University of California
Santa Barbara

# Bulk Comptonization by Turbulence in Black Hole Accretion Discs

A dissertation submitted in partial satisfaction
of the requirements for the degree

Doctor of Philosophy

in

Physics

by

Jason Kaufman

Committee in charge:

> Professor Omer Blaes, Chair
> Professor Lars Bildsten
> Professor Crystal Martin

March 2018

The Dissertation of Jason Kaufman is approved.

_________________________________________

Professor Lars Bildsten

_________________________________________

Professor Crystal Martin

_________________________________________

Professor Omer Blaes, Committee Chair

January 2018

Bulk Comptonization by Turbulence in Black Hole Accretion Discs

Copyright © 2018

by

Jason Kaufman

## Permissions and Attributions

Parts of this dissertation have appeared in:

1. Kaufman J., Blaes O. M., 2016, MNRAS, 459, 1790K

2. Kaufman J., Blaes O. M., Hirose S., 2017, MNRAS, 467, 1734K

3. Kaufman J., Blaes O. M., Hirose S., 2018, MNRAS (accepted)

# Acknowledgements

We thank Yan-Fei Jiang for providing us with data from the AGN shearing box simulations of Jiang et al. (2016) and with data from global disc simulations. We thank Shane Davis for his Monte Carlo code (Davis et al., 2009), which we modified in order to incorporate bulk Comptonization as well as cylindrical and spherical grids. We thank Shigenobu Hirose for providing us with data from simulation 110304a. This work was supported by NASA Astrophysics Theory Program grant NNX13AG61G and the International Space Science Institute (ISSI) in Bern.

I also must thank Crystal Martin for a very helpful discussion that we had in July 2016. At the time I was concerned that it may turn out that I could not model bulk Comptonization accurately enough to make detailed comparisons with observations. Crystal pointed out that even if this were the case it may still be useful to show how bulk Comptonization depends on various accretion disc parameters. I quickly gravitated to this approach, which led to the principal results presented in Chapter 4.

Most importantly, I give a personal thank you to Omer Blaes for his unwavering support and patience, and for countless discussions.

# Curriculum Vitæ
### Jason Kaufman

**Education**

| | |
|---|---|
| 2018 | Ph.D. in Physics, University of California, Santa Barbara, CA. Adviser: Omer Blaes |
| 2014 | M.A. in Physics, University of California, Santa Barbara, CA. |
| 2011 | M.A.St. in Applied Maths and Theoretical Physics, University of Cambridge, Cambridge, UK. |
| 2009 | B.S. in Engineering Sciences (Electrical), Yale University, New Haven, CT. |

**Publications**

Kaufman J., Blaes O. M., Hirose S., 2018, MNRAS (accepted)

Kaufman J., Blaes O. M., Hirose S., 2017, MNRAS, 467, 1734K

Kaufman J., Blaes O. M., 2016, MNRAS, 459, 1790K

**Teaching assistant positions**

| | |
|---|---|
| Fall 2011 | Physics 3/3L: Basic Physics (waves, sound, electrostatics) |
| Winter 2013 | Physics 101: Math Methods (complex variables, Fourier series and transforms, Laplace transforms, asymptotic series, the gamma function, Green's functions) |
| Spring 2013 | Physics 141: Optics |
| Spring 2017 | Physics 131: Gravitation and Relativity |
| Fall 2017 | Physics 2: Basic Physics (rotation, simple harmonic motion, gravitation, fluid mechanics, thermodynamics) |

**Abstract**

Bulk Comptonization by Turbulence in Black Hole Accretion Discs

by

Jason Kaufman

Radiation pressure dominated accretion discs may have turbulent velocities that exceed the electron thermal velocities. Bulk Comptonization by the turbulence may therefore dominate over thermal Comptonization in determining the emergent spectrum. We discuss how to self-consistently resolve and interpret this effect in calculations of spectra of radiation MHD simulations. In particular, we show that this effect is dominated by radiation viscous dissipation and can be treated as thermal Comptonization with an equivalent temperature. We investigate whether bulk Comptonization may provide a physical basis for warm Comptonization models of the soft X-ray excess in AGN. We characterize our results with temperatures and optical depths to make contact with other models of this component. We show that bulk Comptonization shifts the Wien tail to higher energy and lowers the gas temperature, broadening the spectrum. More generally, we model the dependence of this effect on a wide range of fundamental accretion disc parameters, such as mass, luminosity, radius, spin, inner boundary condition, and $\alpha$. Because our model connects bulk Comptonization to one dimensional vertical structure temperature profiles in a physically intuitive way, it will be useful for understanding this effect in future simulations run in new regimes. We also develop a global Monte Carlo code to study this effect in global radiation MHD simulations. This code can be used more broadly to compare global simulations with observed systems, and in particular to investigate whether magnetically dominated discs can explain why observed high Eddington accretion discs appear to be thermally stable.

# Contents

# Chapter 1

# Introduction

Electron scattering is one of the most important processes in determining the emergent spectrum from models of optically thick accretion discs around black holes and neutron stars. Electron scattering opacity generally dominates absorption opacity in the atmospheres of the innermost regions of such discs (Shakura & Sunyaev, 1973). In the case where coherent scattering is a good approximation, the resulting local thermal spectrum of some annulus in the disc is generally harder than a blackbody with the same effective temperature, due to incomplete thermalization at the scattering photosphere. However, Compton (incoherent) scattering in the disc surface layers by thermal electrons can reduce this spectral hardening by increasing the energy exchange between the photons and the plasma (Shimura & Takahara, 1995; Davis et al., 2005; Davis, Done & Blaes, 2006). For those unfamiliar with the basics of Compton scattering, we give a very brief overview in Appendix A.

In addition to the disc atmosphere itself, many models invoke a powerful corona above the disc consisting of high temperature or non-thermal electrons that Compton upscatter disc photons to produce the energetically significant hard X-rays that exist in certain classes of active galactic nuclei and in certain black hole X-ray binary accretion states (e.g. Haardt & Maraschi 1991; Svensson & Zdziarski 1994). These hard X-rays in turn interact with the relatively cool disc atmosphere to produce reflection spectra that are widely observed in many black hole sources (e.g. Lightman & White 1988; Ross & Fabian 1993).

In this work we explore turbulent Comptonization, which is the effect of bulk Comptonization by turbulence on photon spectra (Socrates, Davis & Blaes, 2004; Socrates, 2010). By bulk Comptonization, therefore, we will usually mean turbulent Comptonization, though we also briefly discuss bulk Comptonization by the background shear and bulk Comptonization by arbitrary velocity fields. In sources with radiation pressure dominated accretion flows, bulk velocities may exceed thermal electron velocities, a phenomenon first pointed out in Socrates, Davis & Blaes (2004). The turbulent speeds $v_{\text{turb}}$ on the outer scale of a magnetohydrodynamic (MHD) turbulent cascade will be of order the Alfvén speed, and the ratio of this to the root mean square electron thermal velocity is therefore

$$\frac{v_{\text{turb}}}{\langle v_{\text{th}}^2 \rangle^{1/2}} \sim \left( \frac{P_{\text{mag}}}{P_{\text{rad}}} \right)^{1/2} \left( \frac{P_{\text{rad}}}{P_{\text{gas}}} \right)^{1/2} \left( \frac{m_{\text{e}}}{m_{\text{p}}} \right)^{1/2}. \tag{1.1}$$

Here $P_{\text{gas}}$, $P_{\text{rad}}$, and $P_{\text{mag}}$ are the gas, radiation, and magnetic pressures, respectively, and $m_{\text{e}}/m_{\text{p}}$ is the ratio of the electron to proton mass. Stratified shearing box simulations of magnetorotational turbulence generally have disc atmospheres that are supported by magnetic fields rather than thermal pressure (Miller & Stone, 2000; Hirose, Krolik & Stone, 2006; Hirose, Krolik & Blaes, 2009; Guan & Gammie, 2011; Jiang, Stone & Davis, 2014a). Hence the first factor generally exceeds unity in an otherwise radiation pressure dominated disc. Bulk speeds on the outer scale of the turbulence will therefore exceed the electron thermal speeds whenever the radiation pressure to gas pressure ratio exceeds the ratio of the proton to electron mass ratio, and even smaller depending on how magnetically supported is the disc atmosphere. In this regime, bulk Comptonization by the turbulence may dominate thermal Comptonization in determining the shape of the spectrum emitted by a local patch of the disc.

This regime is commonly reached for near-Eddington accretion on black holes of all mass scales. Indeed, the inner disc solution of the standard geometrically thin model of Shakura & Sunyaev (1973) gives vertically averaged radiation to gas pressure ratios of approximately

$$\frac{P_{\rm rad}}{P_{\rm gas}} \sim 10^7 \alpha^{1/4} \left(\frac{M}{10^8 M_\odot}\right)^{1/4} \eta^{-2} \left(\frac{L}{L_{\rm Edd}}\right)^2 \left(\frac{R}{R_{\rm g}}\right)^{-21/8}, \qquad (1.2)$$

where $\alpha$ is the ratio of turbulent stress to thermal pressure, $M$ is the black hole mass, $\eta \equiv L/(\dot{M}c^2)$ is the radiative efficiency of the disc as a whole, $L/L_{\rm Edd}$ is the luminosity in Eddington units, and $R/R_{\rm g}$ is the radius in the disc in units of the gravitational radius $R_{\rm g} \equiv GM/c^2$. Hence the radiation to gas pressure ratio of the innermost disc will generally exceed the proton to electron mass ratio for near-Eddington accretion even for stellar mass black holes, and certainly for supermassive black holes. On the other hand, energy exchange between the photons and the plasma is generally dominated by true absorption opacity in standard disc models for the most supermassive black holes (Laor & Netzer, 1989; Hubeny et al., 2001), which may reduce bulk Comptonization by turbulence in these sources.

Comptonization by bulk motions in the accretion flow has also been considered by others. Blandford & Payne (1981a,b) considered bulk Comptonization in converging flows and shocks. Starting from this seminal work, bulk Comptonization by radial flows has been calculated in detail by numerous authors (Payne & Blandford, 1981; Colpi, 1988; Titarchuk, Mastichiadis & Kylafis, 1997; Psaltis, 2001; Niedźwiecki & Zdziarski, 2006). Kawashima et al. (2012) included bulk Comptonization in their Monte Carlo calculations of photon spectra from radiation MHD simulations of super-Eddington accretion flows, and found that it produced significant spectral hardening which resembled spectra of ultra-luminous X-ray sources. Here we focus on smaller scale bulk Comptonization by turbulence within the disc atmosphere itself. Turbulent Comptonization has also been invoked in other areas of astrophysics. Zel'dovich, Illarionov & Sunyaev (1972) and Chan & Jones (1975) used then current limits on cosmic microwave background temperature anisotropies to constrain possible turbulent energy on cosmological scales prior to re-combination. Thompson (1994) considered Comptonization by Alfvénic turbulence in a relativistic outflow as a model for the spectrum of gamma-ray bursts.

We approach the study of bulk Comptonization from several angles. In Chapter 2 we investigate analytically how photon spectra produced by turbulent Comptonization depend on properties of the turbulence itself. The material in this chapter is based on an updated version of Kaufman & Blaes (2016). We show that bulk Comptonization actually corresponds to two different physical processes, ordinary work done by radiation pressure and radiation viscous dissipation, and are due to terms that are first and second order in the velocity field, respectively. We discuss why we expect radiation viscous dissipation to be dominant over work done by radiation pressure in determining the emergent spectrum of accretion disc atmospheres. To study radiation viscous dissipation, we first use the Helmholtz theorem to decompose the velocity field into a divergenceless component and a curl-free (compressible) component. For the divergenceless component, bulk Comptonization is due to radiation viscous dissipation alone and can be treated as thermal Comptonization with an equivalent "wave" temperature. For statistically homogeneous turbulence it is simply a weighted sum over the power present at each scale in the turbulent cascade. Scales with wavelengths that are short relative to the photon mean free path contribute fully to the wave temperature, while scales with wavelengths that are long relative to the photon mean free path are significantly downweighted and contribute negligibly. The fact that the wave temperature downweights modes with long wavelengths is physically intuitive because for these modes electron velocity differences between subsequent photon scatterings are significantly smaller. To confirm our physical intuition, we also define a heuristic wave temperature that is proportional to the average square velocity difference between subsequent photon scatterings. We show that our heuristic wave temperature is in fact very similar to the exact wave temperature, in agreement with our intuition. Since the wave temperature increases as the photon mean free path increases, in real accretion discs we expect the wave temperature to be negligible deep inside the photosphere and increase significantly near it. We therefore expect bulk Comptonization to be dominated by a region just inside the photosphere. We discuss how to self-consistently resolve and interpret turbulent Comptonization in spectral calculations for radiation MHD simulations of high Eddington accretion flows.

In Chapter 3 we study bulk Comptonization directly by computing spectra of radiation MHD simulations with Monte Carlo post-processing simulations. We focus specifically on the contribution of bulk Comptonization to the soft X-ray excess in AGN, which refers to the part of the spectrum below 1keV that lies above the power law fit to the hard (2-10keV) X-rays. The material in this chapter is based on Kaufman, Blaes, and Hirose (2017). This is a "top down" approach, as opposed to the "bottom up" approach in Chapter 2. We calculate spectra both taking into account and not taking into account bulk velocities using scaled data from radiation MHD shearing box simulations. We characterize our results with temperatures and optical depths to make contact with other warm Comptonization models of the soft excess. We chose our fiducial mass, $M = 2 \times 10^6 M_\odot$, and accretion rate, $L/L_{\rm Edd} = 2.5$, to correspond to those fit to the super-Eddington narrow-line Seyfert 1 (NLS1) REJ1034+396. The temperatures, optical depths, and Compton $y$ parameters we find broadly agree with those fit to REJ1034+396. We discuss how the effect of bulk Comptonization is to shift the Wien tail to higher energy and lower the gas temperature, broadening the spectrum.

In Chapter 4 we use our heuristic definition of the wave temperature from Chapter 2 to simplify and generalize the bulk Comptonization model presented in Chapter 3 in order to develop greater physical insight into this process and explore a larger parameter space. The material in this chapter is based on Kaufman, Blaes, and Hirose (2018). We model the dependence of bulk Comptonization on fundamental accretion disc parameters, such as mass, luminosity, radius, spin, inner boundary condition, and $\alpha$. In addition to constraining warm Comptonization models, our results can help distinguish contributions from bulk Comptonization to the soft X-ray excess from those due to other physical mechanisms, such as absorption and reflection. By linking the time variability of bulk Comptonization to fluctuations in the disc vertical structure due to MRI turbulence, our results show that observations of the soft X-ray excess can be used to study disc turbulence in the radiation pressure dominated regime. Because our model connects bulk Comptonization to one dimensional vertical structure temperature profiles in a physically intuitive way, it will be useful for understanding this effect in future simulations run in new regimes.

Finally, in Chapter 5, we explore bulk Comptonization in global radiation MHD simulations with global Monte Carlo post-processing simulations. We obtain a global Monte Carlo code by modifying our Monte Carlo shearing box code. We perform Monte Carlo spectral calculations for two radiation MHD simulations of black hole accretion discs with $M = 5 \times 10^8 M_\odot$. For one, $L \sim 0.08 L_{\mathrm{Edd}}$, and for the other, $L \sim 0.2 L_{\mathrm{Edd}}$. We find that there is no statistically significant difference between spectra computed with and without the turbulent velocities included, either overall or in any narrow range of radii, both for the $L = 0.08 L_{\mathrm{Edd}}$ and $L = 0.2 L_{\mathrm{Edd}}$ simulations. This is not surprising given our shearing box results, which indicate that turbulent Comptonization becomes relevant closer to $L = L_{\mathrm{Edd}}$. We discuss how to pursue these questions in future work. We also discuss broader applications of our global Monte Carlo code, such as exploring whether magnetically dominated discs can explain why observed high Eddington accretion discs appear to be thermally stable.

We summarize our results in Chapter 6. Even though the chapters are ordered chronologically, each chapter is largely self-contained. For example, Chapter 2 presents a detailed analysis of the equations underlying bulk Comptonization, but the important parts of these ideas are summarized when they are invoked in Chapters 3 and 4.

# Chapter 2

# Theory of bulk Comptonization by turbulence

## 2.1 Introduction

In this chapter we study analytically how photon spectra produced by turbulent Comptonization depend on properties of the turbulence itself, and how to resolve and interpret this effect in radiation MHD simulations. In section 2.2 we show that the macroscopic physical origins of turbulent Comptonization energy exchange are work due to radiation pressure and viscous dissipation due to the radiation viscous stress tensor, and we discuss why this requires us to treat divergenceless turbulence separately from turbulence with non-zero divergence. In section 2.3 we discuss the consequences of this for correctly implementing radiative transport in simulations, and derive the appropriate radiation energy equation in both lab and fluid frame variables. In section 2.4 we address the conjecture of Socrates, Davis & Blaes (2004) that turbulent Comptonization can be treated as thermal Comptonization with an equivalent "wave" temperature critically dependent on the photon mean free path. We show this is true only for divergenceless turbulence, derive the exact wave temperature with an analytic solution of the radiative transfer equation, and use this result to discuss how the wave temperature depends on the power spectrum of the turbulence. To provide physical insight, we also perform an intuitive, heuristic calculation of the wave temperature which well approximates the analytic solution. In section 2.5 we consider bulk Comptonization by turbulence with non-zero divergence. We show that Comptonization by turbulence whose wavelengths are short relative to the photon mean free path can be treated as thermal Comptonization with an equivalent temperature given by the full turbulent power. In the limit of extremely optically thick turbulence, we show how the evolution of local photon spectra can be understood in terms of compression and expansion of the strongly coupled photon and gas fluids. In section 2.6 we discuss how to apply our results to real, spatially stratified accretion disc atmospheres, and we summarize our findings in section 2.7.

## 2.2 General considerations of turbulent Comptonization

In order to determine how photon spectra produced by turbulent Comptonization depend on properties of the turbulence itself, it is useful to first understand how the frequency-integrated radiation variables couple to the gas. In particular, we show that the resulting energy exchange terms correspond to the work done by radiation pressure and radiation viscous dissipation, and discuss the major consequences of this. We limit our consideration in this work to non-relativistic velocities. The major results of this section are summarized in Table 2.1.

Before proceeding, we define terms and quantities that will be used repeatedly. We denote the characteristic photon mean free path $\lambda_{\mathrm{p}} = (n_{\mathrm{e}}\sigma_{\mathrm{T}})^{-1}$, where $n_{\mathrm{e}}$ is the electron density and $\sigma_\mathrm{T}$ is the Thomson cross section. We denote the typical length scale for bulk velocity variations $\lambda \equiv 2\pi/k$, such as the wavelength if there is a well-defined spatial period. Unless otherwise stated, by the terms optically thin and thick we mean $\lambda_\mathrm{p} \gg \lambda_\mathrm{max}$ and $\lambda_\mathrm{p} \ll \lambda_\mathrm{min}$, where $\lambda_\mathrm{min}$ and $\lambda_\mathrm{max}$ are the minimum and maximum length scales in the turbulent cascade, respectively, not referring to the optical depth that a photon would need to travel to escape the medium.

| Component name | Divergenceless/ Incompressible/ Transverse | Curl-free/ Compressible/ Longitudinal |
|---|---|---|
| Mode decomposition | Transverse | Longitudinal |
| Continuity equation | $\frac{D}{Dt}\rho = 0$ | $\frac{D}{Dt}\rho = -\rho\nabla\cdot\mathbf{v}$ |
| Non-zero rate of strain tensor components | $D_{ij}$ | $D_{ij},$ $\frac{1}{3}\delta_{ij}\nabla\cdot\mathbf{v}$ |
| Non-zero energy exchange terms | $P^{ij}_{\text{vis,shear}}D_{ij}$ | $P^{ij}_{\text{vis,shear}}D_{ij},$ $P_1\nabla\cdot\mathbf{v},$ $P_0\nabla\cdot\mathbf{v}$ |
| Order of energy exchange terms | $(v/c)^2$ | $(v/c)^2$, $(v/c)^2$, and $v/c$, respectively |
| Non-zero radiation viscous stress tensor components | $P^{ij}_{\text{vis,shear}}$ | $P^{ij}_{\text{vis,shear}},$ $P_1\delta^{ij}$ |
| Non-zero radiation viscous stress tensor components in the optically thick limit | $-2\mu D_{ij}$ | $-2\mu D_{ij}$ and $-\zeta\nabla\cdot\mathbf{v}\delta^{ij}$, respectively |

Table 2.1: Summary of contributions to bulk Comptonization by the divergenceless and curl-free components of the velocity field

Net energy exchange due to bulk Comptonization is simply the net energy exchange between gas mechanical energy and radiation. Inside the photosphere, the mechanical energy per unit volume rate of change due to energy exchange with the radiation is

$$\phi = P^{ij}\partial_i v_j, \qquad (2.1)$$

where $P^{ij}$ is the lab frame radiation pressure tensor. This can also be written as

$$\phi = P\nabla \cdot \mathbf{v} + P^{ij}_\mathrm{vis,shear}D_{ij}, \qquad (2.2)$$

where $P = P^{ii}/3$ is the trace of the radiation pressure tensor, $P^{ij}_\mathrm{vis,shear} = P^{ij} - P\delta^{ij}$ is the radiation viscous shear stress tensor, and

$$D_{ij} = \frac{1}{2}\left(\partial_i v_j + \partial_j v_i\right) - \frac{1}{3}\nabla \cdot \mathbf{v}\delta_{ij} \qquad (2.3)$$

is the velocity shear tensor. We see that the energy exchange is separated into two pieces, one due to only the diverging part of the velocity field and another due to the shearing part in the presence of a radiation viscous shear stress tensor. The first piece has contributions from two effects, ordinary work done by radiation pressure, and radiation viscous dissipation. The former effect is first order in velocity since it is due to the contribution to $P$ that is zeroth order in velocity, which we will denote $P_0$. Energy exchange due to viscous effects, on the other hand, is second order in the velocity, as $P_{\text{vis,shear}}^{ij}$ and the relevant contribution to $P$ must themselves be at least first order in velocity since they are a consequence of the velocity field. We will denote the contribution to $P$ that is first order in the velocity field $P_1$. The total energy exchange can then be written

$$\phi = P_0 \nabla \cdot \mathbf{v} + P_1 \nabla \cdot \mathbf{v} + P_{\text{vis,shear}}^{ij} D_{ij}, \qquad (2.4)$$

So far we have decomposed the symmetric part of the rate of strain tensor into two components, the shearing part $D_{ij}$ and the divergence part $\frac{1}{3} \nabla \cdot \mathbf{v} \delta_{ij}$. These two components are fundamentally distinct because under a rotation each component transforms into itself. In other words, if one part is zero then it remains zero in the rotated frame, and if it is non-zero then it remains non-zero in the rotated frame. Note that this is not true of, for example, the off-diagonal components of the strain rate tensor (or any other rank two Euclidean tensor); observers in frames that differ by a rotation may disagree on whether there are off-diagonal components.

But this decomposition cannot be applied to the velocity field itself. In other words, one cannot decompose any velocity field into two parts, one with only $D_{ij}$ non-zero and the other with only $\frac{1}{3} \nabla \cdot \mathbf{v} \delta_{ij}$ non-zero. Instead, according to the Helmholz theorem, the velocity field can be decomposed into a divergenceless component and a curl-free component. Since according to the mass continuity equation the Langrangian derivative of the density is zero if and only if the velocity field is divergenceless, the divergenceless and curl-free components can also be called the incompressible and compressible components, respectively. And since these components themselves can be decomposed into transverse and longitudinal sinusoidal modes, respectively, we can also refer to them as transverse and longitudinal components, respectively.

We now discuss how the velocity field components relate to the strain rate tensor components. The divergenceless component of the velocity field contributes to only the shearing part $D_{ij}$, while the curl-free (compressible) component contributes to both the shearing part $D_{ij}$ *and* the diverging part $\frac{1}{3}\nabla \cdot \mathbf{v}\delta_{ij}$. As a result, energy exchange due to bulk Comptonization by a divergenceless velocity field is non-zero if and only if the radiation viscous shear stress tensor $P^{ij}_{\text{vis,shear}}$ is non-zero. Energy exchange due to bulk Comptonization by the curl-free (compressible) component is non-zero if either the shear stress tensor $P^{ij}_{\text{vis,shear}}$ or the scalar radiation pressure $P$ is non-zero. For example, the bulk Comptonization energy exchange due to a transverse sinusoidal wave $\mathbf{v} = v_0 \sin(kz)\hat{\mathbf{x}}$ is entirely due to the term $P^{ij}_{\text{vis,shear}}D_{ij}$, while energy exchange due to a longitudinal sinusoidal wave $\mathbf{v} = v_0 \sin(kz)\hat{\mathbf{z}}$ is due to both terms, $P^{ij}_{\text{vis,shear}}D_{ij}$ and $P\nabla \cdot \mathbf{v}$. And since we recall that the latter term includes both ordinary work done by radiation pressure $P_0\nabla \cdot \mathbf{v}$ and radiation viscous dissipation $P_1\nabla \cdot \mathbf{v}$, there are in total three ways that a longitudinal sinusoidal wave results in bulk Comptonization energy exchange. Therefore, to calculate the energy exchange due to a divergenceless velocity field we only need to find $P^{ij}_{\text{vis,shear}}$, but to calculate the energy exchange due to the curl-free component we need to find both $P^{ij}_{\text{vis,shear}}$ and $P_1$.

We mentioned that the radiation viscous stress tensor components $P^{ij}_{\text{vis,shear}}$ and $P_1$ are first order in the velocity field, but we have yet to discuss the specific contributions to these by the divergenceless and curl-free velocity field components. In the optically thick limit, for example, the coefficients of shear viscosity $\mu$ and bulk viscosity $\zeta$ are defined such that the radiation viscous stress tensor is given by

$$P^{ij}_{\text{vis}} = -\zeta\nabla \cdot \mathbf{v}\delta_{ij} - 2\mu D_{ij}. \tag{2.5}$$

We note that these are called the dynamic coefficients of viscosity. The kinematic coef-ficients of viscosity are defined by dividing the dynamic coefficients by the mass density $\rho$. The individual components of the stress tensor are therefore

$$P^{ij}_{\text{vis,shear}} = -2\mu D_{ij} \tag{2.6}$$

and

$$P_1 = -\zeta \nabla \cdot \mathbf{v}. \tag{2.7}$$

Since the divergenceless component of the velocity field contributes only to $D_{ij}$, we now also see that (at least in the optically thick limit) it contributes only to $P^{ij}_{\text{vis,shear}}$, whereas the curl-free component of the velocity field contributes to both $P^{ij}_{\text{vis,shear}}$ and $P_1$ since it contributes to both $D_{ij}$ and $\nabla \cdot \mathbf{v}$.

In section 2.4 we show that $P^{ij}_{\text{vis,shear}}$ for a divergenceless velocity field with sinusoidal mode decomposition $\mathbf{v} = \sum_{\mathbf{k}} \mathbf{v_k}$ in a closed, periodic box with sufficiently small escape probability is given by

$$P^{ij}_{\text{vis,shear}} = -2 \sum_{\mathbf{k}} \mu_{\mathbf{k}} D_{ij,\mathbf{k}}, \tag{2.8}$$

where $\mu_{\mathbf{k}}$ and $D_{ij,\mathbf{k}}$ are the dynamic viscosity and strain rate tensor, respectively, of the mode with wave vector $\mathbf{k}$, and we calculate $\mu_{\mathbf{k}}$ in terms of $\mathbf{k}$. We note that since equation (2.6) is valid only when the velocity field varies on only optically thick length scales, our result generalizes this equation. This is critical because we will see that bulk Comptonization is in fact dominated by length scales that are either optically thin or marginally optically thin, not optically thick. We check that in the optically thick limit our result agrees with the radiation viscosity coefficient for scattering. We discuss bulk Comptonization by the curl-free component in section 2.5, but we have fewer closed-form results for this component because in this case the order $v/c$ effect is intertwined non-trivially with the viscous, order $v^2/c^2$ effect. Instead, we focus on developing physical intuition into the underlying equations.

Socrates, Davis & Blaes (2004) conjectured that turbulent Comptonization can be treated as thermal Comptonization with an equivalent "wave" temperature critically dependent on the photon mean free path. This is physically intuitive for divergenceless turbulence since in this case energy exchange is entirely due to radiation viscous dissipation and is therefore second order in velocity. In section 2.4 we prove this conjecture for divergenceless turbulence in a periodic box with sufficiently small escape probability and derive the exact expression for the wave temperature. Since pressure work, on the other hand, is an effect that is first order in velocity, and since Comptonization by a velocity field with non-zero divergence is a combination of pressure work and radiation viscous dissipation, it is not surprising that in this case bulk Comptonization cannot be treated as thermal Comptonization, as we show in section 2.5.

But in the optically thin limit, i.e. when the mean free path is significantly larger than the largest length scale in the turbulence, energy exchange that is first order in velocity vanishes since photons are equally likely to downscatter as they are to upscatter. Bulk Comptonization by a velocity field with non-zero divergence is then solely due to radiation viscous dissipation, and in section 2.5 we show that it may be treated as thermal Comptonization.

In the optically thick case, i.e. when the mean free path is significantly smaller than the smallest scale in the turbulence, the lowest order energy exchange is the work done by radiation pressure to compress the gas, since it is first order in velocity and since radiation viscous effects are suppressed. Socrates, Davis & Blaes (2004) assumed that effects first order in velocity always vanish on average for turbulent eddies, but in the optically thick limit photons trapped in a converging (diverging) region undergo systematic upscattering (downscattering). In the extremely optically thick limit in which the photon and gas fluids are strongly coupled, velocity convergence corresponds to compression in which gas mechanical energy is transferred locally to the photons. In section 2.5 we show that in this process a locally thermal photon distribution remains thermal and only changes temperature, completely analogous to the evolution of the cosmic microwave background radiation under the expansion of the Universe. Unlike energy exchange due to viscous dissipation, this process is reversible. The effect of this process on the emergent spectrum of the disc will depend primarily on how effectively photons are able to escape from such regions to the observer.

## 2.3 Resolving energy exchange due to bulk Comptonization in radiation MHD simulations

Self-consistent radiation MHD simulations of turbulent, radiation pressure dominated accretion flows now exist, both in local vertically stratified shearing box geometries (Hirose, Krolik & Blaes, 2009; Blaes et al., 2011; Jiang, Stone & Davis, 2013) and in global simulations (Ohsuga & Mineshige, 2011; Takeuchi, Ohsuga & Mineshige, 2013; Jiang, Stone & Davis, 2014b; McKinney et al., 2014; Sadowski et al., 2013). Although these simulations use frequency-integrated equations, the emergent radiation spectrum can be computed, including the effects of bulk Comptonization, using post-processing Monte Carlo simulations. Indeed, this has already been done by Kawashima et al. (2012). However, in order for such calculations to be self-consistent, the frequency-integrated radiation MHD equations used in the simulations themselves must include energy exchange due to bulk Comptonization. We now discuss the consequences of the macroscopic physical origins of such energy exchange detailed in section 2.2 for ensuring this effect is captured in simulations. We then proceed to derive the appropriate frequency-integrated source terms due to Compton scattering for the gas and radiation energy equations in both lab frame and fluid frame variables. Using these results, we discuss the extent to which bulk Comptonization is captured by existing radiation MHD simulation codes.

The decomposition of bulk Comptonization energy exchange into pressure work and radiation viscous dissipation shows that radiation MHD schemes that neglect contributions to the viscous stress tensor that are first order in velocity cannot capture bulk Comptonization energy exchange due to a shearing velocity field or any optically thin velocity field with non-zero divergence. As these effects are second order in velocity, we also note that a necessary, but not sufficient, condition for capturing these effects is inclusion of energy terms second order in velocity. Without such terms, turbulence in this form, instead of exchanging energy with photons, will eventually cascade down to the gridscale (or viscous or resistive scale if the code has explicit viscosity or resistivity), and increase the internal energy of the gas. Gas internal energy may then be exchanged with photons through thermal Comptonization. The omission of viscous dissipation by radiation therefore does not prevent the eventual transfer of turbulent energy to radiation, but it may have other physical effects that can in turn affect radiation spectra.

To derive the appropriate frequency-integrated source terms due to Comptonization, we start with the zeroth moment of the radiative transfer equation, correct to order $v^2/c^2$ and $\epsilon/m_{\mathrm{e}}c^2$ (Psaltis & Lamb, 1997),

$$
\begin{aligned}
\frac{1}{n_{\mathrm{e}}\sigma_{\mathrm{T}}} \left( \frac{1}{c}\frac{\partial n}{\partial t} + \partial_i n^i \right) = &\; \frac{1}{\epsilon^2}\frac{\partial}{\partial \epsilon}\left( \epsilon^3 \left( \frac{\epsilon}{m_{\mathrm{e}}c^2}n + \left( \frac{k_{\mathrm{B}}T_{\mathrm{e}}}{m_{\mathrm{e}}c^2} + \frac{1}{3}\frac{v^2}{c^2} \right) \epsilon\frac{\partial}{\partial \epsilon}n \right.\right. \\
&\; \left.\left. + \frac{3}{4}\frac{\epsilon}{m_{\mathrm{e}}c^2}\left( n^2 - n^i n^i + n^{ij}n^{ij} - n^{ijk}n^{ijk} \right) + \frac{v_i}{c}n^i \right)\right) \\
&\; + \left( \frac{18}{5} + \frac{17}{5}\epsilon\frac{\partial}{\partial \epsilon} + \frac{11}{20}\epsilon^2\frac{\partial^2}{\partial \epsilon^2} \right)\left( n^{ij}\frac{v_i v_j}{c^2} - \frac{v^2}{3c^2}n \right).
\end{aligned} \tag{2.9}
$$

Here $\epsilon$ is the photon energy, and the various angle-averaged moments are defined in terms of the energy and direction $(\hat{\ell})$ dependent photon occupation number $n(\epsilon, \hat{\ell})$ by

$$n(\epsilon) \equiv \oint d\Omega\, n(\epsilon, \hat{\ell}),$$

$$n^i(\epsilon) \equiv \oint d\Omega\, \ell^i n(\epsilon, \hat{\ell}),$$

$$n^{ij}(\epsilon) \equiv \oint d\Omega\, \ell^i \ell^j n(\epsilon, \hat{\ell}),$$

$$\text{and} \quad n^{ijk}(\epsilon) \equiv \oint d\Omega\, \ell^i \ell^j \ell^k n(\epsilon, \hat{\ell}). \tag{2.10}$$

In principle, the energy equation is obtained by writing equation (2.9) in terms of moments of the specific intensity and then integrating over all frequencies. Unfortunately, we cannot integrate over terms multiplied by $\epsilon/m_e c^2$ without prior knowledge of the spectrum. For the purpose of simulations, then, we make two approximations. First, we observe that the fractional energy change per scattering off of non-relativistic electrons is small, so that only regions inside the photosphere contribute to Comptonization. Since the stimulated scattering terms are already order $\epsilon/m_e c^2$ and in these regions departures from isotropy are small, we make the following approximation for these terms:

$$n^2 - n^i n^i + n^{ij} n^{ij} - n^{ijk} n^{ijk} \approx \frac{4}{3} n^2. \tag{2.11}$$

Second, we assume that the spectrum can be approximated by a Bose-Einstein distribution with temperature $T_r$. With these approximations, equation (2.9) yields

$$\partial_t E + \partial_i F^i = n_e \sigma_T c \left( -\left(\frac{v_i}{c}\right) \frac{F^i}{c} + \left(\frac{v}{c}\right)^2 E + \left(\frac{v_i}{c}\right)\left(\frac{v_j}{c}\right) P^{ij} + 4 k_B \left(\frac{T_e - T_r}{m_e c^2}\right) E \right). \tag{2.12}$$

This is the correct energy equation, but in order for it to capture second order energy exchange, the substituted value of $F^i$ must be calculated with a moment closure scheme that does not neglect contributions to the radiation viscous stress tensor that are first order in velocity. For example, it is not adequate to calculate $F^i$ by substituting $P^{ij} = (E/3)\delta^{ij}$ into the first moment equation.[1] This is equivalent to flux-limited diffusion in the diffusion regime, such as that implemented in Hirose, Krolik & Blaes (2009). These do, however, capture energy exchange due to pressure work in the optically thick regime. To show this, we substitute into equation (2.12) the standard closure relation,

$$F^i = -\frac{c}{3n_\mathrm{e}\sigma_\mathrm{T}}\partial_i E + \frac{4}{3}v_i E, \tag{2.13}$$

which gives

$$\partial_t E + \partial_i \left(-\frac{c}{3n_\mathrm{e}\sigma_\mathrm{T}}\partial_i E + v_i E\right) = -\frac{1}{3}E\partial_i v_i + n_\mathrm{e}\sigma_\mathrm{T}c\left(4k_\mathrm{B}\left(\frac{T_\mathrm{e} - T_\mathrm{r}}{m_\mathrm{e}c^2}\right)E\right). \tag{2.14}$$

We see that energy exchange that is second order in velocity is not present. Furthermore, we see that energy exchange due to a converging velocity field is indeed the work done by radiation pressure to compress the gas, $-(1/3)E\partial_i v_i \approx -P\partial_i v_i$.

The M1 closure scheme (Levermore, 1984), implemented in, e.g., Sadowski et al. (2013), also captures first order energy exchange but not second order energy exchange. This scheme assumes that there exists a frame in which $P^{ij} = \delta^{ij}E/3$. The lab frame radiation pressure tensor can then be expressed in terms of the energy density and flux (Sadowski et al., 2013):

$$P^{ij} = \left(\frac{1-\xi}{2}\delta^{ij} + \frac{3\xi - 1}{2}f^i f^j\right)E, \tag{2.15}$$

---

[1] However, since the pressure term in the energy equation is already second order in velocity, for the purposes of capturing bulk Comptonization energy exchange it is acceptable to make the approximation $P^{ij} \approx (E/3)\delta^{ij}$ here.

where $f^i = F^i/E$ and

$$\xi = \frac{3 + 4f^i f^i}{5 + 2\sqrt{4 - 3f^i f^i}}. \tag{2.16}$$

To show that second order energy exchange is not captured, we consider the case of a non-zero radiation viscous shear stress tensor due to a non-relativistic velocity field in an otherwise homogeneous medium. The lowest order contribution to the flux must be first order in velocity. In this scheme, then, the radiation viscous shear stress tensor is zero to first order in velocity, and hence second order energy exchange, which requires a contribution that is first order in velocity (equation 2.2), is not captured.

Another way to understand why both flux-limited diffusion and the M1 closure scheme fail to capture second order energy exchange is to observe that they both bridge generically optically thick conditions with optically thin conditions, while optically thin turbulence does not fall into either category. In optically thin turbulence, the turbulence length scales are optically thin ($\lambda_{\max} \ll \lambda_{\mathrm{p}}$), but conditions are otherwise optically thick ($\lambda_{\mathrm{p}} \nabla E/E \ll 1$), since we are far enough inside the photosphere that photons must scatter many times before escaping. It seems that only a more sophisticated approach, such as explicitly solving the transfer equation as done by Jiang, Stone & Davis (2013), can capture this effect.

We note that Sadowski et al. (2015) add an artificial viscosity to the M1 closure scheme in order to address a numerical problem associated with artificial shocks in their simulations. They assume a kinematic radiation viscosity given by

$$\nu_{\mathrm{s}} = 0.1 \left( \frac{E}{\rho c^2} \right) \lambda_{\mathrm{p}} c, \tag{2.17}$$

which, as they acknowledge, underestimates the actual viscosity in the optically thick limit by a factor of $27/80$ (equation 2.79).

For completeness, we also write equation (2.12) in terms of fluid frame radiation variables, indicated by subscript zero:

$$\partial_t \left( E_0 + 2 \left( \frac{v_i}{c} \right) \left( \frac{F_0^i}{c} \right) \right) + \partial_i \left( F_0^i + v_i E_0 + v_j P_0^{ij} \right) =$$
$$n_e \sigma_T c \left( - \left( \frac{v_i}{c} \right) \frac{F_0^i}{c} + 4 k_B \left( \frac{T_e - T_r}{m_e c^2} \right) \right). \tag{2.18}$$

Since

$$\partial_t \left( 2 \frac{v_i}{c} \left( \frac{F_0^i}{c} \right) \right) \sim v_i \partial_i \left( 2 \left( \frac{v_i}{c} \right) \left( \frac{F_0^i}{c} \right) \right) \ll \partial_i F_0^i, \tag{2.19}$$

equation (2.18) simplifies to

$$\partial_t E_0 + \partial_i \left( F_0^i + v_i E_0 + v_j P_0^{ij} \right) = n_e \sigma_T c \left( - \left( \frac{v_i}{c} \right) \frac{F_0^i}{c} + 4 k_B \left( \frac{T_e - T_r}{m_e c^2} \right) \right). \tag{2.20}$$

## 2.4   Comptonization by divergenceless turbulence

Socrates, Davis & Blaes (2004) (hereafter S04) conjectured that Comptonization by turbulence can be treated as thermal Comptonization by solving the Kompaneets equation with an equivalent "wave" temperature $T_w$. They heuristically derived an approximate value for the wave temperature by reasoning that turbulent modes with wavelengths greater than $\lambda_p$ would contribute negligibly since for these modes photons encounter minimal electron velocity differences between subsequent scatterings. This gives

$$T_w \approx \int_{k=2\pi/\lambda_p}^{\infty} T_{tot}(k) dk, \tag{2.21}$$

where $T_{tot}(k)$ is the temperature distribution corresponding to the total electron kinetic energy distribution $E(k)$ at wavenumber $k$. That is, $T_{\text{tot}}(k)$ satisfies

$$\frac{3}{2} n_{\text{e}} k_{\text{B}} T_{\text{tot}}(k) = E(k). \tag{2.22}$$

Equation (2.21) generalizes S04 equation (8),

$$T_{\text{w}}(\lambda_{\text{p}}) \approx T_{\text{w}}(\lambda_0) \left( \frac{\lambda_{\text{p}}}{\lambda_0} \right)^{2/3}, \tag{2.23}$$

which gives $T_{\text{w}}$ for a Kolmogorov spectrum, $E(k) \propto k^{-5/3}$, with maximum wavelength $\lambda_0$. Equation (2.21) is a weighting scheme of the form

$$T_{\text{w}} = \int_0^\infty f(k) T_{\text{tot}}(k) dk, \tag{2.24}$$

with the weighting function $f(k)$ given by

$$f_{\text{S04}}(k) = \begin{cases} 1, k \geq 2\pi/\lambda_{\text{p}} \\ 0, k < 2\pi/\lambda_{\text{p}}. \end{cases} \tag{2.25}$$

This shows explicitly that this scheme simply gives full weight to wavelengths less than $\lambda_{\text{p}}$ and zero weight to wavelengths greater than $\lambda_{\text{p}}$. For a periodic velocity field, since the modes are discrete, equation (2.24) is more clearly written

$$T_{\text{w}} = \sum_{\mathbf{k}} f(k) T_{\text{tot},\mathbf{k}}, \tag{2.26}$$

where $T_{\text{tot},\mathbf{k}}$ is the temperature of the mode with wave vector $\mathbf{k}$. That is,

$$\frac{3}{2} k_{\text{B}} T_{\text{tot},\mathbf{k}} = \frac{1}{2} m_{\text{e}} \left\langle v_{\mathbf{k}}^2 \right\rangle, \tag{2.27}$$

which gives

$$k_{\rm B} T_{\rm w} = \sum_{\bf k} \frac{1}{3} m_{\rm e} \left\langle v_{\bf k}^2 \right\rangle f(k). \tag{2.28}$$

For the remainder of this section we use equation (2.26) since it is more useful for applications to radiation MHD simulations, but note that all results also hold for equation (2.24), as this is just the continuum limit. We also define $\tau_k = 1/\lambda_{\rm p} k = \lambda/2\pi\lambda_{\rm p}$, the optical depth divided by $2\pi$ across a mode with wavenumber $k$. Equation (2.25), for example, can then be written,

$$f_{\rm S04}(k) = \begin{cases} 1, \tau_k \leq 1/2\pi \\ 0, \tau_k > 1/2\pi. \end{cases} \tag{2.29}$$

In section 2.4.1 we perform a detailed analysis of the equations underlying bulk Comptonization for divergenceless velocity fields in a periodic box with a small escape probability. To start, we show that it can in fact be characterized by a wave temperature, which in terms of the viscous stress tensor $P_{\rm vis,shear}^{ij}$ is given by

$$k_{\rm B} T_{\rm w} = \frac{-2\lambda_{\rm p} m_{\rm e} c}{E} \left( P_{\rm vis,shear}^{ij} \left( \partial_i v_j + \partial_j v_i \right) \right). \tag{2.30}$$

We see that the wave temperature varies spatially and is proportional to the rate of energy exchange per unit volume between the gas and the radiation (section 2.2), $P_{\rm vis,shear}^{ij} D_{ij}$. We show that $P_{\rm vis,shear}^{ij}$ is given by

$$P_{\rm vis,shear}^{ij} = -\frac{4\lambda_{\rm p} E}{3c} \sum_{\bf k} \tau_k^2 f(k) \left( \partial_i v_{j,{\bf k}} + \partial_j v_{i,{\bf k}} \right), \tag{2.31}$$

where the weighting function $f(k)$ is given by

$$f(k) = \frac{2}{\tau_k} \left( \frac{1}{Q(\tau_k)} - \frac{1}{\tau_k} \right), \tag{2.32}$$

and where

$$Q(\tau_k) = \tau_k - \frac{3}{4} \tau_k^3 \left( \frac{2}{3} + \tau_k^2 - \tau_k(1 + \tau_k^2) \tan^{-1} \left( \frac{1}{\tau_k} \right) \right). \tag{2.33}$$

The limiting cases of $f(k)$ are

$$f(k) = \begin{cases} 1 & \text{if } \tau_k \to 0 \\ \frac{2}{9\tau_k^2} & \text{if } \tau_k \to \infty. \end{cases} \tag{2.34}$$

The expression for the wave temperature in terms of the velocity field is therefore

$$k_{\mathrm{B}} T_{\mathrm{w}} = \frac{\lambda_{\mathrm{p}}^2 m_{\mathrm{e}}}{6} \left( \partial_i v_j + \partial_j v_i \right) \sum_{\mathbf{k}} \tau_k^2 f(k) \left( \partial_i v_{j,\mathbf{k}} + \partial_j v_{i,\mathbf{k}} \right). \tag{2.35}$$

We then show that for statistically homogeneous turbulence we can take the spatial average of the wave temperature, so that we obtain (homogeneous) thermal Comptonization with a temperature given by

$$k_{\mathrm{B}} T_{\mathrm{w}} = \left\langle \frac{-2\lambda_{\mathrm{p}} m_{\mathrm{e}} c}{E} P_{\mathrm{vis,shear}}^{ij} \left( \partial_i v_j + \partial_j v_i \right) \right\rangle. \tag{2.36}$$

In terms of the velocity field this is

$$k_{\mathrm{B}} T_{\mathrm{w}} = \left\langle \frac{\lambda_{\mathrm{p}}^2 m_{\mathrm{e}}}{6} \left( \partial_i v_j + \partial_j v_i \right) \sum_{\mathbf{k}} \tau_k^2 f(k) \left( \partial_i v_{j,\mathbf{k}} + \partial_j v_{i,\mathbf{k}} \right) \right\rangle. \tag{2.37}$$

Since the modes are sinusoidal, equation (2.37) simplifies to equation (2.28). We plot $f_{\mathrm{S04}}(k)$ and $f(k)$ in Figure 2.1. Since log scaling is used, the curve for $f_{\mathrm{S04}}(k)$ disappears for $\lambda > \lambda_{\mathrm{p}}$. We see that $f_{\mathrm{S04}}(k)$ does roughly approximate $f(k)$. In particular, modes with wavelengths significantly longer than the photon mean free path are significantly downweighted and therefore likely contribute negligibly to the wave temperature.

In section 2.4.2, to provide physical insight into the exact solution, we define a heuristic wave temperature given by

$$\frac{3}{2} k_{\mathrm{B}} T_{\mathrm{w,heur}} = \frac{1}{4} m_{\mathrm{e}} \left\langle (\Delta \mathbf{v})^2 \right\rangle, \tag{2.38}$$

where $\left\langle (\Delta \mathbf{v})^2 \right\rangle$ is the average square velocity difference between subsequent photon scatterings. Our heuristic wave temperature, therefore, also downweights modes with wavelengths greater than the photon mean free path, like the approximate wave temperature in S04. We show that $T_{\mathrm{w,heur}}$ differs from $T_{\mathrm{w}}$ only in its weighting function, which is given by

$$f_{\mathrm{heur}}(k) = 1 - \tau_k \tan^{-1} \left( \frac{1}{\tau_k} \right). \tag{2.39}$$

The limiting cases are

$$f_{\mathrm{heur}}(k) = \begin{cases} 1 & \text{if } \tau_k \to 0 \\ \frac{1}{3\tau_k^2} & \text{if } \tau_k \to \infty. \end{cases} \tag{2.40}$$

We also plot $f_{\mathrm{heur}}(k)$ in Figure 2.1 and see that it is remarkably close to the exact solution. The exact wave temperature is therefore approximately given by our heuristic wave temperature, and so the wave temperature can be thought of as describing the average square velocity difference between subsequent photon scatterings in the box. Not only is this result useful for developing an intuitive understanding of bulk Comptonization by second order (i.e. $v^2/c^2$) terms, but it forms the basis of our model in Chapter 4.

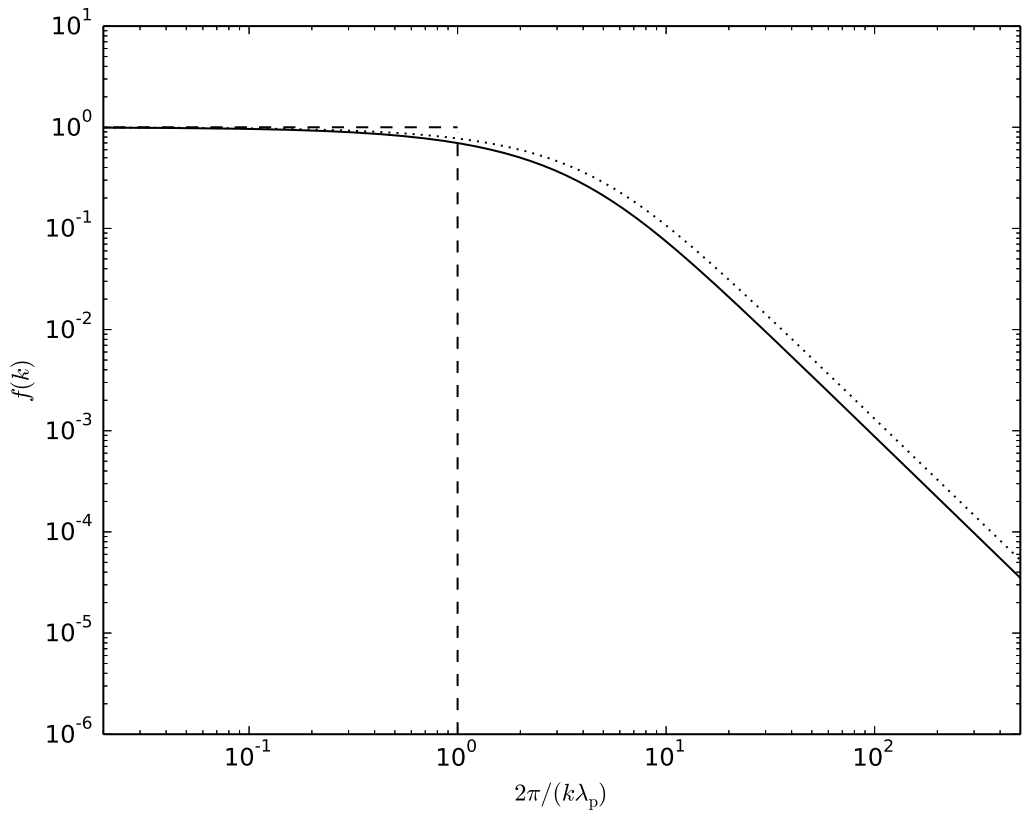

Figure 2.1: Dependence of the mode weighting function on the mode wavelength (in units of the photon mean free path), for statistically homogeneous, divergenceless turbulence. The solid line shows our formal solution, equation (2.32), the dotted line shows our heuristic solution, equation (2.39), and the dashed line shows the weighting from S04, equation (2.25).

Finally, in section 2.4.3 we use the correct expression for $T_{\mathrm{w}}$ to discuss how the wave temperature depends on the power spectrum of the turbulence and determine which turbulent wavelengths contribute most to Comptonization.

## 2.4.1   The exact wave temperature

To derive equations (2.30), (2.31) and (2.28), we start with the zeroth moment of the radiative transfer equation, equation (2.9), correct to order $v^2/c^2$ and $\epsilon/m_{\mathrm{e}}c^2$, and develop our solution in several stages. First (Part 1), in order to contextualize our solution and develop physical insight, we present the formal procedure for simplifying equation (2.9) in the generically optically thick limit. In this limit we discuss (Part 2) the $v = 0$ case, which is inhomogeneous thermal Comptonization (for generically optically thick conditions). In particular, we define the saturated and unsaturated wavelength limits of thermal Comptonization, which are important because we will take limits analogous to these when studying bulk Comptonization. Next (Part 3), we show why bulk Comptonization, by contrast, corresponds to a limiting case of equation (2.9) that differs subtly from the generically optically thick limit. We derive equations (2.30) and (2.31), which show that bulk Comptonization can be reduced to inhomogeneous thermal Comptonization. Finally (Part 4), we discuss various limits of bulk Comptonization and derive equation (2.28) in the case of statistically homogeneous turbulence.

**Part 1: Simplifying the zeroth moment equation in the generically optically thick limit**

In the generically optically thick limit $n^i, n^{ijk} \ll n$ and $n^{ij} \approx n\delta^{ij}/3$, and so equation (2.9) simplifies to

$$\lambda_{\mathrm{p}} \left( \frac{1}{c}\partial_t n + \partial_i n^i \right) = \frac{1}{\epsilon^2}\partial_\epsilon \left( \epsilon^3 \left( \frac{\epsilon}{m_{\mathrm{e}}c^2} \left( n + n^2 \right) \; + \left( \frac{k_{\mathrm{B}}T_{\mathrm{e}}}{m_{\mathrm{e}}c^2} + \frac{v^2}{3c^2} \right) \epsilon\partial_\epsilon n + \frac{v_i}{c}n^i \right) \right). \tag{2.41}$$

But since it turns out that bulk Comptonization corresponds to a limit that differs from the generically optically thick one only subtly, we first define the generic one more rigorously. This limit is defined to be one in which the minimum length scale in the problem $\lambda_{\min}$ is significantly greater than the mean free path $\lambda_{\mathrm{p}}$. In this limit, any term is large relative to $\lambda_{\mathrm{p}}$ multiplied by the term's spatial derivative. We can therefore attempt to expand the angle dependent occupation number $n(\hat{\ell})$ in terms of its zeroth moment $n$ by treating spatial derivatives as order $\lambda_{\mathrm{p}}/\lambda_{\min}$. Since $v/c \ll 1$, we define this to be an expansion in two parameters, $\lambda_{\mathrm{p}}/\lambda_{\min}$ and $v/c$. We note that terms of order $\epsilon/m_{\mathrm{e}}c^2$ are second order in velocity since for Comptonized photons $\epsilon \simeq m_{\mathrm{e}}v^2$. To first order, the expansion can be written

$$n(\hat{\ell}) = n(\hat{\ell})_{0,0} + n(\hat{\ell})_{1,0} + n(\hat{\ell})_{0,1}, \tag{2.42}$$

where the subscripts denote the orders of terms in $\lambda_{\mathrm{p}}/\lambda_{\min}$ and $v/c$, respectively.

To construct this expansion, we begin with the radiative transfer equation, equation (A1) of Psaltis & Lamb (1997), correct to order $v^2/c^2$ and $\epsilon/m_{\mathrm{e}}c^2$. Since this expansion must be constructed recursively, we begin by finding the solution to zeroth order in both parameters. At this order, equation (A1) of Psaltis & Lamb (1997) reduces to

$$0 = n_{0,0} - n_{0,0}(\hat{\ell}) + \frac{3}{4}\ell^i\ell^j\left(n_{0,0}^{ij} - \frac{1}{3}\delta^{ij}n_{0,0}\right). \qquad (2.43)$$

The solution to this is

$$n_{0,0}(\hat{\ell}) = n_{0,0}, \qquad (2.44)$$

from which it follows that to this order $n(\hat{\ell}) = n$, $n^i = 0$ and $n^{ij} = n\delta^{ij}/3$. Before proceeding, we observe that since we have now shown that $n(\hat{\ell}) = n + O(\lambda_{\mathrm{p}}/\lambda_{\mathrm{min}}, v/c)$, it follows that equation (2.41) is simply equation (2.9) to second order in $\lambda_{\mathrm{p}}/\lambda_{\mathrm{min}}$ and $v/c$. We note that this includes terms of order $(\lambda_{\mathrm{p}}/\lambda_{\mathrm{min}})(v/c)$.

Next, we solve for $n_{1,0}(\hat{\ell})$, which is the order $\lambda_{\mathrm{p}}/\lambda_{\mathrm{min}}$ correction to $n_{0,0}(\hat{\ell})$. We again use equation (A1) of Psaltis & Lamb (1997), which reduces to

$$\lambda_{\mathrm{p}}\ell^i\partial_i n_{0,0} = n_{1,0} - n_{1,0}(\hat{\ell}) + \frac{3}{4}\ell^i\ell^j\left(n_{1,0}^{ij} - \frac{1}{3}\delta^{ij}n_{1,0}\right). \qquad (2.45)$$

The solution is

$$n_{1,0}(\hat{\ell}) = -\lambda_{\mathrm{p}}\ell^i\partial_i n_{0,0}, \qquad (2.46)$$

from which it follows that to this order $n(\hat{\ell}) = n - \lambda_{\mathrm{p}}\ell^i\partial_i n$, $n^i = -\lambda_{\mathrm{p}}\partial_i n/3$, and $n^{ij} = n\delta^{ij}/3$.

Finally, we solve for $n_{0,1}(\hat{\ell})$. To this order derivatives of $v$ are still negligible since they are cross terms of order $(\lambda_{\mathrm{p}}/\lambda_{\mathrm{min}})(v/c)$. We note that this assumption will not hold for bulk Comptonization which is why we will have to construct a slightly different

| Limit | Criteria for $\lambda$ | Criteria for $y_\lambda$ | Timescales | Temperature |
|---|---|---|---|---|
| Saturated wavelength | $\lambda_{\min} \gg \sqrt{\frac{m_e c^2}{k_B T_e}} \lambda_p$ | $y_{\lambda_{\min}} \gg 1$ | $t_e \ll t_d$ | $T_e$ |
| Unsaturated wavelength | $\lambda_{\max} \ll \sqrt{\frac{m_e c^2}{k_B T_e}} \lambda_p$ | $y_{\lambda_{\max}} \ll 1$ | $t_d \ll t_e$ | $\langle T_e \rangle$ |
| Unsaturated for $\lambda < \lambda_{\text{unsat}}$ | $\lambda_{\text{unsat}} \ll \sqrt{\frac{m_e c^2}{k_B T_e}} \lambda_p$ | $y_{\lambda_{\text{unsat}}} \ll 1$ | $t_d \ll t_e$ for $\lambda \leq \lambda_{\text{unsat}}$ | $\langle T_e \rangle_{\lambda_{\text{unsat}}}$ |

Table 2.2: Summary of important limits for inhomogeneous thermal Comptonization

expansion in Part 3. Equation (A1) of Psaltis & Lamb (1997) now reduces to

$$0 = n_{0,1} - n_{0,1}(\hat{\ell}) - \ell^i v_i \epsilon \partial_\epsilon n_{0,0} + \frac{3}{4} \ell^i \ell^j \left( n_{0,1}^{ij} - \frac{1}{3} \delta^{ij} n_{0,1} \right). \qquad (2.47)$$

The solution is

$$n_{0,1}(\hat{\ell}) = -l^i v_i \epsilon \partial_\epsilon n_{0,0}, \qquad (2.48)$$

from which it follows that to this order $n(\hat{\ell}) = n - \lambda_p \ell^i \partial_i n - \ell^i v_i \epsilon \partial_\epsilon n$, $n^i = -\lambda_p \partial_i n/3 - v_i \epsilon \partial_\epsilon n/3$, and $n^{ij} = n \delta^{ij}/3$. By plugging the expression for $n^i$ into equation (2.41) we could solve for the zeroth moment $n$ to second order in $\lambda_p/\lambda_{\min}$ and $v/c$. We therefore have derived a closed set of equations for $n$ in the generically optically thick limit.

## Part 2: The saturated and unsaturated wavelength limits of thermal Comptonization

Before proceeding to study bulk Comptonization, we first develop several important ideas by discussing inhomogeneous thermal Comptonization. We summarize these in Table 2.2.

Not including source and sink terms, inhomogeneous thermal Comptonization is described by equation (2.41) with $v_i = 0$, which gives

$$\lambda_{\rm p}\left(\frac{1}{c}\partial_t n + \partial_i n^i\right) = \frac{1}{\epsilon^2}\partial_\epsilon\left(\epsilon^3\left(\frac{\epsilon}{m_{\rm e}c^2}\left(n + n^2\right) + \left(\frac{k_{\rm B}T_{\rm e}}{m_{\rm e}c^2}\right)\epsilon\partial_\epsilon n\right)\right). \qquad (2.49)$$

For a constant temperature the problem is spatially homogeneous so $\partial_i n^i = 0$, which gives the famous Kompaneets equation,

$$\frac{\lambda_{\rm p}}{c}\partial_t n = \frac{1}{\epsilon^2}\partial_\epsilon\left(\epsilon^3\left(\frac{\epsilon}{m_{\rm e}c^2}\left(n + n^2\right) + \left(\frac{k_{\rm B}T_{\rm e}}{m_{\rm e}c^2}\right)\epsilon\partial_\epsilon n\right)\right). \qquad (2.50)$$

In what we define as the saturated and unsaturated wavelength limits, solutions to the inhomogeneous equation can be understood in terms of solutions to the Kompaneets equation. We define these limits by the timescales for spatial diffusion and photon energy change. We note that by saturated and unsaturated wavelengths, therefore, we are *not* referring to the resultant spectra, but rather characterizing the length scales over which the temperature varies. We avoid referring to these as long and short wavelength limits since in this work these phrases generally refer instead to the optically thick and thin wavelength limits, respectively.

Diffusion results from the term $\partial_i n^i$, and since $n^i \sim \lambda_{\rm p}\partial_i n$, the time it take photons to diffuse across a length scale $\lambda$ is

$$t_{\rm d} = \frac{\lambda^2}{\lambda_{\rm p}c}. \qquad (2.51)$$

Meanwhile, the timescale for photon energy change is

$$t_{\rm e} = \frac{m_{\rm e}c\lambda_{\rm p}}{k_{\rm B}T_{\rm e}}. \qquad (2.52)$$

In the saturated wavelength limit, the temperature varies slowly enough in space that photons change energy much faster than they can diffuse. In this limit the minimum length scale $\lambda_{\min}$ for variations in the temperature must satisfy $t_e \ll t_d$, which gives

$$\lambda_{\min} \gg \sqrt{\frac{m_e c^2}{k_B T_e}} \lambda_p. \tag{2.53}$$

Since in this limit the spatial diffusion term $\partial_i n^i$ is negligible, the resulting photon energy distribution is determined by solving the Kompaneets equation at each point separately. We note that if we assign a Compton $y$ parameter $y_\lambda$ to the length scale $\lambda$ (not to be confused with the $y$ parameter associated with the resultant spectrum), given by

$$y_\lambda = \frac{4 k_B T_e}{m_e c^2} \left( \frac{\lambda}{\lambda_p} \right)^2, \tag{2.54}$$

then the saturated wavelength limit is simply the limit in which $y_\lambda \gg 1$ for the minimum wavelength $\lambda_{\min}$.

In the unsaturated wavelength limit, photons diffuse much faster than they can change energy so spatial variations in the occupation number $n$ are negligible (by contrast, in the saturated wavelength limit the variations themselves may be significant but happen on longer length scales). The maximum length scale $\lambda_{\max}$ for variations in the temperature must satisfy $t_d \ll t_e$, which gives

$$\lambda_{\max} \ll \sqrt{\frac{m_e c^2}{k_B T_e}} \lambda_p. \tag{2.55}$$

In this limit we can take the spatial average of equation (2.49), which gives

$$\frac{\lambda_p}{c} \partial_t n = \frac{1}{\epsilon^2} \partial_\epsilon \left( \epsilon^3 \left( \frac{\epsilon}{m_e c^2} \left( n + n^2 \right) + \left( \frac{k_B \langle T_e \rangle}{m_e c^2} \right) \epsilon \partial_\epsilon n \right) \right). \tag{2.56}$$

This is just equation (2.50), the Kompaneets equation, with the temperature given by the spatially averaged temperature $\langle T_e \rangle$. We note that the unsaturated wavelength limit is simply the limit in which $y_\lambda \ll 1$ for the maximum wavelength $\lambda_{\max}$.

If neither the criterion for the saturated limit nor the criterion for the unsaturated limit is satisfied, then to simplify the problem it may be helpful to take a spatial average over the largest length scale that is unsaturated. In this case the process remains inhomogeneous thermal Comptonization but with a temperature field that is possibly simpler, $\langle T_e \rangle_{\lambda_{\mathrm{unsat}}}$, where $\lambda_{\mathrm{unsat}}$ denotes that the average is taken over the largest unsaturated length scale.

We note that since the optical depth of a length scale is given by $\lambda/\lambda_p$, a length scale can be optically thick yet have an unsaturated wavelength as defined here. On the other hand, although in principle a length scale can be optically thin yet have a saturated wavelength, this cannot be the case in the non-relativistic limit (in which we are working), since $k_B T_e / m_e c^2 \ll 1$.

## Part 3: Simplifying the zeroth moment equation in the case of bulk Comptonization

We now proceed to derive equations (2.30) and (2.31) for small escape probability, which show that bulk Comptonization by divergenceless turbulence can be treated as inhomogeneous thermal Comptonization.

For bulk Comptonization we cannot assume that the minimum length scale for variations in the velocity field is large relative to the mean free path $\lambda_p$. Since the solution to the radiative transfer equation depends on the velocity field, this means we cannot assume that spatial derivatives of the zeroth moment $n$ are small, either, and so it appears that we cannot construct an expansion of the form developed in Part 1. But fortunately we can circumvent this obstacle in the following way. We can construct a nearly identical expansion assuming that only higher moments of $n(\hat{\ell})$ (but not the zeroth moment $n$) depend on the velocity field and therefore vary on small length scales. In this case derivatives of $n$ will still be first order in $\lambda_p/\lambda_{\min}$, but derivatives of $v$ will be zeroth order in $\lambda_p/\lambda_{\min}$ since we are allowing $v$ to vary on arbitrarily small length scales. We note that now $\lambda_{\min}$ is still the minimum length scale for variations in $n$, but it may be significantly larger than the minimum length scale for variations in $v$. The resulting solution will then be valid as long as it turns out to be consistent with these assumptions.

To zeroth order the solution is identical to the generically optically thick limit (Part 1), so to this order $n(\hat{\ell}) = n$, $n^i = 0$ and $n^{ij} = n\delta^{ij}/3$. Then, since $n(\hat{\ell}) = n + O(\lambda_p/\lambda_{\min}, v/c)$, equation (2.9) again simplifies to equation (2.41) to second order in $\lambda_p/\lambda_{\min}$ and $v/c$.

Before proceeding with the expansion, we use the first moment of the transfer equation, Psaltis & Lamb (1997) equation (35), to write equation (2.41) in a more physically revealing way. We multiply the first moment equation by $v_i$, so that in steady state, to second order it becomes

$$\lambda_p \left(\frac{v_i}{c}\right) \partial_j n^{ij} = -\frac{v_i}{c} n^i - \frac{1}{3}\frac{v^2}{c^2}\epsilon \partial_\epsilon n. \qquad (2.57)$$

Substituting equation (2.57) into equation (2.41) gives

$$\lambda_p \left(\frac{1}{c}\partial_t n + \partial_i n^i + \partial_i \left(\left(\frac{v_j}{c}\right) n^{ij}\right)\right) = \frac{1}{\epsilon^2}\partial_\epsilon \left(\epsilon^3 \left(\frac{\epsilon}{m_e c^2}\left(n + n^2\right)\right.\right.$$
$$\left.\left. + \left(\frac{k_B T_e}{m_e c^2}\right)\epsilon\partial_\epsilon n + \lambda_p \left(\partial_i \frac{v_j}{c}\right) n^{ij}\right)\right). \qquad (2.58)$$

This is just the inhomogeneous thermal Comptonization equation, equation (2.49), with an advection term $\partial_i \left(\left(\frac{v_j}{c}\right) n^{ij}\right)$ and a bulk Comptonization contribution to the temper-

ature given by

$$k_{\mathrm{B}}T_{\mathrm{w}} = \frac{\lambda_{\mathrm{p}}m_{\mathrm{e}}c}{2\epsilon\partial_\epsilon n}\left(\partial_i v_j + \partial_j v_i\right)n^{ij}. \tag{2.59}$$

Multiplying both sides by $\epsilon^4\partial_\epsilon n$ and integrating over energy gives equation (2.30). We note, however, that this contribution can be regarded as a temperature only if it turns out to be independent of the photon energy $\epsilon$, which we will show.

We now finish constructing the expansion for $n(\hat{\ell})$. This will allow us to find $n^{ij}$ and evaluate equation (2.59). To first order in $\lambda_{\mathrm{p}}/\lambda_{\min}$ the solution is again identical to the generically optically thick limit (Part 1), so to this order $n(\hat{\ell}) = n - \lambda_{\mathrm{p}}\ell^i\partial_i n$, $n^i = -\lambda_{\mathrm{p}}\partial_i n/3$, and $n^{ij} = n\delta^{ij}/3$. However, the solution for $n_{0,1}(\hat{\ell})$ is different, since this time derivatives of $v$ are not negligible. Equation (A1) of Psaltis & Lamb (1997) now reduces to

$$\lambda_{\mathrm{p}}\ell^i\partial_i n_{0,1} = n_{0,1} - n_{0,1}(\hat{\ell}) - \ell^i v_i\epsilon\partial_\epsilon n_{0,0} + \frac{3}{4}\ell^i\ell^j\left(n_{0,1}^{ij} - \frac{1}{3}\delta^{ij}n_{0,1}\right). \tag{2.60}$$

If the density is constant and the velocity field is divergenceless with sinusoidal mode decomposition

$$\mathbf{v} = \sum_{\mathbf{k}}\mathbf{v_k}, \tag{2.61}$$

then the solution is given by (Appendix B)

$$n_{0,1}^{ij} = \frac{\lambda_{\mathrm{p}}\epsilon\partial_\epsilon n_{0,0}}{3c}\sum_{\mathbf{k}}\tau_k^2 f(k)\left(\partial_i v_{j,\mathbf{k}} + \partial_j v_{i,\mathbf{k}}\right), \tag{2.62}$$

where $f(k)$ is given by equation (2.32). To this order, the full second moment is therefore

$$n^{ij} = \frac{1}{3}n\delta^{ij} + \frac{\lambda_{\mathrm{p}}\epsilon\partial_{\epsilon}n}{3c}\sum_{\mathbf{k}}\tau_k^2 f(k)\left(\partial_i v_{j,\mathbf{k}} + \partial_j v_{i,\mathbf{k}}\right).  \qquad (2.63)$$

Frequency integrating and subtracting off the scalar radiation pressure then gives equation (2.6). We evaluate equation (2.59) by plugging in equation (2.63) to get equation (2.35). This confirms that the temperature is independent of the photon energy $\epsilon$.

Before proceeding, we must check that the expansion for the zeroth moment $n$ is consistent with our assumption that it does not depend explicitly on the velocity field and therefore does not vary over short length scales. We recall that to first order $n = n_{0,0} + n_{1,0} + n_{0,1}$. The terms $n_{0,0}$ and $n_{1,0}$ do not depend on the velocity field by definition since they are zeroth order in $v$. For the third term, we have $n_{0,1} = \delta^{ij}n_{0,1}^{ij} = 0$ (since the velocity field is divergenceless) so our expansion is self-consistent. The physical interpretation of this result is that the velocity field affects the angular distribution but not the energy distribution of photons on short length scales. This is because variations in the photon energy distribution are washed out by spatial diffusion on length scales of order the photon mean free path.

Since the derivation of equation (2.63) assumes that any escape probability term $p_{\mathrm{e}}n$ added to equation (2.60) is negligible, we must derive a constraint on $p_{\mathrm{e}}$. This term must be small compared to the term in the radiative transfer equation that sets the diffusion time scale. The diffusion term comes from the term $\lambda_{\mathrm{p}}\ell^i\partial_i n(\hat{\ell})$, which is approximated by

$$\sim \lambda_{\mathrm{p}}\ell^i\partial_i\left(n - \lambda_{\mathrm{p}}\ell^i\partial_i n\right)  \qquad (2.64)$$

The diffusion timescale is set by the second derivative term,

$$\ell^i \ell^j \lambda_{\mathrm{p}}^2 \partial_i \partial_j n \sim \left( \frac{\lambda_{\mathrm{p}}}{\lambda_{\max}} \right)^2 n, \qquad (2.65)$$

which gives the condition

$$p_{\mathrm{e}} \ll \left( \frac{\lambda_{\mathrm{p}}}{\lambda_{\max}} \right)^2. \qquad (2.66)$$

We note that equation (2.63) is the solution for a time-independent velocity field, which is typically what Monte Carlo spectral simulations assume. Such solutions may appear to be unphysical (though still good approximations) since velocity fields typically evolve on the flow timescale

$$t_{\mathrm{f}} \sim \lambda/v, \qquad (2.67)$$

which is often significantly smaller than the diffusion timescale. But we can always choose a mode decomposition of traveling waves, and for each wave separately the solution in the frame of the wave is given by equation (2.63). To obtain the general solution in the lab frame, we need to transform equation (2.63) for each mode separately back to the lab frame and then sum the results. But to order $v/c$ the solutions are invariant under such transformations when written in the form of equation (2.63), so it turns out that equation (2.63) gives the correct solution even for a time-dependent divergence-less velocity field. It follows, therefore, that in a periodic box with a sufficiently small escape probability, time-independent Monte Carlo simulations correctly capture bulk Comptonization by divergenceless turbulence. In a vertically stratified atmosphere other time-dependent effects complicate the problem, but such simulations should still capture this effect reasonably well.

**Part 4: Important limiting cases of bulk Comptonization**

For bulk Comptonization the wave temperature given by equation (2.35) simplifies in different ways depending on what limit is taken. In order to develop physical insight we now discuss several important limits in detail. In particular, we show that to derive equation (2.28) we must take either the unsaturated limit or the limit of statistically homogeneous turbulence. We summarize our results in Table 2.3.

**The unsaturated, statistically homogeneous, and marginally optically thin limits**  In the unsaturated limit (Part 2) the diffusion timescale is significantly shorter than the energy timescale so that spatial variations in the zeroth moment of the occupation number $n$ are negligible. As a result, we can spatially average equation (2.30), which gives equations (2.36), (2.37), and (2.28). The criterion for this limit is given by equation (2.55), where $\lambda_{\max} = \lambda_{T,\max}$, the maximum lengthscale for variations in the wave temperature, not the underlying velocity field. If it is not straightforward to calculate $\lambda_{T,\max}$, then it is safest to set it equal to the box size.

We note that the maximum wavelength for variations in the velocity field may not be a good approximation for $\lambda_{T,\max}$, so it is not safe to make this approximation in equation (2.55). For example, consider a velocity field composed of two optically thin modes with wavenumbers $k + \Delta k$ and $k - \Delta k$ such that $\Delta k/k \ll 1$. The resulting velocity field is the familiar beats pattern, given by

$$
\begin{aligned}
v(x) &= v_0 \sin\left((k + \Delta k)x\right) + v_0 \sin\left((k - \Delta k)x\right) \\
&= 2v_0 \cos(\Delta k x) \sin(kx)
\end{aligned}
\tag{2.68}
$$

| Limit | Criteria | Type of thermal Comptonization equivalent to | $k_B T_w$ |
|---|---|---|---|
| None | - | inhomogeneous | $\frac{1}{6} m_e \lambda_p^2 \left( \partial_i v_j + \partial_j v_i \right) \sum_{\mathbf{k}} \tau_k^2 f(k) \left( \partial_i v_{j,\mathbf{k}} + \partial_j v_{i,\mathbf{k}} \right)$ |
| Unsaturated | $\lambda_{T,\max} \ll \sqrt{\frac{m_e c^2}{k_B T_w}} \lambda_p$ | homogeneous | $\sum_{\mathbf{k}} \frac{1}{3} m_e \left\langle v_{\mathbf{k}}^2 \right\rangle f(k)$ |
| Statistically homogeneous turbulence | Random mode phases | homogeneous | $\sum_{\mathbf{k}} \frac{1}{3} m_e \left\langle v_{\mathbf{k}}^2 \right\rangle f(k)$ |
| Unsaturated, optically thin | $\lambda_{T,\max} \ll \sqrt{\frac{m_e c^2}{k_B T_w}} \lambda_p$, $\lambda_{\max} \ll \lambda_p$ | homogeneous | $\frac{1}{3} m_e \langle v^2 \rangle$ |
| Statistically homogeneous turbulence, optically thin | Random mode phases, $\lambda_{\max} \ll \lambda_p$ | homogeneous | $\frac{1}{3} m_e \langle v^2 \rangle$ |
| Unsaturated, optically thick | $\lambda_{T,\max} \ll \sqrt{\frac{m_e c^2}{k_B T_w}} \lambda_p$, $\lambda_{\min} \gg \lambda_p$ | homogeneous | $\sum_{\mathbf{k}} \frac{2}{27 \tau_k^2} m_e \left\langle v_{\mathbf{k}}^2 \right\rangle$, or equivalently $\frac{1}{27} m_e \lambda_p^2 \left\langle \left( \partial_i v_j + \partial_j v_i \right)^2 \right\rangle$ |
| Statistically homogeneous turbulence, optically thick | Random mode phases, $\lambda_{\min} \gg \lambda_p$ | homogeneous | $\sum_{\mathbf{k}} \frac{2}{27 \tau_k^2} m_e \left\langle v_{\mathbf{k}}^2 \right\rangle$, or equivalently $\frac{1}{27} m_e \lambda_p^2 \left\langle \left( \partial_i v_j + \partial_j v_i \right)^2 \right\rangle$ |
| Marginally optically thin* | $\lambda_{\max} < 10 \lambda_p$ | inhomogeneous | $\sum_{\mathbf{k}} \frac{1}{3} m_e \left\langle v_{\mathbf{k}}^2 \right\rangle f(k)$ |
| Optically thick | $\lambda_{\min} \gg \lambda_p$ | inhomogeneous | $\frac{1}{27} m_e \lambda_p^2 \left( \partial_j v_i + \partial_i v_j \right)^2$ |

Table 2.3: Summary of important limits for bulk Comptonization by a divergence-less velocity field in a closed box. $\lambda_T$ denotes the lengthscale for variations in the wave temperature, not the underlying velocity field. If $\lambda_{T,\max}$ is not known then it should be set equal to the box size. *For the marginally optically thin case, the mode decomposition is taken locally over a smaller box size $\sim 10 \lambda_p$.

and the temperature is, for $\Delta k/k \ll 1$,

$$k_B T_w = \frac{4}{3} m_e v_0^2 \cos^2(\Delta k x) \sin^2(k x). \tag{2.69}$$

We see that even though the maximum wavelength of the velocity field is only $2\pi/(k + \Delta k)$, the resulting temperature varies on the significantly longer length scale $2\pi/\Delta k$ (this is intuitive but can be verified by plotting the Fourier series coefficients), so in this case $\lambda_{max} \sim 2\pi/\Delta k$. Therefore, in the criterion for the unsaturated limit given by equation (2.55) we should set $\lambda_{max}$ equal to the box size unless we have directly calculated the maximum wavelength of the temperature field.

Even if the criterion for the unsaturated limit is not satisfied for the box overall, variations in the zeroth moment $n$ will still be negligible in the case of statistically homogeneous turbulence. The reason for this is as follows. For variations in the temperature with $\lambda \sim \lambda_p$, equation (2.55) is trivially satisfied. To study optically thick variations, we begin by observing that for statistically homogeneous turbulence the phases of modes are random so that modes with wavelength $\sim \lambda$ are unlikely to cause significant inhomogeneities on scales larger than $\lambda$. Significant inhomogeneities on the scale $\lambda$ must therefore be due to modes with wavelengths greater than or equal to $\lambda$. But we see from equation (2.34) that the wave temperature is downweighted by $\lambda_p^2/\lambda^2$ for optically thick modes, so for a mode with amplitude $v_0$ the resulting wave temperature is $k_B T_w \sim m_e v_0^2 \lambda_p^2/\lambda^2$. If we plug this into the criterion given by equation (2.55) we get $\lambda \ll c\lambda/v$, which we see is satisfied for all $\lambda$. Inhomogeneities on the scale $\lambda$ are therefore unsaturated for all scales $\lambda$ and so variations in the zeroth moment $n$ are negligible for statistically homogeneous turbulence. This argument can be summarized as follows. For optically thin length scales the diffusion timescale is much smaller than the energy timescale, and as the number of scatterings across a length scale increases (increasing the diffusion timescale), the wave temperature decreases by the same factor (increasing the energy timescale) so that all length scales remain unsaturated.

In the unsaturated, optically thin limit, $f(k) = 1$, so equation (2.28) simplifies to

$$k_{\mathrm{B}} T_{\mathrm{w}} = \frac{1}{3} m_{\mathrm{e}} \left\langle v^2 \right\rangle = k_{\mathrm{B}} T_{\mathrm{tot}}. \tag{2.70}$$

This result also holds in the limit of optically thin, statistically homogeneous turbulence. In the unsaturated, optically thick limit, $f(k) = 2/9\tau_k^2$ so equation (2.28) simplifies to

$$k_{\mathrm{B}} T_{\mathrm{w}} = \sum_{\mathbf{k}} \frac{2}{27\tau_k^2} m_{\mathrm{e}} \left\langle v_{\mathbf{k}}^2 \right\rangle. \tag{2.71}$$

From equation (2.37) we see that in this limit the wave temperature can also be written in a way that is independent of the mode decomposition, as

$$k_{\mathrm{B}} T_{\mathrm{w}} = \frac{1}{27} \lambda_{\mathrm{p}}^2 m_{\mathrm{e}} \left\langle (\partial_i v_j + \partial_j v_i)^2 \right\rangle. \tag{2.72}$$

This result also holds in the limit of optically thick, statistically homogeneous turbulence.

If the criterion for the unsaturated limit is not satisfied (and the velocity field is not statistically homogeneous turbulence), then just as in the case of inhomogeneous thermal Comptonization (Part 2) it may be helpful to spatially average the wave temperature over a smaller length scale that is unsaturated. In the beats pattern discussed above, for example, even if the maximum length scale $2\pi/\Delta k$ for variations in the temperature is not unsaturated we can spatially average the wave temperature over the length scale $2\pi/k$ which gives

$$k_{\mathrm{B}} \left\langle T_{\mathrm{w}} \right\rangle_{2\pi/k} = \frac{2}{3} m_{\mathrm{e}} v_0^2 \cos^2(\Delta k x). \tag{2.73}$$

Alternatively, in such a case we could divide the box into smaller boxes, each with length $2\pi/k$, and derive the wave temperature in each box separately. In this case we would find that each box has only one mode with wavelength $2\pi/k$ and amplitude $2v_0 \cos(\Delta k x)$, so that the resultant wave temperature is still given by equation (2.73). We see, therefore, that two different ways of simplifying the problem yield the same result. The latter approach is perhaps nicer since it results in a single, unsaturated mode in each box, but we have to be careful when generalizing it to more complicated velocity fields. In principle the smaller box length should still be significantly greater than the maximum wavelength in the velocity field in order to ensure that the resulting mode decomposition is representative of the velocity field nearby. But if the minimum box length that satisfies this criterion happens not to be unsaturated then this does not help to simplify the problem. In practice it is a good approximation to choose the smaller box length equal to $\sim 10\lambda_{\rm p}$ since such a length scale is small enough to remain unsaturated, and modes with larger wavelengths are significantly downweighted by equation (2.32) so their contribution to the wave temperature is likely negligible anyway. This yields a spatially varying wave temperature given by

$$k_{\rm B} T_{\rm w, 10\lambda_{\rm p}} = \sum_{\mathbf{k}} \frac{1}{3} m_{\rm e} \left\langle v_{\mathbf{k}}^2 \right\rangle f(k),\qquad(2.74)$$

where, unlike the wave temperature given by equation (2.28), the mode decomposition is taken locally over the smaller box length $10\lambda_{\rm p}$. We define this to be the marginally optically thin limit since it holds when the contribution of optically thick modes to the wave temperature is negligible.

**The optically thick limit and radiation viscosity**  In the optically thick limit equation (2.35) simplifies to

$$k_{\mathrm{B}} T_{\mathrm{w}} = \frac{1}{27} m_{\mathrm{e}} \lambda_{\mathrm{p}}^2 \left( \partial_j v_i + \partial_i v_j \right)^2 , \tag{2.75}$$

in agreement with the "heating temperature" in Chan & Jones (1975).

Furthermore, in this limit bulk Comptonization can be described by a coefficient of kinematic viscosity $\nu_{\mathrm{s}}$. This coefficient is usually defined by

$$P^{ij}_{\mathrm{vis,shear}} = -\nu_{\mathrm{s}} \rho \left( \partial_i v_j + \partial_j v_i \right) , \tag{2.76}$$

where $\rho$ is the fluid mass density. According to equations (2.76) and (2.31), although we cannot in general define a kinematic viscosity coefficient, for any single velocity mode the kinematic radiation viscosity is given by

$$\nu_{\mathrm{s,k}} = \frac{4}{3} \tau_k^2 f(k) \left( \frac{E}{\rho c^2} \right) \lambda_{\mathrm{p}} c, \tag{2.77}$$

so that the radiation viscous shear stress tensor can be written

$$P^{ij}_{\mathrm{vis,shear}} = -\rho \sum_{\mathbf{k}} \nu_{\mathrm{s,k}} \left( \partial_i v_{j,\mathbf{k}} + \partial_j v_{i,\mathbf{k}} \right) . \tag{2.78}$$

We note that according to equation (2.77) the viscosity coefficient for a single mode, which is proportional to $\tau_k^2 f(k)$, is greatest in the optically thick limit and goes to zero in the optically thin limit. This is surprising since the corresponding wave temperature, which is proportional to $f(k)$, is greatest in the optically thin limit and goes to zero in the optically thick limit. The reason for this is that the wave temperature is related to the viscosity coefficient by two factors of $\tau_k$, one due to the derivative in equation (2.76) and the other due to the derivative in equation (2.30). In other words, since the dissipation is due the product of the viscosity coefficient with the square of the velocity shear tensor, the viscosity coefficient and the temperature have different limiting behaviors.

In the optically thick limit (i.e. $\lambda_\mathrm{p} \ll \lambda_\mathrm{min}$), the kinematic viscosity is independent of $\mathbf{k}$,

$$\nu_\mathrm{s} = \frac{8}{27} \left( \frac{E}{\rho c^2} \right) \lambda_\mathrm{p} c,  \tag{2.79}$$

so that in this limit the kinematic viscosity is well-defined for an arbitrary (divergenceless) velocity field. This coefficient for the viscosity was first derived by Masaki (1971), and it differs by a factor of $\frac{10}{9}$ from the more commonly cited value, $\nu_a = \frac{4}{15} \left( \frac{E}{\rho c^2} \right) c \lambda_\mathrm{p}$, in, e.g., Weinberg (1971), Weinberg (1972), and Mihalas & Mihalas (1984). The reason for the discrepancy is that the more commonly cited value, first derived by Thomas (1930), assumes pure absorption, while equation (2.79) is correct for pure scattering (Masaki 1971; Straumann 1976).

## 2.4.2   A heuristic wave temperature

To provide physical insight into our analytic solution given by equation (2.28), we now find a heuristic, approximate expression for $T_\mathrm{w}$ with a simple physical model motivated by ideas put forth in S04. To do this, we first consider the wave temperature of a single mode with wave vector $\mathbf{k}$, which we denote $T_\mathrm{w}(\lambda_\mathrm{p}, \mathbf{k})$. S04 suggested that photons can only sample turbulent velocities on scales $\lambda \leq \lambda_\mathrm{p}$, since longer turbulent wavelengths will advect photons back and forth with the flow without allowing the photons to "sample" their velocities. A rough interpretation of this reasoning leads to

$$T_\mathrm{w,rough}(\lambda_\mathrm{p}, \mathbf{k}) = f_\mathrm{S04}(k) T_\mathrm{tot,\mathbf{k}}.  \tag{2.80}$$

If one then assumes that the wave temperature of an arbitrary field is the sum of the wave temperatures of its modes, then equation (2.21) follows. But a more subtle interpretation of this reasoning suggests that $T_{\mathrm{w}}(\lambda_{\mathrm{p}}, \mathbf{k})$ is determined by the second moment of the distribution of velocity differences between subsequent scatterings, $\langle (\Delta \mathbf{v})^2 \rangle$. In the long wavelength limit, velocity differences between subsequent scatterings are negligible and so the wave power does not contribute to Comptonization. In the short wavelength limit, velocity differences allow photons to sample the full power of the wave. This model of Comptonization suggests that we define

$$T_{\mathrm{w,heur}}(\lambda_{\mathrm{p}}, \mathbf{k}) = f_{\mathrm{heur}}(k) T_{\mathrm{tot}, \mathbf{k}}, \tag{2.81}$$

where

$$f_{\mathrm{heur}}(k) \propto \langle (\Delta \mathbf{v})^2 \rangle. \tag{2.82}$$

Before proceeding, we note that defining $\langle (\Delta \mathbf{v})^2 \rangle$ is potentially tricky because the distribution of $\Delta \mathbf{v}$ for a photon is dependent on its current location, in effect introducing correlations into subsequent $\Delta \mathbf{v}$'s. In other words, subsequent $\Delta \mathbf{v}$'s are not independent. But if the escape probability is low enough, a condition we quantify below, then the set of $\Delta \mathbf{v}$'s encountered by a photon before it escapes is indistinguishable from a set of $\Delta \mathbf{v}$'s independently drawn from the position-averaged $\Delta \mathbf{v}$ distribution. The order of $\Delta \mathbf{v}$'s is different in the two cases, but the total photon energy change does not depend on the order because the fractional photon energy change per scattering is small for $v^2/c^2 \ll 1$. With these potential problems accounted for, we proceed to calculate $f_{\mathrm{heur}}(k)$.

First we find the proportionality constant between $f_{\mathrm{heur}}(k)$ and $\langle (\Delta \mathbf{v})^2 \rangle$ by evaluating both sides of equation (2.81) in the short wavelength limit. In this limit, the full wave power must contribute, so $f_{\mathrm{heur}}(k) \to 1$. To evaluate $\langle(\Delta\mathbf{v})^2\rangle$ in this limit, let $f(\mathbf{v})$ be any normalized distribution of velocities. Then,

$$
\begin{aligned}
\langle(\Delta\mathbf{v})^2\rangle &= \int (\mathbf{v_2} - \mathbf{v_1})^2 f(\mathbf{v_1}) f(\mathbf{v_2}) d\mathbf{v_1} d\mathbf{v_2} \\
&= 2\left(\langle v^2\rangle - \langle\mathbf{v}\rangle^2\right) = 2\sigma_{\mathbf{v}}^2.
\end{aligned}
\tag{2.83}
$$

Therefore,

$$
f_{\mathrm{heur}}(k) = \frac{\langle(\Delta\mathbf{v})^2\rangle}{2\sigma_{\mathbf{v}}^2}.
\tag{2.84}
$$

For sinusoidal modes, $\langle\mathbf{v}\rangle = 0$ so $\sigma_{\mathbf{v}}^2 = \langle v^2\rangle$ and

$$
f_{\mathrm{heur}}(k) = \frac{\langle(\Delta\mathbf{v})^2\rangle}{2\langle v^2\rangle}.
\tag{2.85}
$$

We now calculate $\langle(\Delta\mathbf{v})^2\rangle$ for the position-averaged $\Delta\mathbf{v}$ for a single, divergenceless (i.e. transverse) mode with wavelength $\lambda$,

$$
\mathbf{v} = \mathbf{v_0}\sin\left(\frac{2\pi}{\lambda}z\right).
\tag{2.86}
$$

Let $P_{\Delta\mathbf{r}}(\Delta\mathbf{r})$ be the probability density that a photon travels a displacement $\Delta\mathbf{r}$ between scatterings. Then, at a given position $\mathbf{r}$,

$$
\langle(\Delta\mathbf{v})^2\rangle_{\mathbf{r}} = \int \left(\Delta\mathbf{v}(\Delta\mathbf{r}, \mathbf{r})\right)^2 P_{\Delta\mathbf{r}}(\Delta\mathbf{r}) d^3\Delta\mathbf{r}.
\tag{2.87}
$$

Averaging over all positions in a volume $V$, this is

$$
\langle(\Delta\mathbf{v})^2\rangle = \frac{1}{V}\int \left(\Delta\mathbf{v}(\Delta\mathbf{r}, \mathbf{r})\right)^2 P_{\Delta\mathbf{r}}(\Delta\mathbf{r}) d^3\Delta\mathbf{r} d^3\mathbf{r}.
\tag{2.88}
$$

For a single mode, equation (2.86) gives

$$\Delta \mathbf{v} = \mathbf{v_0} \sin \left( \frac{2\pi}{\lambda} \left( z + \Delta z \right) \right) - \mathbf{v_0} \sin \left( \frac{2\pi}{\lambda} z \right) \tag{2.89}$$

and

$$\langle (\Delta \mathbf{v})^2 \rangle = \frac{1}{\lambda} \int_0^\lambda dz \int d^3 \Delta \mathbf{r} \left( \Delta \mathbf{v}(\Delta z, z) \right)^2 P_{\Delta \mathbf{r}} \left( \Delta \mathbf{r} \right). \tag{2.90}$$

The probability that a photon with mean free path $l$ travels a distance between $s$ and $s + ds$ is $P_s \left( s \right) ds = (1/l) \mathrm{e}^{-s/l} ds$. Let $\mu = \cos \theta$, where $\theta$ is the angle between the photon propagation direction and the $z$-axis, so that $\Delta z = s\mu$. Then, expressing $\Delta \mathbf{r}$ in spherical polar coordinates and invoking axisymmetry about the $z$-axis, equation (2.90) becomes

$$\langle (\Delta \mathbf{v})^2 \rangle = \frac{v_0^2}{2l\lambda} \int_0^\lambda dz \int_{-1}^1 d\mu \int_0^\infty ds \left( \sin \left( \frac{2\pi}{\lambda} \left( z + s\mu \right) \right) \right.$$
$$\left. - \sin \left( \frac{2\pi}{\lambda} z \right) \right)^2 \mathrm{e}^{-s/l}. \tag{2.91}$$

This is easily evaluated by performing the integral over $z$ first, giving

$$\langle (\Delta \mathbf{v})^2 \rangle = v_0^2 \left( 1 - \tau_k \tan^{-1} \left( \frac{1}{\tau_k} \right) \right). \tag{2.92}$$

$$= 2 \left\langle v^2 \right\rangle \left( 1 - \tau_k \tan^{-1} \left( \frac{1}{\tau_k} \right) \right). \tag{2.93}$$

Equation (2.85) then gives equation (2.39). By comparison, the exact solution for a single mode is determined by equation (2.32). Our heuristic result is plotted in Figure 2.1, and is remarkably close to the exact solution. In particular, it is a much better approximation than the rough weighting function, equation (2.25), also shown in Figure 2.1. Our model based on the second moment of the velocity difference distribution therefore captures the essential physics of Comptonization by a single mode.

Before proceeding to define $T_{\mathrm{w,heur}}$ for an arbitrary velocity field, we quantify the condition that the escape probability be low enough, presupposed by our derivation of $\langle (\Delta \mathbf{v})^2 \rangle$. Since the distribution of velocity differences, as a function of position, repeats every quarter wavelength, our results should be valid provided photons travel a distance $\Delta z$ in the $z$ direction that is greater than $\lambda_{\max}/4$ before escaping. For an (optically thick) random walk, $\Delta z \sim (N/3)^{1/2} \lambda_{\mathrm{p}}$, where $N$ is the average number of scatterings. Since $N = 1/p_{\mathrm{e}} - 1$, where $p_{\mathrm{e}}/t_{\mathrm{C}}$ is the escape probability per unit time during the average time $t_{\mathrm{C}}$ between subsequent scattering events,

$$p_{\mathrm{e}} < \left( \frac{3}{16} \left( \lambda_{\max}/\lambda_{\mathrm{p}} \right)^2 + 1 \right)^{-1} \simeq \frac{16}{3} \left( \lambda_{\mathrm{p}}/\lambda_{\max} \right)^2 \tag{2.94}$$

for optically thick modes. Up to a factor of order unity, this agrees with equation (2.66).

We now use our model to compute $T_{\mathrm{w,heur}}$ for an arbitrary velocity field in terms of $T_{\mathrm{w,heur}}(\lambda_{\mathrm{p}}, \mathbf{k})$. Proceeding analogously to the single mode case, we define

$$T_{\mathrm{w,heur}} = \frac{\langle (\Delta \mathbf{v})^2 \rangle}{2 \langle v^2 \rangle} T_{\mathrm{tot}}, \tag{2.95}$$

where $T_{\mathrm{tot}} = \sum_{\mathbf{k}} T_{\mathrm{tot},\mathbf{k}}$ is the temperature corresponding to the average kinetic energy of the electrons due to the velocity field. We note that since $\frac{3}{2} k_{\mathrm{B}} T_{\mathrm{tot}} = \frac{1}{2} m_{\mathrm{e}} \langle v^2 \rangle$ this is equivalent to equation (2.38). To simplify this, we compute

$$
\begin{aligned}
\langle (\Delta \mathbf{v})^2 \rangle &= \frac{1}{V} \int \left( \sum_{\mathbf{k}} \Delta \mathbf{v_k}(\Delta \mathbf{r}, \mathbf{r}) \right)^2 P_{\Delta \mathbf{r}}(\Delta \mathbf{r}) d^3 \Delta \mathbf{r} d^3 \mathbf{r} \\
&= \frac{1}{V} \int \left( \sum_{\mathbf{k}} (\mathbf{v_k}(\mathbf{r} + \Delta \mathbf{r}) - \mathbf{v_k}(\mathbf{r})) \right)^2 P_{\Delta \mathbf{r}}(\Delta \mathbf{r}) d^3 \Delta \mathbf{r} d^3 \mathbf{r} \\
&= \frac{1}{V} \int \sum_{\mathbf{k}} (\mathbf{v_k}(\mathbf{r} + \Delta \mathbf{r}) - \mathbf{v_k}(\mathbf{r}))^2 P_{\Delta \mathbf{r}}(\Delta \mathbf{r}) d^3 \Delta \mathbf{r} d^3 \mathbf{r} \\
&= \sum_{\mathbf{k}} \langle (\Delta \mathbf{v_k})^2 \rangle.
\end{aligned}
\tag{2.96}
$$

To get from line 2 to line 3 we made use of the orthogonality for distinct sinusoidal modes. That is, for two distinct modes, $\mathbf{v_k}$ and $\mathbf{v_{k'}}$, $\mathbf{k} \neq \mathbf{k'}$, and any displacement $\Delta \mathbf{r}$,

$$
\int \mathbf{v_k}(\mathbf{r} + \Delta \mathbf{r}) \cdot \mathbf{v_{k'}}(\mathbf{r}) d^3 \mathbf{r} = 0.
\tag{2.97}
$$

Then, since $\frac{3}{2} k_{\mathrm{B}} T_{\mathrm{tot},\mathbf{k}} = \frac{1}{2} m_{\mathrm{e}} \langle v_{\mathbf{k}}^2 \rangle$ and $\frac{3}{2} k_{\mathrm{B}} T_{\mathrm{tot}} = \frac{1}{2} m_{\mathrm{e}} \langle v^2 \rangle$, equations (2.85), (2.95), and (2.96) give

$$
T_{\mathrm{w,heur}} = \sum_{\mathbf{k}} f_{\mathrm{heur}}(k) T_{\mathrm{tot},\mathbf{k}},
\tag{2.98}
$$

or, alternatively,

$$
T_{\mathrm{w,heur}} = \sum_{\mathbf{k}} T_{\mathrm{w,heur}}(\lambda_{\mathrm{p}}, \mathbf{k}).
\tag{2.99}
$$

In terms of the velocity field, this is

$$
k_{\mathrm{B}} T_{\mathrm{w,heur}} = \sum_{\mathbf{k}} \frac{1}{3} m_{\mathrm{e}} \langle v_{\mathbf{k}}^2 \rangle f_{\mathrm{heur}}(k).
\tag{2.100}
$$

Equation (2.100) is the same as the exact solution for the wave temperature, equation (2.28), except that here the heuristic weighting function is used. Note that in our heuristic derivation it is the orthogonality of distinct modes that allows us to express the wave temperature of an arbitrary velocity field as a sum over the wave temperatures of its modes. Unsurprisingly, orthogonality was used analogously to prove equation (2.26). Therefore, our model based on the second moment of the velocity difference distribution captures the essential physics of Comptonization by statistically homogeneous, divergenceless turbulence.

### 2.4.3  The dependence of the wave temperature on the turbulence power spectrum

We now analyze the dependence of $T_{\mathrm{w}}$ on the power spectrum of the turbulence. For the remainder of this section, we write $k$ in units of $1/\lambda_{\mathrm{p}}$ for clarity (i.e. $k \ll 1$ and $k \gg 1$ denote optically thick and thin scales, respectively). If the turbulence is completely optically thin on all scales ($k_{\min} \gg 1$), then $T_{\mathrm{w}} = T_{\mathrm{tot}}$, independent of the energy spectral index, $p$. However, if some scales in the turbulent cascade are optically thick, then $p$ will affect $T_{\mathrm{w}}$.

For the case where all scales in the turbulence are optically thick ($k_{\max} \ll 1$), equation (2.32) implies that

$$f(k) = \frac{2}{9}k^2.$$  (2.101)

Integrating this over an energy spectrum $T_{\mathrm{tot}}(k) \propto k^{-p}$ then gives

$$\frac{T_{\mathrm{w}}}{T_{\mathrm{tot}}} = \left(\frac{2}{9}\right)\frac{1-p}{3-p}\left(\frac{k_{\max}^{3-p} - k_{\min}^{3-p}}{k_{\max}^{1-p} - k_{\min}^{1-p}}\right).$$  (2.102)

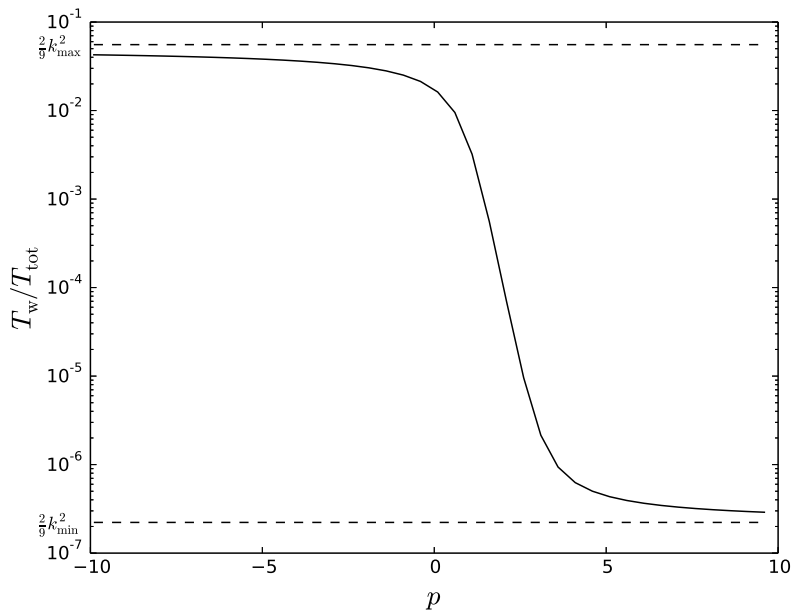

Figure 2.2: The dependence of $T_\mathrm{w}/T_\mathrm{tot}$ on $p$ (calculated exactly) for $k_\mathrm{min} = 0.001$, $k_\mathrm{max} = 0.5$, where $k$ is in units of $1/\lambda_\mathrm{p}$. Note that $T_\mathrm{w}/T_\mathrm{tot}$ approaches $\frac{2}{9}k_\mathrm{max}^2$ and $\frac{2}{9}k_\mathrm{min}^2$ for $p \ll 1$ and $p \gg 3$, respectively, as expected for a broad power spectrum with $k_\mathrm{max} < 1$.

For a broad power spectrum, i.e. $k_\mathrm{min}/k_\mathrm{max} \ll 1$, this simplifies to

$$\frac{T_\mathrm{w}}{T_\mathrm{tot}} = \frac{1}{9} \times \begin{cases} 2k_\mathrm{max}^2, & p \ll 1 \\ k_\mathrm{max}^{4/3}k_\mathrm{min}^{2/3}, & p = 5/3 \\ 2k_\mathrm{min}^2, & p \gg 3. \end{cases} \qquad (2.103)$$

This is illustrated in Figure 2.2. Note that $T_\mathrm{w}/T_\mathrm{tot}$ drops significantly for $p > 1$, because then the energy bearing modes are on the largest scales. These are the most optically thick and therefore the most downweighted in their contribution to bulk Comptonization.

We next analyze whether $T_\mathrm{w}/T_\mathrm{tot}$, for a given spectral index $p$ and range of modes, $k_\mathrm{min} < k < k_\mathrm{max}$, is dominated by small or large scales. In other words, we examine which turbulent modes in a given spectrum contribute most to bulk Comptonization. The relative contribution of a scale with wavenumber $k$ is

$$T_{\mathrm{w}}(\lambda_{\mathrm{p}}, k)dk \sim T_{\mathrm{w}}(\lambda_{\mathrm{p}}, k)k$$
$$\sim f(k)T_{\mathrm{tot}}(k)k$$
$$\sim k^{q-p+1}, \tag{2.104}$$

where, from equation (2.32), $q = 2$ for optically thick ($k \ll 1$) scales, and $q = 0$ for optically thin ($k \gg 1$) scales. Now consider an underlying power spectrum with some $k_{\min}$ and $k_{\max}$. We see that for $p < 1$ the exponent in equation (2.104) is always positive, and so small scales contribute most to bulk Comptonization, regardless of $k_{\min}$ and $k_{\max}$. This is physically intuitive; for $p < 1$, the turbulent power is concentrated on small scales. Since the weighting factor $f(k)$ also favors small scales, they of course contribute most. For $p > 3$, the exponent is always negative, and so large scales always contribute most. In this case, the turbulent power is so concentrated on large scales that they contribute more even though $f(k)$ favors small scales.

For $1 < p < 3$, we first consider the part of the spectrum with $k \gg 1$ (if it exists). Since $q = 0$, small scales contribute more than large scales for these modes. Now consider the part of the spectrum with $k \ll 1$ (if it exists). Here, large scales contribute more than small scales. Therefore, looking at the entire power spectrum, it is intermediate scales that contribute most, assuming it is broad enough to include regions of both small and large $k$. If it is not sufficiently broad, then whether small or large scales contribute most depends on $k_{\min}$ and $k_{\max}$ relative to $k \approx 1$ (the optically thin to thick transition wavenumber).

These results are depicted in Figure 2.3. The curve in this figure shows the values of $p$ and $k$ such that the derivative of $k_{\mathrm{B}}T_{\mathrm{w}}(\lambda_{\mathrm{p}}, k) = kf(k)T_{\mathrm{tot}}(k)$ is zero, using the full analytic expression for $f(k)$ from equation (2.32). To connect this figure to our discussion, draw a horizontal line from $k_{\min}$ to $k_{\max}$ at a given value of $p$. If the line lies in the lower (upper) region, then small (large) scales contribute most. If the line straddles the two regions, then for the part of the line that lies in the lower region small scales contribute most, and for the part that lies in the upper region large scales contribute the most. In this case, then, for the entire spectrum it is the scales that straddle the curve which contribute most. Note that for $p < 1$ and $p > 3$ a spectrum can never straddle the curve, whereas for a Kolmogorov spectrum ($p = 5/3$), e.g., it can, if $k_{\min} < 3 < k_{\max}$.

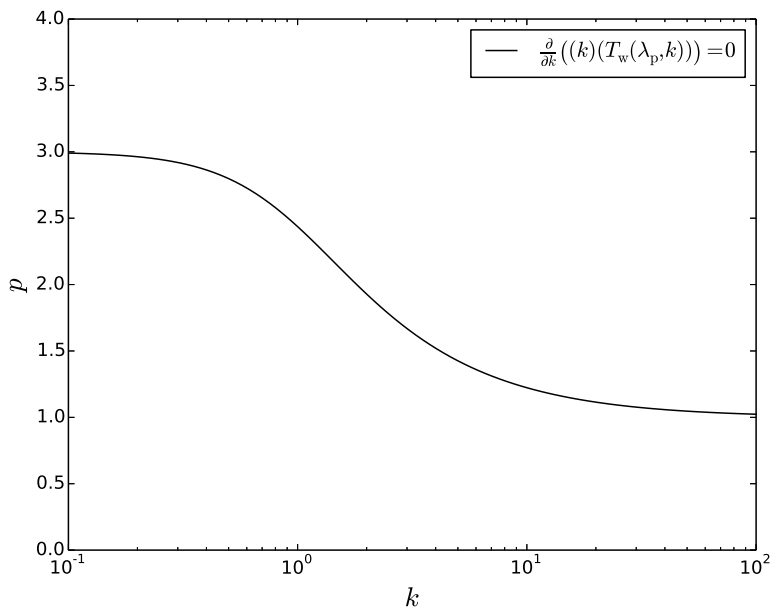

Figure 2.3: Regions in $k - p$ space in which bulk Comptonization is dominated by small (lower region) and large (upper region) scales, as determined by the sign of $\frac{d}{dk}((k)(T_w(\lambda_p, k)))$.

## 2.5 Comptonization by turbulence with non-zero divergence

Socrates, Davis & Blaes (2004) conjectured that Comptonization by turbulence can be treated as thermal Comptonization by solving the Kompaneets equation with an equivalent "wave" temperature. In section 2.4 we proved this under certain conditions, one of which is that the turbulence be divergenceless. In this section we investigate Comptonization by velocity fields with non-zero divergence, an effect that usually cannot be treated as thermal Comptonization alone, with the aim of understanding how it impacts radiation spectra in generic, turbulent regions of stratified accretion disc atmospheres.

Because in this case effects that are second order in velocity are non-trivially intertwined with those that are first order in velocity, it is harder to find closed-form solutions. To develop physical intuition we focus on two limiting cases. The trivial case is the optically thin case, i.e. when the mean free path is significantly larger than the largest length scale in the turbulence, $\lambda_p/\lambda_{max} \gg 1$. Electron velocities "sampled" by photons are uncorrelated and so Compton scattering should depend on only the total spatial average distribution of electron velocities. This is, therefore, the one case where Comptonization by a velocity field with non-zero divergence can be treated as thermal Comptonization, by solving the Kompaneets equation with $T_w$ given by

$$\frac{3}{2}k_B T_w = \frac{3}{2}k_B T_{tot} = \frac{1}{2}m_e \left\langle v^2 \right\rangle. \qquad (2.105)$$

We note that in this limit $T_w$ is independent of $\lambda_p$. Energy exchange that is first order in velocity vanishes since photons are equally likely to downscatter as they are to upscatter. Bulk Comptonization is then solely due to radiation viscous dissipation (see section 2.2). As viscous effects are second order in velocity, it is unsurprising that they can be characterized by a temperature. We also note that in this limit the wave temperature is the same as that for a divergenceless velocity field, equation (2.34). Optically thin bulk Comptonization is therefore a single phenomenon that depends on only the mean square speed of the velocity field.

To arrive at this result with a more formal approach, we start with the zeroth moment of the radiative transfer equation, equation (2.9). In the limit of optically thin turbulence the radiation variables must be homogeneous and isotropic, so that $n^i = 0 = n^{ijk}$, and $n^{ij} = (1/3)n\delta^{ij}$. Then, averaging equation (2.9) over the largest scale $\lambda_{\max}$ gives

$$\frac{\lambda_{\mathrm{p}}}{c}\frac{\partial n}{\partial t} = \frac{1}{m_{\mathrm{e}}c^2\epsilon^2}\frac{\partial}{\partial\epsilon}\left(\epsilon^4\left(n + n^2 + \left(k_{\mathrm{B}}T_{\mathrm{e}} + \frac{1}{3}m_{\mathrm{e}}\langle v^2\rangle\right)\frac{\partial}{\partial\epsilon}n\right)\right). \qquad (2.106)$$

This is the Kompaneets equation, with the contribution from the velocity field to the Comptonization temperature given by equation (2.105).

In the optically thick case, i.e. when the photon mean free path is significantly smaller than the smallest scale in the turbulence, the lowest order energy exchange is the work done by radiation pressure to compress the gas, since it is first order in velocity and since radiation viscous effects are suppressed (section 2.4). We focus on the extremely optically thick case, which we define as the limit in which photon diffusion is negligible relative to photon advection, so that the photon and gas fluids are strongly coupled. If we define $\psi_\epsilon = \epsilon^2 n$, the photon number density at energy $\epsilon$, and $\psi = \int \psi_\epsilon d\epsilon$, the total photon number density, then the advection and diffusion fluxes are given by

$$\mathbf{F}_{\mathrm{a}} = \mathbf{v}\psi \qquad (2.107)$$

and

$$\mathbf{F}_{\mathrm{d}} = -\frac{1}{3}\lambda_{\mathrm{p}} c \nabla \psi, \qquad (2.108)$$

respectively. The extremely optically thick limit is then given by

$$\frac{\lambda_{\min}}{\lambda_{\mathrm{p}}} \gg \frac{c}{v}. \qquad (2.109)$$

In this case, velocity convergence corresponds to compression in which gas mechanical energy is transferred locally to the photons. We expect that photons with wavelength $\lambda_\gamma$ are effectively compressed at a rate given by the velocity difference across $\lambda_\gamma$,

$$\frac{d\lambda_\gamma}{dt} = \frac{1}{3}\lambda_\gamma \nabla \cdot \mathbf{v}, \qquad (2.110)$$

so that, for example, a locally thermal photon distribution remains thermal and only changes temperature, completely analogous to the evolution of the cosmic microwave background radiation under the expansion of the Universe. This is equivalent to a fractional energy change per scattering given by

$$\frac{\lambda_{\mathrm{p}}}{c}\frac{1}{\epsilon}\frac{d\epsilon}{dt} = -\frac{\lambda_{\mathrm{p}}\nabla \cdot \mathbf{v}}{3c}, \qquad (2.111)$$

since

$$\frac{d\epsilon}{d\lambda_\gamma} = -\frac{\epsilon}{\lambda_\gamma}. \qquad (2.112)$$

We now confirm that equation (2.111) correctly describes extremely optically thick Compton scattering in a converging or diverging flow, both by providing a heuristic argument and by deriving it from the radiative transfer equation.

Before proceeding, we note that the study of photon upscattering by a converging velocity field can be traced back to Blandford & Payne (1981a,b) and Payne & Blandford (1981), who, along with later authors, made detailed spectral calculations for specific velocity fields in shocks and spherically accreting systems. In fact, equation (2.111) can equivalently be stated as the upscattering timescale

$$t_{\mathrm{up}}^{-1} = \frac{1}{3}\nabla \cdot \mathbf{v} \tag{2.113}$$

given in Blandford & Payne (1981a). In this section, by contrast, we have been investigating how this effect manifests itself locally in a generic, turbulent region of a stratified disc atmosphere, with the goal of resolving and interpreting it in spectral calculations of radiation MHD simulations. We have been focusing on the extremely optically thick case, in which the photon and gas fluids are strongly coupled, because the physics is both relevant and intuitive. In the moderately optically thick case, on the other hand, i.e.

$$1 \ll \frac{\lambda_{\mathrm{min}}}{\lambda_{\mathrm{p}}} \sim \frac{c}{v}, \tag{2.114}$$

such as photon upscattering in a radiation pressure dominated shock (Blandford & Payne, 1981b), diffusion competes with advection so that photon distributions at neighboring fluid elements mix. Photon upscattering in such a converging flow may not be viewed as simply the compression of a photon fluid strongly coupled to the gas, and photon upscattering in which a photon thermal distribution is not preserved can occur.

To heuristically derive equation (2.111), consider a disturbance converging in the $z$

direction given by

$$\mathbf{v} = -\alpha z \hat{z}. \tag{2.115}$$

For a 3D random walk, the average distance between scatterings traveled by a photon in the direction of convergence is $\lambda_{\mathrm{p}}/3$. Since the fractional energy change per scattering for low energy photons is $v/c$, at $z = 0$ this gives

$$\frac{\Delta\epsilon}{\epsilon} = \frac{-\lambda_{\mathrm{p}}\partial_z v_z}{3c}, \tag{2.116}$$

in agreement with equation (2.111).

We now derive equation (2.111) with the radiative transfer equation, equation (2.9). If we (1) omit stimulated scattering terms to facilitate comparison with simulations, (2) substitute in the standard closure relations for the first moment in the optically thick limit,

$$n^i = -\frac{v_i}{3c}\epsilon\frac{\partial n}{\partial \epsilon} - \frac{1}{3}\lambda_{\mathrm{p}}\partial_i n \tag{2.117}$$

and

$$n^{ij} = \frac{1}{3}n\delta^{ij}, \tag{2.118}$$

and (3) substitute in the photon number density $\psi_\epsilon$, then the radiative transfer equation to second order in velocity and first order in $\epsilon/m_e c^2$ becomes

$$
\begin{aligned}
\frac{\lambda_\mathrm{p}}{c}\partial_t\psi_\epsilon = {} & -\frac{\lambda_\mathrm{p}}{c}\nabla\cdot\left(\mathbf{v}\psi_\epsilon - \frac{1}{3}\lambda_\mathrm{p}c\nabla\psi_\epsilon\right) \\
& -\partial_\epsilon\left(\epsilon\left(\frac{-\lambda_\mathrm{p}\nabla\cdot\mathbf{v}}{3c}\right)\psi_\epsilon\right) - \partial_\epsilon\left(\epsilon\left(\frac{4k_\mathrm{B}T_\mathrm{e}-\epsilon}{m_e c^2}\right)\psi_\epsilon\right) \\
& +\partial_\epsilon^2\left(\epsilon^2\left(\frac{k_\mathrm{B}T_\mathrm{e}}{m_e c^2}\right)\psi_\epsilon\right).
\end{aligned}
\tag{2.119}
$$

Neglecting stimulated scattering, this is Blandford & Payne (1981a) equation (18), cast in the physically revealing form of a Fokker-Plank equation. The terms inside the divergence operator correspond to spatial drift (i.e. photon advection) and spatial diffusion, respectively. The next term corresponds to energy drift due to photon upscattering (downscattering) in the presence of a converging (diverging) velocity field. The remaining terms correspond to energy drift and diffusion due to thermal Comptonization. Note that even though Blandford & Payne (1981a) start with a zeroth moment equation correct only to first order in $v/c$, their resulting equation is the same because with the standard closure relation (equation 2.117) the second order terms cancel. The fractional energy change per scattering given by equation (2.111) follows from the bulk upscattering term. In the extremely optically thick limit the spatial diffusion term is negligible, and so photons are advected with the velocity field and upscatter according to equation (2.111). The effect of this process on the emergent spectrum of the disc will depend primarily on how effectively photons are able to escape from converging (or diverging) regions to the observer.

We note that this effect may be very sensitive to the time dependence of the velocity field. This is important because usually post-processing Monte Carlo simulations invoke time-independent atmospheres with the assumption that they approximately capture the effects of interest. For example, consider Comptonization in a converging region. A time-independent velocity field results in the accumulation of photons and subsequent upscattering to high energies at the point of zero velocity. If the region is near the photosphere, the emergent spectrum will be strongly upscattered. But in a time-dependent velocity field photons have a limited time to upscatter before converging regions become diverging regions, and so the effect on the emergent spectrum is significantly different.

## 2.6 Discussion

In order to develop physical intuition we have focused on relatively simple cases, such as the periodic box with escape probability. We now discuss how to apply these results to bulk Comptonization by turbulence in real accretion discs.

Real accretion flows are spatially stratified. Simulations of magnetorotational turbulence generally indicate that this turbulence dominates the fluid velocities in the midplane regions of the accretion flow. This turbulence is largely incompressible (divergenceless), although it generally excites compressible spiral acoustic waves (Heinemann & Papaloizou, 2009a,b). Sufficiently far from the midplane, magnetic forces always dominate thermal pressure forces, and support the flow vertically against the tidal gravitational field of the compact object. Such regions are dominated by Parker instability dynamics, and exhibit considerable compressive behavior (i.e. the flow has non-zero divergence) with significant density fluctuations (Blaes, Hirose & Krolik, 2007).

We expect terms that are second order in the velocity field to contribute most to turbulent Comptonization for two principal reasons. First, we observe that bulk Comptonization is significant only when bulk velocities exceed thermal velocities, and so second order terms cause photon upscattering but not downscattering. First order terms, on the other hand, can cause either upscattering or downscattering depending on whether the velocity field is converging or diverging, respectively. Therefore, for a turbulent velocity field we expect Comptonization by first order terms to be negligible on average. Second, MHD turbulence is generally incompressible (divergenceless) except near the photosphere, and first order terms vanish for an incompressible velocity field. Near the photosphere conditions become optically thin which suppresses the first order effect (section 2.5).

In the presence of vertical stratification we can still apply our results for statistically homogeneous turbulence to *local* regions of the atmosphere. For example, we can apply our analytic solution for the wave temperature, equation (2.28), to a local region by finding the decomposition of the turbulence in terms of sinusoidal modes in that region. This is equivalent to the spatially varying wave temperature given by equation (2.74). In section (2.4.1) we referred to this as taking the marginally optically thin limit. This neglects contributions to the wave temperature by longer wavelength modes, but since these modes are downweighted by a factor proportional to $\lambda_{\mathrm{p}}^2/\lambda^2$ it is a good approximation. In the same way, we could apply equation (2.38), which heuristically defines the wave temperature in terms of the mean square velocity difference, to a local region. This is in effect the approach we take in Chapter 4. We note that when applied to the entire velocity field (rather than just the divergenceless part), the heuristically defined wave temperature should capture all viscous energy exchange (i.e. bulk Comptonization due to terms second order in velocity), not just viscous dissipation arising from the divergenceless part.

Once the wave temperature is locally defined, we are left with an inhomogeneous thermal Comptonization problem. Since for $\lambda_{\mathrm{p}} \ll \lambda$ the wave temperature is downweighted by a factor proportional to $\lambda_{\mathrm{p}}^2/\lambda^2$, we expect the wave temperature to be negligible deep inside the (scattering) photosphere. At the photosphere, on the other hand, Comptonization is negligible because even though the wave temperature may be substantial, photons scatter too few times before escaping to change energy significantly. The contribution to bulk Comptonization is therefore dominated by an optically thick region near the photosphere.

## 2.7   Summary

Bulk Comptonization energy exchange is due to ordinary work done by radiation pressure to compress the gas as well as radiation viscous dissipation. These effects are due to terms that are first and second order in the velocity field, respectively. In general these effects are intertwined non-trivially.

According to the Helmholtz theorem, we can decompose a velocity field into a divergenceless component and curl-free (compressible) component. For the divergenceless component, bulk Comptonization is due to radiation viscous dissipation alone and can be treated as thermal Comptonization with an equivalent "wave" temperature. In terms of the viscous shear stress tensor and the velocity shear tensor the wave temperature is given by equation (2.30):

$$k_{\mathrm{B}} T_{\mathrm{w}} = \frac{-2\lambda_{\mathrm{p}} m_{\mathrm{e}} c}{E} \left( P_{\mathrm{vis,shear}}^{ij} \left( \partial_i v_j + \partial_j v_i \right) \right).$$

If we decompose the (divergenceless part of the) velocity field into sinusoidal modes with wave vectors $\mathbf{k}$, the viscous shear stress tensor is given by equation (2.31),

$$P_{\mathrm{vis,shear}}^{ij} = -\frac{4\lambda_{\mathrm{p}} E}{3c} \sum_{\mathbf{k}} \tau_k^2 f(k) \left( \partial_i v_{j,\mathbf{k}} + \partial_j v_{i,\mathbf{k}} \right),$$

where $\tau_k = \lambda/2\pi\lambda_p$ is the optical depth over $2\pi$ of a mode with wavenumber $k$. The function $f(k)$ is a weighting function given by equation (2.32) which goes to unity for optically thin modes and downweights optically thick modes by a factor $2/9\tau_k^2$. The expression for the wave temperature in terms of the velocity field is therefore given by

equation (2.35):

$$k_{\mathrm{B}}T_{\mathrm{w}} = \frac{\lambda_{\mathrm{p}}^2 m_{\mathrm{e}}}{6} \left(\partial_i v_j + \partial_j v_i\right) \sum_{\mathbf{k}} \tau_k^2 f(k) \left(\partial_i v_{j,\mathbf{k}} + \partial_j v_{i,\mathbf{k}}\right).$$

For statistically homogeneous turbulence we can take the spatial average of the wave temperature, which gives equation (2.28):

$$k_{\mathrm{B}}T_{\mathrm{w}} = \sum_{\mathbf{k}} \frac{1}{3} m_{\mathrm{e}} \left\langle v_{\mathbf{k}}^2 \right\rangle f(k). \tag{2.120}$$

For statistically homogeneous turbulence, therefore, the wave temperature is simply a weighted sum over the power present at each scale in the turbulent cascade. Scales with wavelengths that are short relative to the photon mean free path contribute fully to the wave temperature, while scales with wavelengths that are long relative to the photon mean free path are significantly downweighted and contribute negligibly.

The fact that the wave temperature downweights modes with wavelengths longer than the photon mean free path is physically intuitive because for these modes electron velocity differences between subsequent photon scatterings are significantly smaller. To confirm our physical intuition, we also define a heuristic wave temperature by equation (2.38):

$$\frac{3}{2} k_{\mathrm{B}}T_{\mathrm{w,heur}} = \frac{1}{4} m_{\mathrm{e}} \left\langle (\Delta \mathbf{v})^2 \right\rangle.$$

Here, $\left\langle (\Delta \mathbf{v})^2 \right\rangle$ is the average square velocity difference between subsequent photon scatterings. We find that $T_{\mathrm{w,heur}}$ is also given by equation (2.120) but with a slightly different weighting function $f_{\mathrm{heur}}(k)$ given by equation (2.39). The function $f_{\mathrm{heur}}(k)$ goes to unity for optically thin modes and downweights optically thick modes by a factor $1/3\tau_k^2$. Both $f(k)$ and $f_{\mathrm{heur}}(k)$ are plotted in Figure 2.1. The function $f_{\mathrm{heur}}(k)$ well approximates $f(k)$, which confirms that the wave temperature can be intuitively understood in terms of the electron velocity differences between subsequent photon scatterings.

For the curl-free (compressible) component of the velocity field, bulk Comptonization is due to both radiation viscous dissipation and ordinary work done by radiation pressure. The part due to radiation viscous dissipation can be understood using the physical intuition we developed from studying the divergenceless component. If the turbulence is optically thin, i.e. if all wavelengths are significantly shorter than the photon mean free path, then the full power contributes to viscous dissipation. In this limit work done by radiation pressure is negligible, so bulk Comptonization can be treated as thermal Comptonization with $(3/2)k_B T_w = (1/2)m_e \langle v^2 \rangle$.

If the turbulence is optically thick then viscous dissipation is suppressed since electron velocity differences between subsequent photon scatterings are small. In this limit, therefore, the first order effect is dominant, and the effect on photon spectra is analogous to the effect of the work done by radiation pressure on the gas. Just as the gas gains or loses internal energy depending on whether the gas is either compressing or expanding, respectively, so do the photons. Whether the gas is compressing or expanding depends on whether the sign of $-\nabla \cdot \mathbf{v}$ is positive or negative, respectively. Photons upscatter when $-\nabla \cdot \mathbf{v}$ is positive and downscatter when it is negative. The effect of this process on the emergent spectrum, however, depends on how effectively photons are able to escape from such regions to the observer.

We expect that radiation viscous dissipation will be dominant over work done by radiation pressure in determining the emergent spectrum in accretion disc atmospheres. Since the latter effect can give rise to either photon upscattering or downscattering depending on the sign of $-\nabla \cdot \mathbf{v}$, we expect it to be negligible on average for statistically homogeneous turbulence. In addition, it is in principle most significant in optically thick regions, but in these regions turbulence is generally dominated by divergenceless (in-compressible motions). Furthermore, such regions are deeper inside the photosphere and therefore have less impact on the emergent spectrum.

To calculate the wave temperature, equations (2.28) and (2.38) should be applied to a local region of an accretion disc atmosphere in which the turbulence is statistically homogeneous. The turbulence is not statistically homogeneous over the entire atmosphere since the vertical structure is spatially stratified. Since the wave temperature increases as the photon mean free path increases, we expect the wave temperature to be negligible deep inside the photosphere and increase significantly near it. We therefore expect bulk Comptonization to be dominated by a region just inside the photosphere.

In order for radiation MHD simulations to properly account for energy exchange due to turbulent Comptonization so that post-processing Monte Carlo simulations of photon spectra are self-consistent, they must include energy terms second order in velocity and use a moment closure scheme that correctly captures contributions to the radiation stress tensor that are first order in velocity. The appropriate energy equation source terms in lab and fluid frame variables are given by equations (2.12) and (2.20), respectively. Flux-limited diffusion and the M1 closure scheme are insufficient because they neglect the lowest order contribution to the radiation stress tensor.

Modeling turbulent Comptonization ultimately requires detailed analysis of radiation MHD simulations and post-processing Monte Carlo simulations. By exploring how photon spectra produced by turbulent Comptonization depend on the properties of the turbulence itself, we have laid the groundwork necessary to make sure this effect is both captured and correctly interpreted in these simulations.

# Chapter 3

# The contribution of bulk Comptonization to the soft X-ray excess in AGN

## 3.1 Introduction

The soft X-ray excess in AGN spectra is the component below 1keV that lies on top of the extrapolation of the best fitting 2-10keV power law (Singh et al., 1985; Arnaud et al., 1985; Vasudevan et al., 2014). The dependence of effective temperature on mass and accretion rate in optically thick accretion disc models (Shakura & Sunyaev 1973, hereafter SS73) is $T_{\text{eff}} \sim (\dot{m}/M)^{1/4}$, where $\dot{m} = \dot{M}/\dot{M}_{\text{Edd}}$. We therefore expect intrinsic disc emission to contribute to the soft excess most in narrow-line Seyfert Is (NLS1), which are comparatively low mass ($\sim 10^6 M_\odot$), near-Eddington sources. In the most luminous regions of NLS1 discs the temperature is greater than the hydrogen ionization energy, so electron scattering is the dominant opacity. The color temperature is therefore greater than the effective temperature, which augments the expected contribution to the soft excess in these sources. While the soft excess is particularly prominent in NLS1s, the expected disc contribution is insufficient to account for it (Done et al. 2012, hereafter D12). In broad-line Seyferts, which are lower Eddington ratio sources, the intrinsic disc emission does not extend to high enough energies to contribute at all, and so in these sources the entire soft excess must originate elsewhere.

One class of models for the soft excess invokes warm Comptonization. In this picture, a warm ($k_B T_e \sim 0.2$ keV) medium with moderate optical depth upscatters photons from a cool, optically thick disc. Magdziarz et al. (1998), for example, fit the soft excess of the broad-line Seyfert 1 NGC 5548 with $k_B T_e = 0.3$keV, $\tau = 30$. In this case, they pictured the medium as a transition region between the accretion disc and an inner hot, geometrically thick flow. In other studies the medium is a warm layer above the inner regions of the disc. For example, Janiuk et al. (2001) fit the soft excess of the quasar PG 1211+143 with $k_B T_e = 0.4$keV, $\tau = 10$. Dewangan et al. (2007) fit two NLS1s, Ark 564 and Mrk 1044, with $k_B T_e = 0.18$keV, $\tau = 45$, and $k_B T_e = 0.14$keV, $\tau = 45$, respectively. Jin et al. (2009) fit the super-Eddington ($L/L_{\rm Edd} = 2.7$) NLS1 RXJ0136.9-3510 with $k_B T_e = 0.28$keV, $\tau = 12$. Mehdipour et al. (2011) fit the broad-line Seyfert 1 Mrk 509 with $k_B T_e = 0.2$keV, $\tau = 17$. More recently, D12 constructed the XSPEC model OPTXAGNF for the soft excess, which uses the disc spectrum at the outer coronal radius as the seed photon source and, for the purpose of energy conservation, models the warm medium as part of the disc atmosphere. D12 fit the super-Eddington ($L/L_{\rm Edd} = 2.4$) NLS1 REJ1034+396 with $k_B T_e = 0.23$keV, $\tau = 11$. Since then, this model has been applied to several sources, such as the NLS1 II Zw 177 (Pal et al., 2016), for which they found $k_B T_e \sim 0.2$keV, $\tau \sim 20$.

Warm Comptonization models fit the spectra well, but the minimal variation of the fitted electron temperature with black hole mass and accretion rate (e.g. Gierlinski & Done 2004) motivated alternative models based on discrete atomic features. In reflec-tion models, photons from the hot ($\sim 100$ keV) corona are reflected and relativistically blurred by the inner regions of the accretion disc (e.g. Crummy et al. 2006; Ross & Fabian 2005). In ionized absorption models, high velocity winds originating from the accretion disc absorb and reemit photons from the hot corona (Gierlinski & Done, 2004). While these models naturally predict the minimal variation in the soft excess temperature, they typically require extreme parameters to sufficiently smear the discrete atomic features on which they are based. Reflection models, for example, require near maximal spin black holes (e.g. Crummy et al. 2006), and the original absorption models require unrealistically large wind velocities (Schurch & Done, 2007). More complex absorption models circumvent this difficulty, but they lack predictive power (e.g. Middleton et al. 2009). Other proposed explanations for the soft excess include magnetic reconnection (Zhong & Wang, 2013) and Comptonization by shock-heated electrons (Fukumura et al., 2016). Because warm Comptonization, reflection, and absorption all fit the spectra adequately (e.g. Middleton et al. 2009), solving this problem requires variability and multiwavelength studies (e.g. Mehdipour et al. 2011; Vasudevan et al. 2014).

Because optically thick disc models predict that disc emission associated with NLS1s already extends into the soft X-rays, in these sources warm Comptonization could be due to modifications to the vertical structure that occur in this regime. For example, warm Comptonization may be due to turbulence in the disc (Socrates, Davis & Blaes 2004; Chapter 2), if bulk electron velocities exceed thermal electron velocities. For the alpha disk model (SS73),

$$\frac{\langle v_{\text{turb}}^2 \rangle}{\langle v_{\text{th}}^2 \rangle} \sim \alpha \left( \frac{m_e}{m_{\text{p}}} \right) \left( \frac{P_{\text{rad}}}{P_{\text{gas}}} \right), \tag{3.1}$$

so we expect turbulent Comptonization to be important in the extremely radiation pressure dominated regime. Since the ratio of radiation to gas pressure increases with mass and accretion rate, turbulent Comptonization should be most relevant for supermassive black holes accreting at near-Eddington rates, such as NLS1s. In this regime, therefore, turbulent Comptonization could provide a physical basis for the construction of warm Comptonization models. By connecting the observed temperature and optical depth to the disc vertical structure, this could help solve the problem of the soft excess and also shed light on the properties of MHD turbulence. In broad-line Seyferts, which have lower Eddington ratios, the ratio of radiation to gas pressure is too small for turbulent Comptonization to be significant, so if warm Comptonization is present it must originate elsewhere. In these sources, it is unlikely that warm Comptonization could be due to modifications to the intrinsic disc atmosphere physics, because the thermal spectrum falls off at energies significantly below the soft X-rays.

In Chapter 2 we outlined the fundamental physical processes underlying bulk Comptonization by turbulence in accretion disc atmospheres. In this chapter we model the effect of bulk Comptonization on disc spectra using data from radiation MHD simulations (Hirose, Krolik & Blaes, 2009), including both turbulent Comptonization and Comptonization by the background shear. We parametrize this effect by temperature and optical depth in order to make contact with observations fit by other warm Comptonization models. In particular, we compare our results to the temperature and optical depth fit to REJ1034+396 (D12), a super-Eddington NLS1 with an unusually large soft excess. The structure of this chapter is as follows. In section 3.2 we describe our model in detail. In section 3.3 we describe our results, and in section 3.4 we discuss them. Finally, we summarize our findings in section 3.5.

| Simulation | $M/M_\odot$ | $L/L_{\mathrm{Edd}}$ | $r$ |
|------------|-------------|----------------------|-----|
| 110304a    | 6.62        | 1.68                 | 30  |
| OPALR20    | $5 \times 10^8$ | 0.03             | 40  |

Table 3.1: Shearing box simulation parameters

## 3.2 Modeling bulk Comptonization

### 3.2.1 Overview

In order to facilitate comparisons with warm thermal Comptonization models of the soft X-ray excess, we seek to characterize the contribution of bulk Comptonization with a temperature and an optical depth. To do this, we use data from radiation MHD shearing box simulations to compute spectra both including and excluding bulk velocities. Since our simulation data is limited, we use a scheme to scale data from a simulation run with a particular radius, mass, and accretion rate to different sets of these parameters. We describe this scheme in section 3.2.2. In this work we use data from simulation 110304a, which is similar to simulations 1112a and 1226b (Hirose, Krolik & Blaes, 2009), but has a lower surface density, $\Sigma = 2.5 \times 10^4 \mathrm{g\ cm^{-2}}$, which results in a higher radiation to gas pressure ratio. The parameters of interest for 110304a are given in Table 3.1. We note that all numerical radii in this chapter are in units of the gravitational radius $R_{\mathrm{g}} = GM/c^2$ of the black hole.

We calculate the spectrum at a given timestep using Monte Carlo post-processing simulations. For this work, we chose the 140 orbit timestep at random. The details of our Monte Carlo implementation of bulk Compton scattering are in Appendix C. To isolate the effect of the turbulence alone, we also calculate spectra without the background shear. The background shear is modeled by simply including the background Keplerian velocity field, although this is not ideal (see section 5.4.1). To model an entire accretion disc we calculate spectra at multiple radii. We discuss our choice of radii in section 3.2.3. The flux obtained at a particular radius corresponds to an Eddington ratio. If our scaling scheme were perfect, the corresponding Eddington ratios at the other radii would be the same by construction. We correct for minor discrepancies by normalizing the other spectra so that their corresponding Eddington ratios are the same.

We transport the spectra computed with bulk velocities at multiple radii to infinity and superpose the results to obtain the final, observed spectrum. We choose a viewing angle of 60°. At this angle the gravitational redshift approximately cancels the Doppler blueshift (D12, Zhang et al. 1997), which allows us to use a Newtonian transport code. We chose this method because it is easy to include the propagation of error bars, but we verified that our results are unchanged when a fully relativistic Kerr spacetime transport code (Agol, 1997) is used instead.

The spectra computed without bulk velocities are used as seed photon sources for a warm Comptonizing medium characterized solely by a uniform temperature and optical depth. We implement this by solving the Kompaneets equation at each radius. We then transport the resultant spectra to infinity to obtain the observed spectrum. We fit the observed spectrum Comptonized by the warm medium to the observed spectrum computed with bulk velocities by adjusting the temperature and optical depth. We explore the effect of varying the outer radius, $r_{cor}$, of the warm Comptonizing medium on the goodness of fit parameter, $\chi^2/\nu$, and select the radius for which this parameter is minimized.

To provide insight into the physics of bulk Comptonization, we also perform spectral calculations in which the simulation data are truncated at the effective photosphere and the emissivity is zero everywhere except in the cells at the base. Since we expect bulk Comptonization to be dominated by the contribution from photons emitted at the effective photosphere, we expect the resulting temperature and optical depth to be nearly unchanged. We discuss this point more in section 3.4.2.

### 3.2.2    Scalings for radiation MHD shearing box simulation data

In this section we derive a scheme to scale data from a radiation MHD simulation run with a particular radius, mass, and accretion rate to a different set of these parameters. We first observe that the construction of an appropriate scheme is made possible by the fact that the density, temperature, and velocity profiles show considerable self-similarity across a wide range of simulation parameters. For example, in Figures 3.1 and 3.3 we compare the density and bulk velocity profiles from the 140 orbits timestep of 110304a, which is the basis of this work, with those from a snapshot of OPALR20 (Jiang et al., 2016), a simulation run in an entirely different regime (Table 3.1). The bulk temperature is defined by $(3/2)k_{\mathrm{B}}T_{\mathrm{bulk}} = (1/2)m_e v^2$. We note that $T_{\mathrm{tot}}$ in Chapter 2 is just the average value of $T_{\mathrm{bulk}}$ over some region. Subscript "c" denotes midplane values. The variable $z$ is the distance from the midplane and the scale height $h$ is the value of $z$ for which $\rho/\rho_{\mathrm{c}} = 1/e$. The profiles nearly coincide, and even the discrepancy between the density profiles at large $z/h$ is likely just due to a temporary fluctuation at 140 orbits. At 180 orbits, for example, there is no discrepancy (Figure 3.2). This self-similarity is perhaps an even more robust phenomenon than the difference in simulation parameters alone would indicate since the inclusion of the iron opacity bump in OPALR20 is a non-trivial effect. In particular, the thermal stability of OPALR20 depends on the inclusion of this effect (Jiang et al., 2016), whereas it is now believed that the thermal stability in 110304a is a result of the narrow box size in the radial direction and is therefore artificial (Jiang, Stone & Davis, 2013). Despite this caveat as well as the fact that the mass parameter for OPALR20 is closer to our regime of interest, we chose 110304a for this work because the photospheres are better resolved, a decisive advantage for the purpose of computing spectra.

Because of self-similarity, we primarily need to scale the midplane values for the profiles of interest and the scale height. Analogous to the derivation of the standard $\alpha$-disc scalings in the radiation pressure dominated regime (SS73), we derive scalings in terms of the shearing box surface density $\Sigma$, the vertical epicyclic frequency $\Omega_z$, and the shear $\partial_x v_y$. The integrated hydrostatic equilibrium equation for a density profile with scale height $h$ and midplane radiation pressure $P_c$ is

$$P_c = \frac{1}{4}\Omega_z^2\Sigma h.$$ (3.2)

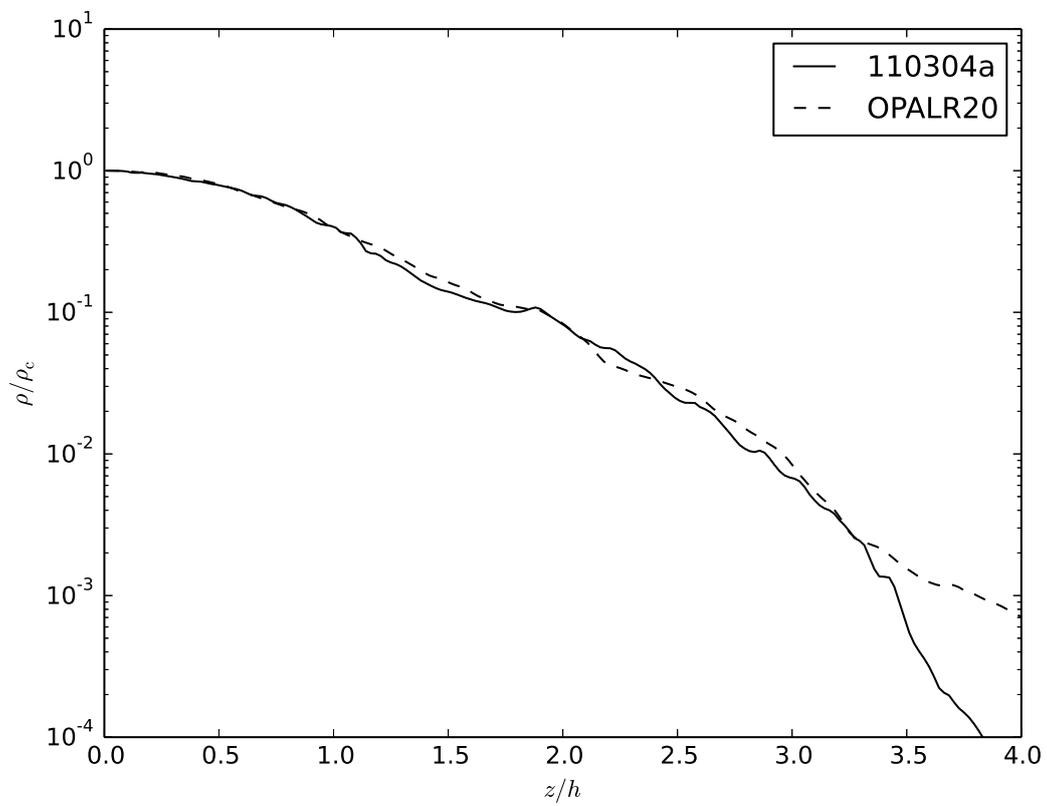

Figure 3.1: Normalized shearing box density profiles. The timestep for 110304a is 140 orbits.

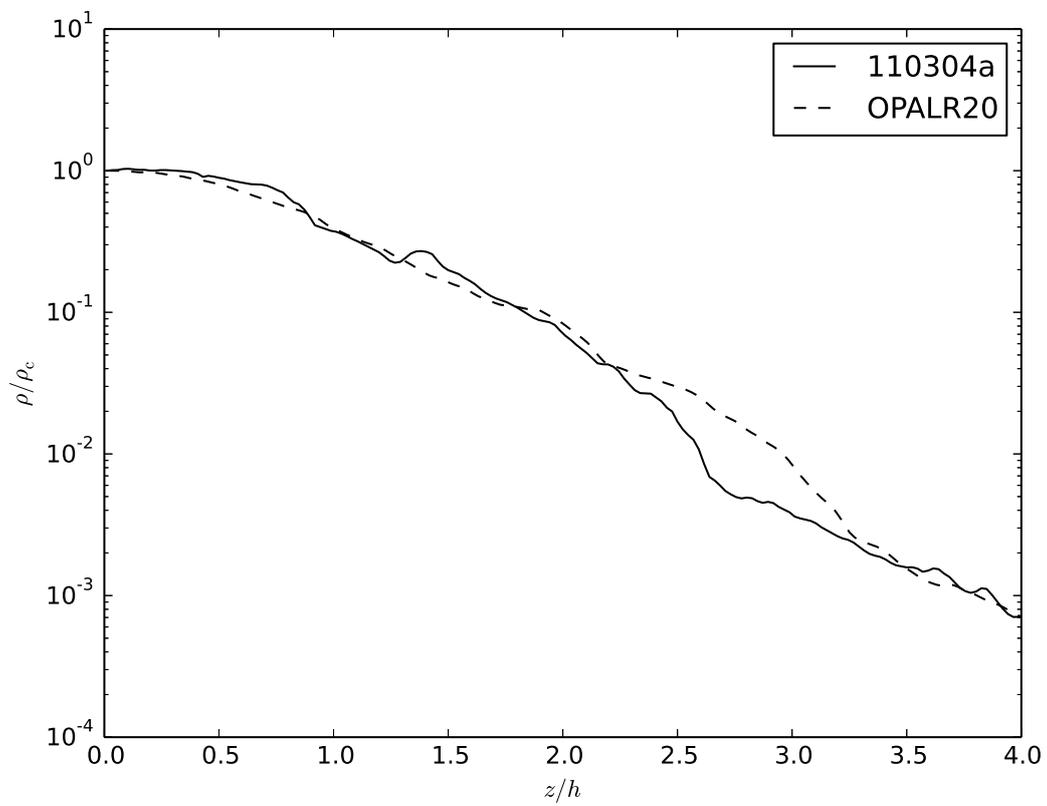

Figure 3.2: Normalized shearing box density profiles. The timestep for 110304a is 140 orbits.

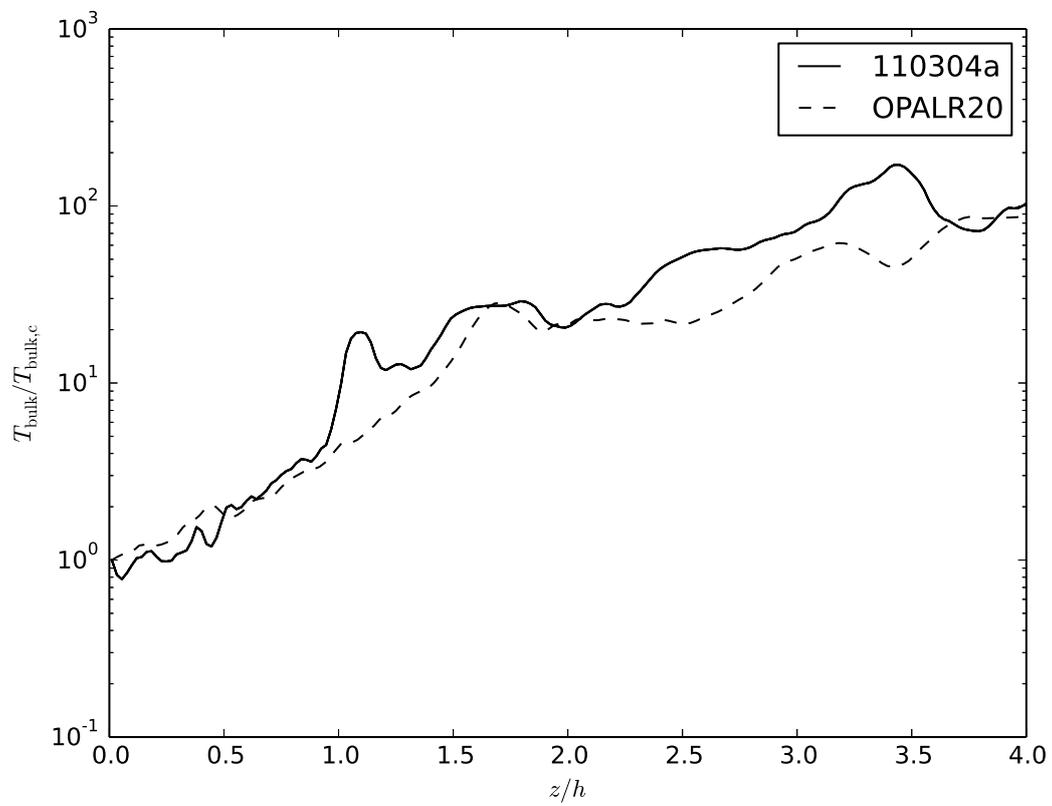

Figure 3.3: Normalized shearing box bulk temperature profiles. The timestep for 110304a is 140 orbits.

The thermal equilibrium equation, given the radiation flux $F$ and the midplane turbulent stress $\tau_c$ is

$$F = (\partial_x v_y)\tau_c h.$$ (3.3)

The stress prescription is

$$\bar{\tau} = \alpha\bar{P},$$ (3.4)

which for a profile that decays with scale height $h$ is equivalent to

$$\tau_c = \alpha P_c.$$ (3.5)

The radiative diffusion equation with the opacity given by $\kappa$ is

$$F = \frac{2cP_c}{\kappa\Sigma}.$$ (3.6)

Equations (3.3), (3.5), and (3.6) give the scale height scaling:

$$\left(\frac{h}{h_0}\right) = \left(\frac{\alpha}{\alpha_0}\right)^{-1} \left(\frac{\kappa}{\kappa_0}\right)^{-1} \left(\frac{\partial_x v_y}{\partial_x v_{y,0}}\right)^{-1} \left(\frac{\Sigma}{\Sigma_0}\right)^{-1}. \tag{3.7}$$

Since we intend to scale to the lower mass ($\sim 10^6 M_\odot$), high Eddington ratio regime, the opacity remains dominated by electron scattering so we set $\kappa/\kappa_0 = 1$. Equations (3.2) and (3.7) give the midplane pressure scaling:

$$\left(\frac{P_c}{P_{c,0}}\right) = \left(\frac{\alpha}{\alpha_0}\right)^{-1} \left(\frac{\kappa}{\kappa_0}\right)^{-1} \left(\frac{\Omega_z}{\Omega_{z,0}}\right)^2 \left(\frac{\partial_x v_y}{\partial_x v_{y,0}}\right)^{-1}. \tag{3.8}$$

Below we will also need the flux scaling:

$$\left(\frac{F}{F_0}\right) = \left(\frac{\alpha}{\alpha_0}\right)^{-1} \left(\frac{\kappa}{\kappa_0}\right)^{-2} \left(\frac{\Omega_z}{\Omega_{z,0}}\right)^2 \left(\frac{\partial_x v_y}{\partial_x v_{y,0}}\right)^{-1} \left(\frac{\Sigma}{\Sigma_0}\right)^{-1}. \tag{3.9}$$

For the purpose of calculating spectra, the profiles of interest are the density, the gas temperature, the turbulent velocity, and the shear velocity. The midplane density is trivially given by

$$\left(\frac{\rho_c}{\rho_{c,0}}\right) = \left(\frac{\Sigma}{\Sigma_0}\right) \left(\frac{h}{h_0}\right)^{-1}. \tag{3.10}$$

Since the gas temperature is coupled to the radiation temperature, the scaling for the midplane gas temperature follows directly from equation (3.8). To find the turbulent velocity scaling, we define $\beta$ as follows:

$$\frac{1}{2} \left\langle \rho v^2 \right\rangle = \beta \tau. \tag{3.11}$$

| Variable | Ratio |
|---|---|
| $h_{\text{scaled}}/h$ | 0.9 |
| $T_{\text{g,c,scaled}}/T_{\text{g,c}}$ | 1.0 |
| $T_{\text{bulk,c,scaled}}/T_{\text{bulk,c}}$ | 0.9 |

Table 3.2: Ratios of variables predicted using 110304a data to variables measured in OPALR20, taking into account $\alpha/\alpha_0 = 2.38$.

The midplane turbulent velocity scaling is then

$$\frac{\langle v_c^2 \rangle}{\langle v_{c,0}^2 \rangle} = \left(\frac{\alpha}{\alpha_0}\right)^{-1} \left(\frac{\beta}{\beta_0}\right) \left(\frac{\kappa}{\kappa_0}\right)^{-2} \left(\frac{\Omega_z}{\Omega_{z,0}}\right)^2$$
$$\left(\frac{\partial_x v_y}{\partial_x v_{y,0}}\right)^{-2} \left(\frac{\Sigma}{\Sigma_0}\right)^{-2}. \tag{3.12}$$

To test these scalings, we scale the midplane values and the scale height from 110304a to the simulation parameters of OPALR20 and then divide by the actual midplane values and the scale height in OPALR20 (Table 3.2). We assume $\beta/\beta_0 = 1$. Taking into account the empirical turbulent stress ratio $\alpha/\alpha_0 = 2.38$, we see that the resulting ratios are all near unity, and that our scalings therefore capture the essential physics in the shearing box. This is even more remarkable given that our scalings only take into account Thomson scattering and radiation diffusion, while the iron opacity bump and vertical advection are non-trivial effects in OPALR20.

The density and turbulent velocity profiles follow directly from equations (3.7), (3.10), and (3.12), but the pressure profile, which determines the gas temperature profile, is non-trivial. The density profile is

$$\rho(z) = \left(\frac{\Sigma}{\Sigma_0}\right) \left(\frac{h}{h_0}\right)^{-1} \rho_0 \left(h_0 z/h\right). \tag{3.13}$$

The turbulent velocity profile is

$$v(z) = \left(\frac{\alpha}{\alpha_0}\right)^{-1/2} \left(\frac{\beta}{\beta_0}\right)^{1/2} \left(\frac{\kappa}{\kappa_0}\right)^{-1} \left(\frac{\Omega_z}{\Omega_{z,0}}\right) \left(\frac{\partial_x v_y}{\partial_x v_{y,0}}\right)^{-1} \left(\frac{\Sigma}{\Sigma_0}\right)^{-1} v_0(h_0 z/h). \quad (3.14)$$

But scaling the radiation pressure profile by adjusting only the scale height and the overall normalization is too simplistic a scheme for the purpose of calculating spectra because near the photosphere the flux begins to free stream and is no longer carried by radiative diffusion. In such a scheme, therefore, the profile will be least accurate in the region that it is most important. This difficulty can be addressed by imposing a boundary condition at the photosphere. Inside the photosphere,

$$P_{\mathrm{ph,in}} \sim T_{\mathrm{ph,in}}^4 \sim (f_{\mathrm{cor}} T_{\mathrm{ph,out}})^4 \sim f_{\mathrm{cor}}^4 F, \quad (3.15)$$

where $f_{\mathrm{cor}}$ is determined by the physics at the photosphere. For example, if the opacity is dominated by coherent scattering and the boundary condition is imposed at the effective photosphere, then $f_{\mathrm{cor}} = f_{\mathrm{col}}$, the color correction. The scaling for $P_{\mathrm{ph,in}}$ is then

$$P_{\mathrm{ph,in}} = \left(\frac{f_{\mathrm{cor}}}{f_{\mathrm{cor,0}}}\right)^4 \left(\frac{F}{F_0}\right) P_{\mathrm{ph,in,0}}. \quad (3.16)$$

The simplest scheme that imposes this boundary condition is given by

$$P(z) = P_{\mathrm{ph,in}} + \left(\frac{P_c}{P_{c,0}}\right) (P_0(h_0 z/h) - P_0(h_0 z_{\mathrm{ph}}/h)), \quad (3.17)$$

which we formally derive in Appendix D1. We recall that $P_c/P_{c,0}$ is given by equation (3.8). Since the pressure at the photosphere is always orders of magnitude smaller than the midplane pressure, we find that

$$P(0) \approx \left( \frac{P_{\mathrm{c}}}{P_{\mathrm{c},0}} \right) P_0 \left( 0 \right), \qquad (3.18)$$

so that this scheme is self-consistent. Inside the photosphere the gas temperature is coupled to the radiation temperature, so in this region the gas temperature profile is then given by

$$T_{\mathrm{g,in}}^4(z) = T_{\mathrm{g,ph}}^4 + \left( \frac{P_{\mathrm{c}}}{P_{\mathrm{c},0}} \right) \left( T_{\mathrm{g},0}^4 \left( h_0 z / h \right) - T_{\mathrm{g},0}^4 \left( h_0 z_{\mathrm{ph}} / h \right) \right), \qquad (3.19)$$

where

$$T_{\mathrm{g,ph}}^4 = \left( \frac{P_{\mathrm{ph,in}}}{P_{\mathrm{ph,in},0}} \right) T_{\mathrm{g,ph},0}^4. \qquad (3.20)$$

In order that the gas temperature profile be continuous, the scaling outside the photosphere is given by

$$T_{\mathrm{g,out}}^4 \left( z \right) = \left( \frac{P_{\mathrm{ph,in}}}{P_{\mathrm{ph,in},0}} \right) T_{\mathrm{g},0}^4 \left( z_{\mathrm{ph},0} + h_0(z - z_{\mathrm{ph}})/h \right). \qquad (3.21)$$

Finally, we also need the scaling for the shear velocity profile, which is trivially given by

$$v_{\mathrm{s}} \left( x \right) = \left( \frac{\partial_x v_y}{\partial_x v_{y,0}} \right) \left( \frac{h}{h_0} \right) v_{\mathrm{s},0} \left( h_0 x / h \right). \qquad (3.22)$$

We define $z_{\mathrm{ph}}$ to be where the scattering optical depth $\tau_{\mathrm{s}} = 1$ (where subscript "s" denotes scattering) and set $f_{\mathrm{cor}}/f_{\mathrm{cor},0} = 1$. Near the photosphere magnetic pressure begins to play a major role in hydrostatic equilibrium (e.g. Blaes, Hirose & Krolik 2007), and near the effective photosphere the gas temperature begins to diverge from the radiation temperature, so we acknowledge that the assumptions underlying our scheme do not reflect the detailed physics in this region. But since our goal is only to calculate spectra, for optical depths $\tau_s \ll 1$ the accuracy of this scheme is not important. We can assess the validity of this scheme in the region $\tau_s \approx 1$ by comparing the flux from spectral calculations with the intended flux given by equation (3.9), or, equivalently, by comparing the corresponding Eddington ratios. In section 3.3.1, we make this comparison for each set of scaling parameters we use and find that they generally agree to within 10%. More importantly, we find that normalizing the spectra at different radii so that their corresponding Eddington ratios match has a negligible impact on the observed spectrum when contrasted with the discrepancies between spectral calculations with and without bulk velocities. In other words, because the potential error is significantly less than the effect we are measuring, our scaling scheme is adequate.

These are the appropriate equations for scaling data to a different set of fundamental shearing box simulation parameters, in particular $\Omega_z$, $\partial_x v_y$, and $\Sigma$. If we substitute in equation (3.9) for $\Sigma$, we can alternatively regard $F$ as a fundamental parameter instead of $\Sigma$. Shearing box scalings in terms of $F$ are given in Appendix D2. This substitution is useful in order to scale to a different set of fundamental accretion disc parameters, since it is straightforward to express $F$ in terms of accretion disc radius, mass, and accretion rate. The scalings for $\Omega_z$, $\partial_x v_y$, and $F$ for both Newtonian and Kerr discs, allowing for a non-zero stress inner boundary condition, are given in Appendix D3. The final scalings for $\rho$, $T_g$, $v$, and $v_s$ in terms of fundamental accretion disc parameters are given in Appendix D4. We only use Kerr scalings for our spectral calculations, but the Newtonian scalings are potentially useful for the purpose of comparing with other works in which Newtonian parameters are used and also for developing physical intuition.

### 3.2.3   Dependence of turbulent Comptonization on radius

To characterize the contribution of turbulent Comptonization, we must model spectra at multiple radii. Our choice of radii is guided by the scaling of the ratio of bulk to thermal electron energies. We estimate this effect for a disc with no spin and a stress-free inner boundary condition with the Newtonian scalings in Appendix D4. The bulk velocity scaling is

$$\left\langle v_{\text{turb}}^2 \right\rangle \sim r^{-3} \left(1 - \sqrt{r_{\text{in}}/r}\right)^2.$$ (3.23)

The photosphere thermal velocity scaling is

$$\left\langle v_{\text{th,ph}}^2 \right\rangle \sim r^{-3/4} \left(1 - \sqrt{r_{\text{in}}/r}\right)^{1/4}.$$ (3.24)

The scaling for the ratio of bulk velocity to thermal velocity at the photosphere is

$$\frac{\left\langle v_{\text{turb}}^2 \right\rangle}{\left\langle v_{\text{th,ph}}^2 \right\rangle} \sim r^{-9/4} \left(1 - \sqrt{r_{\text{in}}/r}\right)^{7/4}.$$ (3.25)

We also calculate the scaling for the ratio of bulk to thermal velocity using the midplane thermal velocity scaling, which is

$$v_{\text{th,c}}^2 \sim r^{-3/8}.$$ (3.26)

The scaling for the ratio is

$$\frac{\left\langle v_{\text{turb}}^2 \right\rangle}{\left\langle v_{\text{th,c}}^2 \right\rangle} \sim r^{-21/8} \left(1 - \sqrt{r_{\text{in}}/r}\right)^2.$$ (3.27)

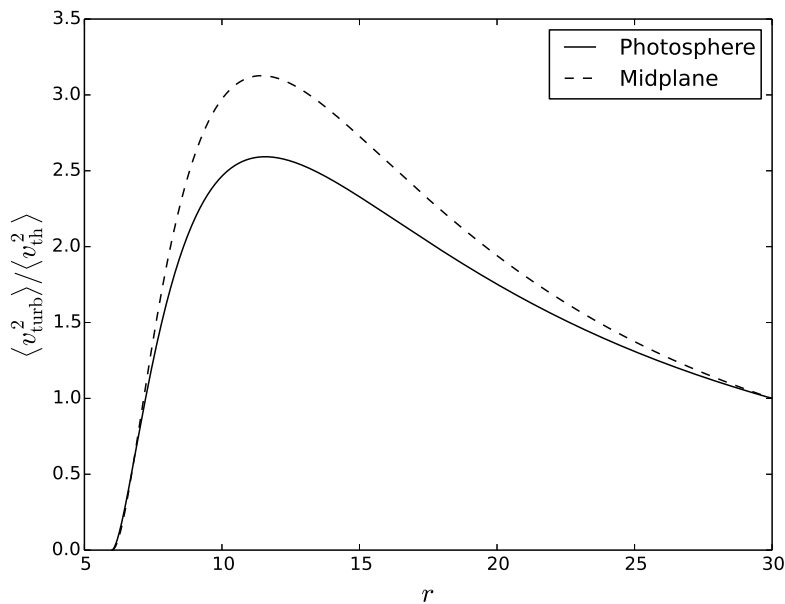

Figure 3.4: Scaling for the relative magnitude of the turbulent velocity for $r_{in} = 6$, normalized to $r = 30$.

We plot equations (3.25) and (3.27) in Figure 3.4, normalized to 30 gravitational radii. We expect that turbulent Comptonization will be most significant between 8 and 20 gravitational radii. We verify this assumption in section 3.3. For our model we choose to compute spectra at 30, 20, 14, 11, 10, 9.5, 9.0, 8.5, and 7.5 gravitational radii. We also run simulations for spin $a = 0.5$, for which $r_{in} = 4.2$. For these we compute spectra at 30, 20, 15, 12, 10, 8, 7, 6, 5.5, and 5 gravitational radii.

## 3.3  Results

We compute the contribution of bulk Comptonization to the soft X-ray excess and characterize our results with a temperature and optical depth. Our fiducial mass, $M = 2 \times 10^6 M_\odot$, and Eddington ratio, $L/L_{Edd} = 2.5$, were chosen to correspond to those of the NLS1 source REJ1034+396 in D12 (Table 3.8). Table 3.4 summarizes our main results.

| Set | Type | $M/M_\odot$ | $L/L_{\rm Edd}$ (target) | a | $\alpha/\alpha_0$ | $v_{\rm turb}$ | $v_{\rm shear}$ |
|---|---|---|---|---|---|---|---|
| a | Full | $2 \times 10^6$ | 2.5 | 0 | 1 | Y | Y |
| a2 | Truncated, emissivity at base | $2 \times 10^6$ | 2.5 | 0 | 1 | Y | Y |
| b | Full | $2 \times 10^6$ | 2.5 | 0 | 1 | Y | N |
| b2 | Truncated, emissivity at base | $2 \times 10^6$ | 2.5 | 0 | 1 | Y | N |
| c | Full | $2 \times 10^6$ | 2.5 | 0 | 2 | Y | Y |
| c2 | Truncated, emissivity at base | $2 \times 10^6$ | 2.5 | 0 | 2 | Y | Y |
| d | Full | $2 \times 10^6$ | 2.5 | 0.5 | 1 | Y | Y |
| d2 | Truncated, emissivity at base | $2 \times 10^6$ | 2.5 | 0.5 | 1 | Y | Y |
| e | Full | $2 \times 10^7$ | 2.5 | 0 | 1 | Y | Y |
| e2 | Truncated, emissivity at base | $2 \times 10^7$ | 2.5 | 0 | 1 | Y | Y |

Table 3.3: Simulation set independent variables

| Set | $M/M_\odot$ | $L/L_{\rm Edd}$ (target) | a | $\alpha/\alpha_0$ | $v_{\rm turb}$ | $v_{\rm shear}$ | $L/L_{\rm Edd}$ (observed) | $k_{\rm B}T_e$ (keV) | $\tau$ | $r_{\rm cor}$ | $y_p$ | $\chi^2/\nu$ |
|---|---|---|---|---|---|---|---|---|---|---|---|---|
| a | $2 \times 10^6$ | 2.5 | 0 | 1 | Y | Y | 2.5 | $0.14 \pm 0.0067$ | $15 \pm 1.4$ | 20 | 0.26 | 1 |
| b | $2 \times 10^6$ | 2.5 | 0 | 1 | Y | N | 2.5 | $0.18 \pm 0.056$ | $11 \pm 4.2$ | 14 | 0.14 | 1.7 |
| c | $2 \times 10^6$ | 2.5 | 0 | 2 | Y | Y | 2.3 | $0.17 \pm 0.012$ | $17 \pm 1.8$ | 20 | 0.38 | 2.3 |
| d | $2 \times 10^6$ | 2.5 | 0.5 | 1 | Y | Y | 2.3 | $0.21 \pm 0.011$ | $12 \pm 0.82$ | 20 | 0.22 | 1.9 |
| e | $2 \times 10^7$ | 2.5 | 0 | 1 | Y | Y | 2.1 | $0.081 \pm 0.0075$ | $24 \pm 4.1$ | 20 | 0.37 | 0.87 |

Table 3.4: Results for full atmosphere spectral calculations

| Set | $M/M_\odot$ | $L/L_{\rm Edd}$ (target) | a | $\alpha/\alpha_0$ | $v_{\rm turb}$ | $v_{\rm shear}$ | $k_{\rm B}T_e$ (keV) | $\tau$ | $r_{\rm cor}$ | $y_p$ | $\chi^2/\nu$ |
|---|---|---|---|---|---|---|---|---|---|---|---|
| a2 | $2 \times 10^6$ | 2.5 | 0 | 1 | Y | Y | $0.14 \pm 0.0065$ | $16 \pm 1.4$ | 30 | 0.26 | 0.67 |
| b2 | $2 \times 10^6$ | 2.5 | 0 | 1 | Y | N | $0.13 \pm 0.013$ | $12 \pm 2.5$ | 20 | 0.15 | 1.3 |
| c2 | $2 \times 10^6$ | 2.5 | 0 | 2 | Y | Y | $0.18 \pm 0.015$ | $14 \pm 1.4$ | 30 | 0.28 | 0.93 |
| d2 | $2 \times 10^6$ | 2.5 | 0.5 | 1 | Y | Y | $0.18 \pm 0.011$ | $14 \pm 1.2$ | 20 | 0.28 | 0.93 |
| e2 | $2 \times 10^7$ | 2.5 | 0 | 1 | Y | Y | $0.074 \pm 0.0040$ | $32 \pm 4.5$ | 20 | 0.57 | 0.52 |

Table 3.5: Results for truncated atmosphere spectral calculations with emissivity only at the base.

| Set | $k_{\rm B}T_e$ (keV) | $\tau$ | $r_{\rm cor}$ | $y_p$ | $\chi^2/\nu$ |
|---|---|---|---|---|---|
| a | 0.14 | 16 | 30 | 0.26 | 1.6 |
| b | 0.13 | 12 | 20 | 0.15 | 2.0 |
| c | 0.18 | 14 | 30 | 0.28 | 2.6 |
| d | 0.18 | 14 | 20 | 0.28 | 1.9 |
| e | 0.074 | 32 | 20 | 0.57 | 1.1 |

Table 3.6: Goodness of fit of parameters derived from truncated atmosphere spectral calculations to observed spectra calculated with the full atmosphere.

The original (unscaled) simulation parameters for 110304a are listed in Table 3.1. Each system is modeled by calculating spectra with and without the bulk velocities at the set of radii discussed in section 3.2.3. The target $L/L_{\mathrm{Edd}}$ is the Eddington ratio that would correspond to the observed flux at 30 gravitational radii if the scaling scheme were exact. The turbulent stress scaling is given by $\alpha/\alpha_0$. In all cases, $\Delta\epsilon = 0$ (Appendix D3), which imposes the stress-free inner boundary condition. The choices of whether or not to include turbulent and shear velocities in the spectral calculations with bulk velocities are indicated by $v_{\mathrm{turb}}$ and $v_{\mathrm{shear}}$, respectively. The Compton $y$ parameter is calculated from the fitted temperature and optical depth. To calculate $\chi^2/\nu$, we first correct for uncertainty in the overall normalization of the data point errors by normalizing them to the standard deviation calculated from the fit for set (a) (shown in Figure 3.5). In section 3.3.1, we discuss the results of each set. To provide physical insight into the physics of bulk Comptonization, we also perform spectral calculations in which the simulation data was truncated at the effective photosphere and the emissivity was set to zero everywhere except in the cells at the base. Table 3.3 summarizes these results, which we discuss in section 3.3.2. For clarity, in Table 3.3 we list the independent variables for all simulation sets.

We note that here we use the generic formula $N = \tau^2$ for the average number of scatterings $N$, whereas in Chapter 4 we use $N = 1.6\tau^2$ which applies specifically to a plane parallel geometry. Therefore, to obtain physical optical depths in this chapter one should divide the fitted optical depths by $\sqrt{1.6}$. Since the $y$ parameter, on the other hand, is always related to the average number of scatterings by

$$y_{\mathrm{p}} = \frac{4k_{\mathrm{B}}T_{\mathrm{e}}}{m_{\mathrm{e}}c^2}N, \tag{3.28}$$

its definition here is the same as in Chapter 4.

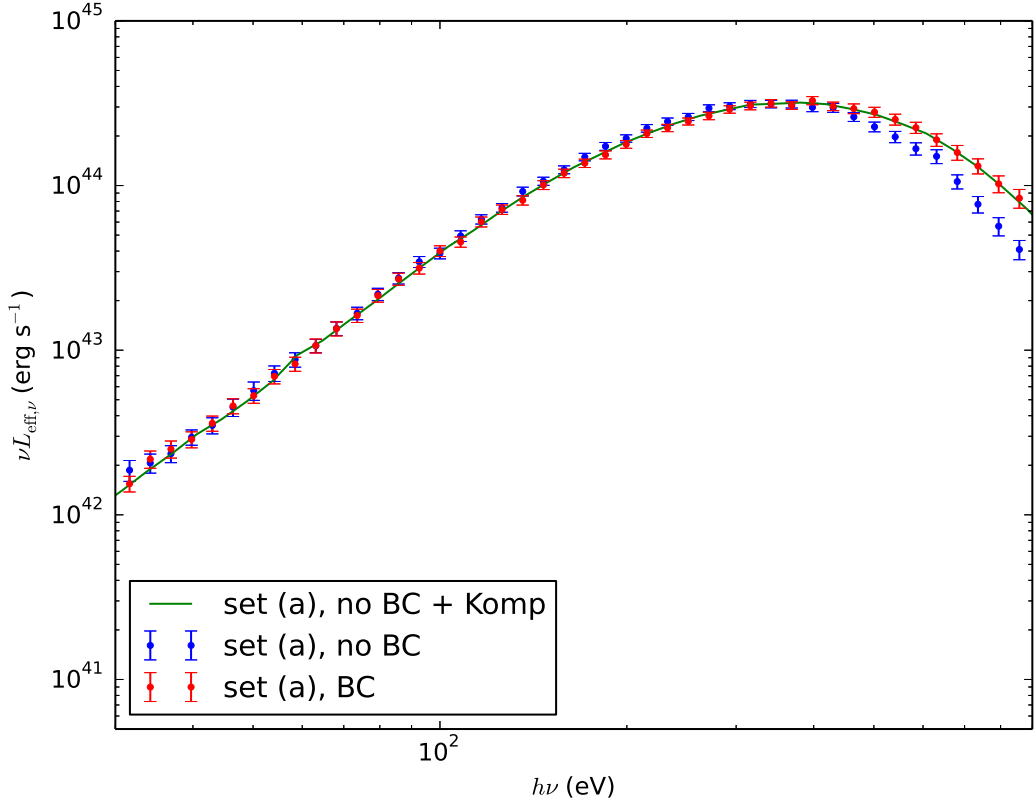

Figure 3.5: Observed disc spectra computed for set (a). BC (bulk Comptonization) means bulk velocities were included. Komp means the zero bulk Comptonization spectrum from each radius for $r \leq r_{\mathrm{cor}}$ was passed through a warm Comptonizing medium with the parameters given in Table 3.4.

### 3.3.1  Full spectral calculations

The observed spectrum for set (a) computed with and without the bulk velocities along with the Kompaneets fit are shown in Figure 3.5. We see that the fit is excellent, which means that bulk Comptonization here is well modeled by thermal Comptonization with a fitted temperature and optical depth. We note that the observed $L/L_{\mathrm{Edd}}$ matches the target $L/L_{\mathrm{Edd}}$, which confirms that our scaling scheme is self-consistent. The required flux normalizations given the flux at 30 gravitational radii are given in Table 3.7. They hardly deviate from unity, which provides another check for the self-consistency of our scalings. In Figure 3.6 we show local spectra at multiple radii for set (a). We see that the spectra passed through the warm Comptonizing medium fit the spectra calculated with bulk velocities for $9.5 \leq r \leq 20$, but overshoot them for $r = 7.5$ and $r = 30$. This confirms that bulk Comptonization is most significant in the region we expected it to be (section 3.2.3). Furthermore, this is consistent with the value we find for $r_{\mathrm{cor}}$, since we expect the best fit to be obtained when the Comptonizing medium is restricted to the region in which bulk Comptonization is most significant.

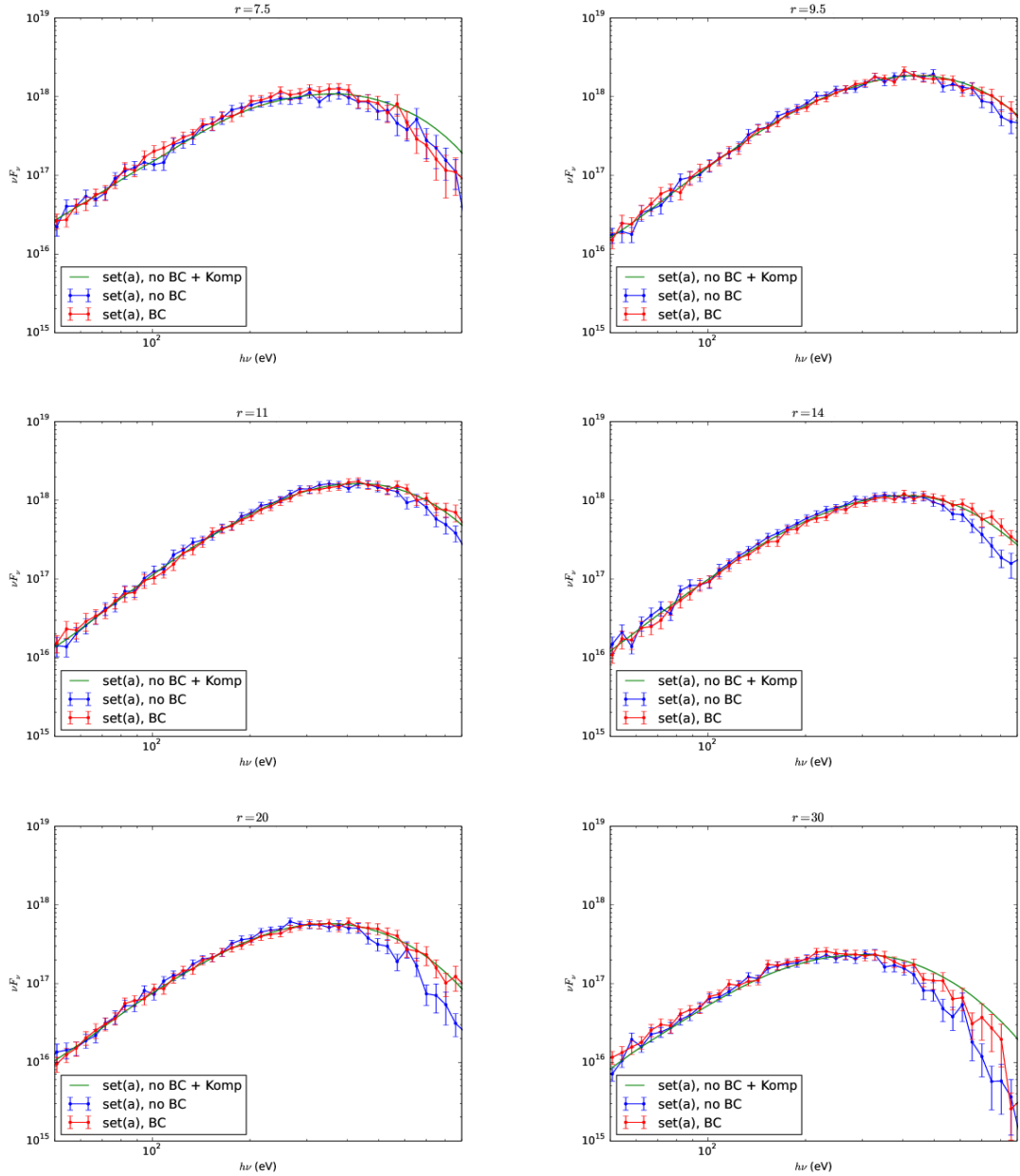

Figure 3.6: Disc spectra at select radii, labeled at the top of each plot, computed for set (a). BC (bulk Comptonization) means bulk velocities were included. Komp means the zero bulk Comptonization spectrum was passed through a warm Comptonizing medium with the parameters given in Table 3.4.

| $r$ | Flux norm (No BC) | Flux norm (BC) |
|-----|-------------------|----------------|
| 30  | 1                 | 1              |
| 20  | 1.04              | 1.10           |
| 14  | 1.04              | 1.15           |
| 11  | 0.99              | 1.06           |
| 10  | 0.95              | 0.96           |
| 9.5 | 0.92              | 0.94           |
| 9.0 | 0.91              | 0.89           |
| 8.5 | 0.90              | 0.87           |
| 7.5 | 1.03              | 1.06           |

Table 3.7: Flux normalizations to the Eddington ratio at $r = 30$ for set (a).

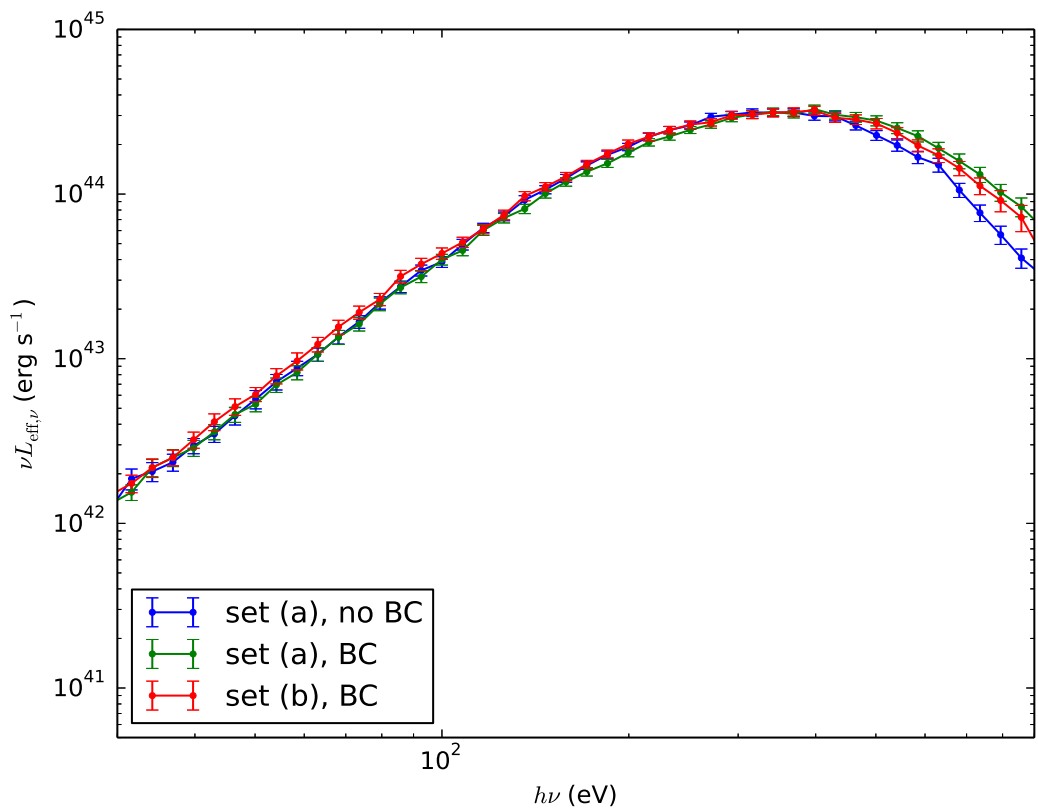

Figure 3.7: Observed disc spectra computed for sets (a) and (b). BC (bulk Comptonization) means bulk velocities were included. Set (a) includes both turbulence and shear. Set (b) includes only turbulence.

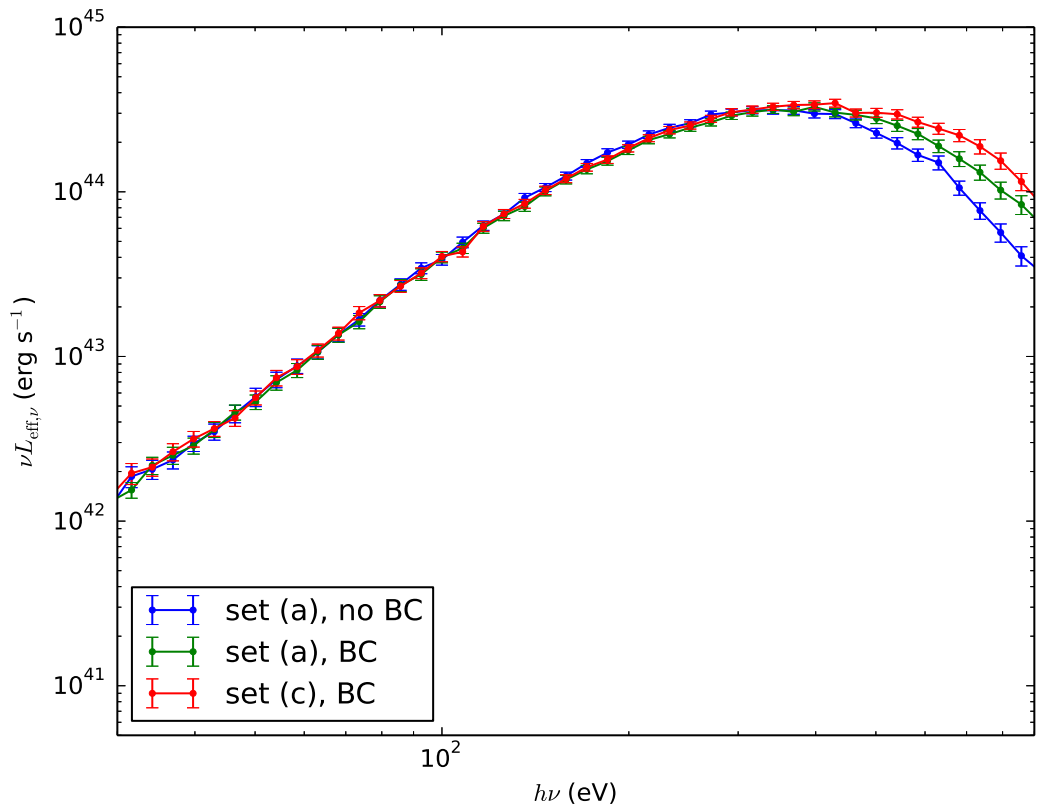

Figure 3.8: Observed disc spectra computed for sets (a) and (c). BC (bulk Comptonization) means bulk velocities were included. For set (a) the turbulent stress scaling $\alpha/\alpha_0$ is 1. For set (c), $\alpha/\alpha_0 = 2$.

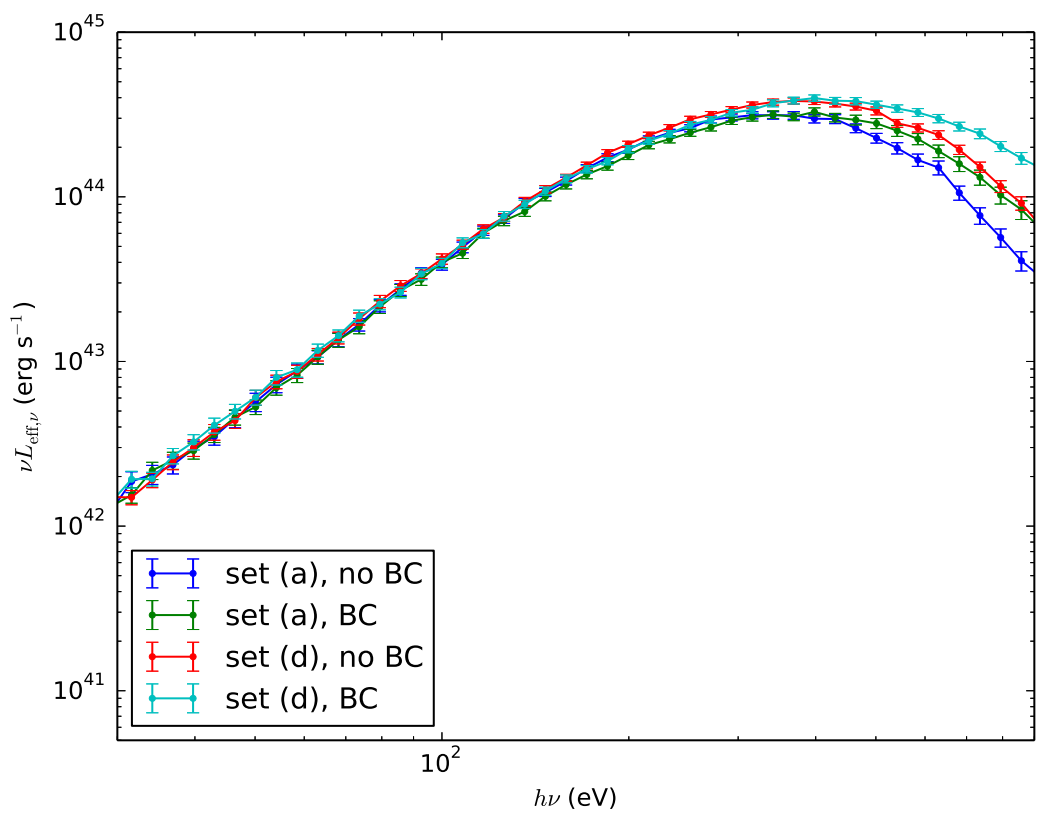

Figure 3.9: Observed disc spectra computed for sets (a) and (d). BC (bulk Comptonization) means bulk velocities were included. For set (a), the spin parameter $a = 0$. For set (d), $a = 0.5$.

For set (b) we calculate spectra without the background shear to isolate the effect of turbulence. The resulting observed spectrum is plotted in Figure 3.7. We see that the spectrum computed without shear lies significantly closer to the spectrum computed with shear than to the spectrum computed without the bulk velocities. This indicates that bulk Comptonization is primarily due to turbulence, not shear.

For set (c) we test the robustness of our results by repeating spectral calculations with a different turbulent stress scaling ratio, $\alpha/\alpha_0 = 2$. For OPALR20 (section 3.2.2), for example, $\alpha/\alpha_0 = 2.38$. The resulting observed spectrum is plotted in Figure 3.8. We see that although the observed spectrum computed with $\alpha/\alpha_0 = 2$ is Comptonized more than the spectrum computed with $\alpha/\alpha_0 = 1$, the effect is not huge. In particular, the fitted temperature and optical depth are only 21% and 13% higher, respectively. Since the turbulent velocity squared scales as $\alpha$ (equation D8), one might expect that the fitted temperature would also scale as $\alpha$, but this neglects the contribution by shear as well as the fact that we are fitting the optical depth along with the temperature rather than holding the optical depth fixed. The magnitude of bulk Comptonization is better indicated by $y_{\mathrm{p}}$. From set (b) we see that for $\alpha/\alpha_0 = 1$, $y_{\mathrm{p}} = 0.14$ for turbulence alone. From sets (a) and (b) we infer that for $\alpha/\alpha_0 = 1$, $y_{\mathrm{p}} = 0.26 - 0.14 = 0.12$ for shear alone. We would expect, therefore, that for $\alpha/\alpha_0 = 2$, $y_{\mathrm{p}} = 2 \times 0.14 + 0.12 = 0.40$, which is very close to the fitted value $y_{\mathrm{p}} = 0.38$.

For set (d) we explore the effect of varying the spin parameter by setting $a = 0.5$. The resulting observed spectrum is plotted in Figure 3.9. As expected, the original spectra computed without bulk velocities are hotter and more luminous for the higher spin parameter since the accretion efficiency is higher. But the effect of bulk Comptonization is comparable. The fitted temperature is slightly higher, but the fitted optical depth is slightly lower, leading to an effect that is nearly the same.

Finally, for set (e) we use a higher mass, $M = 2 \times 10^7 M_\odot$. The fitted temperature is lower, consistent with the dependence of overall accretion disc temperature on mass. But the larger value of $y_p$ indicates that the effect of bulk Comptonization on the spectrum is greater. This is consistent with equation (3.1), since the ratio of radiation to gas pressure increases with mass (SS73).

## 3.3.2    Truncated atmosphere spectral calculations with emissivity only at the base

We expect that bulk Comptonization is predominantly explained by the Comptonization of photons emitted at the effective photosphere. We discuss this in detail in section 3.4.2. To test this picture, we repeat spectral calculations with the parameters given in Table 3.4 but truncate the atmosphere at the effective photosphere and set the emissivity to zero everywhere except in the cells at the base. Table 3.3 summarizes these results.

For these calculations the observed spectra are different, but we expect the effect of bulk Comptonization on the observed spectra to be nearly unchanged. For example, the spectra computed without velocities for sets (a) and (a2), normalized to the total flux of (a), are plotted in Figure 3.10. The spectra coincide at high energies and diverge at low energies since photons emitted from lower temperature regions are omitted in (a2). But the fitted temperatures and optical depths for corresponding sets are very similar, which supports our picture of bulk Comptonization.

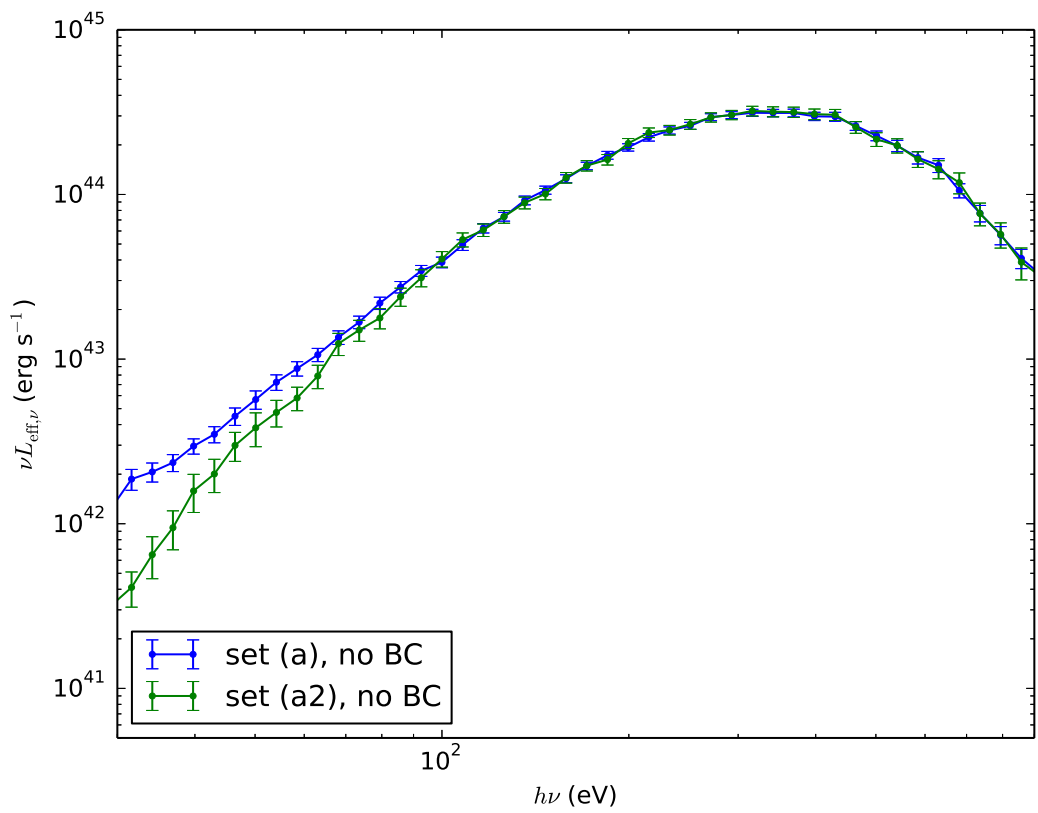

Figure 3.10: Observed disc spectra computed for sets (a) and (a2). In set (a2), the atmosphere is truncated at the effective photosphere and the emissivity is zero everywhere except in the cells at the base.

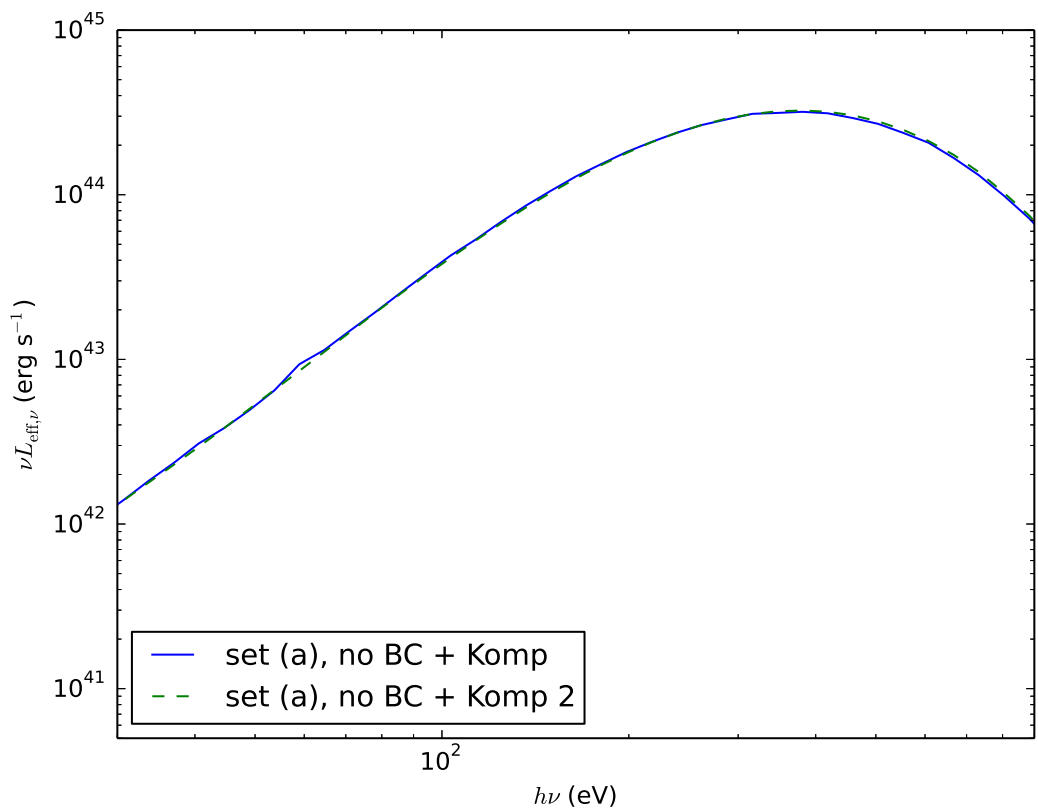

Figure 3.11: Observed disc spectra computed for set (a). BC (bulk Comptonization) means bulk velocities were included. Komp means the zero bulk Comptonization spectrum from each radius for $r \leq r_{\mathrm{cor}}$ was passed through a warm Comptonizing medium with the parameters given in Table 3.4. For Komp 2 the parameters used are those fit to set (a2), given in Table 3.3.

| Source | Model | Reference | $M/M_{\odot}$ | $L/L_{\mathrm{Edd}}$ | $k_{\mathrm{B}}T_{\mathrm{e}}$ (keV) | $\tau$ | $y_{\mathrm{p}}$ |
|--------|-------|-----------|---------------|----------------------|--------------------------------------|--------|------------------|
| REJ1034+396 | OPTXAGNF | D12 | $1.9 \times 10^6$ | 2.4 | $0.23 \pm 0.03$ | $11 \pm 1$ | 0.22 |

Table 3.8: Fits to observed NLS1s

For sets (a) to (e), we also pass the spectra computed without the bulk velocities through a warm Comptonizing medium with the temperatures and optical depths fit to sets (a2) to (e2), respectively, and see whether the results fit the spectra computed with the bulk velocities. For each case we calculate $\chi^2/\nu$ to assess the goodness of fit and list the results in Table 3.6. In Figure 3.11 for set (a) we plot the observed spectrum obtained by this procedure as well as the original fit. We see that the two curves nearly coincide and note that the corresponding values of $\chi^2/\nu$ differ by 0.6. For the other pairs of sets the corresponding values of $\chi^2/\nu$ differ by even less, which again confirms our expectation that bulk Comptonization is due to the Comptonization of photons emitted at the effective photosphere.

## 3.4   Discussion

### 3.4.1   Comparison with REJ1034+396

In NLS1s the Wien tail of the intrinsic disc spectrum contributes to the soft excess (D12). Bulk Comptonization increases the contribution to the soft excess by shifting the Wien tail to higher energy. Since bulk Comptonization increases with accretion rate, we expect this contribution to be greatest in near and super-Eddington sources. In broad-line Seyferts, the ratio of radiation to gas pressure is too low for bulk Comptonization to be significant. We compare our results to the analysis by D12 of REJ1034+396, a super-Eddington NLS1 with an unusually large soft excess. This analysis is summarized in Table 3.8. The comparison is appropriate because the mass and Eddington ratio we chose for our spectral calculations correspond to those fit to REJ1034+396, though we do note that our model for Comptonization is more detailed than the one in D12.[1] We see that the Compton $y$ parameter, $y_p = 0.22$, which characterizes the overall impact of Comptonization on the spectrum, is remarkably similar to the values we found. The fitted temperature and optical depth are also similar to our values. It may be, therefore, that the soft excess is unusually large in this system because of the contribution of bulk Comptonization.

A soft excess is also present in less luminous AGN for which bulk Comptonization is unlikely to be significant, and in general it seems that no single physical effect can fully explain the soft excess in all AGN. Until the contribution to the soft excess by other proposed mechanisms such as reflection and absorption are better understood, it will be difficult to tease out the contribution of bulk Comptonization. But our calculations show that if this can be done then observations of the soft excess can be used to constrain properties of the turbulence as well as other disc parameters.

### 3.4.2   Physical interpretation of results

Comptonization of photons by bulk motions is due to effects both first and second order in velocity (Chapter 2). The mathematics of thermal Comptonization cannot be used to describe first order effects, but in Chapter 2 we showed that for divergenceless (incompressible), statistically homogeneous turbulence it does capture second order effects. The equivalent "wave" temperature for bulk velocities, which is a function of the

---

[1] In particular, in D12 the photon spectrum passed through the warm Comptonizing medium is given by the spectrum at $r_{cor}$ and only the overall normalization varies with radius. This choice was made to minimize computation time.

photon mean free path, is given by

$$k_B T_w = \left\langle \frac{-2\lambda_p m_e c}{E} P^{ij} \left( \partial_i v_j + \partial_j v_i \right) \right\rangle,$$ 
(3.29)

where $E$ is the radiation energy density and $P^{ij}$ is the radiation pressure tensor. Note that only the traceless part of the pressure tensor, which is the shear stress, contributes since this result assumes incompressible motions. We see that the temperature for bulk velocities is proportional to the stress multiplied by the strain rate, which is just the viscous dissipation of bulk motions by the photons.

For our spectral calculations bulk Comptonization is well described by the Kompaneets equation (which describes thermal Comptonization by a single temperature), which suggests that second order effects, not first order effects, are dominant. This may be because MRI turbulence is incompressible and first order effects vanish for incompressible, but not compressible, turbulence (Chapter 2). On the other hand, the photosphere regions are magnetically dominated and show considerable compressible motions because of the Parker instability (Blaes, Hirose & Krolik, 2007), so it seems more likely that first order effects average out.

Assuming second order effects are dominant, we can gain physical insight into the fitted temperatures and optical depths by considering the dependence of the wave temperature on the photon mean free path. The wave temperature is largest when the photon mean free path is long relative to the maximum turbulence wavelength, and is negligible when it is small (Chapter 2). Therefore, Comptonization is only significant in the region near enough to the photosphere that the photon mean free path is comparable to the maximum turbulence wavelength. The resulting Comptonization temperature and optical depth should be the same for all photons emitted below this region. Inside this region, on the other hand, photons emitted nearer the photosphere should have comparatively larger Comptonization temperatures and smaller optical depths. For real disc atmospheres, which are stratified in (gas) temperature, photons contributing to the spectral peak are predominantly emitted at the effective photosphere, which for modest turbulence should be below the region where bulk Comptonization is significant. We therefore expect the resulting Comptonization temperature and optical depth to be unchanged when we truncate the atmosphere at the effective photosphere and set the emissivity equal to zero everywhere except at the base. Our findings confirm this. This is also useful because these spectral calculations run much faster which allows for a more efficient exploration of the disc parameter space.

### 3.4.3   Self-consistency of results with shearing box simulations

We see that when bulk velocities are included in spectral calculations, the observed spectrum is shifted to higher energy. In particular, the Wien tail is shifted right. While this allows us to characterize bulk Comptonization with a temperature and optical depth as a function of accretion disc parameters, to determine the actual impact on disc spectra we must consider whether our spectral calculations are consistent with the underlying shearing box simulations on which they are based.

In section 3.4.2 we showed that bulk Comptonization here is predominantly an effect that is second order in velocity, but the underlying shearing box simulations (Hirose, Krolik & Blaes, 2009) do not include this effect because the flux-limited diffusion approximation is used (Chapter 2). Therefore, according to this picture we expect the spectral calculations without the bulk velocities to be consistent with the flux found in the underlying shearing box simulation. In order to determine the effect of including the bulk velocities on the resulting spectra we must take into account the back-reaction on the vertical structure. Since adding in bulk Comptonization without modifying the vertical structure increases the flux and violates energy conservation, including this effect in the underlying shearing box simulation would lower the gas temperature until energy conservation is restored. Therefore, for significant Comptonization, while the Wien tail shifts to the right, the spectral peak shifts to the left. The overall effect, therefore, is to broaden the spectrum. In practice, if the Comptonization temperature is only slightly higher than the gas temperature then the spectrum will still be broadened but without an obvious leftward shift of the spectral peak.

Because the decrease in gas temperature as well as other changes in the vertical structure may then affect bulk Comptonization, in theory the two should be calculated self-consistently. Another complicating factor is vertical advection of radiation, a velocity dependent effect that increases the number of photons emitted without affecting their energies, which also impacts energy conservation. But as long as bulk Comptonization is a perturbative effect, our fundamental results should hold: Bulk Comptonization broadens the spectrum by lowering the gas temperature and shifting the Wien tail to higher energy such that the total energy is conserved, and the characteristic temperatures and optical depths are given by Table 3.4. Furthermore, our method can be used to explore how bulk Comptonization scales with different parameters such as the mass, accretion rate, spin, turbulent stress scaling, and boundary condition at the innermost stable circular orbit, which we do in Chapter 4.

### 3.4.4 Bulk Comptonization by the background shear

Our results suggest that Comptonization by bulk motions is predominantly due to turbulence, not shear. But since Comptonization by shear is not negligible, here we consider how it differs from Comptonization by turbulence, both in its potential effect on spectra and on the disc vertical structure.

From the perspective of total energy conservation, bulk Comptonization by the background shear at a given radius should have the same impact on the spectrum as turbulent Comptonization. It should shift the Wien tail to the right and decrease the gas temperature, broadening the spectrum. This is because the effective temperature for a steady-state disc at a given radius is strictly fixed by the mass, mass accretion rate, and radius.

But Comptonization by the background shear plays a completely different role in the disc equations than Comptonization by turbulence. For the latter, the stress on the mean fluid flow is still entirely determined by MRI turbulence. For the $\alpha$ prescription, for example, the value of $\alpha$ is still set by the saturation level of the magnetic field and is therefore presumably unchanged. But Comptonization by shear is an additional stress on the mean fluid flow, and would therefore presumably increase $\alpha$. Since Comptonization by shear, at least in the regimes we have explored here, has only a perturbative effect on the spectrum, we expect any increase in $\alpha$ to be small. This is physically intuitive since dissipation by shear can be significant only near the photosphere where the mean free path is larger (see section 3.4.2), whereas dissipation by MRI turbulence is significant throughout the body of the disc.

An interesting consequence of the difference between Comptonization by turbulence and background shear is that they have different effects on the total flux emitted from a shearing box. In a shearing box the density, not the radius, is fixed, and the flux depends on $\alpha$ (equation 3.9). For Comptonization by turbulence, unless the MRI is affected, $\alpha$ is unchanged, and the gas temperature must decrease so that the flux is unchanged. But for Comptonization by the background shear, an additional source of stress on the mean flow is present, which modifies $\alpha$ and allows the flux to change. According to equation (3.9) we would ironically expect the flux to decrease rather than increase, but we should not take this prediction seriously. Comptonization by bulk motions is only significant near the photosphere where predominantly magnetic pressure, not radiation pressure, supports the atmosphere, so a small perturbation to $\alpha$ confined to this region cannot be treated self-consistently by the standard $\alpha$ disc equations.

Of course, in practice it is not well understood what determines $\alpha$; it is possible that even Comptonization by turbulence indirectly affects $\alpha$. It is also possible that Comptonization by the background shear indirectly decreases the saturation level of the magnetic field so that the net effect is to leave $\alpha$ unchanged. Our point is that Comptonization by turbulence and Comptonization by the background shear play different roles in the disc equations and therefore potentially have different effects on the vertical structure.

## 3.5   Summary

We modeled the contribution of bulk Comptonization to the soft X-ray excess in AGN. To do this, we calculated disc spectra both taking into account and not taking into account bulk velocities with data from radiation MHD simulations. Because our simulation data was limited, we developed a scheme to scale the disc vertical structure to different values of radius, mass, and accretion rate. For each parameter set, we characterized our results by a temperature and optical depth in order to facilitate comparisons with other warm Comptonization models of the soft excess. We chose our fiducial mass, $M = 2 \times 10^6 M_\odot$, and accretion rate, $L/L_{\mathrm{Edd}} = 2.5$, to correspond to the values fit by D12 to the super-Eddington narrow-line Seyfert 1 REJ1034+396, which has an unusually large soft excess. Our principal results are as follows.

For zero spin, when Comptonization by both turbulence and the background shear are included, the Compton $y$ parameter we find, $y_{\mathrm{p}} = 0.26$, is close to that found by D12 for REJ1034+396, $y_{\mathrm{p}} = 0.22$. The temperature we find is a bit lower ($k_{\mathrm{B}}T_{\mathrm{e}} = 0.14$keV vs.

$k_B T_e = 0.23\text{keV}$), but the optical depth is higher ($\tau = 15$ vs. $\tau = 11$). For spin $a = 0.5$, the correspondence is remarkable; we find $y_p = 0.22$, $k_B T_e = 0.21\text{keV}$, and $\tau = 12$. We find that bulk Comptonization is primarily due to turbulence, not the background shear (Figure 3.7). Both the fitted temperature and optical depth increase moderately when we double the turbulent stress scaling to $\alpha/\alpha_0 = 2$. When we increase the mass, the fitted temperature decreases, but the $y$ parameter increases. This indicates that bulk Comptonization is more significant, which we expect since the ratio of electron thermal to bulk velocities depends on the ratio of radiation to gas pressure (equation 3.1), which in turn scales with mass (SS73). Our results are given in Table 3.4.

To enforce energy conservation, the impact of bulk Comptonization on disc spectra is to shift the Wien tail to the right while simultaneously lowering the gas temperature, broadening the spectrum. Since we find that bulk Comptonization is well described by the Kompaneets equation, this suggests that it is predominantly an effect second order in velocity (Chapter 2). Knowledge of this is important for self-consistently resolving bulk Comptonization in radiation MHD simulations, since common closure schemes such as flux-limited diffusion do not include this effect (Chapter 2).

The soft excess in general is unlikely due to a single physical mechanism. Other contributing effects, such as reflection and absorption, must be better understood to make precise comparisons of predictions by models of bulk Comptonization with observations. But the fact that our results, based simply on the most naive scalings, are in agreement with observations suggests that at least in the super-Eddington NLS1 regime bulk Comptonization may play a significant role in producing the soft X-ray excess. If so, observations of the soft excess can be directly tied to the properties of MHD turbulence as well as fundamental disc parameters.

# Chapter 4

# A simple framework for modelling the dependence of bulk Comptonization by turbulence on accretion disc parameters

## 4.1 Introduction

In this chapter we simplify and generalize the bulk Comptonization model presented in Chapter 3 in order to develop greater physical insight into this process and explore a larger space of accretion disc parameters. In Chapter 3, bulk Comptonization is modeled by fitting the Comptonization temperature and optical depth parameters to spectra computed with Monte Carlo post-processing of simulations of the turbulence. Here we develop a procedure to infer these Comptonization parameters from the underlying disc vertical structure radiation MHD simulation data without computing spectra. The immediate benefit of this is that we can efficiently explore a larger space of accretion disc parameters. We can also find the time-averaged Comptonization parameters for a given simulation since without computing spectra we can now efficiently calculate Comptonization parameters for multiple timesteps.

More importantly, our model provides a physically intuitive framework for understanding bulk Comptonization. In particular, we show that the variation of this effect with disc parameters can be understood in terms of the vertical gas temperature and "wave" temperature (section 4.2.1) profiles. This allows us to determine the dependence of bulk Comptonization on each accretion disc parameter separately without exhaustively exploring a multiparameter space. We can also probe how various physical effects, such as vertical radiation advection (Blaes et al., 2011; Jiang, Stone & Davis, 2013, 2014b), may impact bulk Comptonization, as well as evaluate how robust our model's predictions are to changes in the disc vertical structure. In particular, although the specific bulk Comptonization parameters that we calculate here result from applying our model to scaled data from the limited shearing box simulation 110304a from Chapter 3, we show that our principal findings regarding how bulk Comptonization scales with fundamental accretion disc parameters is likely to be robust to differences in the disc vertical structure seen in other simulations. Furthermore, understanding this framework should be useful for developing physical intuition in new situations in which some of our particular results may no longer hold, such as shearing box or global disc simulations run in radically different regimes.

The structure of this chapter is as follows. In section 4.2 we describe our model and show why it is effective. In section 4.3 we apply our model to data from radiation MHD simulations. We show how the dependence of bulk Comptonization on shearing box parameters can be understood in terms of one dimensional temperature profiles (section 4.3.2), and then proceed to examine its dependence on each accretion disc parameter individually (section 4.3.3). We estimate bulk Comptonization for an entire disc as well by fixing the radius to the region of maximum luminosity (section 4.3.4). We consider the effect of including radiation advection (section 4.3.5) and discuss the time variability of bulk Comptonization within a given simulation (section 4.3.6). In section 4.4 we discuss our results, and we summarize our findings in section 4.5.

## 4.2    Efficiently modeling bulk Comptonization

### 4.2.1    Overview

Since this work simplifies and generalizes the bulk Comptonization model that we presented in Chapter 3, we begin by summarizing how we calculated the bulk Comptonization temperature and optical depth for a given system. At each radius in an accretion disc, we used radiation MHD stratified shearing box simulation data to calculate spectra both including and excluding velocities. Each spectral computation was performed by running a post-processing Monte Carlo simulation on a simulation data snapshot at a particular epoch in time. Since our data was limited, we developed a scheme to scale the original data to the accretion disc parameters of interest. At each radius in the disc, we used the Kompaneets equation to pass the spectrum computed without velocities through a Comptonizing medium with a given electron temperature and optical depth, and the resulting spectra were superposed to obtain the observed spectrum. Meanwhile, the spectra computed with velocities at each radius were superposed to obtain a different observed spectrum. The temperature and optical depth parameters (which were assumed to be the same at all radii) were adjusted until the two spectra match.

Here we develop a more efficient and physically revealing procedure for calculating the Comptonization temperature and optical depth. We focus on computing these parameters for each radius individually rather than for the whole disc at once, a choice that also allows us to study the dependence of bulk Comptonization on radius. This choice does not limit us to studying individual radii, since bulk Comptonization for a whole disc can be estimated by the Comptonization parameters at the radius where the luminosity is greatest (section 4.3.4). One other difference from Chapter 3 is that we include only turbulent velocities, not shear velocities. This allows us to study the effects of turbulence alone. The preliminary results in Chapter 3 suggest that bulk Comptonization by shear is subdominant to bulk Comptonization by turbulence, but to rigorously calculate the effect of shear near the photosphere will require global simulations. Since the scalings for the shear velocities are nearly identical to the scalings for the turbulent velocities (Chapter 3), this omission should not affect our conclusions regarding the general dependence of bulk Comptonization on accretion disc parameters.

At a given radius, we define the Kompaneets parameters, consisting of the electron temperature $T_K$ and optical depth $\tau_K$, analogously to how the temperature and optical depth are defined for the whole disc in Chapter 3. In other words, spectra are calculated with and without velocities at a given radius, and the spectrum computed without velocities is passed through a Comptonizing medium using the Kompaneets equation. The temperature and optical depth of the medium are adjusted until the resulting spectrum matches the spectrum calculated with velocities. To scale simulation data to different accretion disc parameters we use the scheme developed in Chapter 3.

We show that the Kompaneets parameters are approximated by what we will henceforth refer to as the Comptonization parameters. We define these parameters with an efficient and physically revealing procedure that we outline here and then discuss in greater detail in the following sections. First, we map the bulk velocity grid to a tem-perature grid by defining at each point a bulk "wave" temperature,

$$\frac{3}{2}k_{\mathrm{B}}T_{\mathrm{w}} = \frac{1}{4}m_{\mathrm{e}} \left\langle (\Delta\mathbf{v})^2 \right\rangle_{\mathbf{r}},$$  (4.1)

where $\left\langle (\Delta\mathbf{v})^2 \right\rangle_{\mathbf{r}}$ is the average square velocity difference between subsequent photon scatterings at $\mathbf{r}$. We note that if instead of applying equation (4.1) to the bulk velocities in a region we apply it to the thermal velocity distribution at a particular point, then since

$$\left\langle (\Delta\mathbf{v})^2 \right\rangle = \int (\mathbf{v_2} - \mathbf{v_1})^2 f(\mathbf{v_1}) f(\mathbf{v_2}) \, d\mathbf{v_1} d\mathbf{v_2}$$  (4.2)

$$= 2 \left( \left\langle v^2 \right\rangle - \left\langle \mathbf{v} \right\rangle^2 \right)$$  (4.3)

$$= 2 \left\langle v^2 \right\rangle,$$  (4.4)

we find that

$$\frac{3}{2}k_{\mathrm{B}}T_{\mathrm{w}} = \frac{1}{2}m_{\mathrm{e}} \left\langle v^2 \right\rangle_{\mathbf{r}},$$  (4.5)

as expected. We call this a "wave" temperature and not a "turbulent" temperature because it depends on the power spectrum of the turbulence and is less than the temperature that one usually associates with a turbulent velocity distribution, which is given by $\frac{3}{2}k_{\mathrm{B}}T = \frac{1}{2}m_{\mathrm{e}} \left\langle v^2 \right\rangle_{\mathbf{r}}$, analogous to a thermal temperature. We discuss $\left\langle (\Delta\mathbf{v})^2 \right\rangle_{\mathbf{r}}$ in more detail in section 4.2.2. Next, we horizontally average all simulation variables, including the newly defined wave temperature, to obtain 1D profiles of the data. For the systems of interest, the wave temperature is negligible compared to the gas temperature at the effective photosphere, and it increases going outward so that near the scattering photosphere it may exceed it. We define the Comptonization optical depth $\tau_{\mathrm{C}}$ as the optical depth of the region in which the wave temperature is at least half the gas temperature, a region that we will henceforth refer to as the bulk Comptonization region. We define the Comptonization temperature $T_C$ as a weighted average of the sum of the gas and wave temperatures in this region, given by

$$T_C = \frac{\int_0^{\tau_C} (T_g + T_w)\, \tau d\tau}{\int_0^{\tau_C} \tau d\tau}. \tag{4.6}$$

In the next section we both describe in detail why this procedure approximates the Kompaneets parameters and demonstrate its effectiveness by comparing what it predicts with the results of actual Monte Carlo spectral calculations.

### 4.2.2  Physical justification for the bulk Comptonization model

To justify our definition of the Comptonization parameters, $\tau_C$ and $T_C$, we start with the procedure that defines the Kompaneets parameters and then incrementally simplify it. In the following sections we detail each step of this process. Where appropriate we invoke an accretion disc parameter set for which $M = 2 \times 10^6 M_\odot$ and $L/L_{Edd} = 5$ as a test case. The other parameters are given in Table 4.1. The parameter $a$ is the black hole dimensionless spin parameter, $\Delta\epsilon$ is the change in efficiency for a non-zero torque inner boundary condition (Agol & Krolik, 2000), $\alpha$ is the ratio of vertically integrated stress to vertically integrated pressure, and $\alpha_0$ is the value of $\alpha$ for the original simulation data. We list $\alpha/\alpha_0$ rather than $\alpha$ since it is the former that we can directly adjust with the scaling scheme from Chapter 3. Typically $\alpha_0 \sim 0.01$. We note that these parameters are nearly identical to those of the systems modeled in Chapter 3, and were originally chosen to correspond to those fit by Done et al. (2012) (hereafter D12) to the NLS1 REJ1034+396. The only parameter whose value differs from the value in Chapter 3 is $L/L_{Edd}$, which we set here to 5 rather than 2.5. Since bulk Comptonization increases with $L/L_{Edd}$, we choose a higher value here so that it is easier to see the effectiveness of our approximations in plots of actual spectra. After describing all steps, we demonstrate the effectiveness of the resulting procedure at six different radii for not only the $M = 2 \times 10^6 M_\odot$, $L/L_{\mathrm{Edd}} = 5$ parameter set but for two others as well. One is the same except that $L/L_{\mathrm{Edd}} = 2.5$, the original value in Chapter 3. For the other, $M = 2 \times 10^8 M_\odot$ and $L/L_{\mathrm{Edd}} = 4.2$. All disc parameter sets are given in Table 4.1. As in Chapter 3, all spectra in this section are computed with Monte Carlo post-processing simulations (Davis et al., 2009; Pozdniakov et al., 1983), using data from the 140 orbits timestep of simulation 110304a. We discuss this simulation in more detail in section 4.3.2.

| $M/M_\odot$ | $L/L_{\mathrm{Edd}}$ | $a$ | $\Delta\epsilon$ | $\alpha/\alpha_0$ |
|---|---|---|---|---|
| $2 \times 10^6$ | 5 | 0 | 0 | 2 |
| $2 \times 10^8$ | 4.2 | 0 | 0 | 2 |
| $2 \times 10^6$ | 2.5 | 0 | 0 | 2 |

Table 4.1: Accretion disc parameter sets

## Step 1 - Truncate the simulation data inside the effective photosphere and turn off emissivity above this surface

To begin, we observe that we can modify the defining procedure for calculating the Kompaneets parameters by using simulation data that is truncated inside the effective photosphere and turning off the emissivity everywhere except at this surface, without changing the resulting parameters. The effective photosphere is defined by using the Planck mean opacity. This phenomenon was demonstrated in the context of fitting Kompaneets parameters for an entire disc at once in Chapter 3, but it arises from the fact that it is true for individual radii.

For example, we calculate spectra both with and without velocities at $r = 14$ for the $M = 2 \times 10^6 M_\odot$, $L/L_{\mathrm{Edd}} = 5$ parameter set (Table 4.1). We note that all numerical radii in this chapter are in units of the gravitational radius $GM/c^2$ of the black hole. We also calculate spectra using simulation data that is truncated at the effective photosphere and in which the emissivity is set to zero everywhere except the base. All four resulting spectra are plotted in Figure 4.1. We see that except at very low energies, the two spectra calculated with velocities coincide and the two spectra calculated without velocities coincide.

Two reasons underlie this result. First, the emergent spectrum is dominated by photons originally emitted at or near the effective photosphere since free-free emission depends strongly on density. Second, bulk Comptonization is negligible except near the scattering photosphere (as we show in section 4.2.2) and so its effect does not depend on the precise effective optical depth at which photons are originally emitted. Because all systems in this work meet these conditions, this result is robust. Therefore, modifying the defining procedure for calculating the Kompaneets parameters in this way has a negligible effect on the outcome.

Calculating the temperature and optical depth with this modified method is the first step to simplifying the calculation and developing physical insight into the problem. Not only does this modified procedure run faster, but it shows that in order to model and understand bulk Comptonization we only need to understand the effect of bulk Comptonization on photons emitted at the effective photosphere.

### Step 2 - Map the velocity grid to a "wave" temperature grid

Next, we map the bulk velocity grid to a temperature grid by defining at each point a bulk "wave" temperature given by equation (4.1). This definition is motivated by the results of section 2.4.2. There we showed that for a periodic box with statistically homogeneous turbulence and an escape probability, bulk Comptonization can be treated as thermal Comptonization by solving the Kompaneets equation with a temperature given by $\frac{3}{2}k_B T_w = \frac{1}{4}m_e \left\langle (\Delta \mathbf{v})^2 \right\rangle$, where $\left\langle (\Delta \mathbf{v})^2 \right\rangle$ is the volume average of $\left\langle (\Delta \mathbf{v})^2 \right\rangle_{\mathbf{r}}$, which is in turn given by:

$$\left\langle (\Delta \mathbf{v})^2 \right\rangle_{\mathbf{r}} = \int \left( \Delta \mathbf{v}(\Delta \mathbf{r}, \mathbf{r}) \right)^2 P_{\Delta \mathbf{r}}(\Delta \mathbf{r}) d^3 \Delta \mathbf{r}. \tag{4.7}$$

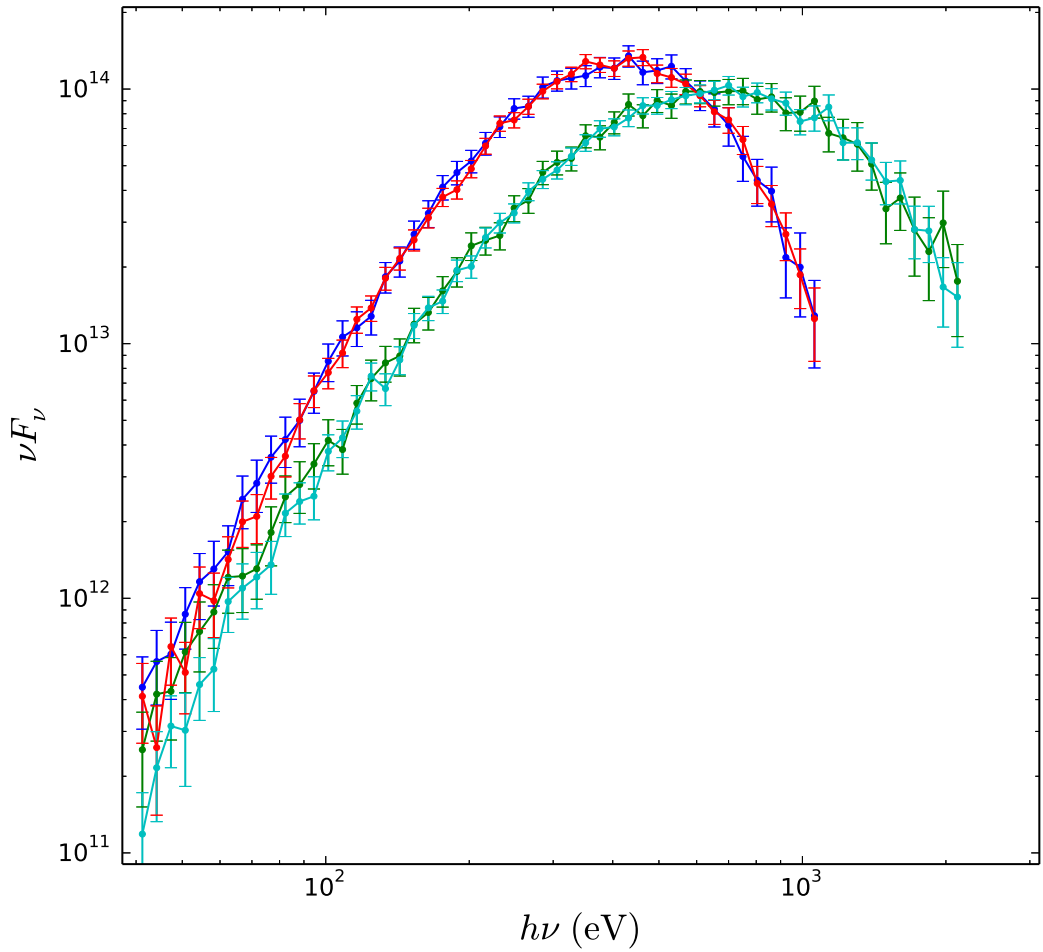

Figure 4.1: Normalized accretion disc spectra at $r = 14$ for the $M = 2 \times 10^6 M_\odot$, $L/L_{\rm Edd} = 5$ parameter set (Table 4.1). The green and cyan curves are computed with velocities, and the blue and red curves are computed without velocities. For the cyan and red curves, the simulation data is truncated inside the effective photosphere and the emissivity is set to zero everywhere except at the effective photosphere.

Here, $\Delta \mathbf{v}(\Delta \mathbf{r}, \mathbf{r})$ is the velocity difference between positions $\mathbf{r}$ and $\mathbf{r} + \Delta \mathbf{r}$, and $P_{\Delta \mathbf{r}}(\Delta \mathbf{r})$ is the probability density that a photon scattering at $\mathbf{r}$ subsequently scatters at $\mathbf{r} + \Delta \mathbf{r}$. Hence $\left\langle (\Delta \mathbf{v})^2 \right\rangle_{\mathbf{r}}$ is the average square velocity difference between subsequent photon scatterings at $\mathbf{r}$. In this work, therefore, equation (4.1) defines a wave temperature at each point instead of taking a volume average and defining it for an entire box.

To develop physical intuition into equation (4.1) it is important to understand the dependence of $\left\langle (\Delta \mathbf{v})^2 \right\rangle_{\mathbf{r}}$ on density. In the high density limit the velocity difference between subsequent photon scatterings is small. In particular, $\left\langle (\Delta \mathbf{v})^2 \right\rangle_{\mathbf{r}}$ is proportional to the square of the mean free path $\lambda_p$ (Chapter 2) so $T_w$ decreases significantly with increasing density. In the low density limit, on the other hand, $\left\langle (\Delta \mathbf{v})^2 \right\rangle_{\mathbf{r}}$ approaches $2\left\langle v^2 \right\rangle$ so that $\frac{3}{2}k_B T_w$ approaches $\frac{1}{2}m_e \left\langle v^2 \right\rangle_{\mathbf{r}}$. We also define the bulk temperature, given by

$$\frac{3}{2}k_B T_{bulk} = \frac{1}{2}m_e v^2. \tag{4.8}$$

We note that $T_{tot}$ in Chapter 2 is just the average of $T_{bulk}$ over some region. Applying equation (4.7) directly to simulation data is somewhat problematic, so we discuss our implementation in detail in Appendix E.

For example, in Figure 4.2 we plot the profile of the density weighted horizontal average of the wave temperature at $r = 14$ for the $M = 2 \times 10^6 M_\odot$, $L/L_{Edd} = 5$ parameter set (Table 4.1). We also plot the gas temperature, $T_g$, the bulk temperature, $T_{\rm bulk}$, and the sum of the gas and wave temperatures. The dashed line on the right denotes the location of the scattering photosphere, which we define as the height at which the Thomson optical depth $\tau_{\rm s} = 1$. We see that the wave temperature significantly increases with decreasing density (that is, moving rightward) as the photon mean free path grows, and is comparable to the bulk temperature only near the scattering photosphere.

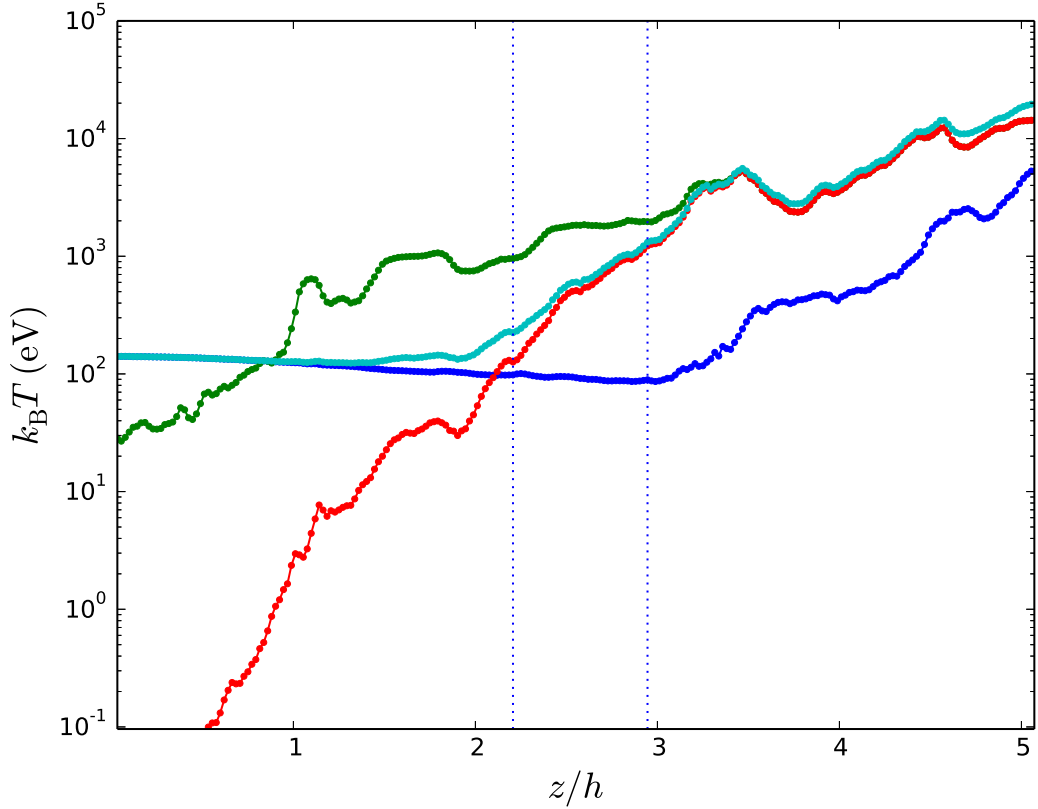

Figure 4.2: Horizontally averaged profiles at $r = 14$ for the $M = 2 \times 10^6 M_\odot$, $L/L_{\rm Edd} = 5$ parameter set (Table 4.1) for the gas temperature $T_{\rm g}$ (blue), bulk temperature $T_{\rm bulk}$ (green), wave temperature $T_{\rm w}$ (red), and sum of gas and wave temperatures (cyan). The dashed lines denote where $\tau_{\rm s} = 1$ and $\tau_{\rm s} = 10$.

We find that photon spectra computed with simulation data in which the velocities are turned off and the wave temperatures are added to the gas temperatures approximate photon spectra computed with the velocities turned on. For example, in Figure 4.3 we plot spectra at $r = 8.5$, 9.5, 11, 14, 20, and 30 for the $M = 2 \times 10^6 M_\odot$, $L/L_{\mathrm{Edd}} = 5$ parameter set (Table 4.1), computed with and without velocities. We also plot spectra computed with data in which the velocities are turned off and the wave temperatures are added to the gas temperatures. We see that these spectra approximate the spectra computed with velocities. In other words, bulk Comptonization can be modeled by thermal Comptonization in which the temperature is given by equation (4.1).

### Step 3 - Horizontally average the simulation data

Once bulk velocities are replaced by wave temperatures, it is straightforward to further simplify the problem by horizontally averaging the simulation data. In order that the effects of bulk Comptonization remain unchanged, the wave temperature data must be density averaged, not volume averaged.

For example, we calculate the spectrum at $r = 14$ for the $M = 2 \times 10^6 M_\odot$, $L/L_{\mathrm{Edd}} = 5$ parameter set (Table 4.1) using data in which the velocities are turned off and the wave temperatures are added to the gas temperatures, as described in section 4.2.2. We repeat this calculation using horizontally, density weighted averaged data and plot both spectra in Figure 4.4. We see that the two spectra coincide. In Figure 4.5 we again plot the spectrum calculated with the unaveraged data as well as a spectrum calculated with horizontally averaged data, except that this time the wave temperatures are computed with a simple spatial horizontal average instead. We see that simple spatially averaging the wave temperatures with no density weighting overestimates bulk Comptonization.

Density averaging improves the accuracy because the time photons spend in a region increases with the region's density. The reason that volume weighting overestimates bulk Comptonization is that the wave temperature is strongly correlated with density. As we discussed in Step 2, the wave temperature decreases with density, and so volume averaging gives too much weight to regions where the wave temperature is larger. Because the gas temperature, on the other hand, is not strongly correlated with density, horizontally volume weighting the gas temperature has a negligible impact on the spectrum. For example, we calculate another spectrum with horizontally averaged data, except that this time the gas temperatures are computed with a simple spatial average instead. We plot the result alongside the spectrum calculated from the unaveraged data in Figure 4.6. We see that the two spectra coincide, which indicates that gas temperature inhomogeneities at a given height are not sufficiently correlated with density inhomogeneities to affect the spectrum.

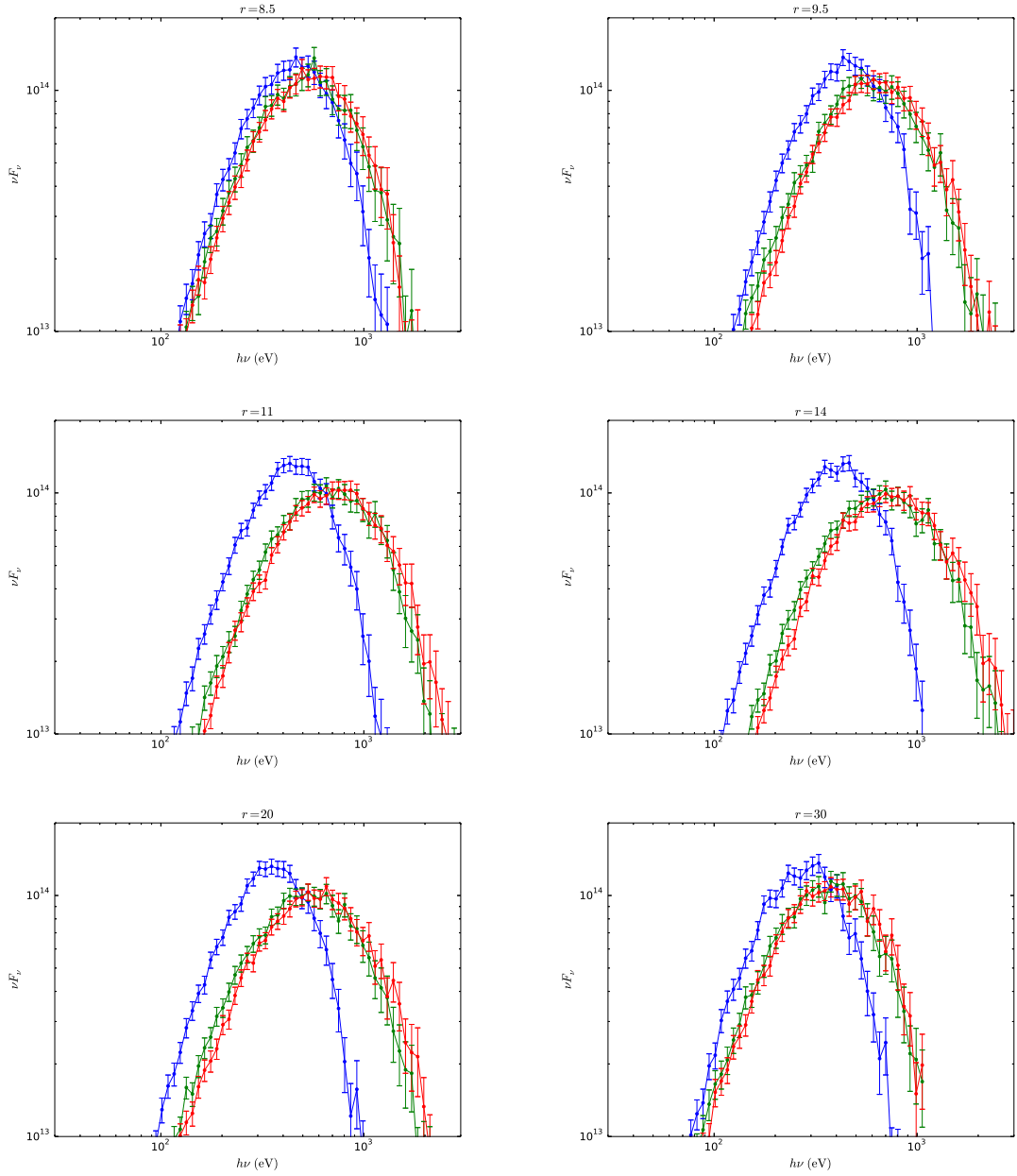

Figure 4.3: Normalized accretion disc spectra at multiple radii for the $M = 2 \times 10^6 M_\odot$, $L/L_{\mathrm{Edd}} = 5$ parameter set (Table 4.1) computed with (green) and without (blue) the velocities. For the red curve, the velocities were not included but the wave temperatures were added to the gas temperatures.

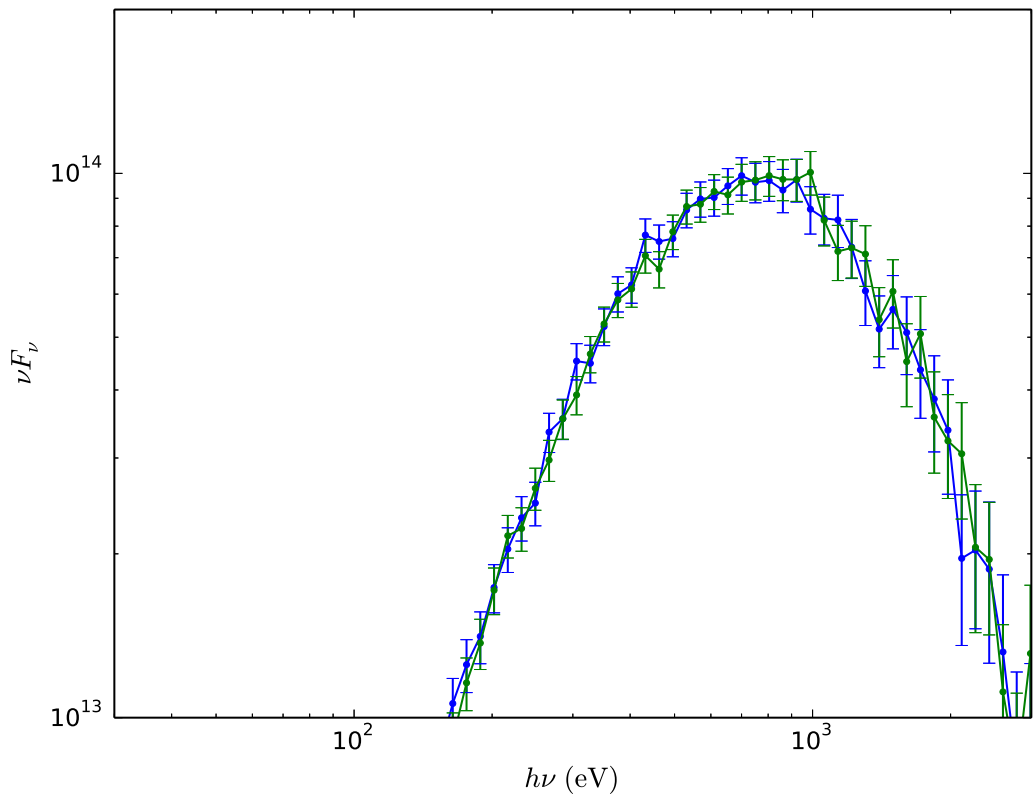

Figure 4.4: Normalized accretion disc spectra at $r = 14$ for the $M = 2 \times 10^6 M_\odot$, $L/L_{\mathrm{Edd}} = 5$ parameter set (Table 4.1) computed by omitting the velocities and instead adding the wave temperatures to the gas temperatures. The blue curve is computed with unaveraged data, and the green curve is computed with horizontally density weighted averaged data.

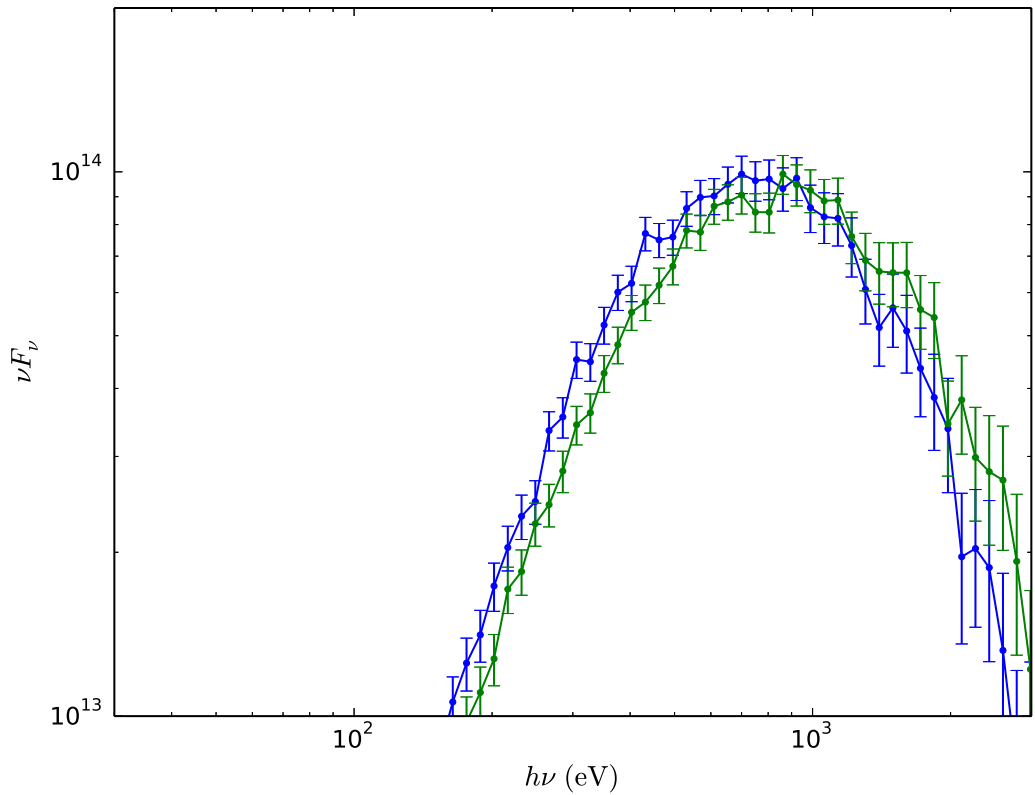

Figure 4.5: Normalized accretion disc spectra at $r = 14$ for the $M = 2 \times 10^6 M_\odot$, $L/L_{\text{Edd}} = 5$ parameter set (Table 4.1) computed by omitting the velocities and instead adding the wave temperatures to the gas temperatures. The blue curve is computed with unaveraged data, and the green curve is computed with horizontally averaged data. For the latter, the wave temperature averages are volume weighted and the other variable averages are density weighted.

The fact that we can use horizontally averaged quantities means we can map a three dimensional problem to a one dimensional problem, an important step to efficiently calculating and understanding bulk Comptonization. Plots of horizontally averaged quantities such as Figure 4.2, first introduced in Step 2, will be of great use in the remainder of this work.

### Step 4 - Solve the 1D inhomogeneous thermal Comptonization problem

By this point, we have modified the original procedure for calculating the Kompaneets parameters by instead calculating spectra with and without adding the wave temperature profile to the gas temperature profile (section 4.2.2), using horizontally averaged data (Step 3) truncated inside the effective photosphere in which velocities are turned off and emission is zeroed everywhere except at the effective photosphere (Step 1). The temperature and optical depth are adjusted until the spectra match.

To further simplify the problem, we first need to understand the effect of thermal Comptonization on photons emitted at the base of an inhomogenous one dimensional medium. We expect that if the optical depth is not too high, so that the average photon energy is always significantly below the local temperature (i.e. the photon spectrum does not saturate), then this process can always be well described by a homogeneous thermal

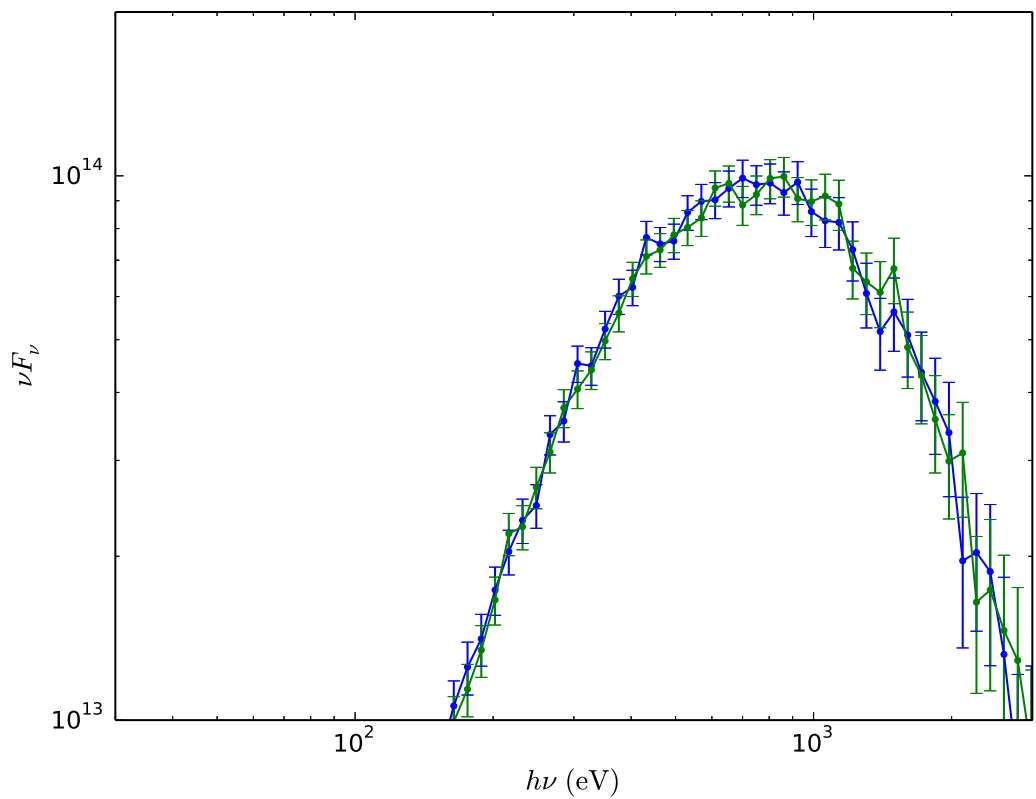

Figure 4.6: Normalized accretion disc spectra at $r = 14$ for the $M = 2 \times 10^6 M_\odot$, $L/L_{\mathrm{Edd}} = 5$ parameter set (Table 4.1) computed by omitting the velocities and instead adding the wave temperatures to the gas temperatures. The blue curve is computed with unaveraged data, and the green curve is computed with horizontally averaged data. For the latter, the gas temperature averages are volume weighted and the other variable averages are density weighted.

| $r$ | 8.5 | 9.5 | 11 | 14 | 20 | 30 |
|-----|-----|-----|-----|-----|-----|-----|
| $T_{1D}$ (eV) | 226 | 304 | 361 | 408 | 344 | 212 |

Table 4.2: Values of $T_{1D}$ for vertical structure data truncated at $\tau_s = 10$ at multiple radii for the $M = 2 \times 10^6 M_\odot$, $L/L_{Edd} = 5$ parameter set (Table 4.1).

Comptonization model. Since the number of scatterings is proportional to the square of the optical depth, the appropriate average scattering temperature should be given by

$$T_{1D} = \frac{\int T \tau d\tau}{\int \tau d\tau}.$$ (4.9)

We test this description of 1D thermal Comptonization at $r = 8.5$, 9.5, 11, 14, 20, and 30 for the $M = 2 \times 10^6 M_\odot$, $L/L_{Edd} = 5$ parameter set (Table 4.1). At each radius, we add the wave temperature to the gas temperature and truncate the data inside $\tau_s = 10$, an optical depth that is large enough to result in significant Comptonization but small enough to prevent the saturation of photon spectra for our purposes. We place a 50eV Planck source at this location and calculate the emergent spectra. At each radius we also use the Kompaneets equation to pass the source through a homogeneous medium with temperature $T_{1D}$ and optical depth $\tau = 10$. We plot the resulting spectra for $r = 14$ in Figure 4.7. Spectra at the other radii illustrate the same effect and are plotted in Appendix F (Figure F1). The value of $T_{1D}$ at each radius is given in Table 4.2. We see that the spectrum calculated using the Kompaneets equation coincides with the spectrum computed directly from the data, confirming that unsaturated, 1D inhomogeneous thermal Comptonization is well modeled by homogeneous thermal Comptonization, with the temperature given by $T_{1D}$.

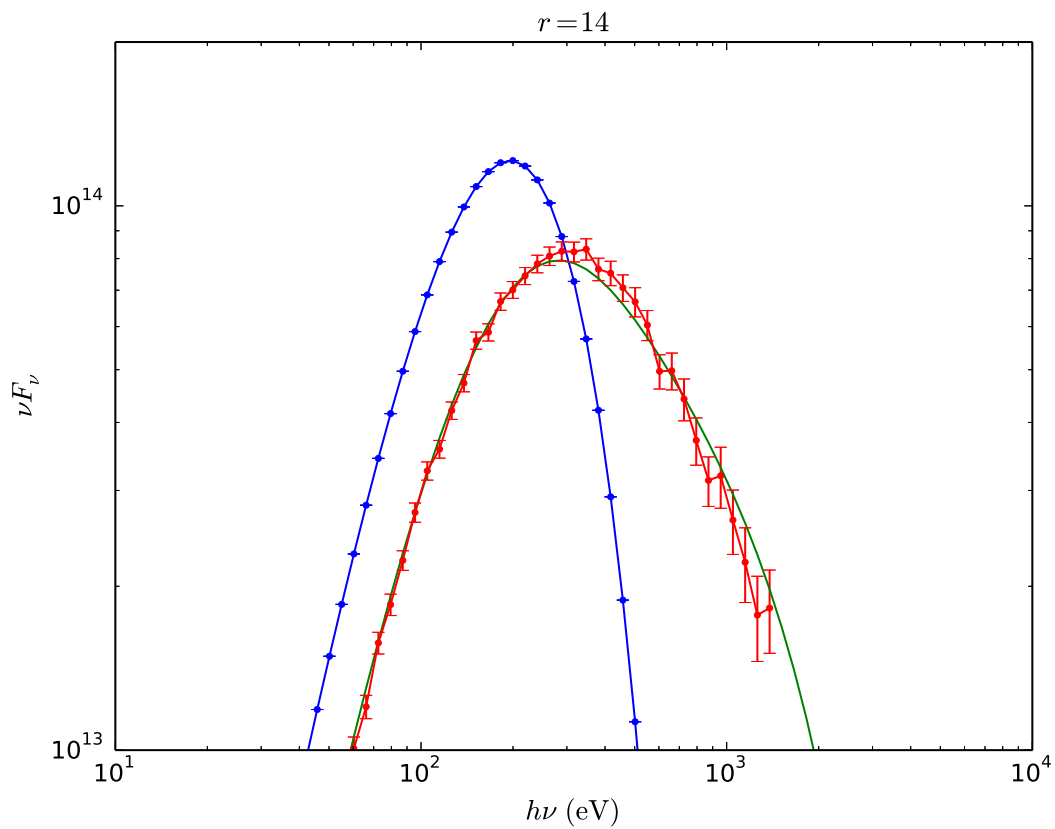

Figure 4.7: Normalized spectrum (red) computed by passing a 50eV Planck source (blue) through vertical structure data truncated at $\tau_s = 10$ at $r = 14$ for the $M = 2 \times 10^6 M_\odot$, $L/L_{\mathrm{Edd}} = 5$ parameter set (Table 4.1). The velocities are zeroed and the wave temperatures are added to the gas temperatures. The green curve is calculated by using the Kompaneets equation to pass the 50eV Planck source through a homogeneous medium with temperature $T_{\mathrm{1D}}$, given in Table 4.2. Spectra at other radii are plotted in Appendix F (Figure F1).

**Step 5 - Use the solution to the 1D thermal Comptonization problem to model bulk Comptonization**

Armed with the results of Step 4, we return to the original problem. We begin by observing that typically in the region between the effective photosphere and the scattering photosphere (that is, where $\tau_{\mathrm{eff}} < 1$ and $\tau_{\mathrm{s}} > 1$), the gas temperature does not vary significantly. The wave temperature, on the other hand, changes rapidly with density. It is negligible compared to the gas temperature at the bottom of the effective photosphere and increases moving outward. Near the scattering photosphere it may exceed the gas temperature, depending on the parameters of the problem. The effect of adding the wave temperature profile to the gas temperature profile, therefore, is to take the spectrum that results from when there is no wave temperature and pass it through a Comptonizing medium of optical depth given by that of the region where the wave temperature is comparable to the gas temperature. We define this to be the region in which the wave temperature is at least half the gas temperature. We refer to it as the bulk Comptonization region and define the Comptonization optical depth parameter $\tau_{\mathrm{C}}$ to be its optical depth. We then define the associated Comptonization temperature parameter $T_{\mathrm{C}}$ by equation (4.9), where $T$ is the sum of the gas and wave temperatures.

For example, we calculate spectra with and without velocities at $r = 8.5$, 9.5, 11, 14, 20, and 30 for the $M = 2 \times 10^6 M_{\odot}$, $L/L_{\mathrm{Edd}} = 5$ parameter set (Table 4.1). At each radius we also use the Kompaneets equation to pass the spectrum computed without velocities through a homogeneous medium with temperature $T_{\mathrm{C}}$ and optical depth $\tau_{\mathrm{C}}$. We plot the resulting spectra for $r = 14$ in Figure 4.8. Spectra at the other radii illustrate the same effect and are plotted in Appendix F (Figure F2). The temperature and optical depth parameters at all radii are given in Table 4.3. We see that the spectrum computed with the Kompaneets equation approximates the spectrum computed with velocities, indicating that the Comptonization parameters $T_C$ and $\tau_C$ approximate the Kompaneets parameters, $T_K$ and $\tau_K$.

| $r$ | 8.5 | 9.5 | 11 | 14 | 20 | 30 |
|---|---|---|---|---|---|---|
| $T_C$ (eV) | 210 | 246 | 251 | 253 | 203 | 149 |
| $\tau_C$ | 11 | 13 | 16 | 17 | 18 | 16 |

Table 4.3: Comptonization temperatures and optical depths at multiple radii for the $M = 2 \times 10^6 M_\odot$, $L/L_{Edd} = 5$ parameter set (Table 4.1).

| $r$ | 8.5 | 9.5 | 11 | 14 | 20 | 30 |
|---|---|---|---|---|---|---|
| $T_C$ (eV) | 75 | 80 | 95 | 90 | 80 | 52 |
| $\tau_C$ | 18 | 24 | 24 | 27 | 25 | 26 |

Table 4.4: Comptonization temperatures and optical depths at multiple radii for the $M = 2 \times 10^8 M_\odot$, $L/L_{Edd} = 5$ parameter set (Table 4.1).

To demonstrate that the effectiveness of the parameters $T_C$ and $\tau_C$ at describing bulk Comptonization is not limited to a narrow mass range, we modify the parameter set by choosing a significantly higher mass, $M/M_\odot = 2 \times 10^8$ (Table 4.1). We again calculate spectra at multiple radii, with and without velocities, and plot the results for $r = 14$ in Figure 4.9. In the same figure we plot the spectrum predicted by the parameters $T_C$ and $\tau_C$ for $r = 14$, given in Table 4.4, and see that the resulting spectrum well approximates the spectrum computed with velocities. Spectra at all radii are plotted in Appendix F (Figure F3).

Finally, we show that the Comptonization parameters $T_C$ and $\tau_C$ well describe bulk Comptonization for the $M = 2 \times 10^6 M_\odot$, $L/L_{Edd} = 2.5$ parameter set (Table 4.1), whose value of $L/L_{Edd}$ is the same as in Chapter 3. The corresponding spectra for $r = 14$ are plotted in Figure 4.10, and the Comptonization parameters are given in Table 4.5. We see that the spectrum predicted by the parameters $T_C$ and $\tau_C$ well approximates the spectrum computed with velocities. Spectra at all radii are plotted in Appendix F (Figure F4).

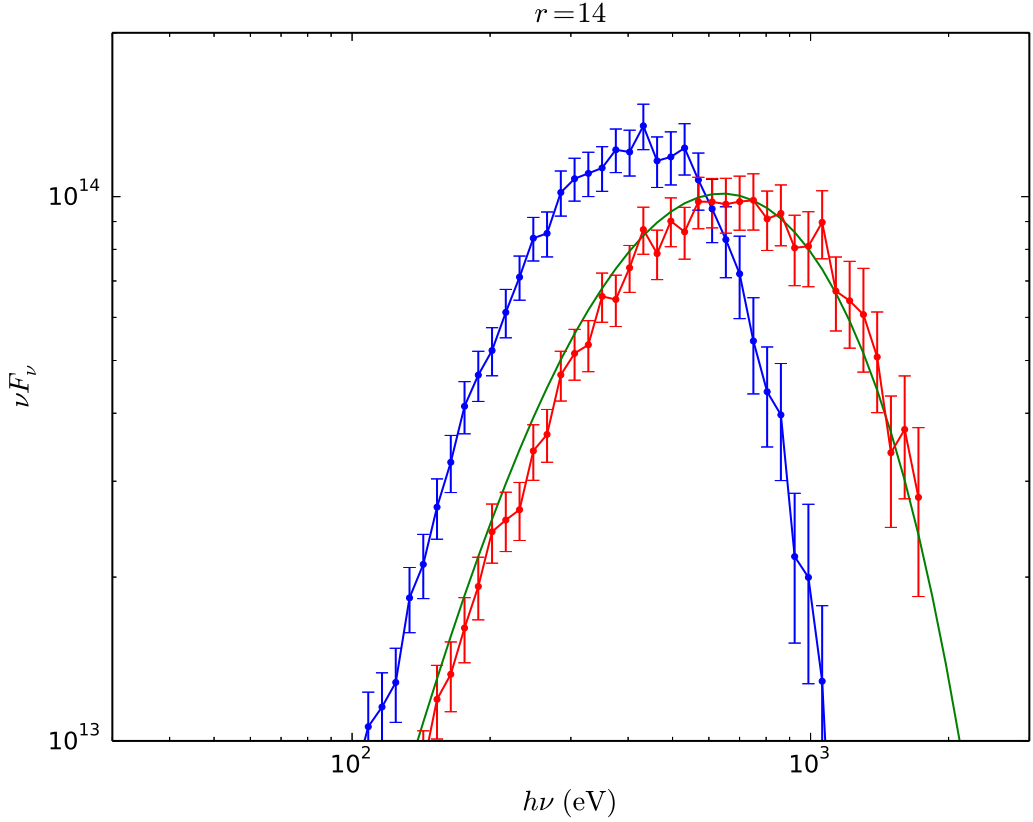

Figure 4.8: Normalized spectra at $r = 14$ for the $M = 2 \times 10^6 M_\odot$, $L/L_{\mathrm{Edd}} = 5$ parameter set (Table 4.1) computed with (red) and without (blue) velocities. The green curve is calculated by using the Kompaneets equation to pass the blue curve through a homogeneous Comptonizing medium with parameters $T_{\mathrm{C}}$ and $\tau_{\mathrm{C}}$, given in Table 4.3. Spectra at other radii are plotted in Appendix F (Figure F2).

| $r$ | 8.5 | 9.5 | 11 | 14 | 20 | 30 |
|---|---|---|---|---|---|---|
| $T_{\mathrm{C}}$ (eV) | 0 | 160 | 162 | 159 | 133 | 100 |
| $\tau_{\mathrm{C}}$ | 0 | 6.7 | 9.5 | 11 | 11 | 8.0 |

Table 4.5: Comptonization temperatures and optical depths at multiple radii for the $M = 2 \times 10^6 M_\odot$, $L/L_{\mathrm{Edd}} = 2.5$ parameter set (Table 4.1).

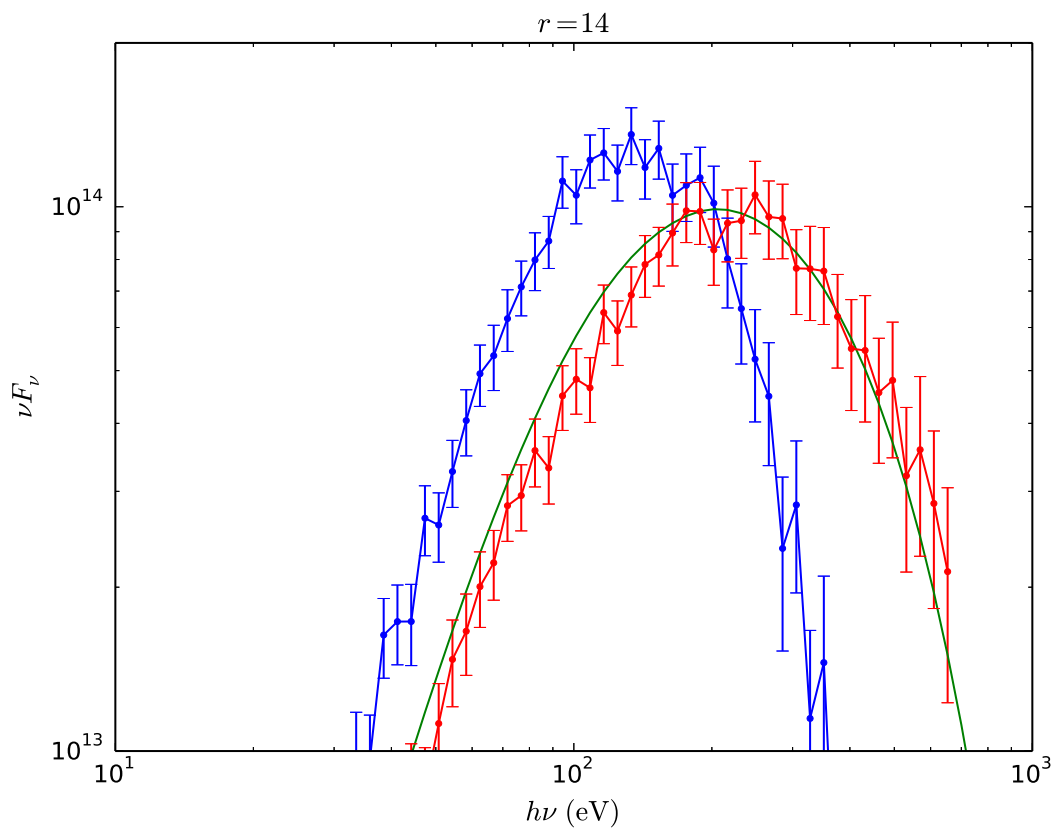

Figure 4.9: Normalized spectra at $r = 14$ for the $M = 2 \times 10^8 M_\odot$, $L/L_{\text{Edd}} = 5$ parameter set (Table 4.1) computed with (red) and without (blue) velocities. The green curve is calculated by using the Kompaneets equation to pass the blue curve through a homogeneous Comptonizing medium with parameters $T_{\text{C}}$ and $\tau_{\text{C}}$, given in Table 4.4. Spectra at other radii are plotted in Appendix F (Figure F3).

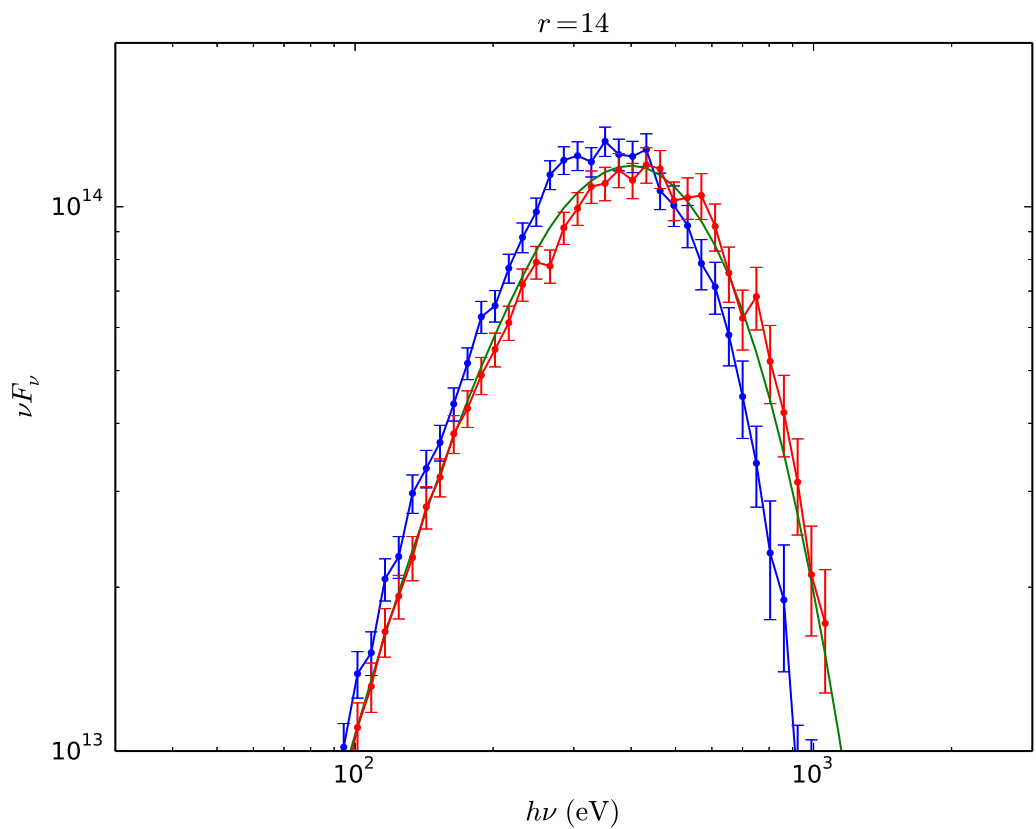

Figure 4.10: Normalized spectra at $r = 14$ for the $M = 2 \times 10^6 M_\odot$, $L/L_{\mathrm{Edd}} = 2.5$ parameter set (Table 4.1) computed with (red) and without (blue) velocities. The green curve is calculated by using the Kompaneets equation to pass the blue curve through a homogeneous Comptonizing medium with parameters $T_{\mathrm{C}}$ and $\tau_{\mathrm{C}}$, given in Table 4.5. Spectra at other radii are plotted in Appendix F (Figure F4).

**Step 6 - Qualify the bulk Comptonization model scope**

In steps 1-5, we justified each step of a process that results in a simplified, physically revealing model for bulk Comptonization by turbulence, and demonstrated the success of this model for multiple radii, masses, and accretion rates. We acknowledge that underlying each step are various assumptions, some of which may not hold over the entire range of accretion disc parameters of interest. This is a limitation only if the sole goal is to reproduce the Kompaneets temperature and optical depth as they are originally defined. But in this work our primary goal is rather to characterize bulk Comptonization in a physically revealing way so that we can easily map out its dependence on a wide range of disc parameters and understand how this dependence itself may change depending on the robustness of certain features in the disc vertical structure. Therefore, each step of this process should be viewed more as a search for parameters that are physically revealing and easily calculated rather than as an attempt to merely speed up the calculation of the Kompaneets parameters.

For example, we may find that in some regimes the calculated Comptonization region optical depth is sufficiently large that photon spectra saturate, which violates an assumption we made in Step 4. In this case, the Comptonization temperature and optical depth will probably differ somewhat from the Kompaneets parameters. But since they would still, by definition, tell us the optical depth of the region in which bulk Comptonization is significant as well as the weighted sum of the gas and wave temperatures in this region, they would still provide a useful characterization of bulk Comptonization.

# 4.3  Results

## 4.3.1  Overview

The independent variables in radiation MHD shearing box simulations are the surface density $\Sigma$, the vertical epicyclic frequency $\Omega_z$, and the strain rate $\partial_x v_y$. Since our simulation data is limited, we use the scheme developed in Chapter 3 to scale data from one set of independent variables to another. This scheme also allows for the variation of $\alpha$ (Shakura & Sunyaev, 1973), defined as the ratio of the vertically integrated total pressure to the vertically integrated total stress.

In section 4.3.2 we calculate the dependence of bulk Comptonization on the four shearing box parameters $\Sigma$, $\Omega_z$, $\partial_x v_y$, and $\alpha$. In particular, we show that these four parameters can in practice be reduced to two parameters, $\Sigma$ and $\alpha^3 \Omega_z$, so that the dependence of bulk Comptonization on shearing box parameters can be illustrated in a single figure with multiple curves. In section 4.3.3 we show the dependence of bulk Comptonization on accretion disc mass, luminosity, radius, spin, and inner boundary condition. To do this, we examine the dependence of the shearing box parameters $\Sigma$ and $\Omega_z$ on these parameters. In section 4.3.4 we estimate bulk Comptonization for an entire disc by setting the radius equal to the value that contributes maximally to the luminosity.

The scaling scheme from Chapter 3 assumes that the radiation energy flux is carried by radiation diffusion, but it does allow for variation in the opacity $\kappa$. In section 4.3.5 we show that vertical radiation advection can be included indirectly by varying $\kappa$, and we examine the effect of this on bulk Comptonization. Finally, in section 4.3.6 we examine how bulk Comptonization is effected by time variability in the simulation data.

### 4.3.2 Dependence of bulk Comptonization on shearing box parameters $\Sigma$, $\Omega_z$, $\partial_x v_y$, and $\alpha$

**Reduction of four shearing box parameters to two**

To simplify the problem, we observe that for Newtonian disc scalings

$$\left( \frac{\partial_x v_y}{\partial_x v_{y,0}} \right) = \left( \frac{\Omega_z}{\Omega_{z,0}} \right) = \left( \frac{M}{M_0} \right)^{-1} \left( \frac{r}{r_0} \right)^{-3/2}. \tag{4.10}$$

For Kerr disc scalings, the strain rate scale factor is nearly equal to the vertical epicyclic frequency scale factor (Chapter 3). To show this, in Figure 4.11 we plot the ratio of these quantities for multiple values of black hole spin. We see that at worst they agree to within $\sim 6\%$. For the purpose of understanding bulk Comptonization, then, we can set these factors equal to each other.

Next, we eliminate the dependence on $\alpha$ by proving the following statements for the bulk Comptonization parameters $T_C$, $\tau_C$, and $y_{p,C}$ (the Compton $y$ parameter): For any constant $k$,

$$\tau_C \left( \Sigma, \Omega_z, k\alpha \right) = \tau_C \left( \Sigma, k^3 \Omega_z, \alpha \right) \tag{4.11}$$

$$T_C \left( \Sigma, \Omega_z, k\alpha \right) = \frac{1}{k} T_C \left( \Sigma, k^3 \Omega_z, \alpha \right) \tag{4.12}$$

$$y_{p,C} \left( \Sigma, \Omega_z, k\alpha \right) = \frac{1}{k} y_{p,C} \left( \Sigma, k^3 \Omega_z, \alpha \right). \tag{4.13}$$

To prove equation (4.11), we first show that for fixed $\Sigma$ the wave temperature scales as the bulk temperature. The wave temperature depends not only on the bulk temperature but on the density via the ratio of each length scale in the turbulence $\lambda_w$ to the photon mean free path $\lambda_p$ (Chapter 2). The scalings for these parameters are $\lambda_w \sim h$ and $\lambda_p \sim \rho^{-1}$, where $h$ is the disc scale height. Since for fixed $\Sigma$ the scaling for the density is given by $\rho \sim h^{-1}$, varying only other parameters leaves the ratio $\lambda_w/\lambda_p$ unchanged and hence the wave temperature scales as the bulk temperature.

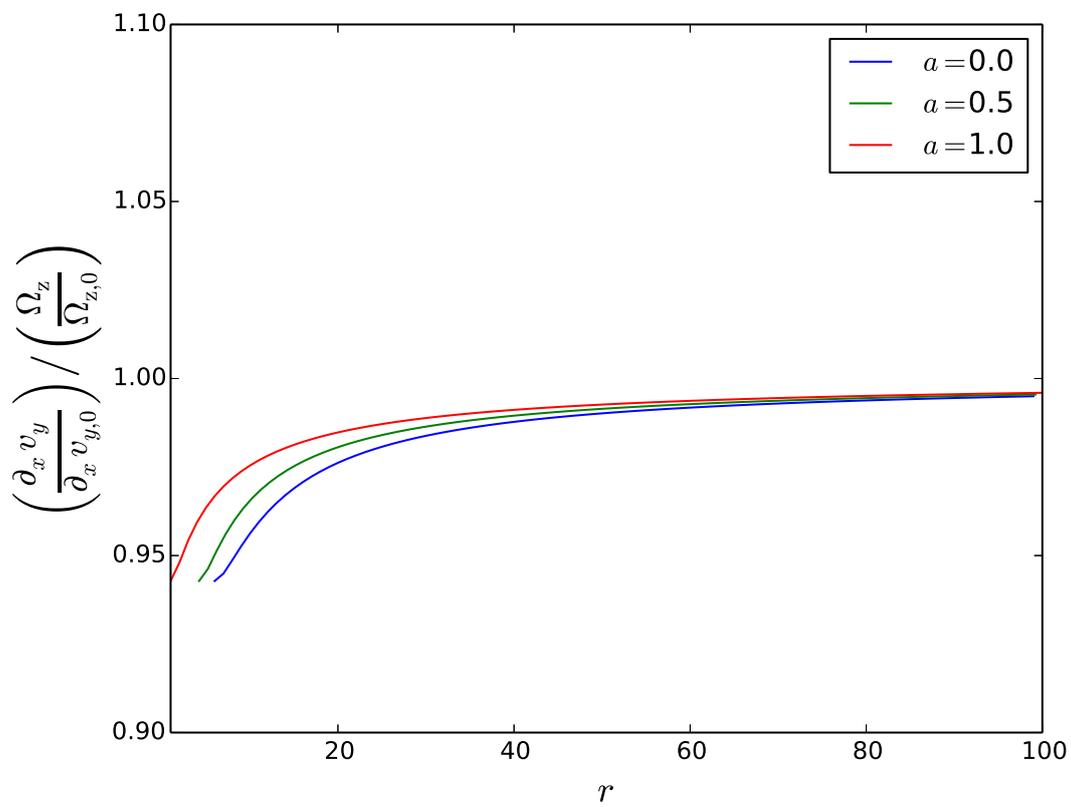

Figure 4.11: Ratio of the strain rate $\partial_x v_y$ scaling to the vertical epicyclic frequency $\Omega_z$ scaling for Kerr discs, for different values of the spin parameter $a$. Note that the maximum deviation from unity is only 6%.

Next, we observe that the ratios of the turbulent velocities to the thermal velocities in the midplane and the photosphere are given by (Chapter 3)

$$\frac{v_{\text{turb}}^2}{v_{\text{th,c}}^2} \sim \left(\alpha^3 \Omega_z\right)^{-1/4} \Sigma^{-2} \tag{4.14}$$

and

$$\frac{v_{\text{turb}}^2}{v_{\text{th,ph}}^2} \sim \left(\alpha^3 \Omega_z\right)^{-1/4} \Sigma^{-7/4}, \tag{4.15}$$

respectively. We see that if we vary the vertical epicyclic frequency inversely to $\alpha^3$, then the ratio of the turbulent kinetic energy to the thermal kinetic energy remains unchanged everywhere. Since for fixed $\Sigma$ the wave temperature is proportional to the gas temperature, it follows that the ratio of the wave temperature to the gas temperature is also everywhere unchanged. Then, since $\tau_C$ is defined as the optical depth of the region in which the wave temperature is comparable to the gas temperature (section 4.2), under these circumstances it can change only if the overall density or the scale height changes. That is,

$$\tau_C \left(\Sigma, k^{-3}\Omega_z, k\alpha\right) = \tau_C \left(\Sigma, \Omega_z, \alpha\right) \left(\frac{\rho_c}{\rho_{c,0}}\right) \left(\frac{h}{h_0}\right). \tag{4.16}$$

But $\rho_c h \sim \Sigma$, and $\Sigma$ is held constant since it is an independent variable, so $\rho_c$ varies inversely to $h$. In other words, for fixed $\Sigma$ the optical depth for any length scale is invariant. (Note that this is the exact same reason that the wave temperature scales as the bulk temperature.) It follows that

$$\tau_{\text{C}}\left(\Sigma, k^{-3}\Omega_z, k\alpha\right) = \tau_{\text{C}}\left(\Sigma, \Omega_z, \alpha\right).$$ (4.17)

Equation (4.11) follows directly from equation (4.17).

To prove equation (4.12), we start with the turbulent and thermal velocity scalings individually (Chapter 3), rather than the ratio of scalings:

$$v_{\text{turb}}^2 \sim \alpha^{-1}\Sigma^{-2}$$ (4.18)

$$v_{\text{th,c}}^2 \sim \left(\alpha^{-1}\Omega_z\right)^{1/4}$$ (4.19)

$$v_{\text{th,ph}}^2 \sim \left(\alpha^{-1}\Omega_z\right)^{1/4}\Sigma^{-1/4}.$$ (4.20)

We observe that if we vary the vertical epicyclic frequency inversely to $\alpha^3$, as we just showed we must do in order to leave $\tau_{\text{C}}$ unchanged, then the scalings for the individual variables are

$$v_{\text{turb}}^2 \sim \alpha^{-1}\Sigma^{-2}$$ (4.21)

$$v_{\text{th,c}}^2 \sim \alpha^{-1}$$ (4.22)

$$v_{\text{th,ph}}^2 \sim \alpha^{-1}\Sigma^{-1/4}.$$ (4.23)

We see that all velocities scale inversely to $\alpha$. Since the Comptonization temperature is defined as a density weighted average of the sum of the wave and gas temperatures (section 4.2), it follows that

$$T_{\mathrm{C}}\left(\Sigma, k^{-3}\Omega_z, k\alpha\right) = \frac{1}{k}T_{\mathrm{C}}\left(\Sigma, \Omega_z, \alpha\right).$$ (4.24)

Equation (4.12) follows directly from equation (4.24). Finally, the definition of the Compton $y$ parameter is

$$y_{\mathrm{p}} = \frac{4k_{\mathrm{B}}T}{m_{\mathrm{e}}c^2}N,$$ (4.25)

where $N$ is the average number of scatterings. For a plane parallel geometry with $\tau > 1$,[1] $N = 1.6\tau^2$, so

$$y_{\mathrm{p,C}} = 1.6\left(\frac{4k_{\mathrm{B}}T_{\mathrm{C}}}{m_{\mathrm{e}}c^2}\right)\tau_{\mathrm{C}}^2,$$ (4.26)

and equation (4.13) follows directly from equations (4.11) and (4.12).

Therefore, for the purpose of understanding bulk Comptonization, we can regard $\Sigma$ and $\alpha^3\Omega_z$ as the fundamental shearing box parameters and $\alpha T_{\mathrm{C}}$, $\tau_{\mathrm{C}}$, and $\alpha y_{\mathrm{p,C}}$ as the Comptonization parameters.

**Dependence of bulk Comptonization on $\Sigma$ and $\Omega_z$**

The original data we use is from ZEUS simulation 110304a (Chapter 3). The shearing box parameters for this simulation are given in Table 4.6. The time-averaged $\alpha$ parameter is $\alpha_0 = 0.01$ (which we do not list in Table 4.6 since it is not an independent variable). These correspond to an accretion disc annulus with parameters given in Table 4.7. The opacities included are electron scattering and free-free. These should be good approximations for the opacities in AGN in the near and super-Eddington regimes of interest in this work. At a given timestep, we scale the data to a range of values of $\Sigma$ and $\Omega_z$ with the scheme in Chapter 3, and then calculate the resulting Comptonization parameters $T_{\mathrm{C}}$, $\tau_{\mathrm{C}}$, and $y_{\mathrm{p,C}}$ with the procedure detailed in section 4.2. We repeat this for 21 timesteps spaced 10 orbital periods apart and plot the time-averaged results in Figure 4.12. The error at each point is estimated by dividing the sample standard deviation by the square root of the number of timesteps. For clarity we add the subscript "fid" to the shearing box parameters of the fiducial system given in Table 4.8 to distinguish them from the original simulation parameters which we denote by the subscript "0". The fiducial shearing box parameters correspond to $r = 20$ for the $M = 2 \times 10^6 M_\odot$, $L/L_{\mathrm{Edd}} = 2.5$ parameter set (Table 4.1). Just as we did in Chapter 3, we chose these parameters to be similar to those fit to the NLS1 REJ1034+396 by D12.

---

[1] Compton scattering is negligible for $\tau < 1$ since $k_{\mathrm{B}}T_{\mathrm{C}} \ll m_{\mathrm{e}}c^2$ for the systems we study here.

| Simulation | $\Omega_{z,0}$ (s$^{-1}$) | $\Sigma_0$ (g cm$^{-2}$) |
|---|---|---|
| 110304a | 186.6 | $2.5 \times 10^4$ |

Table 4.6: Original simulation shearing box parameters

| Simulation | $M/M_\odot$ | $r$ | $L/L_{\mathrm{Edd}}$ | $a$ | $\Delta\epsilon$ |
|---|---|---|---|---|---|
| 110304a | 6.62 | 30 | $\sim 1.7$ | 0 | 0 |

Table 4.7: Accretion disc parameters corresponding to the original simulation shearing box parameters

We can intuitively understand these results by looking at equations (4.18)-(4.20) and the horizontally averaged temperature profiles shown in Figure 4.13. In particular, we show below why $\tau_{\mathrm{C}}$ and $T_{\mathrm{C}}$ strongly increase with increasing $\Sigma^{-1}$, while $\tau_{\mathrm{C}}$ increases weakly and $T_C$ decreases weakly with increasing $\Omega_z^{-1}$. It then follows that since $y_{p,C}$ depends more strongly on $\tau_C$ than on $T_C$, $y_{p,C}$ increases strongly with increasing $\Sigma^{-1}$ and weakly with increasing $\Omega_z^{-1}$. Since $y_p$ is generally used as a proxy for the overall magnitude of Comptonization, we conclude that turbulent Comptonization increases strongly with increasing $\Sigma^{-1}$ and weakly with increasing $\Omega_z^{-1}$. Because of this as well as the fact that $\Omega_z^{-1} \sim M$ and $\Sigma^{-1} \sim L/L_{Edd}$ (which we discuss in section 4.3.3), we treat $\Omega_z^{-1}$ and $\Sigma^{-1}$ as the fundamental parameters rather than $\Omega_z$ and $\Sigma$.

| $(\Omega_{z,\mathrm{fid}}/\Omega_{z,0})^{-1}$ | $(\Sigma_{\mathrm{fid}}/\Sigma_0)^{-1}$ | $\alpha_{\mathrm{fid}}/\alpha_0$ |
|---|---|---|
| $1.6 \times 10^5$ | 4.0 | 2 |

Table 4.8: Fiducial shearing box parameters, corresponding to $r = 20$ for the $M = 2 \times 10^6 M_\odot$, $L/L_{\mathrm{Edd}} = 2.5$ parameter set (Table 4.1).

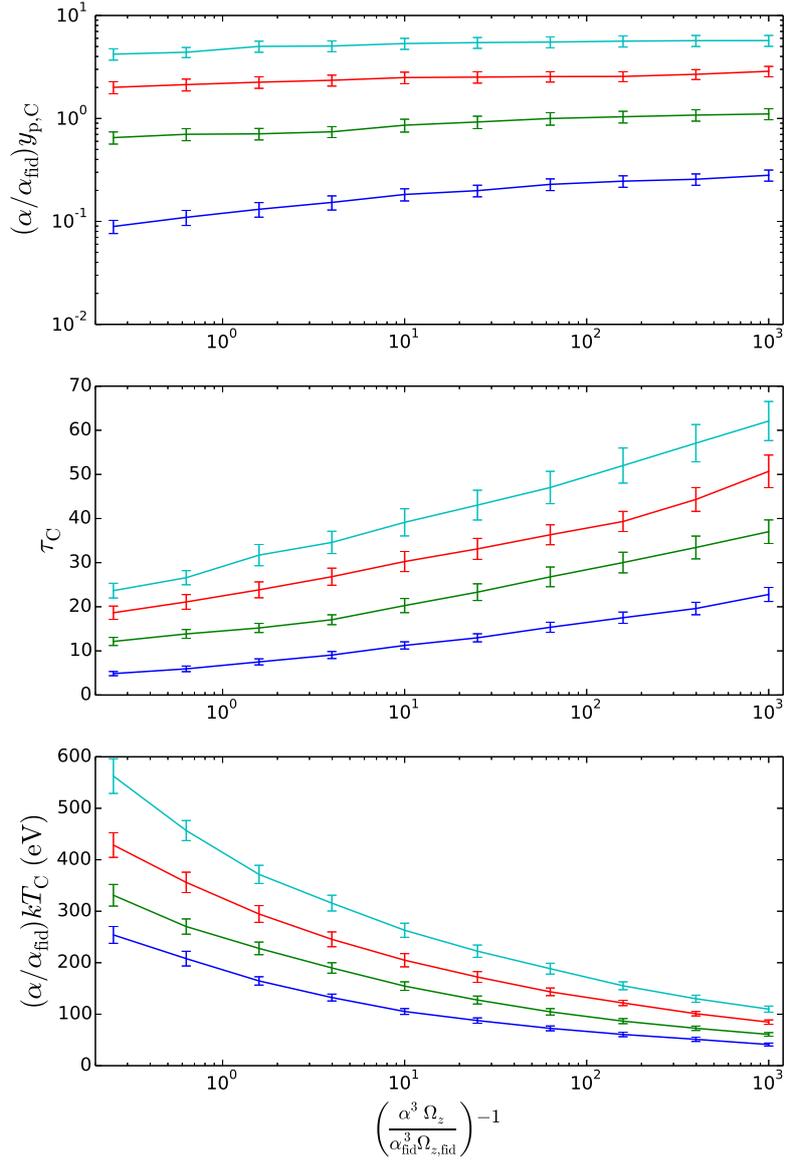

Figure 4.12: Dependence of bulk Comptonization on shearing box parameters. The blue, green, red, and cyan curves correspond to $(\Sigma/\Sigma_{\mathrm{fid}})^{-1} = 1$, 2, 3.3, and 5, respectively.

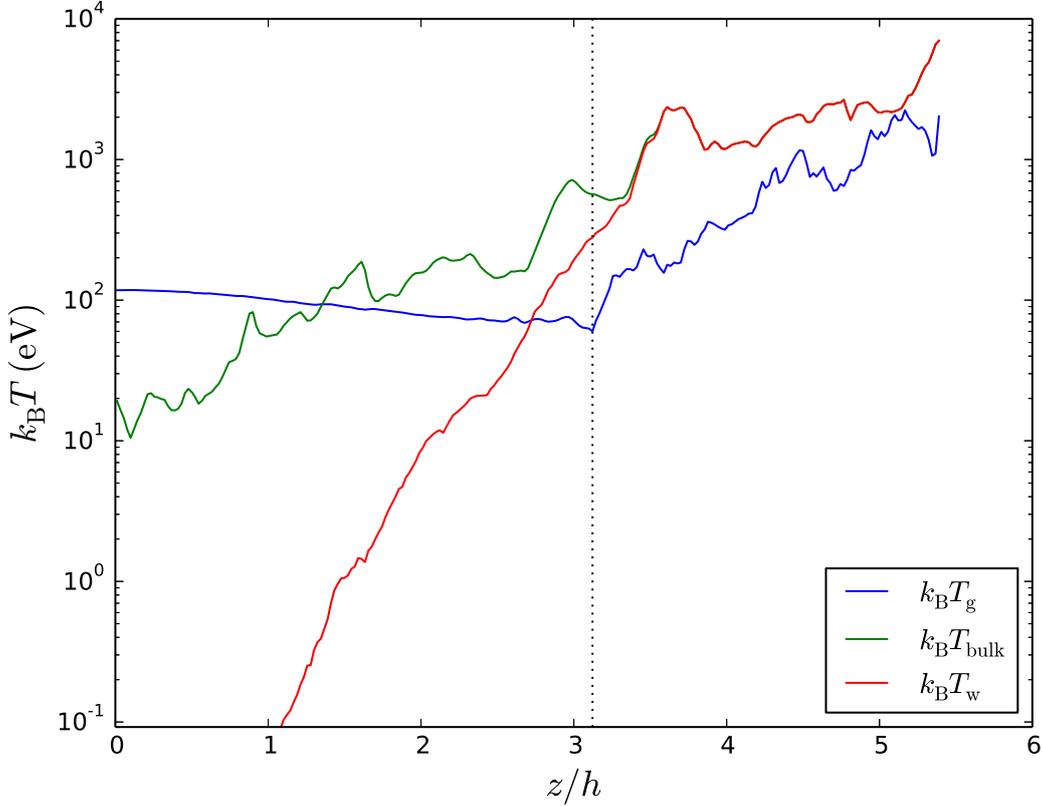

Figure 4.13: Horizontally averaged temperature profiles for the fiducial shearing box parameters, given in Table 4.8, at the 20 orbits timestep. The dashed line denotes where $\tau_s = 1$.

**Dependence of $\tau_C$ on $\Omega_z^{-1}$**    We start by considering the dependence of the bulk Comptonization optical depth $\tau_C$ on $\Omega_z^{-1}$. Equations (4.18)-(4.20) show that the result of varying $\Omega_z^{-1}$ is to multiply the entire gas temperature profile by a constant and leave the bulk temperature profile unchanged. Since the wave temperature scales as the bulk temperature for fixed $\Sigma^{-1}$, the wave temperature profile is also unchanged. Increasing $\Omega_z^{-1}$, therefore, corresponds to moving the gas temperature profile downward in Figure 4.13, increasing the optical depth of the region in which the wave temperature is comparable to the gas temperature, consistent with the results shown in Figure 4.12.

**Dependence of $T_C$ on $\Omega_z^{-1}$**    To understand the dependence of the Comptonization temperature $T_C$ on $\Omega_z^{-1}$, we first need to look at the bulk Comptonization region weighted average gas temperature $T_{C,g}$ and wave temperature $T_{C,w}$, individually. We plot the time-averaged dependence of these two parameters on $\Omega_z^{-1}$ in Figure 4.14. As we already showed, increasing $\Omega_z^{-1}$ moves the gas temperature profile downward in Figure 4.13. Since the gas temperature profile does not spatially vary significantly in this region, equations (4.19) - (4.20) imply that $T_{C,g}$ will be approximately proportional to $\Omega_z^{1/4}$. Typically the gas temperature profile is slightly decreasing at the lower boundary of this region so that the dependence is slightly shallower than $\Omega_z^{1/4}$, which is what we find in Figure 4.14. Since increasing $\Omega_z^{-1}$ has no effect on the wave temperature profile, the only effect of increasing $\Omega_z^{-1}$ on $T_{C,w}$ is to decrease the height of the lower boundary of the bulk Comptonization region. As it decreases, $T_{C,w}$ also decreases since not only does the wave temperature profile decrease with increasing optical depth, but equation (4.6) gives greatest weight to $T_w$ in the region where the optical depth is the largest. Therefore, $T_{C,w}$ also decreases with increasing $\Omega_z^{-1}$ but less so than $T_{C,g}$, as we see in Figure 4.14. Since $T_C = T_{C,g} + T_{C,w}$, $T_C$ also decreases with increasing $\Omega_z^{-1}$ at a rate slightly faster than $T_{C,w}$ but slower than $T_{C,g}$, as we see in Figure 4.12.

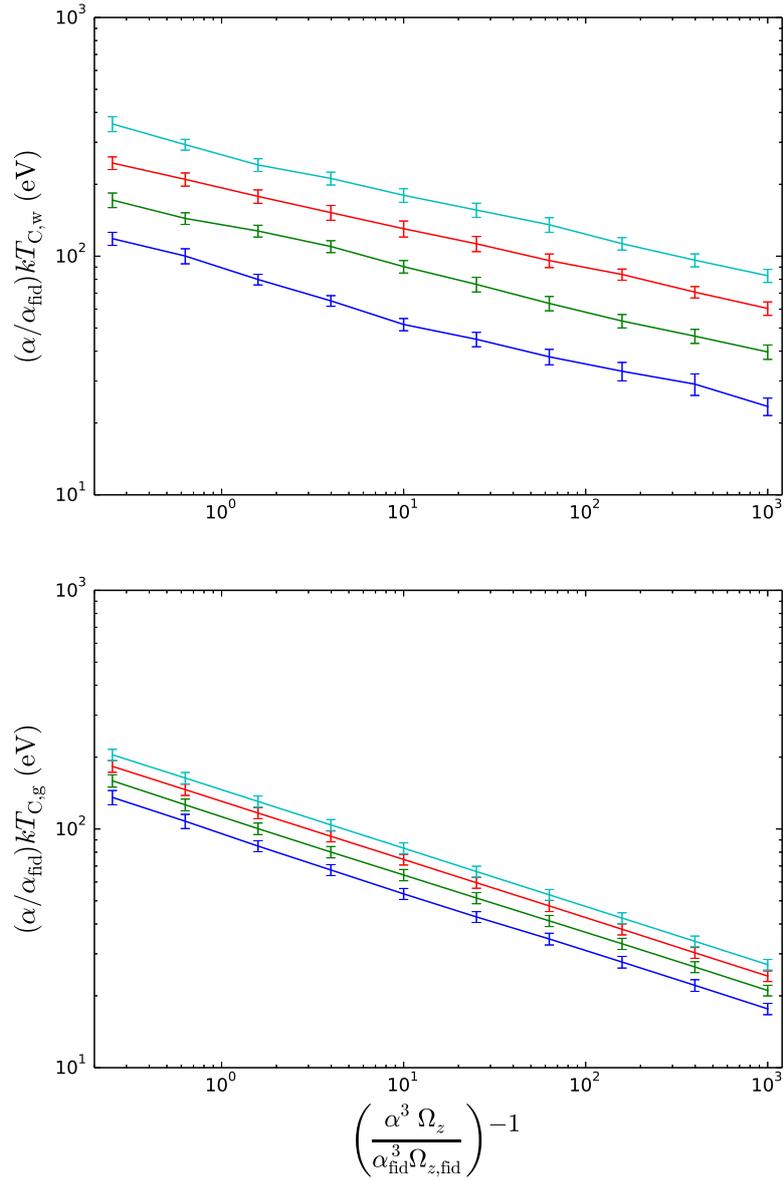

Figure 4.14: Dependence of Comptonization gas and wave temperatures on shearing box parameters. The blue, green, red, and cyan curves correspond to $(\Sigma/\Sigma_{\text{fid}})^{-1} = 1$, 2, 3.3, and 5, respectively.

**Dependence of $\tau_C$ on $\Sigma^{-1}$**    From equations (4.18)-(4.20) we see that $v_{\text{turb}}^2$ depends much more strongly on $\Sigma^{-1}$ than does $v_{\text{th}}^2$, so decreasing $\Sigma^{-1}$ moves the bulk temperature profile downward in Figure 4.13 relative to the gas temperature profile. The wave temperature profile moves downward even more than the bulk temperature profile, since the wave temperature depends on the velocity difference between scatterings, which decreases with decreasing $\Sigma^{-1}$. Therefore, the Comptonization optical depth decreases with decreasing $\Sigma^{-1}$. We expect the dependence of $\tau_C$ on $\Sigma^{-1}$ to be much stronger than its dependence on $\Omega_z^{-1}$ since $v_{\text{turb}}^2 \sim \Sigma^{-2}$ whereas $v_{\text{th}}^2 \sim \Omega_z^{1/4}$, which does not even take into account the fact that the wave temperature depends more strongly on $\Sigma^{-1}$ than does the bulk temperature. These conclusions are consistent with the results shown in Figure 4.12.

**Dependence of $T_C$ on $\Sigma^{-1}$**    To understand the dependence of the Comptonization temperature $T_C$ on $\Sigma^{-1}$, we first look at the dependence of $T_{C,g}$ and $T_{C,w}$, individually. If the gas and wave temperature profiles both decreased proportionally to the same power of $\Sigma^{-1}$, then both $T_{C,g}$ and $T_{C,w}$ would also decrease in proportion to this power of $\Sigma^{-1}$ because the size of the bulk Comptonization region would remain unchanged. But since the wave temperature profile decreases faster than the gas temperature profile, the effect on $T_{C,g}$ and $T_{C,w}$ also depends on other factors, such as the slopes of the gas and wave temperature profiles in the region. Since $v_{\text{turb}}^2 \sim \Sigma^{-2}$ gives a fairly strong dependence on density, we expect $T_{C,w}$ to uniformly decrease with decreasing $\Sigma^{-1}$. But since $v_{\text{th,ph}}^2 \sim \Sigma^{-1/4}$ gives a very weak dependence on density, it is hard to see whether $T_{C,g}$ will increase or decrease with decreasing $\Sigma^{-1}$. Either way, we expect the dependence of $T_{C,g}$ on $\Sigma^{-1}$ to be weaker. Figure 4.14 confirms these expectations. Finally, since $T_C = T_{C,g} + T_{C,w}$, and since by definition the main contribution to $T_C$ is from $T_{C,w}$, we expect $T_C$ to strongly increase with $\Sigma^{-1}$. Figure 4.12 confirms this expectation.

**Dependence of bulk Comptonization on the Reynolds stress fraction**

So far we have assumed that the $\beta$ parameter, defined as the ratio of the vertically integrated Reynolds stress to the vertically integrated total stress, is held constant (Chapter 3). We note that this is not to be confused with the plasma $\beta$, which is the ratio of the plasma pressure to the magnetic pressure. In radiation MHD simulations it is typically found that $\beta \sim 0.2$. We now show how varying $\beta$ affects bulk Comptonization. The turbulent velocity scaling, equation (4.18), becomes (Chapter 3)

$$v_{\mathrm{turb}}^2 \sim \alpha^{-1} \beta \Sigma^{-2}, \tag{4.27}$$

while the thermal velocity scalings, equations (4.19)-(4.20), remain unchanged. Since the dependence of $v_{\mathrm{th,ph}}$ on $\Sigma^{-1}$ is weak, we expect the dependence of bulk Comptonization on $\beta$ to be similar to its dependence on $\Sigma^{-2}$. In Figure 4.15 we plot the dependence of the bulk Comptonization parameters on $\Omega_z^{-1}$ for $\beta/\beta_0 = 1$ and 4. As expected, we see the resulting curves are similar to those in Figure 4.12 corresponding to $(\Sigma/\Sigma_{\mathrm{fid}})^{-1} = 1$ and 2, respectively. We note that unlike the scaling for $\Sigma^{-2}$, the scaling for $\beta$ is restricted to a much narrower range since $\beta \leq 1$. In the rest of this work we suppress the dependence on $\beta$ for clarity.

## 4.3.3 Dependence of bulk Comptonization on accretion disc parameters

Now that we have analyzed in detail the dependence of bulk Comptonization on the shearing box parameters $\Omega_z^{-1}$ and $\Sigma^{-1}$, we proceed by relating these to the underlying

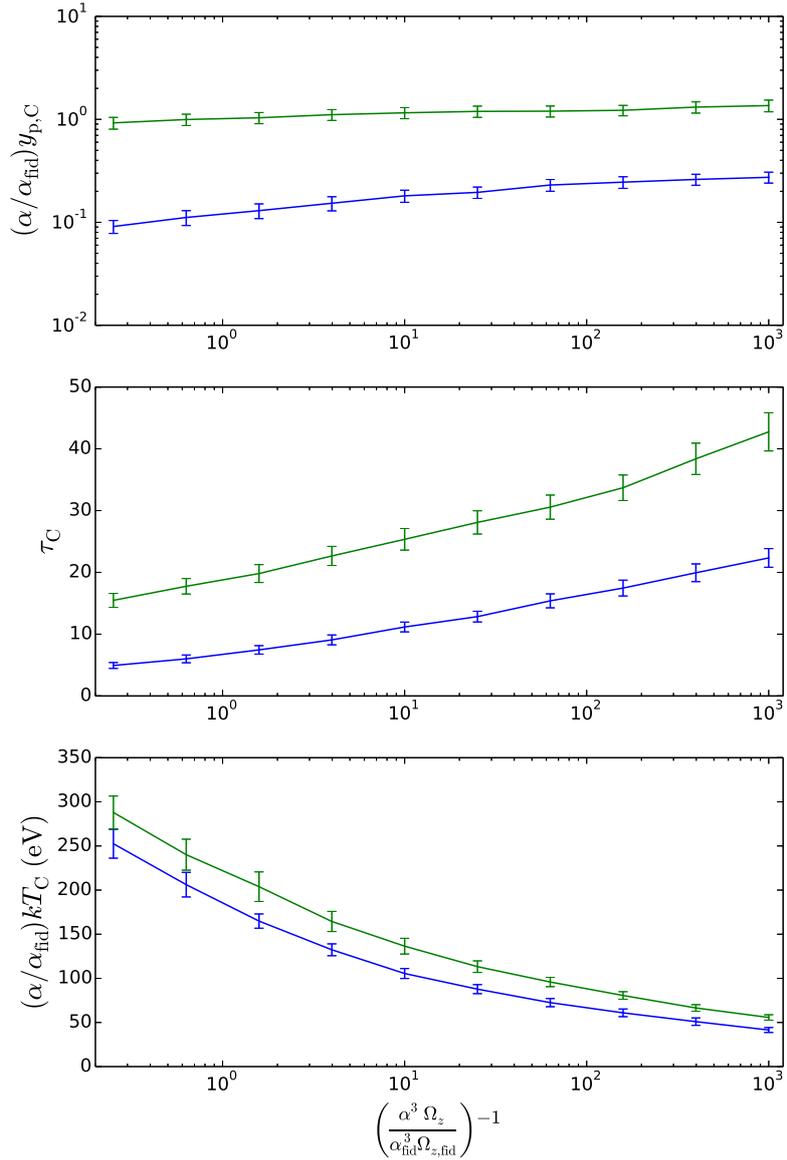

Figure 4.15: Dependence of bulk Comptonization on shearing box parameters for $\beta/\beta_0 = 1$ (blue) and 4 (green). For all curves, $\Sigma = \Sigma_{\mathrm{fid}}$.

accretion disc parameters. First we write $\Sigma^{-1}$ in terms of the local flux $F$ (Chapter 3):

$$\Sigma^{-1} \sim \alpha \Omega_z^{-1} F. \tag{4.28}$$

This says that for fixed $\Omega_z^{-1}$, $\Sigma^{-1}$ is simply proportional to the local flux. Next, we need the scalings for $\Omega_z^{-1}$ and $F$ as functions of the accretion disc parameters. For Newtonian discs, they are

$$\left(\frac{\Omega_z}{\Omega_{z,0}}\right) = \left(\frac{M}{M_0}\right)^{-1} \left(\frac{r}{r_0}\right)^{-3/2} \tag{4.29}$$

and

$$\left(\frac{F}{F_0}\right) = \left(\frac{M}{M_0}\right)^{-1} \left(\frac{r}{r_0}\right)^{-3} \left(\frac{\dot{m}}{\dot{m}_0}\right) \left(\frac{\eta + \Delta\epsilon}{\eta_0 + \Delta\epsilon_0}\right)^{-1}$$
$$\left(\frac{1 - \sqrt{r_{\text{in}}/r} + \left(\sqrt{r_{\text{in}}/r}\right) r_{\text{in}}\Delta\epsilon}{1 - \sqrt{r_{\text{in},0}/r_0} + \left(\sqrt{r_{\text{in},0}/r_0}\right) r_{\text{in},0}\Delta\epsilon_0}\right), \tag{4.30}$$

where $\eta$ is the efficiency assuming a no torque inner boundary condition, $\Delta\epsilon$ is the change in efficiency due to a non-zero torque inner boundary condition (Agol & Krolik, 2000), $r_{\text{in}}$ is the inner radius of the disc, and

$$\dot{m} = L/L_{\text{Edd}}. \tag{4.31}$$

For Kerr discs they are (Chapter 3)

$$\left(\frac{\Omega_z}{\Omega_{z,0}}\right) = \left(\frac{M}{M_0}\right)^{-1} \left(\frac{r}{r_0}\right)^{-3/2} \left(\frac{C}{C_0}\right)^{1/2} \left(\frac{B}{B_0}\right)^{-1/2} \tag{4.32}$$

and

$$\left(\frac{F}{F_0}\right) = \left(\frac{M}{M_0}\right)^{-1} \left(\frac{r}{r_0}\right)^{-3} \left(\frac{\dot{m}}{\dot{m}_0}\right) \left(\frac{\eta + \Delta\epsilon}{\eta_0 + \Delta\epsilon_0}\right)^{-1} \left(\frac{B}{B_0}\right)^{-1}$$
$$\left(\frac{r_{\rm in}^{3/2} B(r_{\rm in})^{1/2} \Delta\epsilon r^{-1/2} + D}{r_{\rm in,0}^{3/2} B(r_{\rm in,0})^{1/2} \Delta\epsilon_0 r_0^{-1/2} + D_0}\right), \tag{4.33}$$

where $B$, $C$, and $D$ are functions of $r$ and the spin parameter $a$, and go to unity for $r \gg r_{\rm in}$. In order that the scalings for both Newtonian and Kerr discs be functions of the same underlying parameters, for Newtonian discs we set $r_{\rm in}$ equal to the innermost stable circular orbit, which is in turn a function of the black hole spin parameter $a$.

We note that since $\dot{m} = L/L_{\rm Edd}$, one should not think of $\dot{m}$ as the mass accretion rate. For example, for fixed mass $M$ and fixed $\dot{m}$, if we vary $\eta + \Delta\epsilon$ (by varying the spin or the inner boundary condition) then the luminosity is unchanged since $L = \dot{m} L_{\rm Edd}$ and $L_{\rm Edd}$ is proportional only to $M$ ($L_{\rm Edd} = 4\pi G M m_{\rm p} c/\sigma_{\rm T}$, so all the other parameters are constants of physics). But the mass accretion rate is NOT unchanged since $\dot{M} = L/(\eta + \Delta\epsilon) c^2$. In other words, when varying other parameters (except for the mass) at fixed $\dot{m}$, one should think of this as varying the mass accretion rate at fixed luminosity.

We will find that it is helpful to reduce the above equations to the following simplified form. For the Newtonian scalings,

$$\Omega_{\rm z}^{-1} \sim M r^{3/2} \tag{4.34}$$

and

$$\Sigma^{-1} \sim \alpha r^{-3/2} \left(L/L_{\rm Edd}\right) (\eta + \Delta\epsilon)^{-1} \left(1 - \sqrt{r_{\rm in}/r} + \left(\sqrt{r_{\rm in}/r}\right) r_{\rm in} \Delta\epsilon\right). \tag{4.35}$$

For the Kerr scalings,

$$\Omega_z^{-1} \sim M r^{3/2} C^{-1/2} B^{1/2} \tag{4.36}$$

and[2]

$$\Sigma^{-1} \sim \alpha r^{-3/2} \left(L/L_{\mathrm{Edd}}\right) \left(\eta + \Delta\epsilon\right)^{-1} B^{-1/2} C^{-1/2} \left(r_{\mathrm{in}}^{3/2} B(r_{\mathrm{in}})^{1/2} \Delta\epsilon r^{-1/2} + D\right). \tag{4.37}$$

We now examine the dependence of the Comptonization parameters $T_{\mathrm{C}}$, $\tau_{\mathrm{C}}$, and $y_{\mathrm{p,C}}$ on the accretion disc parameters.

**Dependence on mass**   The dependence of the bulk Comptonization parameters on mass is straightforward. Since $\Omega_z^{-1}$ is directly proportional to mass and $\Sigma^{-1}$ is independent of mass, the dependence of the bulk Comptonization parameters on mass is identical to their dependence on $\Omega_z^{-1}$. That is, $T_{\mathrm{C}}$ decreases weakly, $\tau_{\mathrm{C}}$ increases weakly, and $y_{\mathrm{p,C}}$ increases weakly with increasing mass. Furthermore, we can immediately regard the $\Omega_z^{-1}$ axis in Figures 4.12 and 4.14 as the mass axis.

**Dependence on luminosity**   The dependence of the bulk Comptonization parameters on $L/L_{\mathrm{Edd}}$ is straightforward. Since $\Sigma^{-1}$ is directly proportional to $L/L_{\mathrm{Edd}}$ and $\Omega_z^{-1}$ is independent of $L/L_{\mathrm{Edd}}$ for both Newtonian and Kerr discs, the dependence of the bulk Comptonization parameters on $L/L_{\mathrm{Edd}}$ is identical to their dependence on $\Sigma^{-1}$. That is, $T_{\mathrm{C}}$, $\tau_{\mathrm{C}}$, and $y_{\mathrm{p,C}}$ all increase strongly with increasing $L/L_{\mathrm{Edd}}$. Furthermore, we can immediately regard the curves corresponding to different values of $\Sigma^{-1}$ in Figures 4.12 and 4.14 as corresponding to different values of $L/L_{\mathrm{Edd}}$. In section 4.5 we reproduce the plots from Figure 4.12 with the independent variables relabeled in order to summarize the dependence of bulk Comptonization on mass and luminosity.

---

[2]The scaling for $\Sigma^{-1}$ differs slightly from that in Chapter 3 since here we have set the strain rate scaling equal to the scaling for the vertical epicyclic frequency, an excellent approximation for our purposes (section 4.3.2).

**Dependence on radius**  Both $\Omega_z^{-1}$ and $\Sigma^{-1}$ depend on $r$. But since all bulk Comptonization parameters depend strongly on $\Sigma^{-1}$ and weakly on $\Omega_z^{-1}$, their dependence on $r$ is almost entirely explained by the dependence of $\Sigma^{-1}$ on $r$. For $r \gg r_{\text{in}}$, $\Sigma^{-1}$ and hence the bulk Comptonization parameters increase with decreasing $r$ for both Newtonian and Kerr discs. For $\Delta\epsilon \ll 1$, $\Sigma^{-1}$ eventually begins to decrease as $r$ approaches $r_{\text{in}}$, after which the bulk Comptonization parameters begin to decrease with decreasing $r$. The precise value of $r$ below which the bulk Comptonization parameters begin to decrease differs slightly from the value of $r$ at which $\Sigma^{-1}$ begins to decrease because the bulk Comptonization parameters also depend on $r$ through $\Omega_z^{-1}$, albeit weakly. For example, in Figure 4.16 we plot the dependence of $\Sigma^{-1}$ and bulk Comptonization on $r$ for $\Delta\epsilon = 0$. We see that the dependence of bulk Comptonization on $r$ is well predicted by the variation in $\Sigma^{-1}$.

If, on the other hand, $\Delta\epsilon$ is large enough, then both $\Sigma^{-1}$ and the bulk Comptonization parameters monotonically increase with decreasing $r$, just as they do for $r \gg r_{\text{in}}$. This holds true for both Newtonian and Kerr discs. For example, in Figure 4.16 we also plot the dependence of $\Sigma^{-1}$ and the Comptonization parameters on $r$ for $\Delta\epsilon = 0.05$. We see that both $\Sigma^{-1}$ and the bulk Comptonization parameters uniformly increase with decreasing $r$.

**Dependence on spin**  For Newtonian discs $\Omega_z^{-1}$ is independent of the spin parameter $a$, and for Kerr discs $\Omega_z^{-1}$ depends on $a$ only for $r$ very close to $r_{\text{in}}$ through the functions $C$ and $B$. But since all bulk Comptonization parameters depend strongly on $\Sigma^{-1}$ and weakly on $\Omega_z^{-1}$, their dependence on $a$ is almost entirely explained by the dependence of

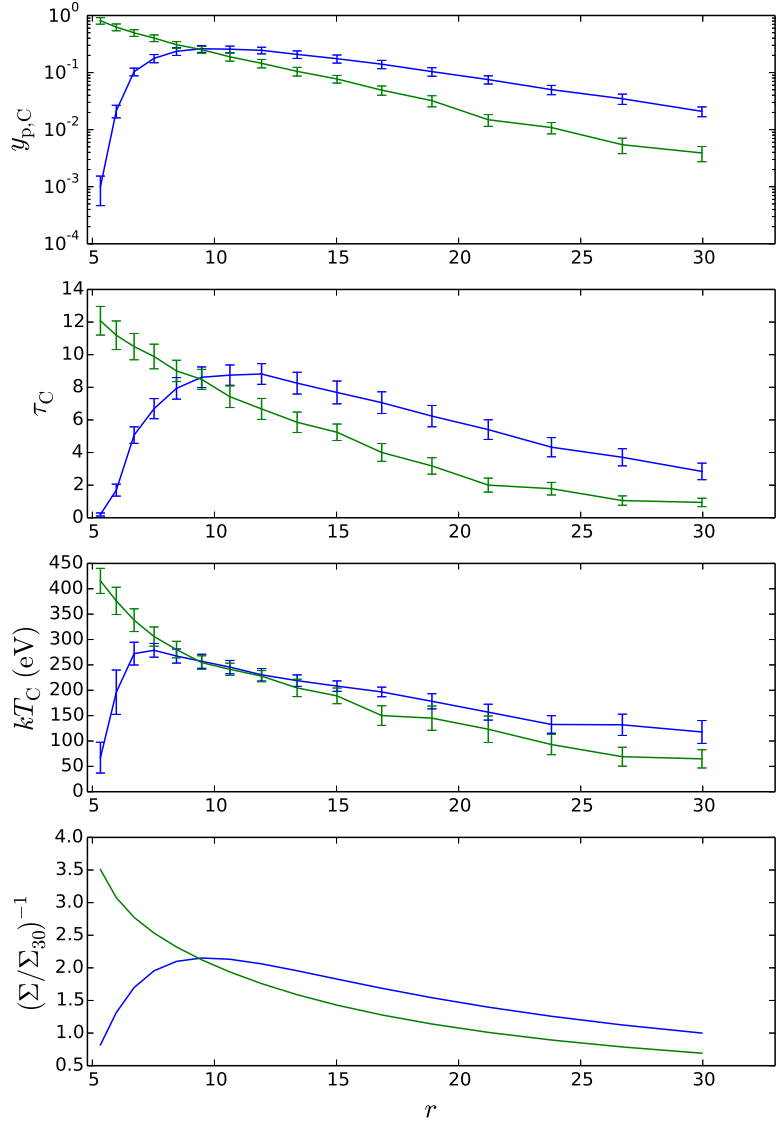

Figure 4.16: Dependence of bulk Comptonization and $\Sigma^{-1}$ on $r$ for $\Delta\epsilon = 0$ (blue) and $\Delta\epsilon = 0.05$ (green). The parameter $\Sigma_{30}$ denotes the surface density at $r = 30$ for $\Delta\epsilon = 0$. The values of the parameters held constant are $M/M_\odot = 2 \times 10^6$, $L/L_{\mathrm{Edd}} = 2.5$, $a = 0.5$, and $\alpha/\alpha_0 = 2$.

$\Sigma^{-1}$ on $a$. For $r \gg r_{\text{in}}$, the dependence of $\Sigma^{-1}$ on $a$ is given by

$$\Sigma^{-1} \sim \alpha r^{-3/2} \left(L/L_{\text{Edd}}\right) \left(\eta + \Delta\epsilon\right)^{-1}, \qquad (4.38)$$

where for both Newtonian and Kerr discs $\eta$ is a monotonically increasing (albeit different) function of $a$. We see that $\Sigma^{-1}$ decreases with increasing $a$. The reason for this is straightforward. We recall that for fixed $\Omega_z^{-1}$, $\Sigma^{-1}$ is proportional to the flux. As the spin and efficiency increase, the flux increases in the inner radii so for fixed luminosity $L/L_{\text{Edd}}$ the flux must decrease at large radii. The bulk Comptonization parameters therefore decrease at large radii with increasing spin. For example, in Figure 4.17 we plot the dependence of flux and bulk Comptonization on spin for $r = 20$, 12, and 7. We see that for $r = 20$, both flux and bulk Comptonization decrease with increasing spin.

For $r$ sufficiently close to $r_{\text{in}}$, on the other hand, since the flux increases with spin, so do the bulk Comptonization parameters. For example, in Figure 4.17 we see that for $r = 7$ flux and bulk Comptonization increase with spin until $a \approx 1$. This is expected because as $a$ approaches 1, $r_{\text{in}}$ approaches 1 and so $r = 7$ is no longer close to $r_{\text{in}}$.

For an intermediate value of $r$ (at which flux does not monotonically increase or decrease with spin), the dependence of the bulk Comptonization parameters on spin can still be understood by simply plotting the flux as a function of spin. For example, in Figure 4.17 we see that for $r = 12$ the dependence of bulk Comptonization on spin tracks the variation in flux.

**Dependence on inner boundary condition**    The dependence of bulk Comptonization on the inner boundary condition is very similar to the dependence on spin. Since $\Sigma^{-1}$ is proportional to the flux for fixed $\Omega_z^{-1}$, the dependence of bulk Comptonization on the inner boundary condition follows the variation in the flux. The inner boundary is pa-rameterized in terms of $\Delta\epsilon$, the change in efficiency due to a non-zero inner torque. Since increasing $\Delta\epsilon$ increases the flux in the inner radii, bulk Comptonization increases with increasing $\Delta\epsilon$ in this region. At large radii, increasing $\Delta\epsilon$ at fixed luminosity decreases the flux so that bulk Comptonization also decreases.

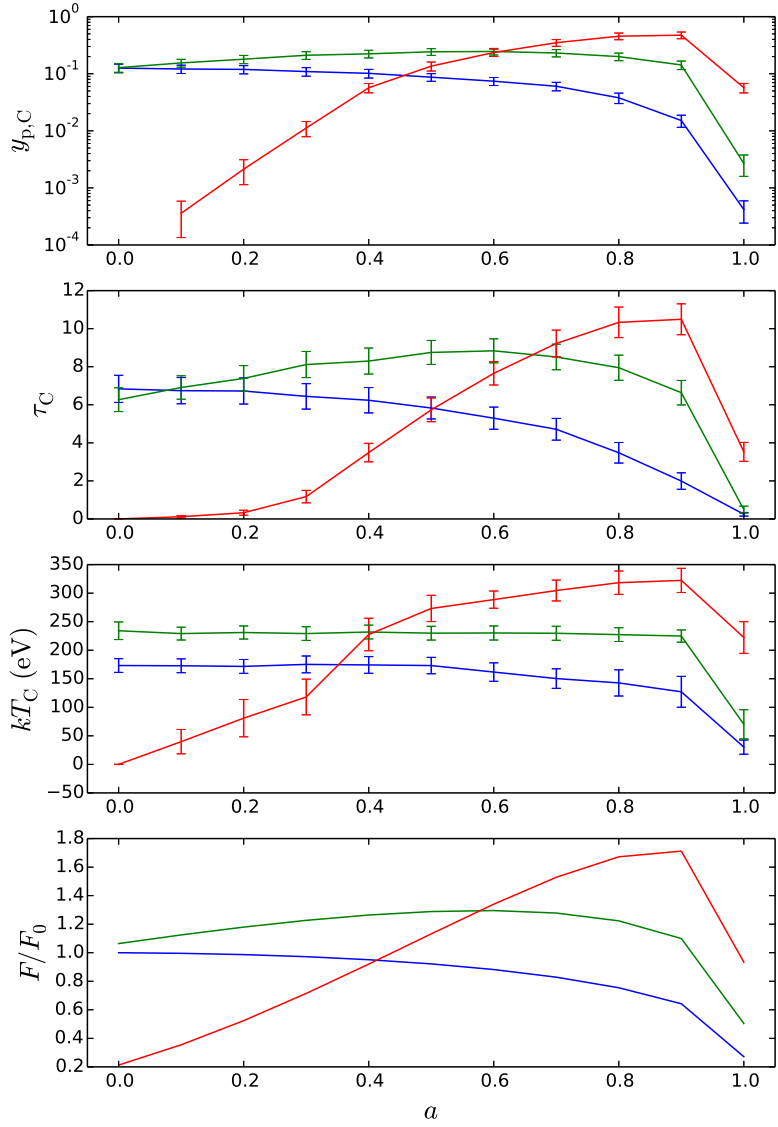

Figure 4.17: Dependence of flux and bulk Comptonization on $a$ for $r = 20$ (blue), $r = 12$ (green), and $r = 7$ (red). The parameter $F_0$ denotes the flux at $a = 0$ for $r = 20$. The values of the parameters held constant are $M/M_\odot = 2 \times 10^6$, $L/L_{Edd} = 2.5$, $\Delta\epsilon = 0$, and $\alpha/\alpha_0 = 2$.

For example, in Figure 4.18 we plot the dependence of flux and bulk Comptonization on $\Delta\epsilon$ at large ($r = 20$) and small ($r = 7$) radii. In both cases we see that bulk Comptonization follows the variation in the flux.

**Dependence on $\alpha$**   For fixed $\Sigma^{-1}$ and $\Omega_z^{-1}$, the variation of bulk Comptonization with $\alpha$, given by equations (4.11)-(4.13), is reflected in Figure 4.12. For $\alpha = \alpha_{\text{fid}}$, these plots are uncomplicated. Multiplying $\alpha$ by a constant $k > 1$ translates each curve for $\tau_{\text{C}}$ to the right on a log scale. For $T_{\text{C}}$ and $y_{\text{p,C}}$, multiplying $\alpha$ by $k$ not only translates each curve to the right but also multiplies each curve by $1/k$. Since we plot $y_{\text{p,C}}$ on a log scale, for this variable multiplying $\alpha$ by $k$ is equivalent to moving each curve to the right and downward.

Alternatively, one can think of multiplying $\alpha$ by a constant $k > 1$ as moving leftward along each curve for $\tau_{\text{C}}$. For a given value of $T_{\text{C}}$ and $y_{\text{p,C}}$, multiplying $\alpha$ by $k$ is equivalent to not only moving leftward but also dividing the resultant value by $k$. To develop physical intuition, we plot the dependence of bulk Comptonization on $\Omega_z^{-1}$ for multiple values of $\alpha$ in Figure 4.19. We see that bulk Comptonization overall decreases moderately with increasing $\alpha$.

But in an accretion disc, we see that $\Sigma^{-1}$ itself is directly proportional to $\alpha$ via the flux. The effect of varying $\alpha$ as an accretion disc parameter, then, affects bulk Comptonization primarily by varying $\Sigma^{-1}$. Since bulk Comptonization increases strongly with $\Sigma^{-1}$, increasing $\alpha$ generally increases bulk Comptonization. For example, in Figure 4.20 we plot the dependence of the flux and bulk Comptonization on $\alpha$. We see that bulk

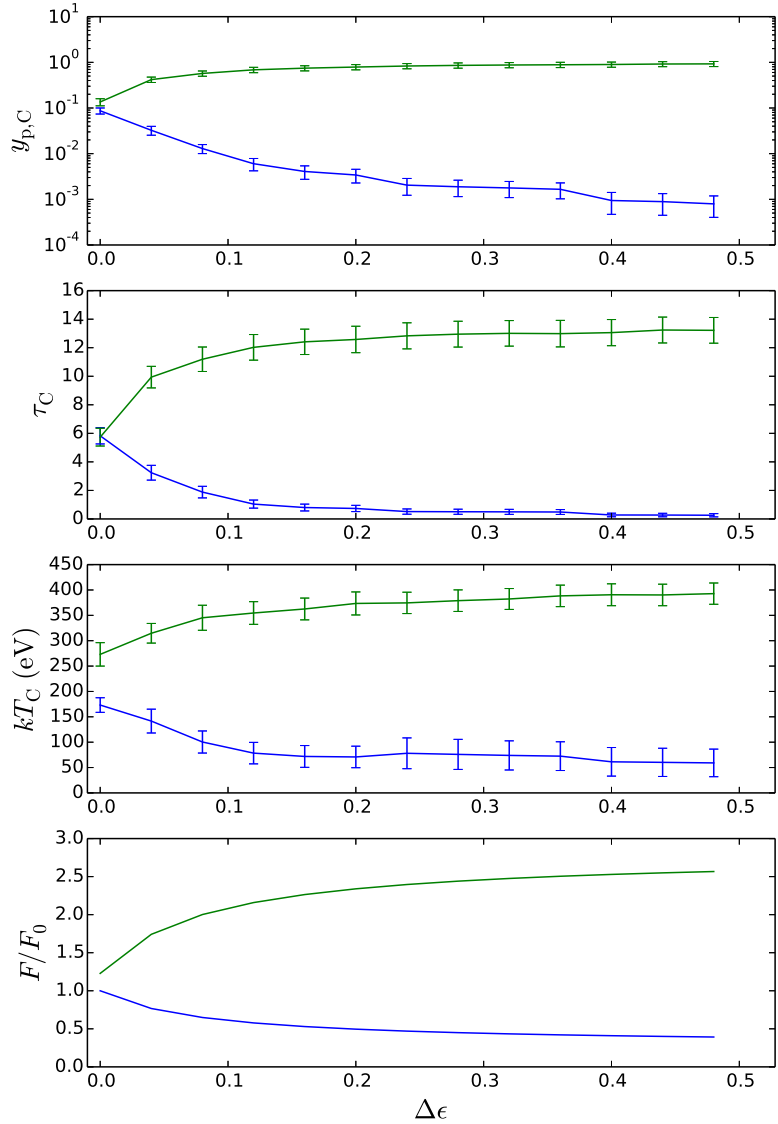

Figure 4.18: Dependence of flux and bulk Comptonization on $\Delta\epsilon$ for $r = 20$ (blue) and $r = 7$ (green). The parameter $F_0$ denotes the flux at $\Delta\epsilon = 0$ for $r = 20$. The values of the parameters held constant are $M/M_\odot = 2 \times 10^6$, $L/L_{\mathrm{Edd}} = 2.5$, $a = 0.5$, and $\alpha/\alpha_0 = 2$.

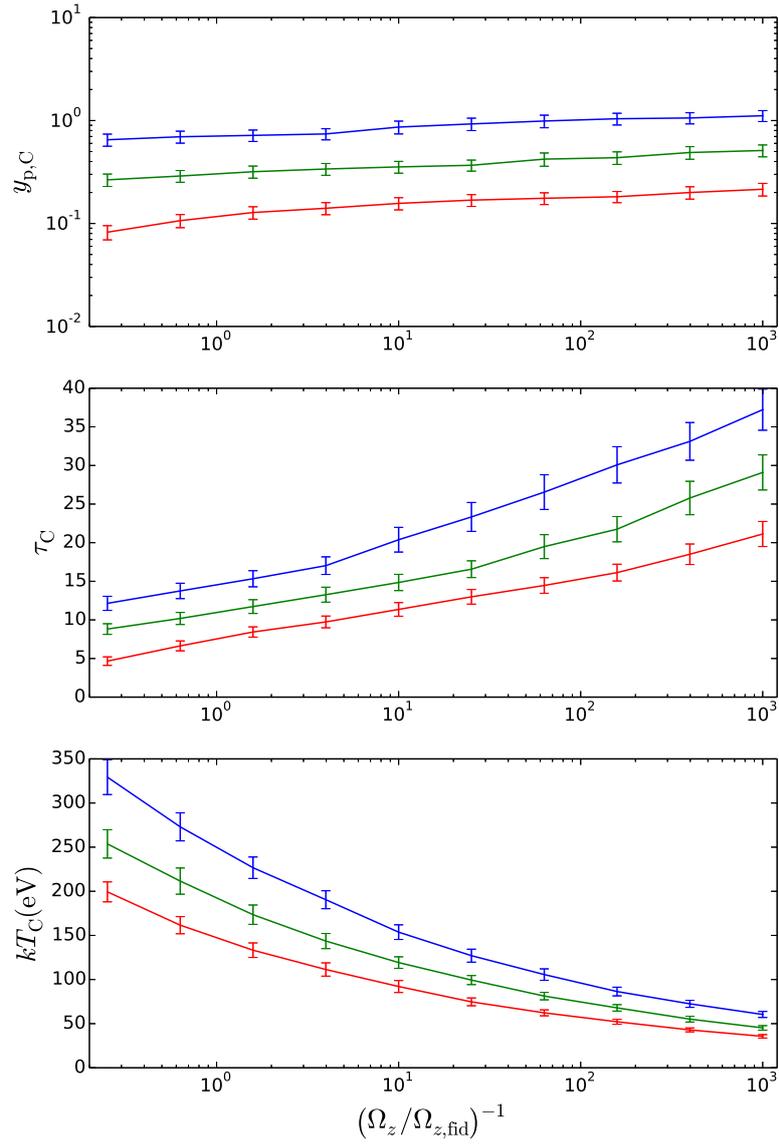

Figure 4.19: Dependence of bulk Comptonization on $\Omega_z^{-1}$ for $\alpha/\alpha_{\mathrm{fid}} = 1$ (blue), 2 (green), and 4 (red). The surface density is $(\Sigma/\Sigma_{\mathrm{fid}})^{-1} = 2$.

Comptonization increases with increasing $\alpha$ for fixed accretion disc parameters, unlike for fixed shearing box parameters.

### 4.3.4 Dependence of bulk Comptonization on accretion disc parameters at the radius of maximum luminosity

To estimate the magnitude of bulk Comptonization for an entire accretion disc, we calculate the bulk Comptonization parameters at the radius $r_{max}$ where the local luminosity is maximized. The luminosity at $r$ is given by

$$L \sim F(2\pi)rdr \sim r^2 F, \qquad (4.39)$$

and $r_{max}$ is determined by maximizing this function with respect to $r$. Since $r_{max}$ is a function only of the spin $a$ and the inner boundary condition parameter $\Delta\epsilon$, the dependence of bulk Comptonization on mass, luminosity, and $\alpha$ at this radius is the same as for fixed $r$, described in the previous section. The dependence on $a$ and $\Delta\epsilon$, however, is different.

**Dependence on spin**    We attempt to determine the dependence of bulk Comptonization on spin by analyzing how the flux depends on spin, as we did earlier. But in this case we have to be careful. Since $r_{max}$ depends on $a$, $r$ is not held constant. Given that

$$\Sigma^{-1} \sim \alpha\Omega_z^{-1}F \sim \alpha M r^{3/2}F, \qquad (4.40)$$

we see that what really matters is the dependence of $r_{max}^{3/2}F$ on $a$. Fortunately, since the dependence of $r^{3/2}F$ on $r$ is qualitatively similar to that of $F$, this does not change our physical intuition. As the spin parameter $a$ increases, $r_{max}$ decreases, and as the region of maximum luminosity becomes smaller, we expect that the flux at each point in this region must increase in order for the total luminosity $L/L_{\mathrm{Edd}}$ to remain the same. In Figure 4.21 we plot the dependence of $r^{3/2}F$ and bulk Comptonization on $a$ at $r_{\max}$. We see that $r^{3/2}F$ and the bulk Comptonization parameters increase with spin, in agreement with our expectations.

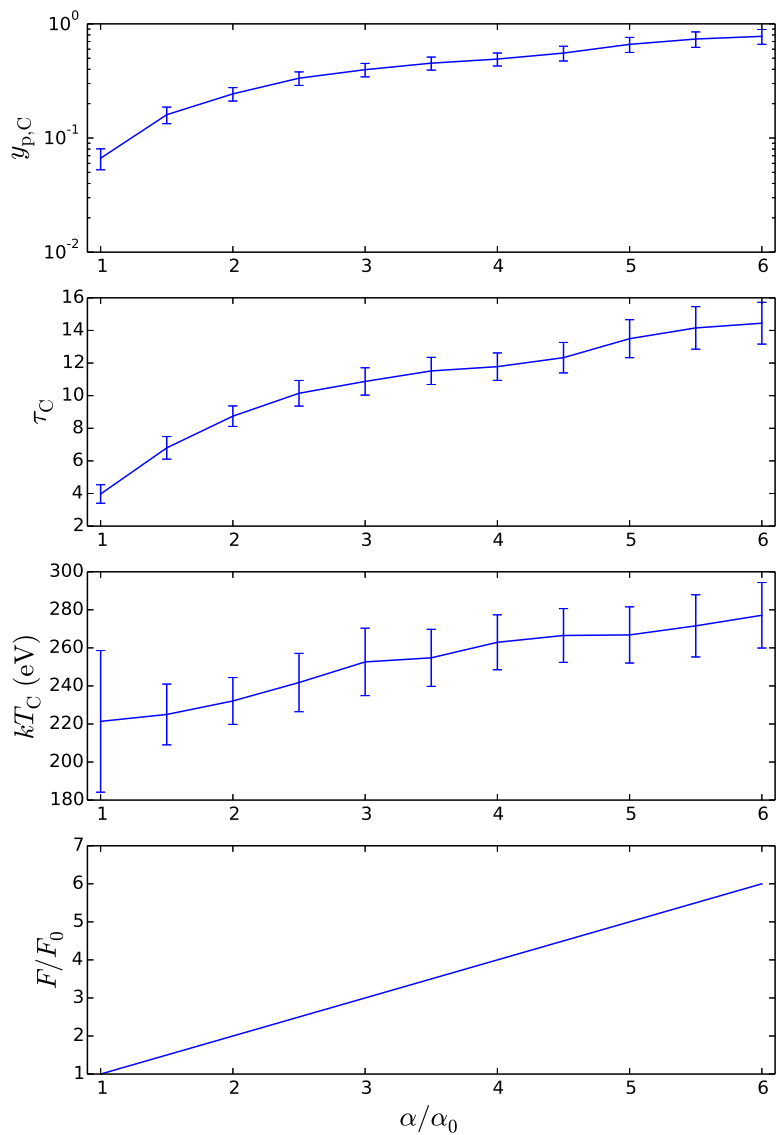

Figure 4.20: Dependence of flux and bulk Comptonization on $\alpha$. The parameter $F_0$ denotes the flux for $\alpha/\alpha_0 = 1$. The values of the parameters held constant are $M/M_\odot = 2 \times 10^6$, $L/L_{\mathrm{Edd}} = 2.5$, $a = 0.5$, and $r = r_{\mathrm{max}} = 11.8$.

We note that for $\Delta\epsilon > 0$, it may be the case that $r_{\max} = r_{\mathrm{in}}$ rather than a value of $r$ at which $dL/dr = 0$. Since $r_{\mathrm{in}}$ decreases with $a$, however, this does not effect our conclusions. For example, in Figure 4.22 we plot the dependence of $r_{\max}$ on $a$ for multiple values of $\Delta\epsilon$. We see that for sufficiently large $\Delta\epsilon$, $r_{\max}$ tracks $r_{\mathrm{in}}$ until $a$ is large enough that the value of $r$ at which $dL/dr$ equals zero is greater than $r_{\mathrm{in}}$. For very large $\Delta\epsilon$, $r_{\max}$ tracks $r_{\mathrm{in}}$ for almost all values of $a$.

**Dependence on inner boundary condition**    The dependence on the inner boundary condition parameter $\Delta\epsilon$ is similar to the dependence on spin. As $\Delta\epsilon$ increases, $r_{\max}$ decreases, and as the region of maximum luminosity becomes smaller, we expect that the flux at each point in this region must increase in order for the total luminosity $L/L_{\mathrm{Edd}}$ to remain the same. Even once $\Delta\epsilon$ is sufficiently large that $r_{\max} = r_{\mathrm{in}}$, we expect the flux at $r_{\mathrm{in}}$ to continue to increase since the increase in efficiency parameterized by $\Delta\epsilon$ should result in an increase in flux at all radii near $r_{\mathrm{in}}$.

In Figure 4.23 we plot the dependence of $r^{3/2}F$ and bulk Comptonization on $\Delta\epsilon$ at $r_{\max}$. We see that $r^{3/2}F$ and the bulk Comptonization parameters increase with $\Delta\epsilon$, in agreement with our expectations. For $\Delta\epsilon$ slightly greater than zero, flux increases both because the flux at all inner radii increases with $\Delta\epsilon$ and because $r_{\max}$ itself decreases towards smaller radii where the flux is larger. For larger values of $\Delta\epsilon$, $r_{\max}$ is fixed to $r_{\mathrm{in}}$, so the flux increases only due to the first effect and therefore increases at a slower rate. In Figure 4.24 we plot the dependence of $r_{\max}$ on $\Delta\epsilon$ for multiple values of $a$. We see that for each curve $r_{max}$ decreases until $r_{max} = r_{in}$.

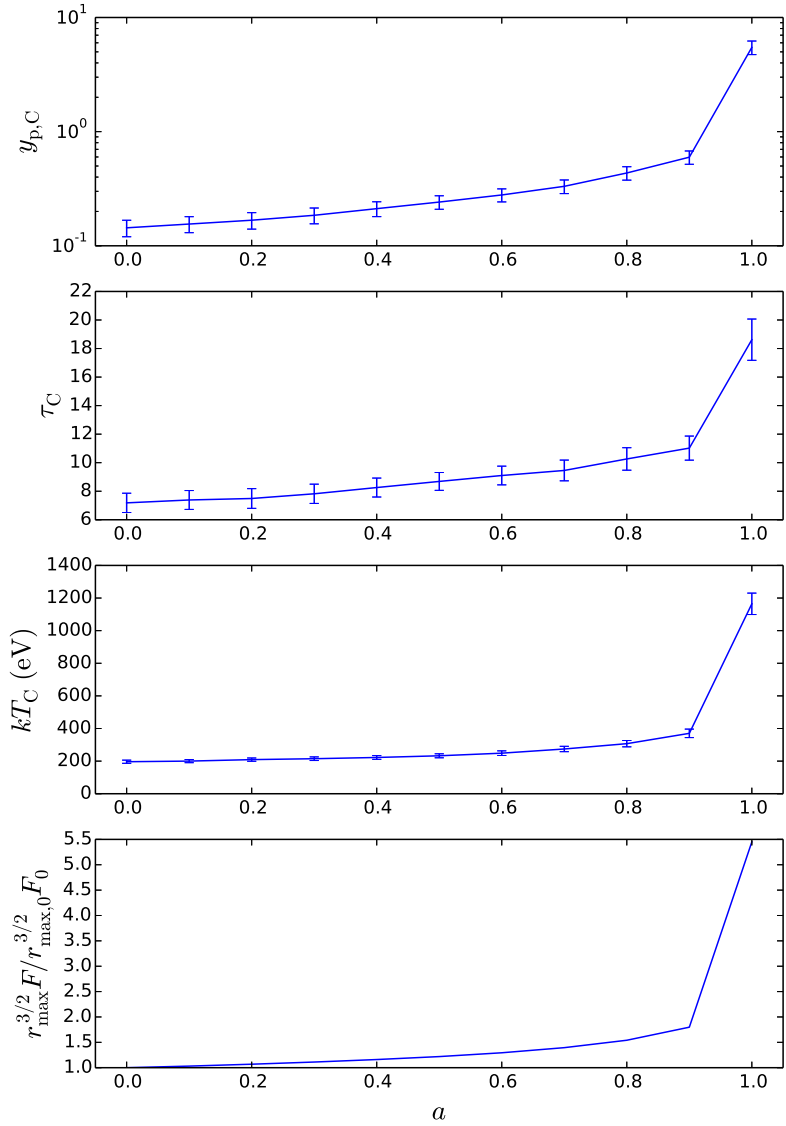

Figure 4.21: Dependence of $\Sigma^{-1}$ and bulk Comptonization on $a$ at the radius where the luminosity is greatest. The subscript zero denotes the value at $a = 0$. The values of the parameters held constant are $M/M_\odot = 2 \times 10^6$, $L/L_{\text{Edd}} = 2.5$, $\Delta\epsilon = 0$, and $\alpha/\alpha_0 = 2$.

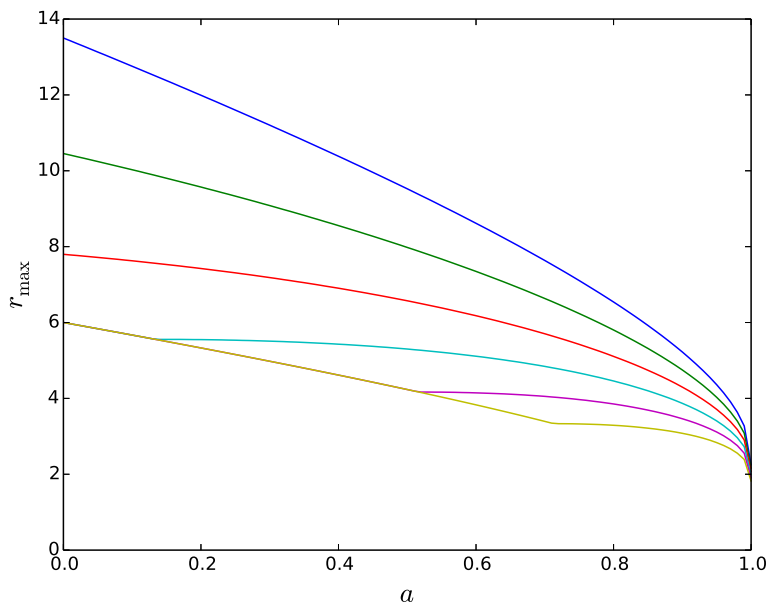

Figure 4.22: Dependence of $r_{max}$ on $a$ for $\Delta\epsilon = 0$ (blue), 0.02 (green), 0.04 (red), 0.06 (cyan), 0.08 (magenta), and 0.1 (yellow).

## 4.3.5   Effect of vertical radiation advection on bulk Comptonization

The scheme from Chapter 3 that we use to scale simulation data assumes that the flux is carried by radiation diffusion. Since we also see substantial vertical radiation advection in some radiation MHD simulations (Blaes et al., 2011), we attempt here to incorporate this process into our analysis of bulk Comptonization. Vertical radiation advection has both a direct and indirect impact on bulk Comptonization. The direct effect is to transport photons through the bulk Comptonization region faster so that they scatter fewer times, reducing bulk Comptonization. But the direct effect is typically negligible since vertical advection is significant only deep inside the photosphere (Blaes et al., 2011), outside the bulk Comptonization region. This is physically intuitive since vertical advection assists radiation diffusion in transporting photons out of the disc in order to maintain thermal equilibrium. Vertical advection is therefore most significant deep inside the photosphere where the photon diffusion time is comparatively large.

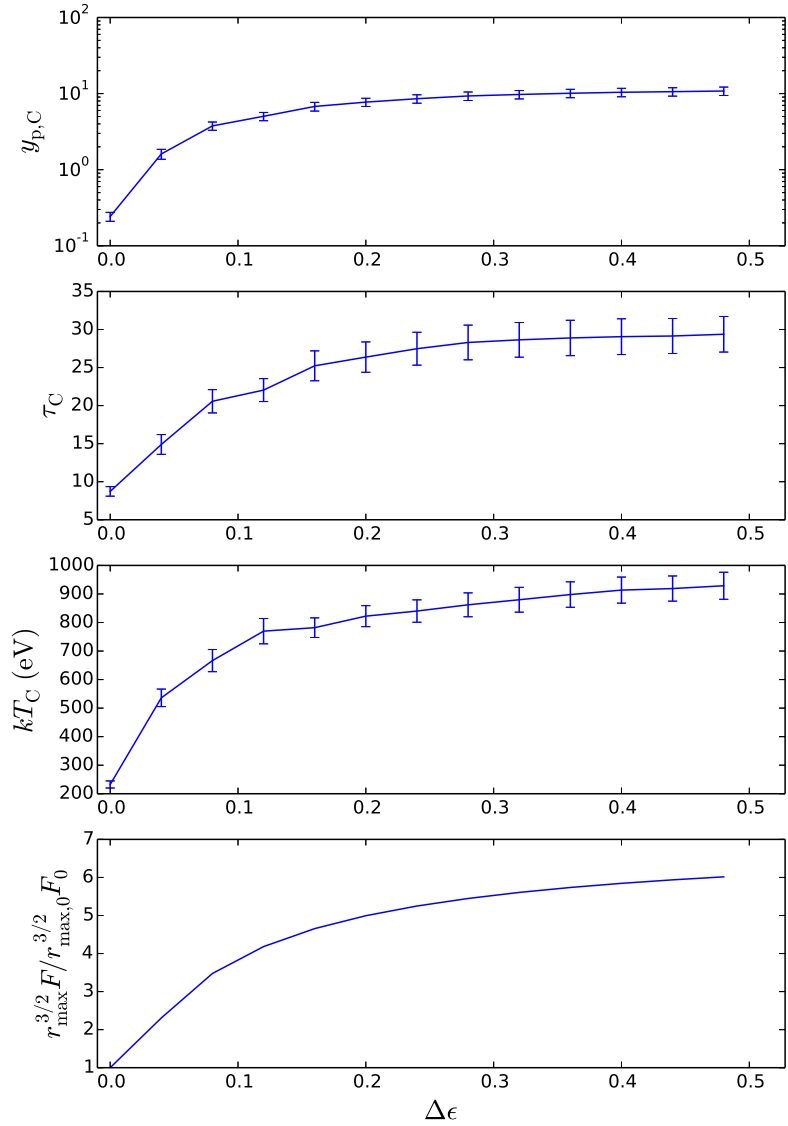

Figure 4.23: Dependence of $\Sigma^{-1}$ and bulk Comptonization on $\Delta\epsilon$ at the radius where the luminosity is greatest. The subscript zero denotes the value at $\Delta\epsilon = 0$. The values of the parameters held constant are $M/M_\odot = 2 \times 10^6$, $L/L_{\rm Edd} = 2.5$, $a = 0.5$, and $\alpha/\alpha_0 = 2$.

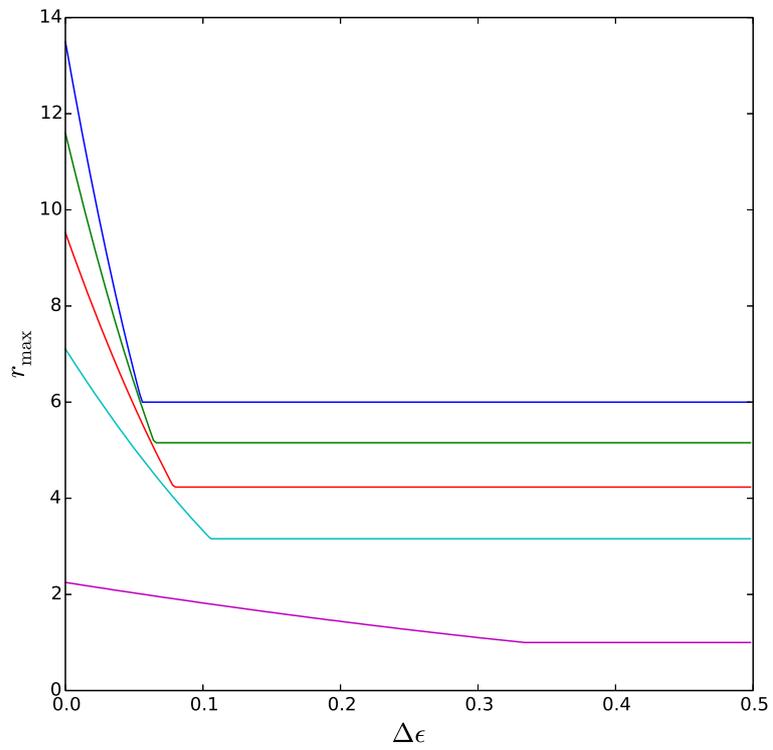

Figure 4.24: Dependence of $r_{\max}$ on $\Delta\epsilon$ for $a = 0$ (blue), 0.25 (green), 0.5 (red), 0.75 (cyan), and 1 (magenta).

The indirect effect of vertical radiation advection on bulk Comptonization, on the other hand, is to modify the underlying vertical structure gas and wave temperature profiles, which in turn either increases or decreases bulk Comptonization depending on whether shearing box parameters or the accretion disc parameters are held constant. In order to study this effect we need to incorporate vertical advection into the scaling scheme. One way to do this is to rederive the shearing box scalings without assuming that the flux is carried by radiation diffusion. We take this approach in Appendix D5. Although this gives physical insight and is necessary to implement a specific model of advection, it is unnecessarily complex for our purpose here. Instead, we begin by simply adding an advection term $F_{\mathrm{a}}$ to the radiation diffusion equation (Chapter 3), which gives

$$F = \frac{2cP_{\mathrm{c}}}{\kappa\Sigma} + F_{\mathrm{a}}. \qquad (4.41)$$

We observe that the only effect of adding $F_{\mathrm{a}}$ at fixed surface density $\Sigma$ is to increase the total flux. Since this is also the equation that introduces the opacity parameter $\kappa$ into the scaling scheme, it follows that the effect of adding $F_{\mathrm{a}}$ is the same as decreasing $\kappa$, as far as our scaling scheme is concerned. Conveniently, we see that the scheme in Chapter 3 already allows for scaling with respect to $\kappa$, even though for clarity we have suppressed the dependence on this parameter until now. We therefore proceed to determine the effect of advection on bulk Comptonization by varying $\kappa$.

**Effect of advection for fixed shearing box parameters**

To determine the effect of including vertical radiation advection on bulk Comptonization for fixed shearing box parameters, we need the turbulent and thermal velocity scalings with $\kappa$ included (Chapter 3):

$$v_{\text{turb}}^2 \sim \kappa^{-2} \alpha^{-1} \Sigma^{-2} \tag{4.42}$$

$$v_{\text{th,c}}^2 \sim \kappa^{-1/4} \left( \alpha^{-1} \Omega_z \right)^{1/4} \tag{4.43}$$

$$v_{\text{th,ph}}^2 \sim \kappa^{-1/2} \left( \alpha^{-1} \Omega_z \right)^{1/4} \Sigma^{-1/4}. \tag{4.44}$$

We see that decreasing $\kappa$ primarily affects the turbulent velocity magnitude. In particular, it moves the bulk and wave temperature profiles upward in Figure 4.13, increasing the size of the bulk Comptonization region. As advection increases at fixed $\Sigma$, therefore, we expect $\tau_C$, $T_C$, and $y_{p,C}$ to increase. For example, in Figure 4.25 we plot the dependence of bulk Comptonization on $\Omega_z^{-1}$ for multiple values of $\kappa$. We see that this result is consistent with our expectations.

**Effect of advection for fixed accretion disc parameters**

To determine the effect of including vertical radiation advection on bulk Comptonization for fixed accretion disc parameters, we write $\Sigma^{-1}$ in terms of the local flux $F$, this time allowing for the variation in $\kappa$ (Chapter 3):

$$\Sigma^{-1} \sim \alpha \kappa^2 \Omega_z^{-1} F. \tag{4.45}$$

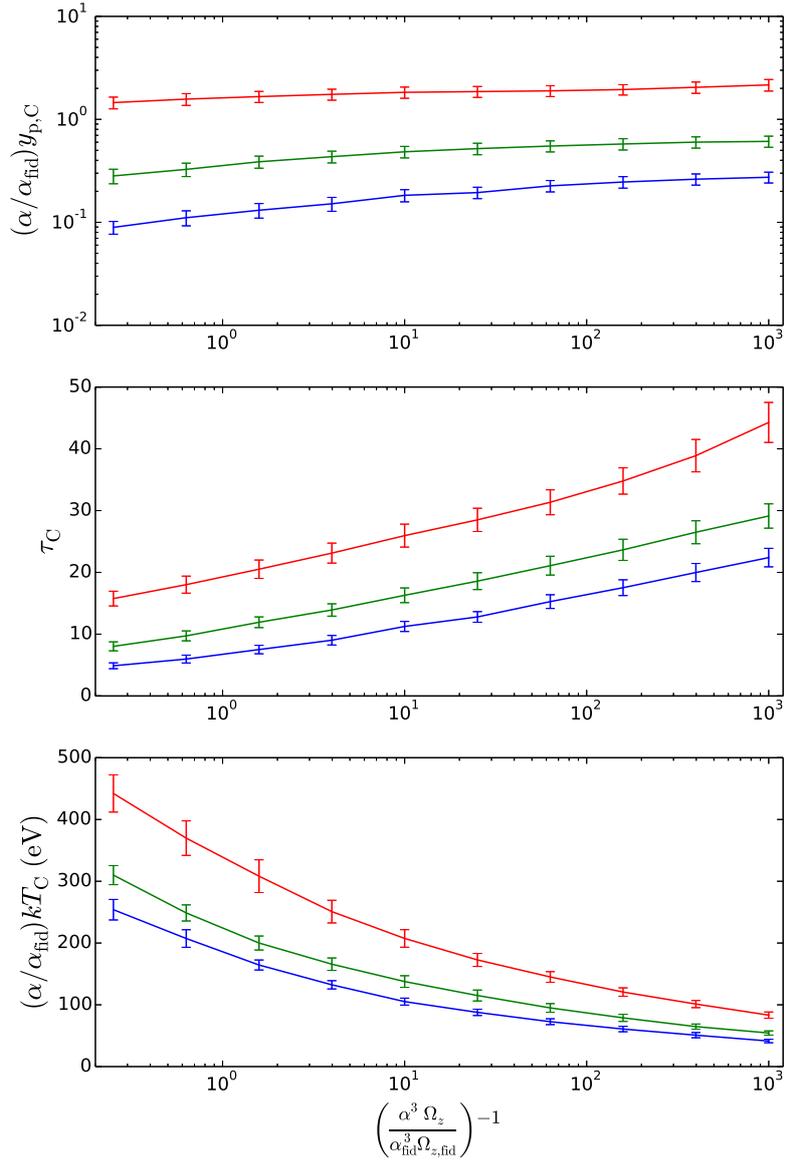

Figure 4.25: Dependence of bulk Comptonization on shearing box parameters for $\kappa = 1$ (blue), 0.75 (green), and 0.5 (red). For all curves, $\Sigma = \Sigma_{\text{fid}}$.

Since neither $\Omega_z^{-1}$ nor $F$ depends on $\kappa$, combining this with equations (4.42)-(4.44) results in the following dependence on $\kappa$:

$$v_{\text{turb}}^2 \sim \kappa^2 \tag{4.46}$$

$$v_{\text{th,c}}^2 \sim \kappa^{-1/4} \tag{4.47}$$

$$v_{\text{th,ph}}^2 \sim 1. \tag{4.48}$$

We see that decreasing $\kappa$ primarily affects the turbulent velocity magnitude, but in the opposite direction to the one in the previous section where the shearing box parameters are fixed. It moves the bulk and wave temperature profiles downward in Figure 4.13, decreasing the size of the bulk Comptonization region. As advection increases, therefore, we expect $\tau_{\text{C}}$, $T_{\text{C}}$, and $y_{\text{p,C}}$ to decrease. For example, in Figure 4.26 we plot the dependence of bulk Comptonization on $\Omega_z^{-1}$ for multiple values of $\kappa$. We see that this result is consistent with our expectations.

## 4.3.6   Time variability of bulk Comptonization

We now explore the effect of the time variability of the vertical structure on bulk Comptonization. We stress that the specific numerical results of this section should not be directly compared to observations of real discs for two primary reasons. First, variability is an inherently global phenomenon which shearing box simulations therefore cannot effectively capture. Second, shearing box simulations with narrow box widths have been found to overestimate variability at a particular radius in the disc, so even not taking into account global phenomena we would expect this analysis to overestimate the variability of bulk Comptonization. The purpose of this section, therefore, is only to demonstrate how the time variability of bulk Comptonization depends on the time variability of the vertical structure profiles. To model the latter will require global disc simulations.

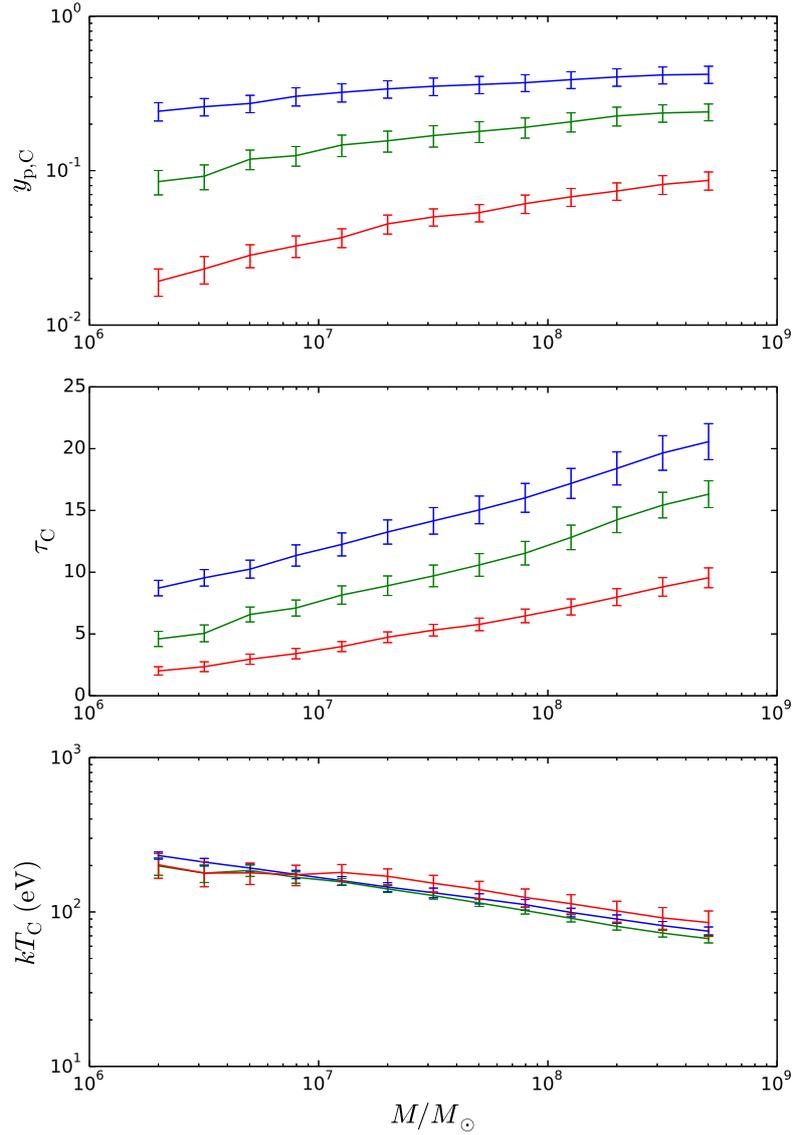

Figure 4.26: Dependence of bulk Comptonization on mass for $\kappa = 1$ (blue), 0.75 (green), and 0.5 (red). The values of the parameters held constant are $M/M_\odot = 2 \times 10^6$, $L/L_{\text{Edd}} = 2.5$, $r = r_{\text{max}} = 11.8$, $a = 0.5$, and $\alpha/\alpha_0 = 2$.

We plot the standard deviation of the bulk Comptonization parameters over the 21 equally spaced timesteps in Figure 4.27. We also plot the standard deviation of the Comptonization gas and wave temperatures individually in Figure 4.28. In order to understand these results, we plot the standard deviations of the time-averaged temperature and density profiles in Figures 4.29 and 4.30, respectively.

The time variation of the density profile alone may cause significant variations in the bulk Comptonization optical depth. For example, in Figure 4.31 at each timestep we plot the Comptonization optical depth $\tau_C$ and the optical depth for a region of fixed size. The bottom of this region is taken to be the point at which the time-averaged gas and wave temperature profiles intersect in Figure 4.29. Variations in the optical depth of this region result in variations in $\tau_C$ that are caused by changes in the density profile alone, rather than changes in the size of the region itself. We see in Figure 4.31 that the overall variance of the optical depth of the region of fixed size is similar to the variance in $\tau_C$. But we also see that the two quantities are only weakly correlated, which means that the variation in the bulk Comptonization optical depth must be due to other factors as well. For example, increasing the density may indirectly *decrease* the bulk Comptonization optical depth by reducing the wave temperature (see section 4.2.2) and therefore the size of the bulk Comptonization region.

Because the spatial variation of the gas temperature profile is so small in the bulk Comptonization region, we expect that its time variation should correlate with the time variation of the bulk Comptonization parameters in a predictable way. We expect that increasing the gas temperature decreases the size of the bulk Comptonization region, thereby increasing the Comptonization wave temperature and decreasing the Comptonization optical depth. In Figure 4.32 at each timestep we plot the Comptonization gas temperature and the Comptonization wave temperature. We see that the two temperatures are strongly correlated in the direction we expect. In Figure 4.33 we plot the Comptonization gas temperature and optical depth. In this case, the correlation is also in the direction we expect, but it is weaker since density variations (among other factors) also play a significant role in determining the Comptonization optical depth.

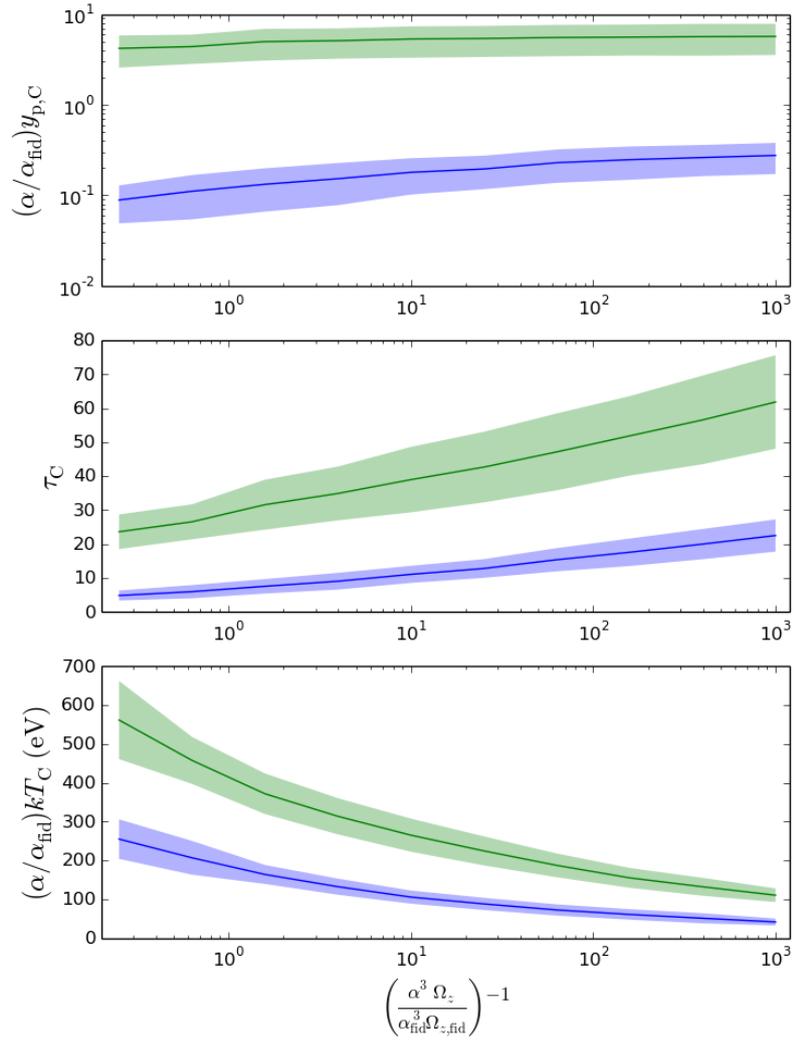

Figure 4.27: Dependence of bulk Comptonization on shearing box parameters for $(\Sigma/\Sigma_{\mathrm{fid}})^{-1} = 1$ (blue) and 5 (green). The shaded region corresponds to points within $0.675\sigma$ (i.e. 50% of the data for a Gaussian distribution) in the time variability.

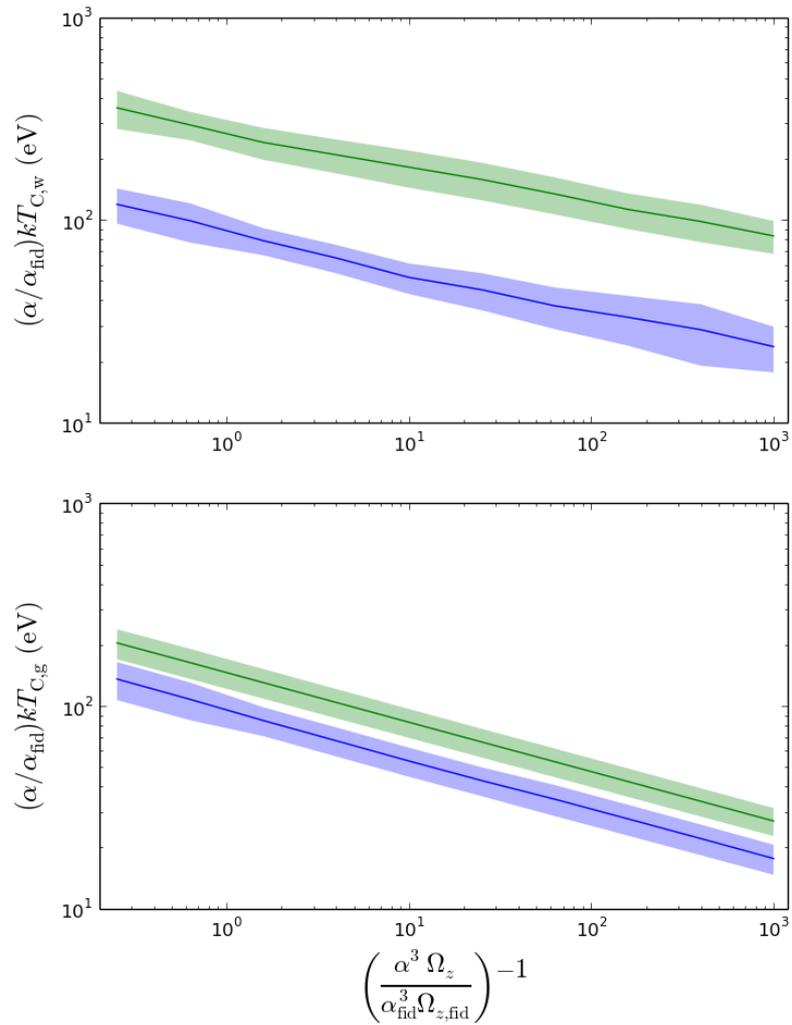

Figure 4.28: Dependence of the Comptonization gas and wave temperatures on shearing box parameters, separately, for $(\Sigma/\Sigma_{\rm fid})^{-1} = 1$ (blue) and 5 (green). The shaded region corresponds to points within $0.675\sigma$ (i.e. 50% of the data for a Gaussian distribution) in the time variability.

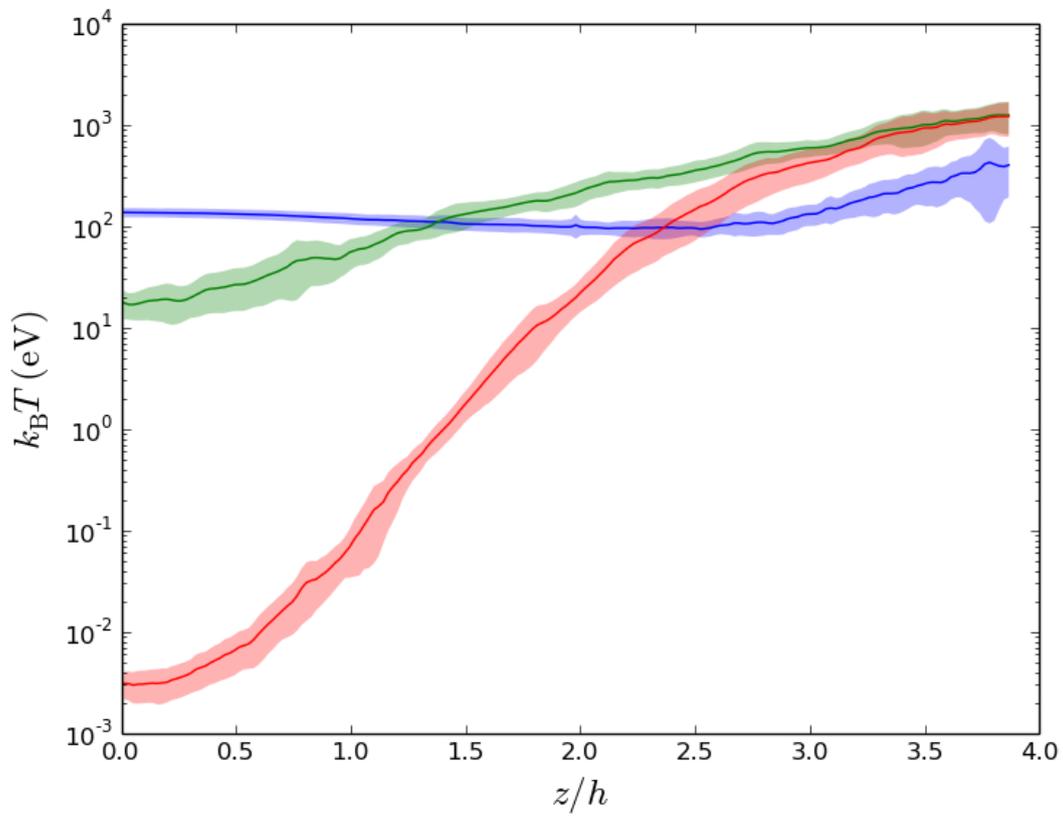

Figure 4.29: Horizontally and time averaged gas (blue), bulk (green), and wave (red) temperature profiles for the fiducial shearing box parameters (Table 4.8). The shaded region corresponds to points within $0.675\sigma$ (i.e. 50% of the data for a Gaussian distribution) in the time variability.

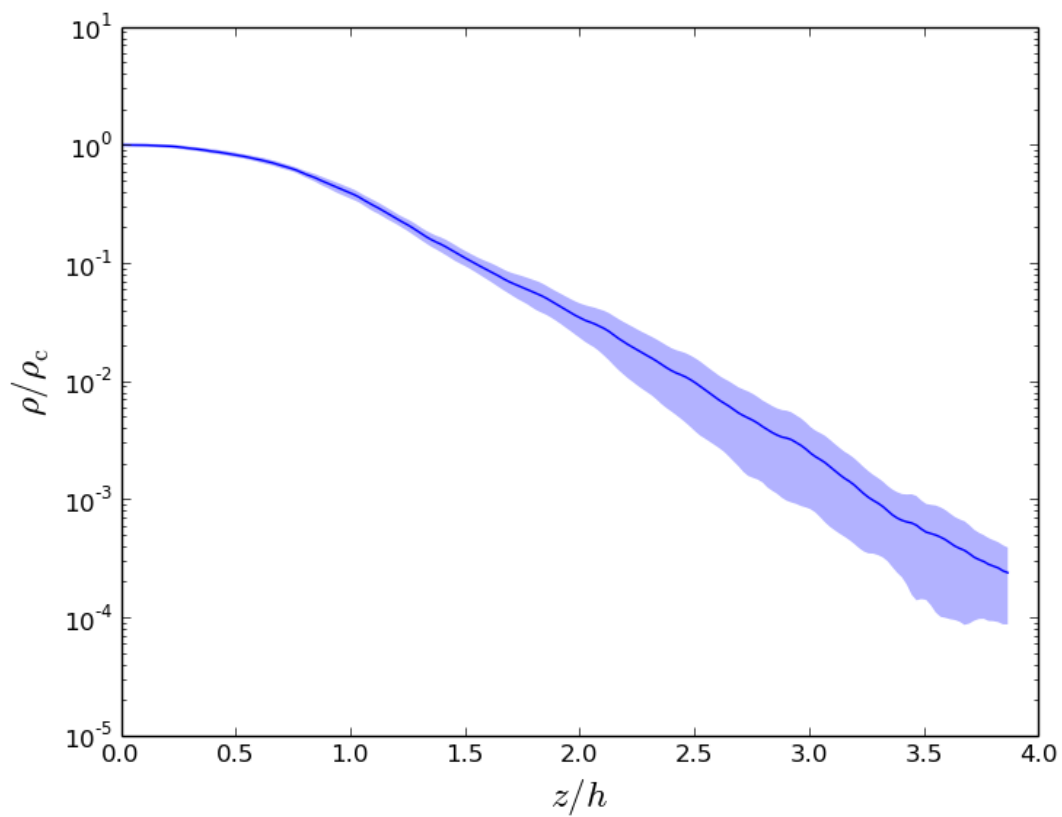

Figure 4.30: Horizontally and time averaged density profile for the fiducial shearing box parameters (Table 4.8). The shaded region corresponds to points within $0.675\sigma$ (i.e. 50% of the data for a Gaussian distribution) in the time variability.

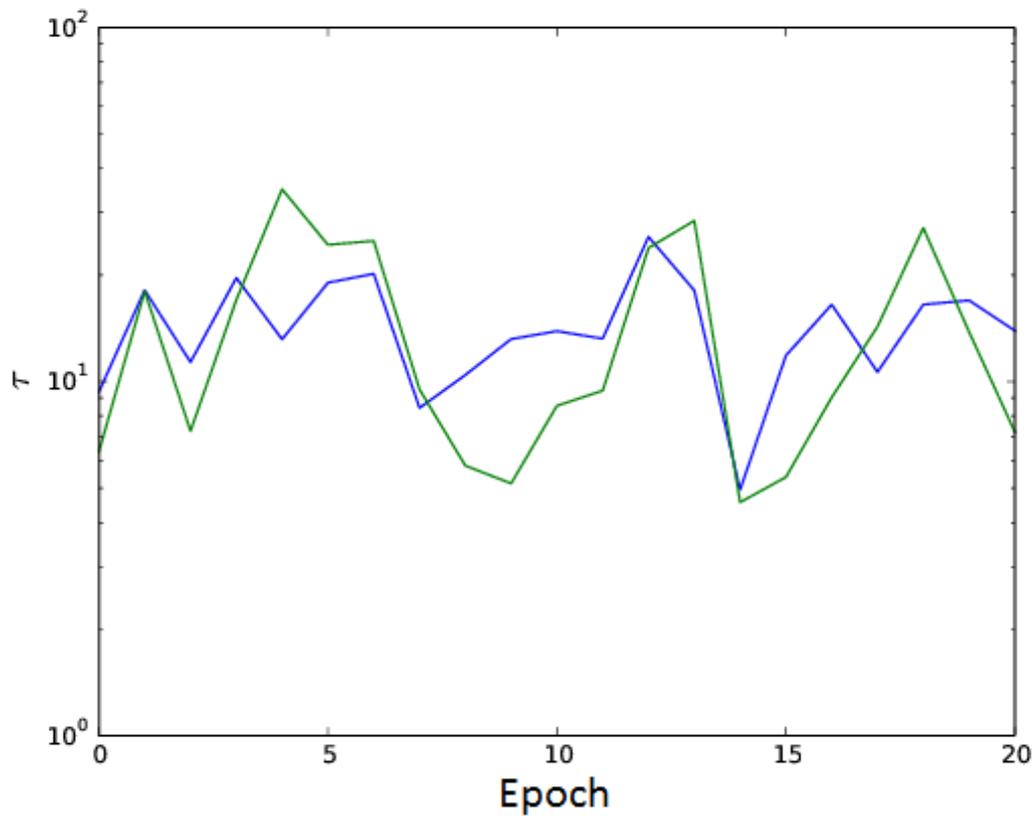

Figure 4.31: Time variability of the Comptonization optical depth $\tau_C$ (blue) and the (normalized) optical depth of a nearly identical region whose physical size is defined to be constant (green). The surface density here is $(\Sigma/\Sigma_{\text{fid}})^{-1} = 2$ and the other parameters are the fiducial shearing box parameters (Table 4.8). The timesteps are spaced 10 orbital periods apart, and each orbital period is 5535 seconds.

We can also estimate the variability of the luminosity powered by bulk Comptonization. The fraction of the luminosity powered by bulk Comptonization is just the total fractional photon energy change, which for unsaturated spectra is approximately

$$\frac{\Delta\epsilon}{\epsilon} \approx e^{y_{p,C}} - 1. \tag{4.49}$$

For $y_{p,C} > 1$ we must check that spectra is unsaturated. For $y_{p,C} \ll 1$ spectra is always unsaturated, and in addition the fractional energy change simplifies to

$$\frac{\Delta\epsilon}{\epsilon} \approx y_{p,C}. \tag{4.50}$$

The variability of the luminosity powered by bulk Comptonization is therefore characterized by the fractional rms (root mean square) $y_{p,C}$, which is the standard deviation divided by the mean of $y_{p,C}$,

$$\text{fractional rms} = \frac{\sigma_{y_{p,C}}}{\langle y_{p,C} \rangle}. \tag{4.51}$$

We plot the fractional rms for $(\Sigma/\Sigma_{\text{fid}})^{-1} = 1$ in Figure 4.34, and see that it is consistent with Figure 4.27. In particular, as the values on the $x$ axis increase from $10^0$ to $4 \times 10^2$, we see that $\sigma_{y_{\mathrm{p,C}}}$ decreases while $\langle y_{\mathrm{p,C}} \rangle$ increases so that the fractional rms increases. Since $\Omega_z^{-1} \propto M$ (section 4.3.3), the fractional rms seems to vary insubstantially with mass. We note that the fiducial shearing box parameters (Table 4.8) correspond to $r = 20$ for the $M = 2 \times 10^6 M_\odot$, $L/L_{\mathrm{Edd}} = 2.5$ parameter set (Table 4.1). Since $r \approx 20$ in the region of the disc that contributes most to the luminosity for this parameter set (i.e. for $a = 0$, $\Delta\epsilon = 0$), Figure 4.34 also characterizes the variability of the luminosity powered by bulk Comptonization for the entire accretion disc.

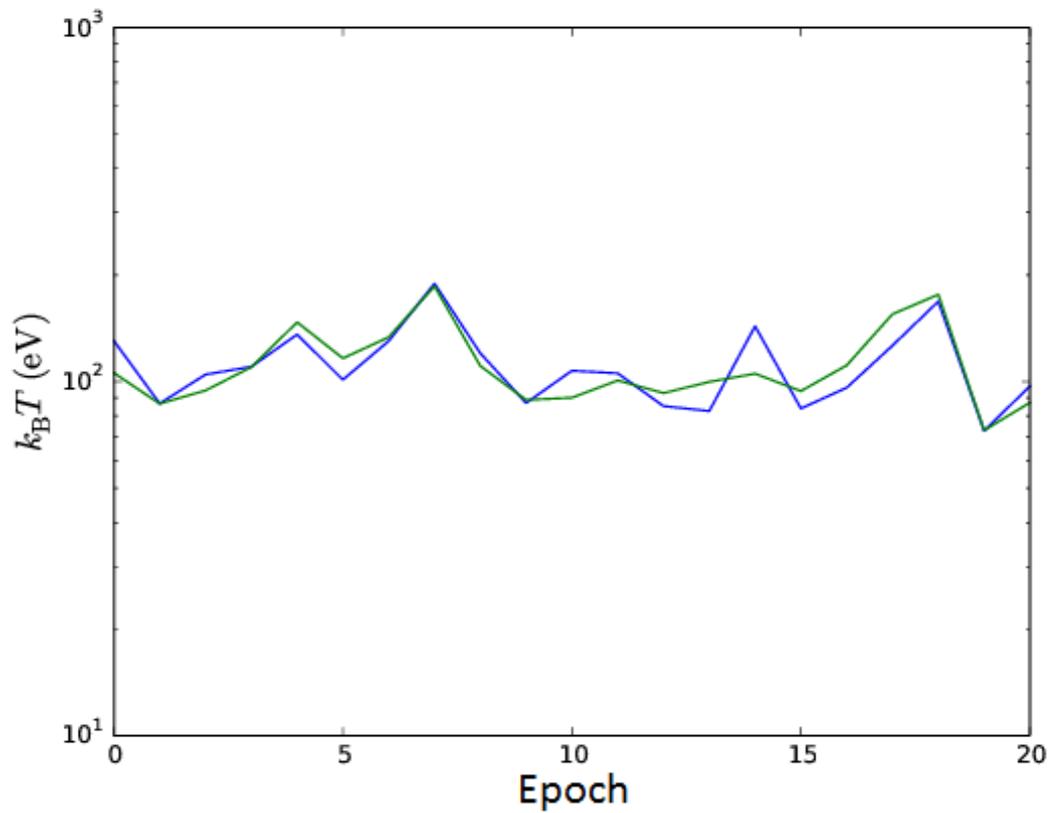

Figure 4.32: Time variability of the Comptonization gas (blue) and wave (green) temperatures, normalized to the average Comptonization gas temperature. The surface density here is $(\Sigma/\Sigma_{\mathrm{fid}})^{-1} = 2$ and the other parameters are the fiducial shearing box parameters (Table 4.8). The timesteps are spaced 10 orbital periods apart, and each orbital period is 5535 seconds.

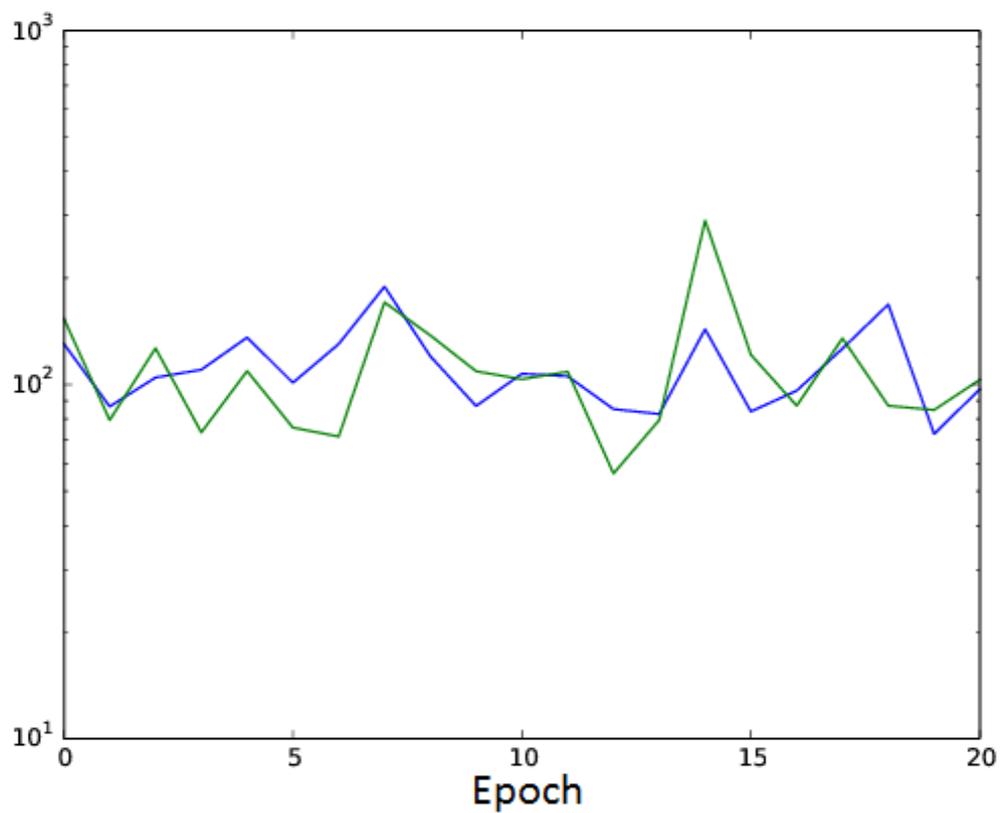

Figure 4.33: Time variability of the Comptonization gas temperature $T_{\mathrm{g,C}}$ (blue) and inverse optical depth $\tau_{\mathrm{C}}^{-1}$ (green), normalized to the average Comptonization gas temperature. The surface density here is $(\Sigma/\Sigma_{\mathrm{fid}})^{-1} = 2$ and the other parameters are the fiducial shearing box parameters (Table 4.8). The timesteps are spaced 10 orbital periods apart, and each orbital period is 5535 seconds.

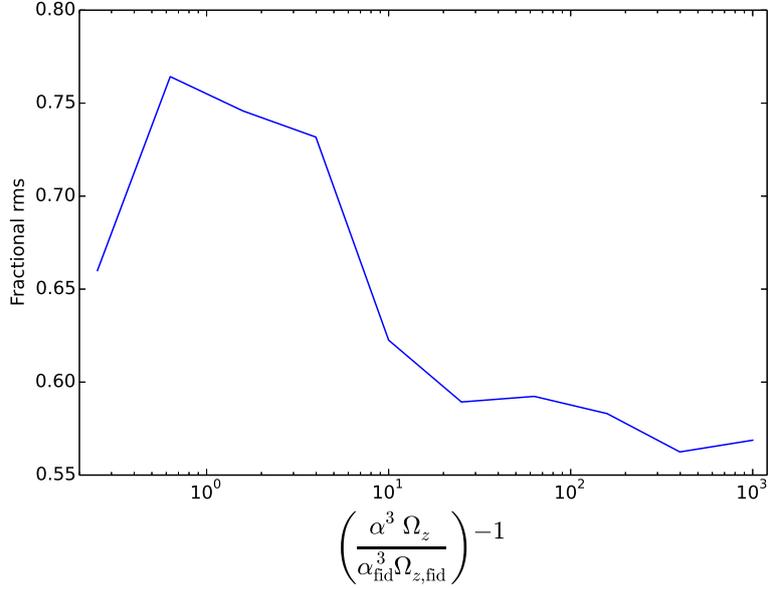

Figure 4.34: Dependence of fractional rms on shearing box parameters for $(\Sigma/\Sigma_{\mathrm{fid}})^{-1} = 1$.

We note that the time-averaged bulk Comptonizaton parameters are not equal to the bulk Comptonization parameters computed from the time-averaged temperature profiles. For example, in Figure 4.35 we plot the bulk Comptonization parameters computed with the time-averaged data. To do this, we first time average the gas and wave temperature profiles and then compute the bulk Comptonization parameters. In the same figure we also plot the time-averaged parameters, originally plotted in Figure 4.12. We see that the parameters computed from the time-averaged profiles significantly overestimate the time-averaged parameters. This result is important because it means that the time-averaged profiles, while often useful, should not directly be used to model bulk Comptonization.

This work is based on only 21 simulation snapshots spaced 10 orbital periods apart, but we note that by applying our model to a complete set of simulation data one could also calculate how the power spectra of the vertical structure profiles affect that of the fraction of the luminosity powered by bulk Comptonization and other bulk Comptonization parameters.

## 4.4    Discussion

### 4.4.1    Comparison of results with previous work and observations

Aside from the fact that our model implements a simplified version of the procedure used in Chapter 3 to calculate the bulk Comptonization parameters, our approach here differs from the approach in Chapter 3 in two important ways: In Chapter 3 bulk Comptonization is modelled for an entire accretion disc at once rather than at each radius individually, and shear velocities are included in addition to turbulent velocities. Because of these differences, the bulk Comptonization parameters found there depart slightly from those found here. But the dependence of bulk Comptonization on accretion disc parameters detailed in sections 4.3.3 and 4.3.4 is consistent with the results of Chapter 3. In particular, the bulk Comptonization $y$ parameter for the overall disc increases with $\alpha$ and mass, while the bulk Comptonization temperature decreases with increasing mass. And within a given disc, bulk Comptonization is greatest at intermediate radii where the flux is also near maximal.

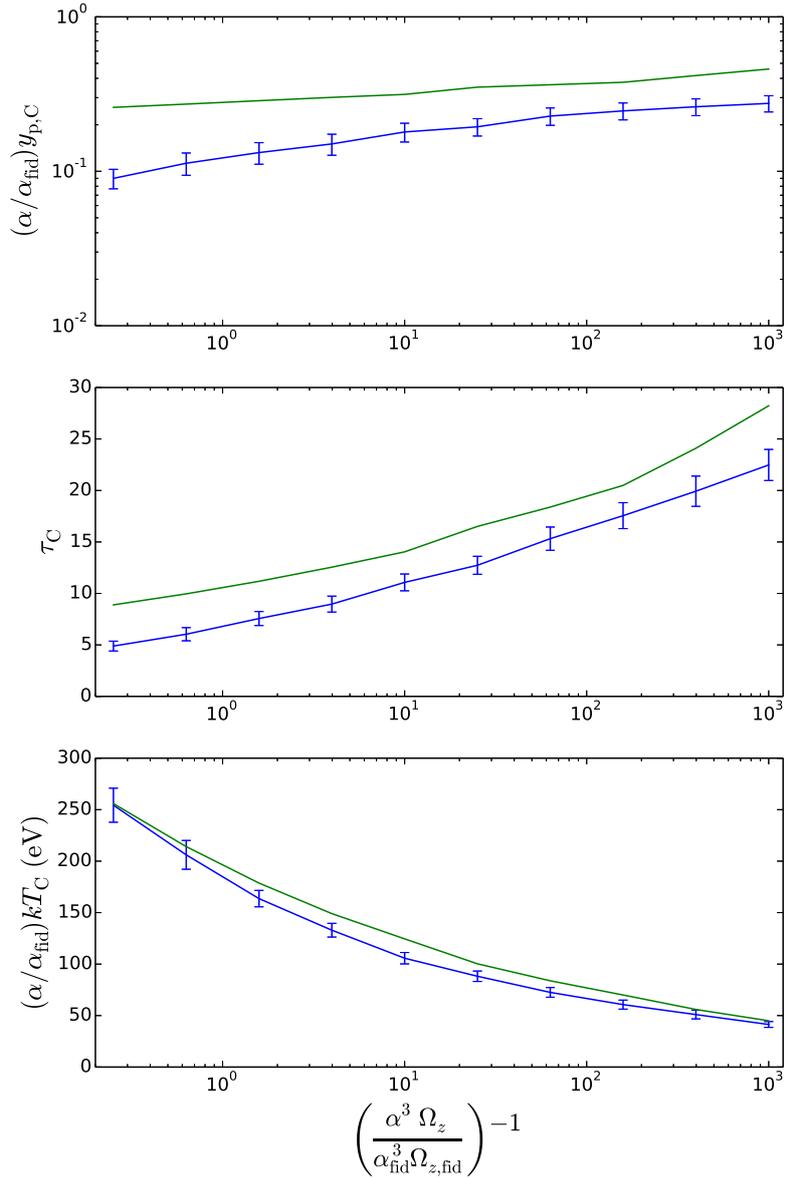

Figure 4.35: Dependence of bulk Comptonization on the shearing box parameters, calculated by either time averaging the Comptonization parameters (blue) or time averaging the vertical structure profiles (green). The surface density is $\Sigma/\Sigma_{\text{fid}} = 1$.

To make contact with observations, in Chapter 3 we modelled bulk Comptonization for a few systems with accretion disc parameters similar to those fit by D12 to REJ1034+396, a narrow-line Seyfert 1 (NLS1) with $L/L_{\rm Edd} = 2.4$. The bulk Comptonization parameters found in Chapter 3 broadly agree with those fit by D12. This agreement suggests that the large soft X-ray excess seen in REJ1034+396 may at least in part be due to bulk Comptonization. By generalizing the results of Chapter 3, our work provides a physical basis for more widely connecting warm Comptonization models of the soft excess to underlying accretion disc parameters.

## 4.4.2 The importance of the disc inner boundary condition and implications for black hole X-ray binaries

An important consequence of our results is that bulk Comptonization is strongly dependent on the disc inner boundary condition parameter, $\Delta\epsilon$. Before proceeding, however, we provide context for the range of $\Delta\epsilon$ since it is not a widely used parameter and we need to have a sense of what it means for it to be large. To start, we observe that since the efficiency for a no torque inner boundary condition, zero spin system is $\eta = 0.057$, any value of $\Delta\epsilon > 0.01$ is relatively large. Even for spin $a = 0.9$, the efficiency with no inner torque is $\eta = 0.16$, so $\Delta\epsilon = 0.1$ corresponds to a substantial physical change.

Another way to understand the effect of $\Delta\epsilon$ is to examine how it affects the disc scalings presented in Chapter 3. We see that this parameter arises in the equations for the flux scalings, reproduced in section 4.3.3, equations (4.30) and (4.33). To understand why $\Delta\epsilon$ appears here, we observe that for $\Delta\epsilon = 0$ in both equations the purpose of the final factor is to ensure that the flux goes to zero as $r$ approaches $r_{\rm in}$ rather than continue to increase as $r^{-3}$. It follows that we can regard $\Delta\epsilon$ large to the extent that it reverses the effects of this factor. For example, for the Newtonian scalings we see that setting $\Delta\epsilon = 1/r_{in}$ removes the dependence on $r$ of this term altogether so that $F \sim r^{-3}$. For zero spin, $r_{in} = 6$ so we should regard $\Delta\epsilon = 0.17$ as very large. For $a = 0.9$, $r_{in} = 2.32$, so the critical value of $\Delta\epsilon$ is 0.43. Therefore, values of $\Delta\epsilon$ anywhere from 0.1 to 0.4 should be viewed as very large, depending on the spin parameter $a$. The Kerr scalings lead to similar conclusions.

As $\Delta\epsilon$ approaches infinity the flux scaling asymptotes to a fixed value rather than continuing to increase. At $r = r_{in}$, we see from the Newtonian scalings that this limit is reached when $\Delta\epsilon \sim 1$. Beyond this point, therefore, bulk Comptonization hardly varies at all with $\Delta\epsilon$. The reason for this is that $\Delta\epsilon$ changes the distribution of flux throughout the disc at a fixed overall luminosity $L/L_{Edd}$. For $\Delta\epsilon \gg 1$, the distribution at the inner radii is fixed and the flux distribution continues to change only for $r \gg r_{in}$.

In Figure 4.36 we plot the dependence of bulk Comptonization on mass for several values of $\Delta\epsilon$ for a system with moderate ($a = 0.5$) spin. We see that all bulk Comptonization parameters strongly increase with increasing $\Delta\epsilon$. Note that this dependence holds only for the region where the disc is brightest, not radii for which $r \gg r_{in}$ (see sections 4.3.3 and 4.3.4).

An important implication of this result is that bulk Comptonization is likely insignificant in black hole X-ray binaries unless the luminosity greatly exceeds Eddington or $\Delta\epsilon$ is large (i.e. $\Delta\epsilon > 0.1$). In Figure 4.36, for which $L/L_{Edd} = 2.5$, for example, we see that for $\Delta\epsilon = 0$ bulk Comptonization is non-existent for $M < 10^5 M_\odot$. For $M = 10 M_\odot$, we see a significant effect only for $\Delta\epsilon > 0.1$.

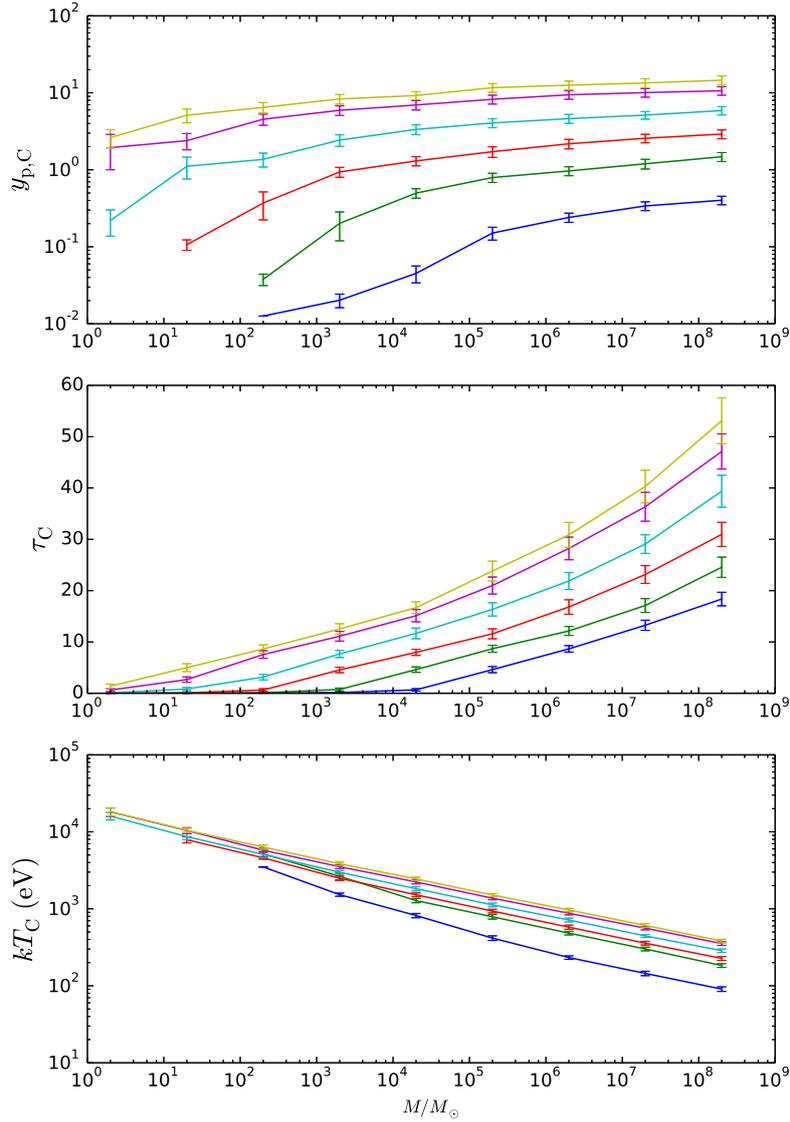

Figure 4.36: Dependence of bulk Comptonization on mass for $\Delta\epsilon = 0$ (blue), 0.03 (green), 0.05 (red), 0.1 (cyan), 0.3 (magenta), and 1 (yellow). The values of the parameters held constant are $L/L_{\mathrm{Edd}} = 2.5$, $a = 0.5$, $r = r_{\mathrm{max}}$, and $\alpha/\alpha_0 = 2$.

### 4.4.3   Robustness of bulk Comptonization results to variations in the disc vertical structure

In section 4.3.2 we showed that the dependence of bulk Comptonization on $\Sigma^{-1}$ and $\Omega_z^{-1}$, plotted in Figure 4.12, can be understood in terms of the shearing box temperature profiles, plotted in Figure 4.13. Since these profiles correspond to scaled data from a single radiation MHD simulation, we need to examine the extent to which our results are robust to changes in the disc vertical structure that may occur in shearing box simulations run in different regimes or global simulations. Certainly the exact values of the bulk Comptonization parameters are sensitive to such changes (section 4.3.6), but we now show that the overall dependence on $\Sigma^{-1}$ and $\Omega_z^{-1}$ (and therefore on the accretion disc parameters) is more robust.

We first consider the dependence of bulk Comptonization on $\Omega_z^{-1}$. In section 4.3.2, using the profiles shown in Figure 4.13, we showed that since only the gas temperature profile varies with $\Omega_z^{-1}$, the Comptonization temperature decreases and the Comptonization optical depth increases with increasing $\Omega_z^{-1}$. There is considerable uncertainty in the shape of the gas temperature profile outside the scattering photosphere, but fortunately the contribution of this region to bulk Comptonization is negligible since the bulk Comptonization temperature is weighted by the optical depth factor $\tau d\tau$. The greatest uncertainty in this analysis, therefore, is the bulk velocity field, which determines the shape of the wave temperature profile. But since the wave temperature is defined to strongly decrease with increasing density (section 4.2.2), we expect that even for significantly different velocity fields the wave temperature profile will increase near the scattering photosphere and that the resulting dependence of the Comptonization parameters on $\Omega_z^{-1}$ will be unchanged. Since our conclusions in section 4.3.2 regarding the dependence of bulk Comptonization on $\Sigma^{-1}$ also rely primarily on the fact that the wave temperature profile strongly increases near the photosphere, we also expect them to be robust to changes in the vertical structure.

In addition to the above concerns, we must also check that as the size of the bulk Comptonization region increases it remains outside the effective photosphere. Otherwise, only part of the bulk Comptonization region will contribute to bulk Comptonization (since photons are emitted at the effective photosphere). However, this condition is likely always satisfied since the size of the bulk Comptonization region increases most significantly as $\Sigma^{-1}$ increases, which simultaneously moves the effective photosphere inward. In particular, for our data we find that the vertical structure becomes effectively thin well before the bulk Comptonization region optical depth is more than a small fraction of the optical depth of the half-thickness of the disc.

Of course, this analysis is still based on the thin disc equations, which assume that $h/r \ll 1$, where $h$ and $r$ are the disc scale height and radius, respectively. This approximation starts to break down when the luminosity approaches a significant fraction of the Eddington luminosity, but this is also when bulk Comptonization starts to become significant. To fully self-consistently study the high Eddington regimes most important for bulk Comptonization, therefore, requires global disc simulations, which we discuss in Chapter 5.

### 4.4.4  Limitations to the scaling scheme parameter range

We now make note of a subtlety that limits the applicability of the scaling scheme from Chapter 3: The scheme can scale data to lower surface densities, but not higher ones. To understand why, we examine how the gas temperature profile scales with decreasing $\Sigma$. First we note that since the gas temperature profile below the photosphere is significantly different from the profile above it, the regions must be scaled separately and then joined together. Next, we observe that the photosphere is not defined to be at a set number of scale heights $h$ away from the midplane but rather at the point at which the scattering optical depth is unity. As a result, as the surface density $\Sigma$ decreases the photosphere moves inward in $z/h$. Since $h$ is the fundamental length scale for variations in the vertical structure and since fewer scale heights of data are needed outside the photosphere, fewer grid cells of data are needed to fill the region outside the photosphere of the scaled disc. This truncated data is scaled appropriately and then joined to the scaled data from above the photosphere. We see, therefore, that scaling to smaller surface densities requires deleting grid cells from the original simulation data. By the same reasoning, this scheme cannot scale to larger surface densities since it would require data from more grid cells than already exist.

It immediately follows that this scheme cannot scale data to any set of accretion disc parameters for which $\Sigma/\Sigma_0 > 1$. In particular, since $\Sigma^{-1}$ is always directly proportional to the luminosity $L/L_{\text{Edd}}$, we can never scale to smaller values of $L/L_{\text{Edd}}$ unless they are offset by simultaneously scaling to, for example, smaller radii or greater $\Delta\epsilon$. We note that this scaling scheme is, therefore, useful for scaling lower Eddington ratio simulations to higher ones, as we do in this work, but not the other way around.

We showed in section 4.3.2 that bulk Comptonization increases strongly with increasing $\Sigma^{-1}$. For the curve in Figure 4.12 with the smallest value of $\Sigma^{-1}$, $\Sigma^{-1} = \Sigma_{\text{fid}}^{-1} = 4\Sigma_0^{-1}$, we see that $0.1 < y_{\text{p,C}} < 0.3$. Therefore, the fact that we cannot scale to values of $\Sigma^{-1}$ smaller than $\Sigma_0^{-1}$ is not a significant limitation since it appears that bulk Comptonization is negligible for such values anyway. But this analysis assumes that the scaling scheme in Chapter 3 is valid over an arbitrarily large parameter range. If we want to scale to a regime with significantly different opacities, for example, then we really should use data from simulations with the relevant opacities included. For example, if the vertical structure is significantly different for sub-Eddington AGN because of changes in the opacities that occur in such regimes, then bulk Comptonization could be larger than we would infer from our analysis of the 110304a simulation data. On the other hand, this seems unlikely given that absorption opacities will substantially increase in this regime.

### 4.4.5   Effect of bulk Comptonization on disc spectra

As we discussed in section 3.4.3, the effect of bulk Comptonization on disc spectra cannot only be to upscatter photons to higher energies because we also must take into account the back-reaction on the disc vertical structure. Since energy conservation fixes the flux as a function of radius and the other accretion disc parameters, we expect that bulk Comptonization will be accompanied by a decrease in the gas temperature at the effective photosphere so that the total emitted flux will remain unchanged. For significant bulk Comptonization, the effect of this is to move the Wien tail to higher energy while moving the spectral peak to lower energy, broadening the spectrum. For moderate bulk Comptonization, the effect of lowering the gas temperature may not translate into a leftward shift of the spectral peak, but the spectrum will still be broadened in such a way that the total flux remains unchanged.

A decrease in the effective photosphere gas temperature is the simplest conceivable back-reaction. This would occur if the only effect of bulk Comptonization on the gas is to remove kinetic energy from the turbulent cascade through radiation viscous dissipation (Chapter 2) so that less kinetic energy is dissipated and converted to gas internal energy. But to self-consistently model this phenomenon, bulk Comptonization must be implemented in the underlying radiation MHD simulations. The shearing box simulations used in this work (Hirose, Krolik & Blaes, 2009), for example, do not include bulk Comptonization since it is primarily a second order effect in velocity (see section 4.4.6), and the flux-limited diffusion approximation does not capture second order effects

(Chapter 2).

## 4.4.6  Effect of the horizontally averaged $z$ component of the velocity field on bulk Comptonization

Bulk Comptonization includes effects that are both first and second order in the velocity field (Chapter 2) and the wave temperature defined in section 4.2.2 captures only the second order effects. The first order effect is non-zero only for compressible modes and is negligible when the photon mean free path is large relative to the mode wavelength. As long as the mean photon energy is less than $4k_B (T_g + T_w)$, the second order effect always results in upscattering, analogous to thermal Comptonization. But the first order effect can result in either upscattering or downscattering depending on whether the velocity field is converging or diverging, respectively (Chapter 2). It follows that only long wavelength compressible modes should result in a non-negligible first order effect, since for shorter wavelength modes either the first order effect is negligible or upscattering in one region is offset by downscattering in another. Therefore, the variations with respect to $z$ of the density weighted, horizontal average of the $z$ component of the velocity field may result in a non-negligible first order effect. As in the case of the wave temperature profile (section 4.2.2), density weighting is appropriate because photons scatter more times in higher density regions. We expect such long wavelength variations to exist since the vertical structure is stratified. For example, in Figure 4.37 we plot this profile at the 140 orbits timestep for the $M = 2 \times 10^6 M_\odot$, $L/L_{\mathrm{Edd}} = 5$ parameter set (Table 4.1) with $r = 14$. In the remainder of this section we show that this effect is discernable but subdominant to the second order effect. We also show that once this effect is taken into account the slight discrepancy in Figure 4.3 between the spectra calculated directly with the turbulence and the spectra calculated by modeling the turbulence with the wave temperature vanishes. We therefore conclude that the wave temperature models the second order effect more accurately than we originally had reason to believe based on the preliminary analysis in section 4.2.2. All data in this section are scaled to the $M = 2 \times 10^6 M_\odot$, $L/L_{\text{Edd}} = 5$ parameter set (Table 4.1). As in section 4.2.2, all spectra and vertical structure profiles correspond to the 140 orbits timestep.

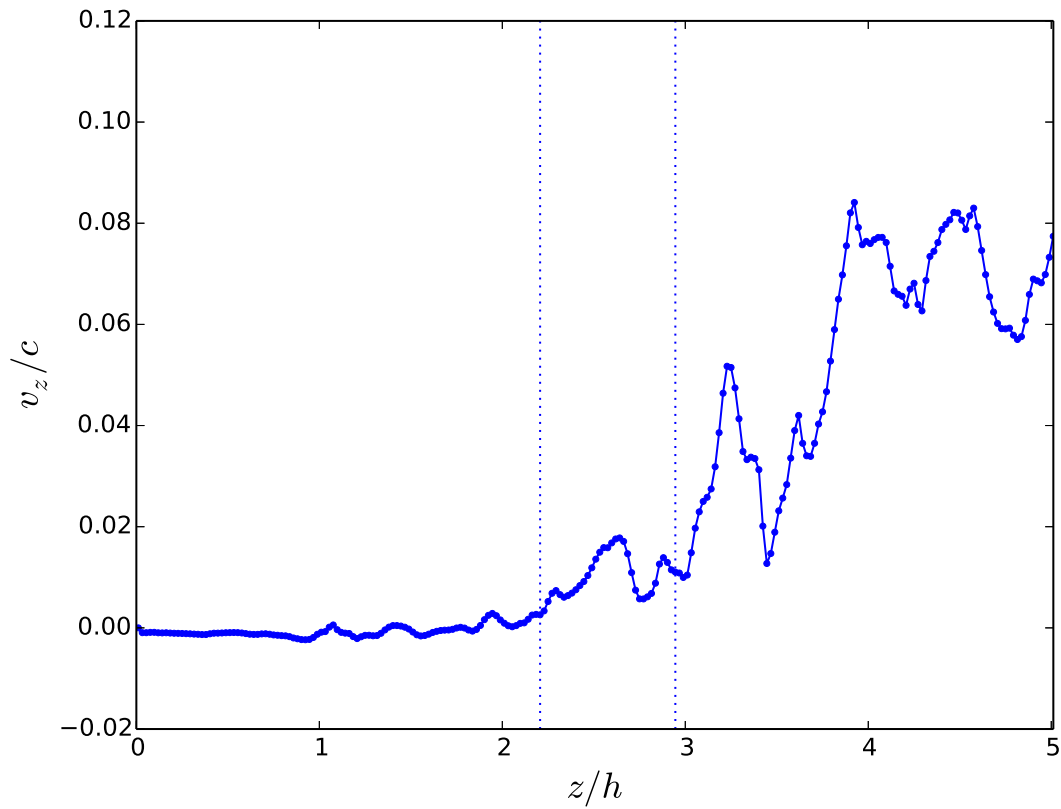

Figure 4.37: Horizontally averaged profile at $r = 14$ for the $M = 2 \times 10^6 M_\odot$, $L/L_{\text{Edd}} = 5$ parameter set (Table 4.1) of the $z$ component of the velocity field. The dashed lines denote where $\tau_s = 1$ and $\tau_s = 10$.

To begin, we calculate spectra with the original velocity field, both with and without subtracting off the horizontally averaged $z$ component, and plot the results for $r = 14$ in Figure 4.38. We see that the spectrum computed with the horizontally averaged $z$ component included is shifted to slightly lower energies. The spectra at the other radii illustrate the same effect. Since any additional second order effect associated with this component can only *increase* upscattering, this energy shift must either be due to the first order effect or vertical radiation advection. As explained in section 4.3.5, vertical radiation advection transports photons through the bulk Comptonization region faster, which decreases the number of photon scatterings in the region and may therefore reduce the overall second order effect. In order to show that the energy shift is predominantly due to the first order effect, not radiation advection, we calculate spectra with uniform temperature profiles both for the case of no velocities and for the case where only the horizontally averaged $z$ component is included, and plot the results for $r = 14$ in Figure 4.39. The spectra at other radii illustrate the same effect. Since a uniform temperature profile with no velocity field has no effect on the base spectrum there is no second order effect, and so adding in the horizontally averaged $z$ component of the velocity field can shift the resulting spectra to lower energies only through the first order effect, not radiation advection. Since the spectrum in Figure 4.39 is shifted by the same amount as in Figure 4.38, we conclude that the original shift is predominantly due to the first order effect, not vertical radiation advection. As a check on this analysis, we repeated the uniform temperature profile spectral calculations but included instead the negative of the horizontally averaged $z$ component of the velocity field, and found that the energy shifts were opposite in direction and equal in magnitude.

We also estimate the energy shift due to the first order effect heuristically and check that the result is consistent with our spectral calculations. The fractional energy change per scattering due to this effect is (Chapter 2)

$$\frac{\Delta\epsilon}{\epsilon} = -\frac{\lambda_{\mathrm{p}}\nabla \cdot \mathbf{v}}{3c}. \tag{4.52}$$

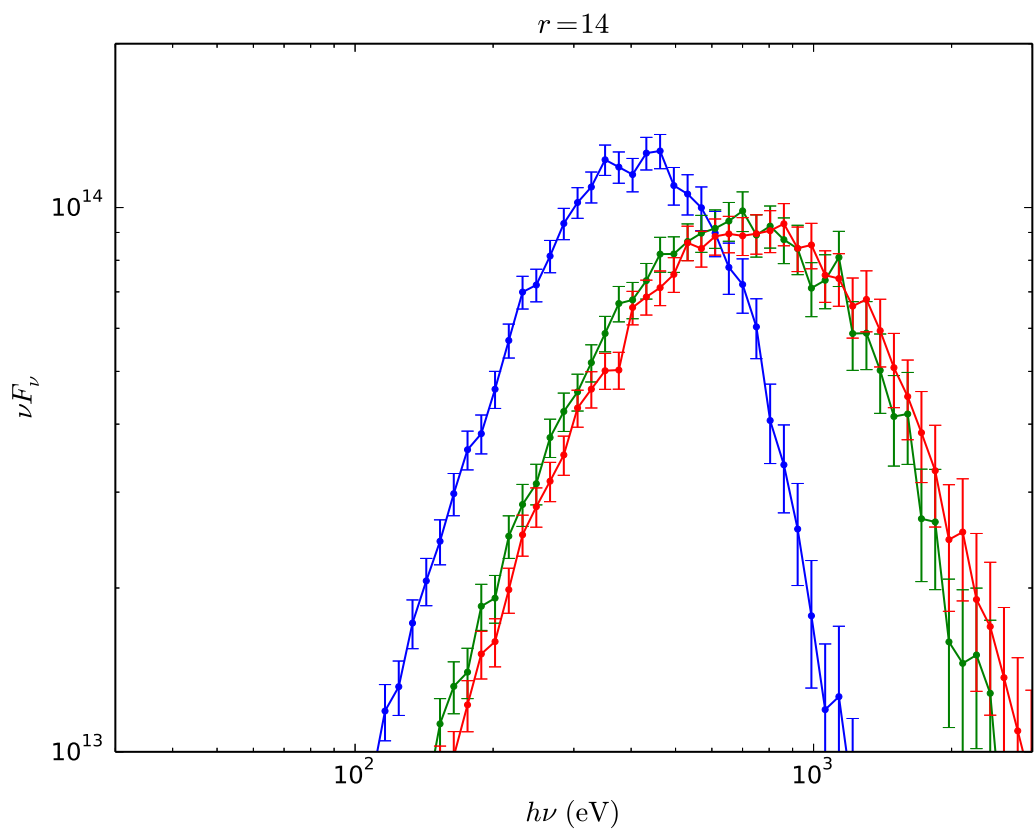

Figure 4.38: Normalized accretion disc spectra at $r = 14$ for the $M = 2 \times 10^6 M_\odot$, $L/L_{\mathrm{Edd}} = 5$ parameter set (Table 4.1) computed with (green) and without (blue) the velocities. For the red curve, the spectrum was computed with velocities but the horizontally averaged $z$ component of the velocity field was subtracted from the total $z$ component.

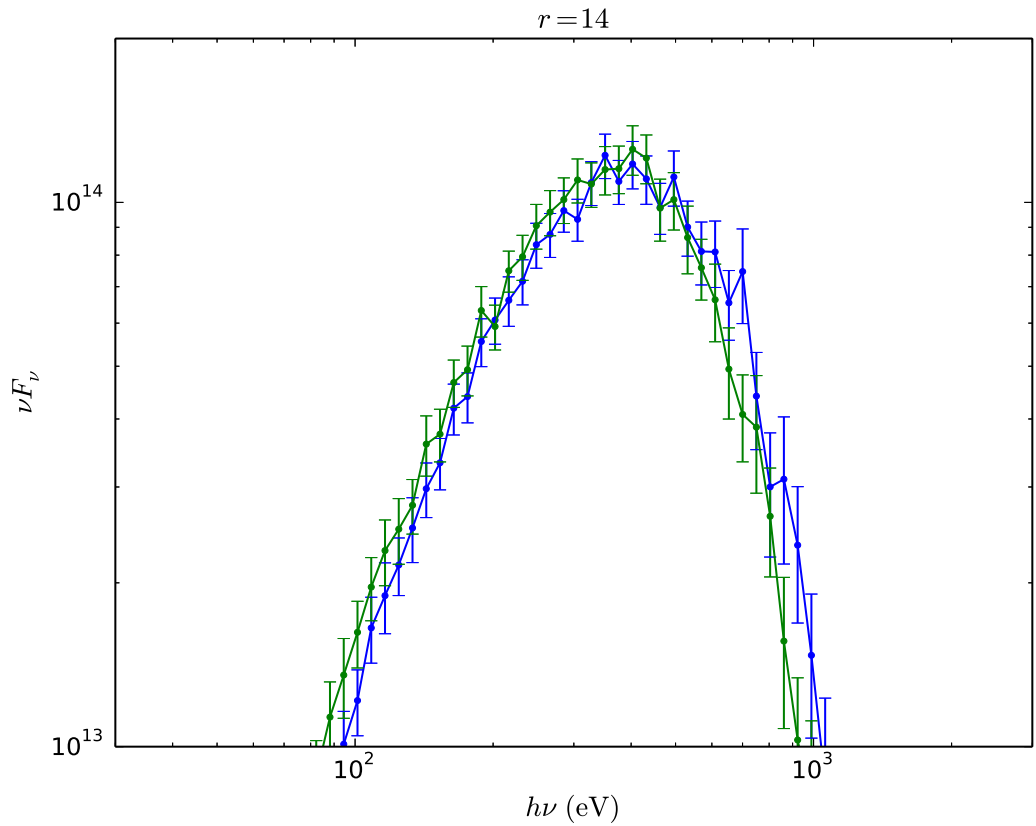

Figure 4.39: Normalized accretion disc spectra at $r = 14$ for the $M = 2 \times 10^6 M_\odot$, $L/L_{\mathrm{Edd}} = 5$ parameter set (Table 4.1), computed with data truncated at $\tau_{\mathrm{s}} = 20$. All gas temperatures were set to the horizontally averaged value at the base. For the blue curve, the velocities were not included, and for the green curve only the horizontally averaged $z$ component of the velocity field was included.

In this case, therefore, the region where this effect is greatest is near the photosphere (Figure 4.37), where we conveniently just confirmed that vertical advection is dominated by diffusion. It follows that the average number of scatterings $dN$ in a region of optical depth $d\tau$ is approximately (section 4.3.2) equal to $1.6(2\tau d\tau)$. The total approximate fractional energy change $f$ in this region is then

$$f = -1 + \lim_{\Delta\tau \to 0} \prod_i \left( 1 - \frac{\lambda_{\mathrm{p}} \nabla \cdot \mathbf{v}}{3c} \right)^{1.6(2\tau_i \Delta\tau_i)} \tag{4.53}$$

$$= -1 + \lim_{\Delta\tau \to 0} \prod_i \exp\left( \ln\left( \left( 1 - \frac{\lambda_{\mathrm{p}} \nabla \cdot \mathbf{v}}{3c} \right)^{1.6(2\tau_i \Delta\tau_i)} \right) \right) \tag{4.54}$$

$$= -1 + \exp\left( \int 1.6 \ln\left( 1 - \frac{\lambda_{\mathrm{p}} \nabla \cdot \mathbf{v}}{3c} \right) 2\tau d\tau \right). \tag{4.55}$$

Since the fractional energy change per scattering is much smaller than unity,

$$f \approx -1 + \exp\left( \int -1.6 \left( \frac{\lambda_{\mathrm{p}} \nabla \cdot \mathbf{v}}{3c} \right) 2\tau d\tau \right). \tag{4.56}$$

In this case we find that at all radii $f \approx -0.1$, consistent with the results in Figures 4.38 and 4.39. We note that if the *total* fractional energy change is also much less than unity, such as in this case, then

$$f \approx \int -1.6 \left( \frac{\lambda_{\mathrm{p}} \nabla \cdot \mathbf{v}}{3c} \right) 2\tau d\tau. \tag{4.57}$$

In order for Monte Carlo calculations to self-consistently capture the first order effect, one must take into account the time-dependent nature of the problem (Chapter 2), either by performing time-dependent simulations or by careful analysis of the results. This is because this effect can result in either upscattering or downscattering depending on whether a region is converging or diverging, and a diverging region will typically evolve into a converging one on the flow timescale, given by

$$t_\mathrm{f} \sim \lambda_\mathrm{p} \tau / v_\mathrm{z}. \tag{4.58}$$

If the region is near the photosphere, such as in the case examined here, then the photons escape the region on the diffusion timescale, which is shorter, and the first order effect will on average broaden the spectrum. If the region is sufficiently deep inside the photosphere that the diffusion timescale,

$$t_\mathrm{d} \sim \lambda_\mathrm{p} \tau^2 / c, \tag{4.59}$$

exceeds the flow timescale, then the flow will change significantly before photons can diffuse very far. This is the case for standing acoustic modes, for example (Blaes et al., 2011). In this case, results from time-independent Monte Carlo simulations can be trusted only if the spectrum is negligibly affected by the upscattering or downscattering in such regions.

In order to capture only second order effects in a Monte Carlo simulation, we can first subtract off the horizontally averaged $z$ component of the velocity field, but we can do this only if its second order effect is negligible. To investigate this, in Figure 4.40 for $r = 14$ we plot the original wave temperature profile along with the wave temperature profile computed by first subtracting off the horizontally averaged $z$ component of the velocity field. We see that the resulting two curves are essentially identical except in a small region where bulk Comptonization is negligible since $\tau_\mathrm{s} \ll 1$. The horizontally averaged $v_\mathrm{z}$ profile contributes negligibly to the wave temperature both because it contributes negligibly to the underlying bulk temperature profile, plotted in Figure 4.41, and because the wave temperature downweights long wavelength variations (section 4.2.2). To confirm that the contribution of the horizontally averaged $v_z$ profile to the second order effect is negligible, we calculate spectra in which we model the turbulence with wave temperatures calculated both with and without including the horizontally averaged $v_z$ and plot the results for $r = 14$ in Figure 4.42. We see that there is no discrepancy between the respective curves, consistent with the wave temperature profiles in Figure 4.40. The spectra at the other radii illustrate the same effect.

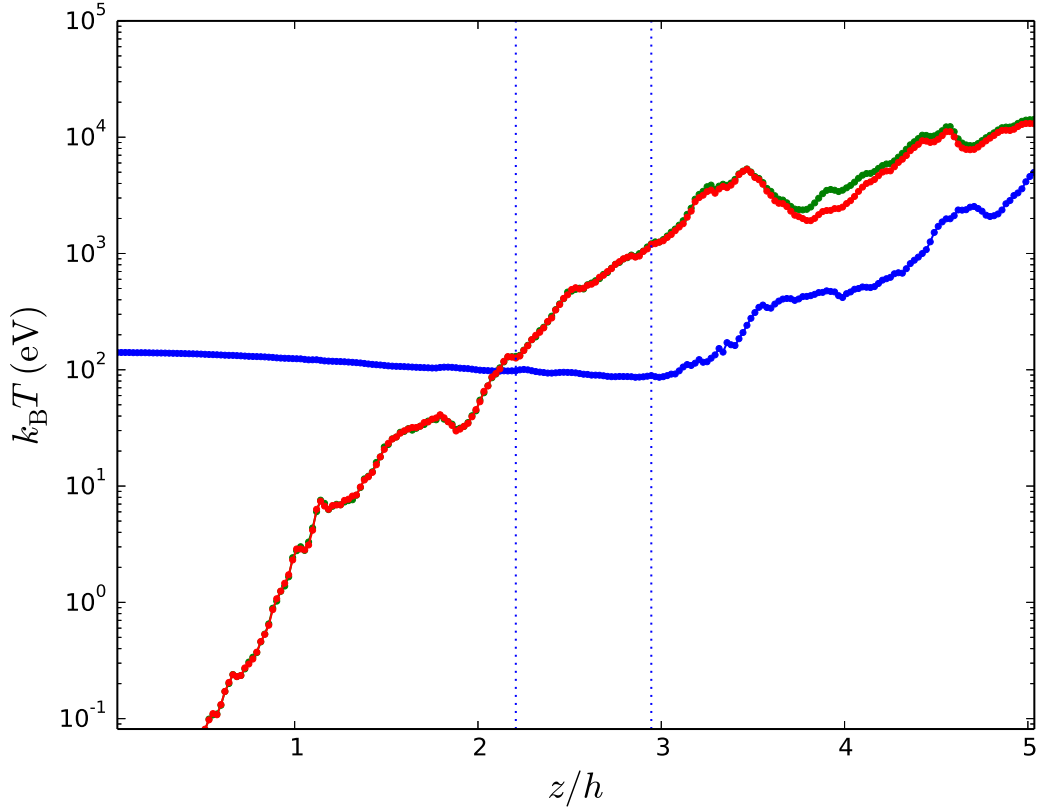

Figure 4.40: Horizontally averaged profiles at $r = 14$ for the $M = 2 \times 10^6 M_\odot$, $L/L_{\mathrm{Edd}} = 5$ parameter set (Table 4.1) for the gas temperature $T_{\mathrm{g}}$ (blue), wave temperature $T_{\mathrm{w}}$ (green), and wave temperature computed by first subtracting off the horizontally averaged $z$ component of the velocity field (red). The dashed lines denote where $\tau_{\mathrm{s}} = 1$ and $\tau_{\mathrm{s}} = 10$.

In section 4.2.2 we showed in Figure 4.3 that spectra computed with data in which the velocities were turned off and the wave temperatures were added to the gas temperatures approximated spectra computed with the velocities. Since the wave temperature captures only second order effects and since the horizontally averaged $z$ component of the velocity field results in a non-negligible first order effect, we expect that when this component is subtracted off the approximation will improve. We perform this comparison in Figure 4.43 for $r = 14$. We see that in this case the approximation is so good that the respective spectra are indistinguishable from each other. The spectra at the other radii illustrate the same effect. This is not only consistent with our prediction but shows that the wave temperature captures second order effects even more accurately than we originally had reason to believe based on the preliminary analysis in section 4.2.2. Given these results, we also expect that spectra computed with the velocities turned off except for the horizontally averaged $z$ component and with the wave temperatures added to the gas temperatures will coincide with spectra computed with the velocities turned on, since both sets of spectra should capture both first and second order effects. We plot these spectra in Figure 4.44 for $r = 14$ and see that they agree with our prediction. The spectra at the other radii illustrate the same effect.

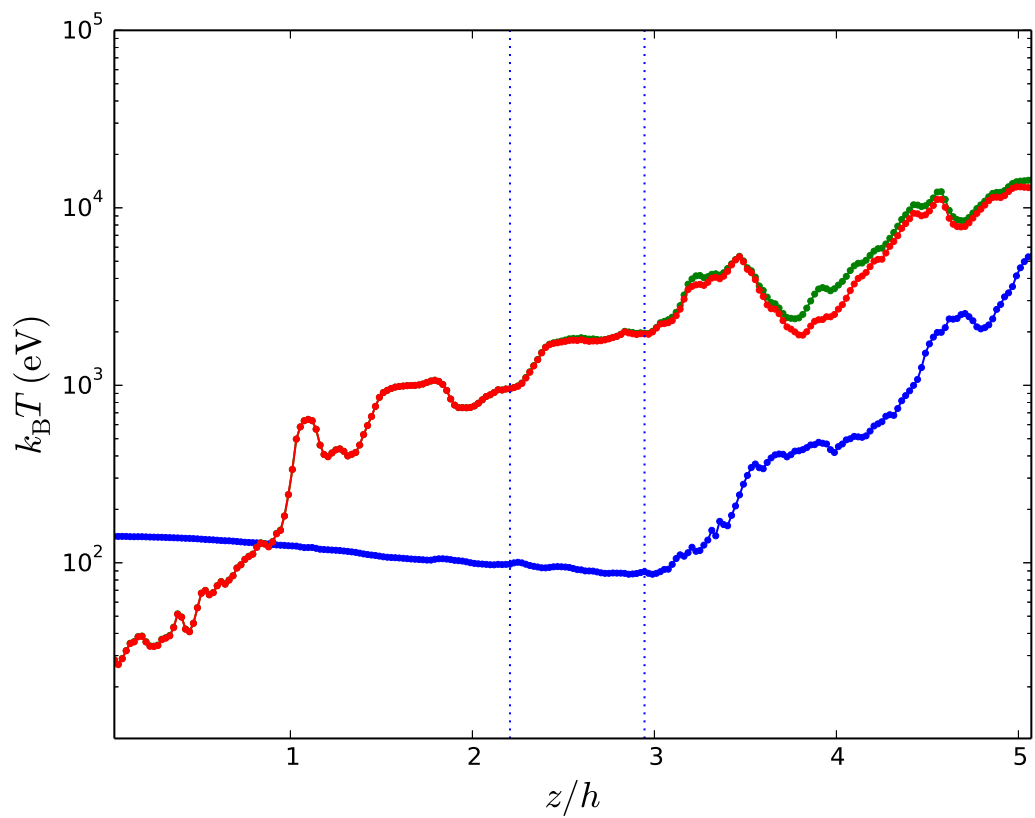

Figure 4.41: Horizontally averaged profiles at $r = 14$ for the $M = 2 \times 10^6 M_\odot$, $L/L_{\mathrm{Edd}} = 5$ parameter set (Table 4.1) for the gas temperature $T_{\mathrm{g}}$ (blue), bulk temperature $T_{\mathrm{bulk}}$ (green), and bulk temperature computed by first subtracting off the horizontally averaged $z$ component of the velocity field (red). The dashed lines denote where $\tau_{\mathrm{s}} = 1$ and $\tau_{\mathrm{s}} = 10$.

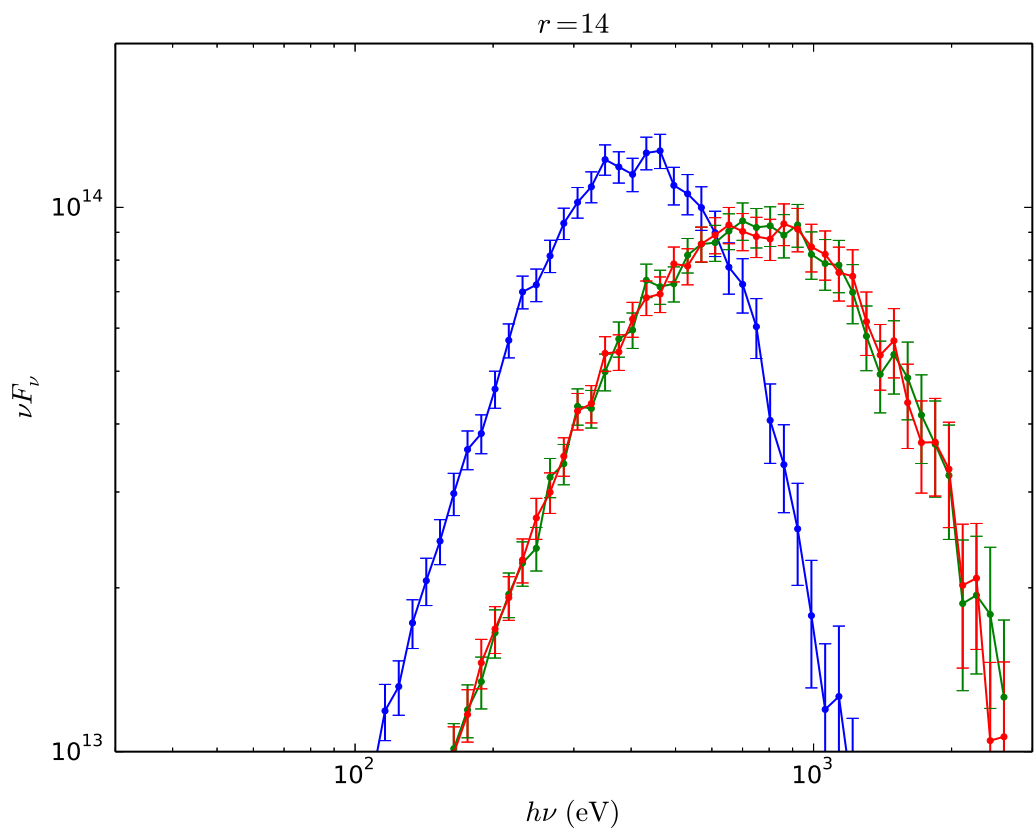

Figure 4.42: Normalized accretion disc spectra at $r = 14$ for the $M = 2 \times 10^6 M_\odot$, $L/L_{\mathrm{Edd}} = 5$ parameter set (Table 4.1) computed without the velocities. For the green and red curves, the wave temperatures were added to the gas temperatures. For the red curve, the wave temperatures were computed by first subtracting off the horizontally averaged $z$ component of the velocity field.

## 4.5   Summary

We have simplified the bulk Comptonization model of Chapter 3 in order to explore a larger space of accretion disc parameters and develop greater physical insight into this phenomenon. Rather than fit the temperature and optical depth to spectra computed with Monte Carlo post-processing simulations, we developed a procedure to calculate the Comptonization temperature and optical depth directly from the underlying vertical structure data (section 4.2). Using this, we plotted the dependence of the Comptonization parameters on the shearing box parameters and showed how these results can be under-stood in terms of the one dimensional temperature profiles (sections 4.3.1 and 4.3.2). We then showed how we can analytically determine the dependence of bulk Comptonization on each accretion disc parameter individually (sections 4.3.3 and 4.3.4). Our principal results are as follows.

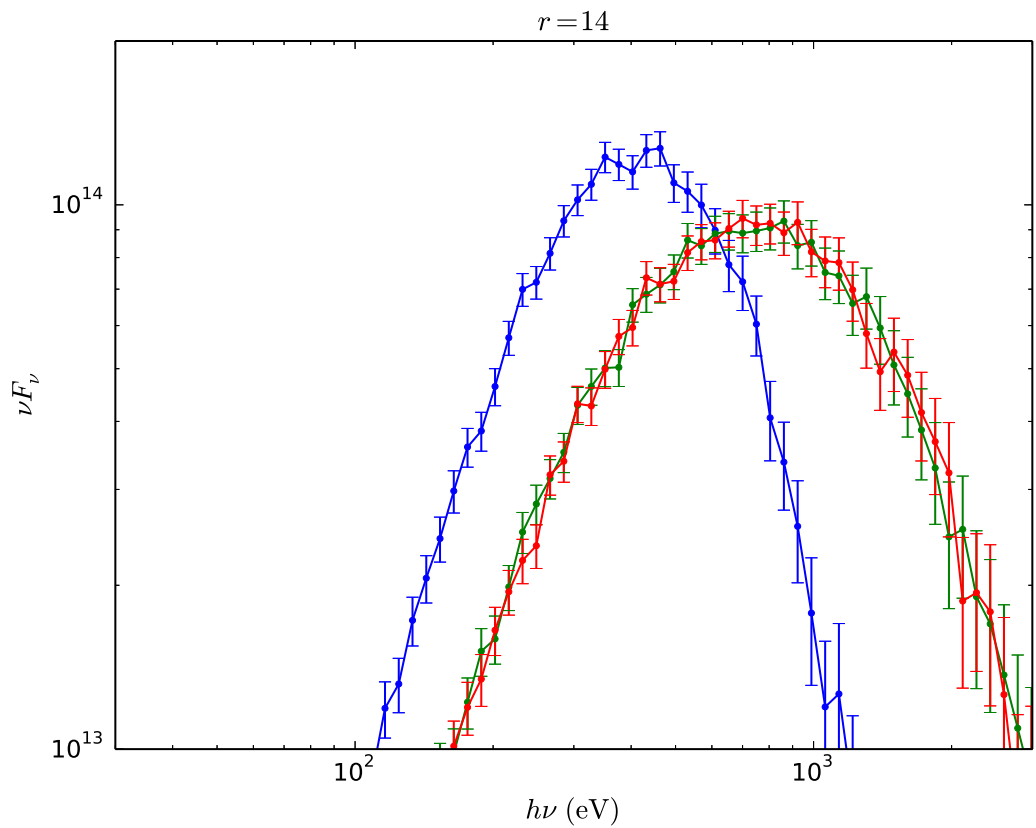

Figure 4.43: Normalized accretion disc spectra at $r = 14$ for the $M = 2 \times 10^6 M_\odot$, $L/L_{\rm Edd} = 5$ parameter set (Table 4.1) computed with (green) and without (blue) the velocities. For the spectrum computed with velocities, the horizontally averaged $z$ component of the velocity field was subtracted from the total $z$ component. For the red curve, the velocities were not included but the wave temperatures were added to the gas temperatures.

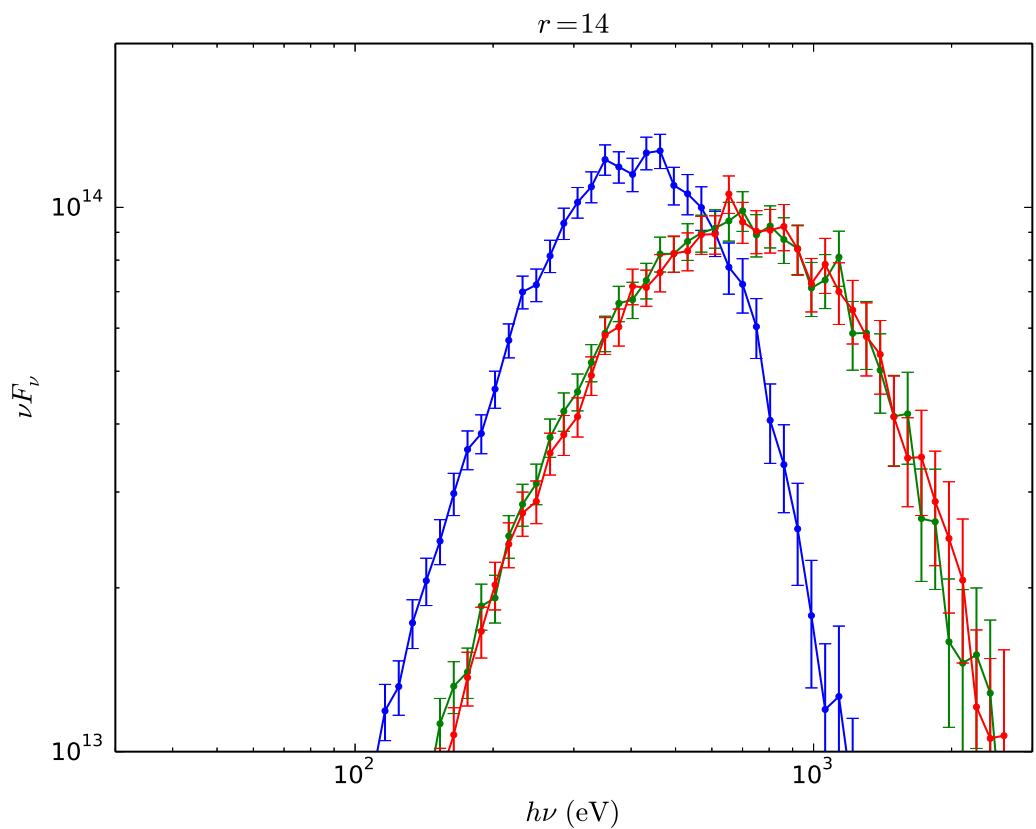

Figure 4.44: Normalized accretion disc spectra at $r = 14$ for the $M = 2 \times 10^6 M_\odot$, $L/L_{\mathrm{Edd}} = 5$ parameter set (Table 4.1) computed with (green) and without (blue) the velocities. For the red curve, the $z$ component of the velocity field was horizontally averaged, the $x$ and $y$ components were set to zero, and the wave temperatures were added to the gas temperatures.

The primary independent variables in a shearing box are the surface density $\Sigma$, the vertical epicyclic frequency $\Omega_z$, and the strain rate, $\partial_x v_y$. We also allow $\alpha$, the ratio of the vertically integrated stress to the vertically integrated total pressure, to vary. For Kerr discs the scalings for the strain rate and vertical epicyclic frequency are always nearly equal (equation 4.10), which leaves three independent parameters. Using the velocity scalings (equations 4.18, 4.19, and 4.20), we showed that the dependence of the Comptonization parameters on $\alpha$ can be subsumed into the other parameters (equations 4.11, 4.12, and 4.13), which reduces the parameter space to two variables, $\Sigma$ and $\Omega_z$.

We plotted the dependence of the bulk Comptonization temperature, optical depth, and $y$ parameter on $\Sigma$ and $\Omega_z$ (Figure 4.12). We showed that these results can be understood by analyzing the one dimensional temperature profiles (Figure 4.13) and the velocity scalings (equations 4.18, 4.19, and 4.20). In particular, the Comptonization optical depth and $y$ parameter increase strongly with increasing $\Sigma^{-1}$ and weakly with increasing $\Omega_z^{-1}$. The Comptonization temperature also increases strongly with increasing $\Sigma^{-1}$, but decreases weakly with increasing $\Omega_z^{-1}$.

To determine the dependence of bulk Comptonization on accretion disc parameters, we write $\Sigma$ in terms of $F$ and then write the scalings for $F$ and $\Omega_z$ in terms of mass, luminosity, radius, spin, and inner boundary condition (section 4.3.3). Since $\Omega_z^{-1}$ is directly proportional to mass, and $\Sigma$ is independent of mass, the dependence of bulk Comptonization on mass is identical to its dependence on $\Omega_z^{-1}$. Similarly, since $\Sigma^{-1}$ is directly proportional to luminosity, and $\Omega_z^{-1}$ is independent of luminosity, the dependence of bulk Comptonization on luminosity is identical to its dependence on $\Sigma^{-1}$. Therefore,

Figure 4.12 also summarizes the dependence of bulk Comptonization on mass and luminosity. Here, for clarity, we reproduce the plots from Figure 4.12 in Figure 4.45 with the independent variables labeled as mass and luminosity.

The dependence of bulk Comptonization on the other accretion disc parameters is inferred by analyzing how they affect $\Sigma^{-1}$ since bulk Comptonization depends much more strongly on $\Sigma^{-1}$ than it does on $\Omega_z^{-1}$. Since $\Sigma^{-1}$ is proportional to the flux $F$ (equation 4.28), we showed that the dependence of bulk Comptonization on the other disc parameters can be understood intuitively in terms of how they effect $F$. In particular, at large radius (i.e. $r \gg r_{\text{in}}$) bulk Comptonization always decreases with increasing radius. At small radius, whether bulk Comptonization increases or decreases with radius depends on the inner boundary condition. Using the same line of reasoning, we showed that bulk Comptonization increases with both spin and the inner boundary condition parameter $\Delta\epsilon$ at small radius ($r \approx r_{\text{in}}$), and decreases with those parameters at large radius. Finally, we showed that bulk Comptonization increases with $\alpha$, since once the accretion disc parameters are substituted in for $\Sigma$ and $\Omega_z$, $\Sigma^{-1}$ itself becomes proportional to $\alpha$ (equation 4.28) and this outweighs the dependence on $\alpha$ discussed earlier.

Next we studied bulk Comptonization for an entire accretion disc by examining how it varies when the radius is fixed to the region of maximum luminosity (section 4.3.4). The dependence of bulk Comptonization on mass, luminosity, and $\alpha$ is unchanged from above since the radius of maximum luminosity does not vary with these parameters. But since this radius does depend on the spin and inner boundary condition parameter $\Delta\epsilon$, the dependence of bulk Comptonization on these parameters required a new treatment. We showed that in this case bulk Comptonization always increases with spin and $\Delta\epsilon$.

In section 4.3.5 we showed that the effect of including vertical radiation advection at a fixed radius in an accretion disc is to decrease bulk Comptonizaton. We discussed how to include advection in our model more formally in Appendix D5.

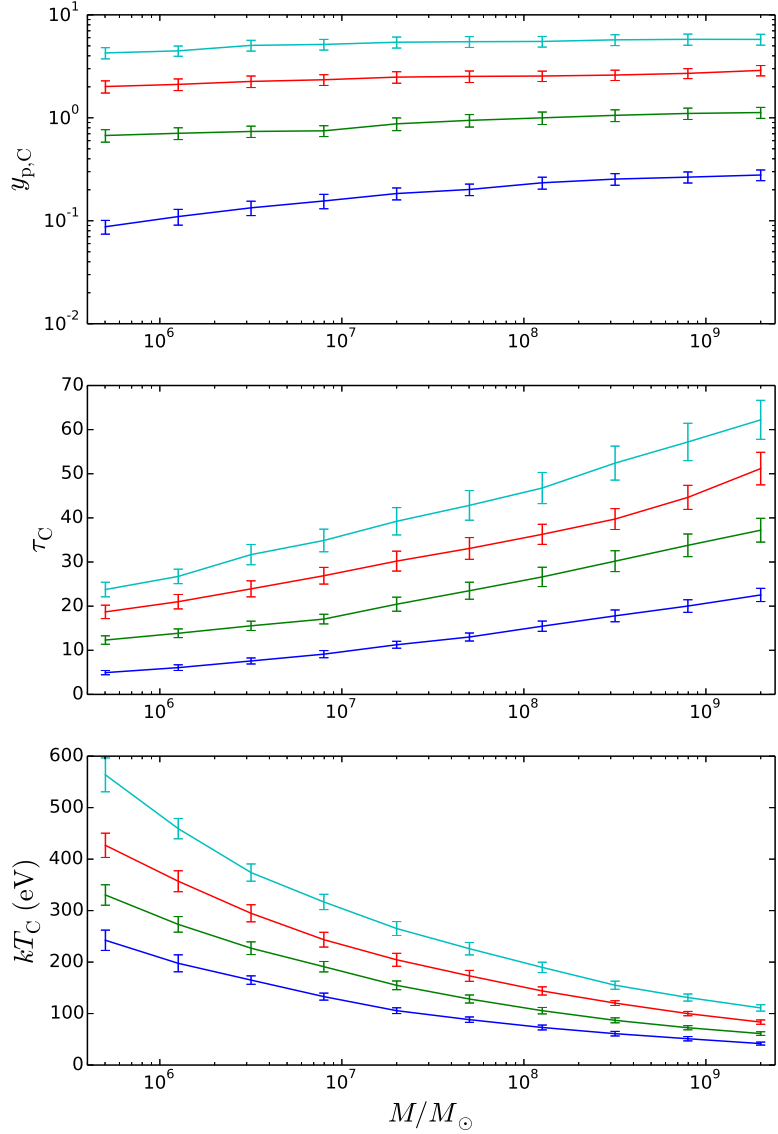

Figure 4.45: Dependence of bulk Comptonization on mass. The blue, green, red, and cyan curves correspond to $L/L_{\mathrm{fid}} = 1$, 2, 3.3, and 5, respectively, where $L_{\mathrm{fid}}/L_{\mathrm{Edd}} = 2.5$. The parameters held constant are $r = 20$, $a = 0$, $\Delta\epsilon = 0$, and $\alpha/\alpha_0 = 2$ (i.e. the other parameters in the $M = 2 \times 10^6 M_\odot$, $L/L_{\mathrm{Edd}} = 2.5$ parameter set (Table 4.1), with $r = 20$). The only difference between this figure and Figure 4.12 is the labeling of the axes.

In section 4.4.1 we pointed out that our results broadly agree with the results of Chapter 3, which in turn agree with the analysis by D12 of the narrow-line Seyfert 1 REJ1034+396.

An important result of this work is that bulk Comptonization is strongly dependent on the disc inner boundary condition (section 4.4.2). In particular, the larger that $\Delta\epsilon$ is, the lower the luminosity can be without bulk Comptonization being negligible. Figure 4.46 summarizes the dependence of bulk Comptonization on mass and luminosity for $\Delta\epsilon = 0.2$. In both Figures 4.45 and 4.46 the radius is fixed to the region where the luminosity is greatest, but for $\Delta\epsilon = 0.2$ this radius corresponds to the innermost stable circular orbit. By comparing these two figures we see that for a given luminosity bulk Comptonization is significantly greater for $\Delta\epsilon = 0.2$. We also showed (Figure 4.36) that bulk Comptonization is negligible in black hole X-ray binaries unless the disc inner boundary condition parameter is very large ($\Delta\epsilon \sim 0.1$) or the luminosity greatly exceeds Eddington.

We expect that in a real disc bulk Comptonization at a given radius will be accompanied by a decrease in the gas temperature at the effective photosphere in order to leave the flux unchanged, which is required by energy conservation (section 4.4.5).

Because our model connects the bulk Comptonization parameters to the disc vertical structure one dimensional temperature profiles in a way that is physically intuitive, it provides a useful framework for understanding bulk Comptonization even in situations in which some of our specific results may not hold, such as shearing box or global radiation MHD simulations run in entirely different regimes.

Since our results outline how bulk Comptonization depends on fundamental accretion disc parameters, an observer who fits the soft X-ray excess with a warm Comptonization model can use them to distinguish contributions to the soft X-ray excess due to bulk Comptonization from those due to other physical mechanisms. In high Eddington sources, this can help provide a physical basis for and therefore constrain warm Comptonization models of the soft excess. Bulk Comptonization is likely insignificant in lower Eddington flows, on the other hand, even though the data show that in these flows the soft excess carries a more significant fraction of the power (Jin et al., 2012; Mehdipour et al., 2011, 2015). Moreover, since bulk Comptonization depends on the properties of MRI turbulence through $\alpha$ (sections 4.3.3 and 4.3.4) and through the time variability of the temperature and density profiles (section 4.3.6), our work indicates that observations of the soft X-ray excess may in turn advance our understanding of disc turbulence in the radiation pressure dominated regime.

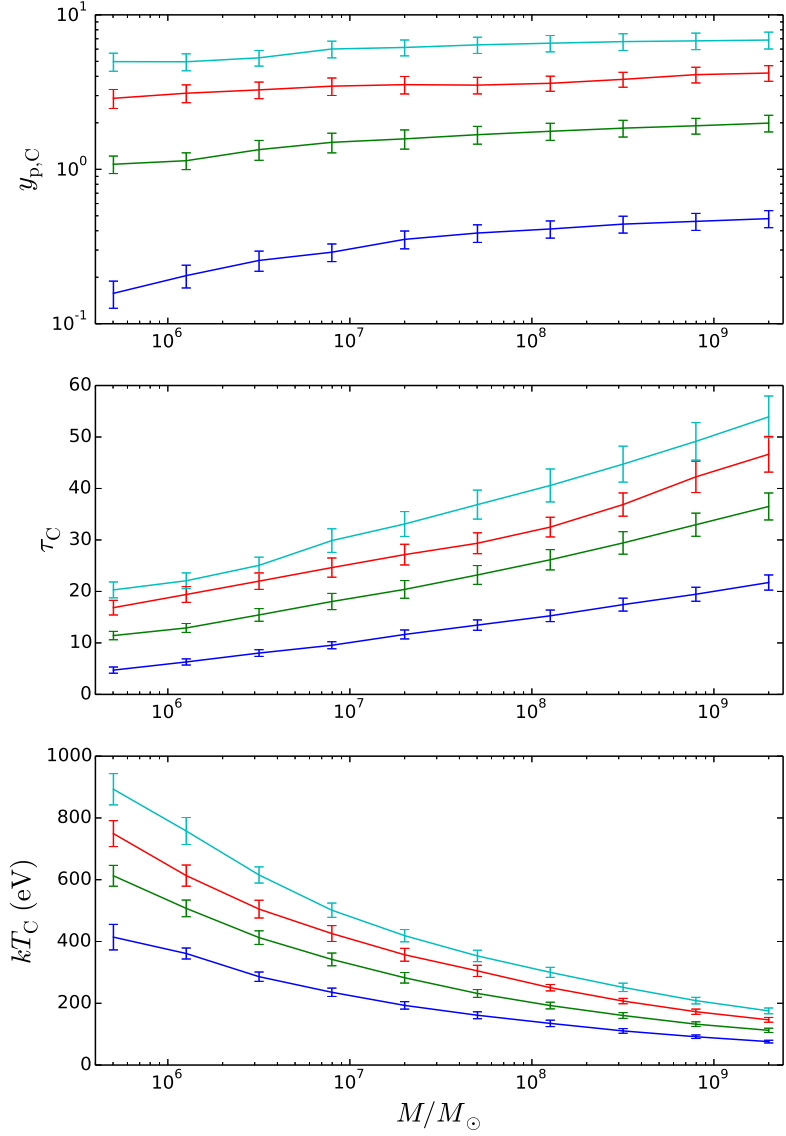

Figure 4.46: Dependence of bulk Comptonization on mass for $\Delta\epsilon = 0.2$. The blue, green, red, and cyan curves correspond to $L/L_{\mathrm{fid}} = 0.25$, $0.5$, $0.75$, and $1$, respectively, where $L_{\mathrm{fid}}/L_{\mathrm{Edd}} = 2.5$. The parameters held constant are $r = 6$, $a = 0$, and $\alpha/\alpha_0 = 2$.

# Chapter 5

# Global Monte Carlo simulations and future work

## 5.1   Introduction

In Chapters 3 and 4 we modeled bulk Comptonization with shearing box simulations. While this approach is effective for developing physical intuition and obtaining preliminary results, it has several limitations. Of course, to the extent that bulk Comptonization turns out to be tied to other global phenomena, its study requires global simulations. But there are also limitations that are problematic specifically for studying bulk Comptonization. The greatest of these is that shearing boxes assume that the disc is geometrically thin, i.e. that $h/r \ll 1$, where $h$ and $r$ are the disc scale height and radius, respectively. This approximation starts to break down when the luminosity approaches a significant fraction of the Eddington luminosity, but according to the results in Chapter 4 this is also when bulk Comptonization starts to become significant. To properly study the high Eddington regimes most important for bulk Comptonization, therefore, requires global disc simulations. Another limitation is that Monte Carlo shearing box simulations can-not properly calculate bulk Comptonization by shear even for thin discs. In Chapter 3 we included the shear flow in the velocity field, but this is problematic because for the periodic boundary condition in the $r$ direction this results in a discontinuous, unphysical velocity field. We discuss the difficulties with implementing shear in Monte Carlo shearing box simulations in section 5.4.1. To properly study bulk Comptonization by shear, therefore, requires global disc simulations. On the other hand, the advantage of shearing box simulations is that the turbulence is better resolved since they focus on a small patch of an accretion disc. Because turbulent Comptonization depends on velocity differences between subsequent photon scatterings (Chapter 2), global disc simulations may underestimate bulk Comptonization if the turbulence is underresolved.

In this chapter we discuss global Monte Carlo simulations of accretion disc spectra and future work. We use these simulations to study bulk Comptonization and to explore spectra of radiation MHD simulations more broadly. This chapter is organized as follows. We first give a broad overview of our implementation of global Monte Carlo simulations (section 5.2). We then discuss preliminary results (section 5.3) and future work (section 5.4).

## 5.2   Implementation

In this section we give an overview of our implementation of the global Monte Carlo code. In Appendix G we describe the code in greater detail (Appendix G1), how to use it (Appendix G2), additional options (Appendix G3), and the problems we used to test it (Appendix G4).

**Photon position coordinates**   The main difference between the shearing box code and the global simulation code is the geometry of the grid cells, which in turn influenced our choice of photon position coordinates. The shearing box position coordinates are the cartesian coordinates $(x, y, z)$ while the global coordinates are $(r, \phi, \theta')$, where $r$ and $\phi$ are the usual spherical coordinates and $\theta' = \pi/2 - \theta$. Note that this differs slightly from the usual spherical coordinates $(r, \theta, \phi)$. The global coordinates are instead a simple generalization of the shearing box coordinates, since in the shearing box limit $\hat{\mathbf{x}} = \hat{\mathbf{r}}$, $\hat{\mathbf{y}} = \hat{\phi}$, $\hat{\mathbf{z}} = \hat{\theta}'$, and $z = 0$ corresponds to $\theta' = 0$.

**Photon unit wave vector components**   In the global code the photon unit wave vector components are still $(k_x, k_y, k_z)$ rather than $(k_r, k_\phi, k_{\theta'})$. We made this choice for two reasons. First, it is more computationally efficient. Photons travel in straight lines between scatterings, along which the components $(k_x, k_y, k_z)$ remain the same while the components $(k_r, k_\phi, k_{\theta'})$ change. Second, $(k_x, k_y, k_z)$ are well-defined everywhere, while $(k_r, k_\phi, k_{\theta'})$ are not well-defined on the poles. Even near the poles we were concerned that this could lead to consequential numerical errors.

**Grid spacing**   The global grid is spaced linearly in $\phi$ and $\theta'$ and logarithmically in $r$. We note that each cell therefore does not occupy the same solid angle. Since $d\Omega = \sin(\theta)d\theta d\phi = \cos(\theta')d\theta d\phi$, each cell occupies a vanishingly small solid angle near the poles. For each cell to occupy the same solid angle the grid could instead be spaced linearly in $\sin(\theta')$, not $\theta'$. Our choice was made out of necessity in order to conform to the radiation MHD simulation data of interest. The $r$ coordinate of the lower $r$ boundary of a cell with indices $(i, j, k)$ we denote $r_i$, and we denote the $\phi$ and $\theta'$ boundary coordinates analogously.

**Boundary conditions**   Photons that cross the lower boundary for the $r$ coordinate escape the grid and are discarded (i.e. they are not binned). Photons that cross the upper boundary for the $r$ coordinate escape the grid and are binned. The boundary condition for the $\phi$ coordinate is periodic. The default boundary condition for the $\theta'$ coordinate is to increment $\phi$ by $\pi$. This reflects the photon's position across the $z$ axis. If the bounds on $\theta'$ are given by $-\pi/2 \leq \theta' \leq \pi/2$, as usual, then this simply propagates the photon in a straight line across the $z$ axis. We discuss the optional escape boundary condition for $\theta'$ in Appendix G3.

**Photon propagation**   The non-rectangular geometry of the cells in the global grid significantly complicates the problem of propagating photons. Unlike in the shearing box case, it is tricky to even figure out the next cell that a photon enters based on its current trajectory, and this is just the analytical aspect of the problem. Due to consequential rounding errors introduced in this process, there is also a numerical aspect. Since the bare propagation algorithm sometimes fails to propagate photons across the correct cell boundary, we had to modify it to robustly correct for such errors. We describe both aspects of this problem in greater detail in Appendix G1.

**Output**   The code outputs the total luminosity distribution $L(\nu, \mu_k)$ as a function of the frequency $\nu$ and the $z$ component of the photon wave vector, $\mu_k = k_z = \cos(\theta_k)$. We note that $\theta_k$ is therefore the polar angle of the photon unit wave vector $\mathbf{k}$, not to be confused with the polar angle $\theta$ of the photon position vector. We define $\phi_k$ analogously. The luminosity of photons within frequency range $d\nu$ and solid angle $d\Omega = d\mu_k d\phi_k$ is therefore given by $L(\nu, \mu_k)d\nu d\mu_k d\phi_k$. The code outputs $L(\nu, \mu_k)$ for each frequency bin and $\mu_k$ bin. If there are $n_\nu$ bins for frequency and $n_\mu$ bins for $k_z$, then the code outputs $n_\nu \times n_\mu$ values of $L(\nu, \mu_k)$. The code also estimates the error $L_{\mathrm{err}}(\nu, \mu_k)$ at each value of the luminosity distribution. We discuss additional output options in Appendix G3.

## 5.3 Preliminary results

We perform Monte Carlo spectral calculations for two radiation MHD simulations of black hole accretion discs with $M = 5 \times 10^8 M_\odot$. For one, $L \sim 0.08 L_{\text{Edd}}$, and for the other $L \sim 0.2 L_{\text{Edd}}$. It is noteworthy that both simulations are magnetically dominated–i.e. the magnetic pressure is greater than the radiation and gas pressures. This was unexpected since for higher Eddington flows we usually expect the radiation to dominate the total pressure. We believe this is due to the initial magnetic field configuration, which consists of two poloidal loops above and below the disc midplane. The disc shear flow converts the initial poloidal field into toroidal field due to flux freezing, magnifying the initial field.

In the original radiation MHD simulation the initial gas density is concentrated in a torus symmetric about the $z$ axis, far from the origin. The accretion disc begins to form as turbulent stresses transport the gas to smaller radii. Once the disc reaches the inner boundary, the innermost regions of the disc begin to form a steady-state flow, followed by regions at larger radius, etc. For a given region we are most interested in the dynamics of the flow after this point in time has been reached since we want to compare it with observations of real systems. For Monte Carlo global simulations, therefore, we use a snapshot of the MHD simulation from a point in time at which the innermost regions of the disc have reached an approximately steady-state flow.

We do not run the global simulation on the entire grid since that would include the original torus of gas, which is not part of the disc. We exclude this region by truncating the grid outside some value of $r$ that we denote $r_{\text{max}}$. For the $L \sim 0.08 L_{\text{Edd}}$ simulation we chose $r_{\text{max}} = 35$, and for the $L \sim 0.2 L_{\text{Edd}}$ simulation we chose $r_{\text{max}} = 40$.

We also must exclude photons that are emitted from the edge of the disc, i.e. the midplane region where $\theta' \sim 0$. Such photons escape only because we have truncated the grid, not because they are part of the actual emitted disc spectrum. In other words, we must exclude photons that are not emitted from the photosphere. To do this, we output the spectrum as a function of the angle $\theta'$ at which photons are emitted (see Appendix G3) and exclude photons emitted in some range $\theta'_1 < \theta' < \theta'_2$.

Another way to exclude photons not emitted from the photosphere is to perform another truncation of the grid at the minimum value of $|\theta'|$ for which the disc is optically thin at all $r$. We then output the luminosity as a function of the radius $r$ at which photons are emitted (see Appendix G3). We ultimately prefer this method because it has the additional advantage of producing the emitted spectrum as a function of radius. It may seem that one could determine this with the prior method by looking at the spectrum as a function of $\theta'$, but the problem is that photons that originate at multiple radii may emerge at the same value of $\theta'$. It is also true that even photons that originate from different radii in the disc may escape the grid at the same value of $r$. But this effect is far smaller since truncating the disc near the photosphere places the escape points much closer to the points of origination. Using this method, we truncate the $L \sim 0.08 L_{\mathrm{Edd}}$ simulation at $|\theta'| = \pi/12$ and the $L \sim 0.2 L_{\mathrm{Edd}}$ simulation at $|\theta'| = \pi/6$. We plot the resulting spectra as a function of radius in Figures 5.1 and 5.2, respectively.

The spectra are noteworthy for several reasons. First, they are poorly fit by thermal spectra–i.e. they are non-thermal. For small radii especially, this is because the high energy tail contains a significant fraction of the power. It seems clear that it is the dramatic increase in the gas temperature in the low density region immediately outside the disc that gives rise to the high energy component. This is particularly notable because it would be desirable to be able to account for non-thermal high energy components in observed AGN spectra with self-consistent radiation MHD simulations. Second, even the low energy part of the spectrum at each radius is non-thermal since the power law is more shallow than the $\nu^2$ Rayleigh-Jeans law (i.e. $\nu L_\nu$ is more shallow than $\nu^3$). While this is also true of a multitemperature blackbody, the range of radii is too small to attribute the spectral shape to this effect, although it may play some role. The fact that these discs are magnetically dominated likely is a critical factor. We discuss these ideas further in section 5.4.1.

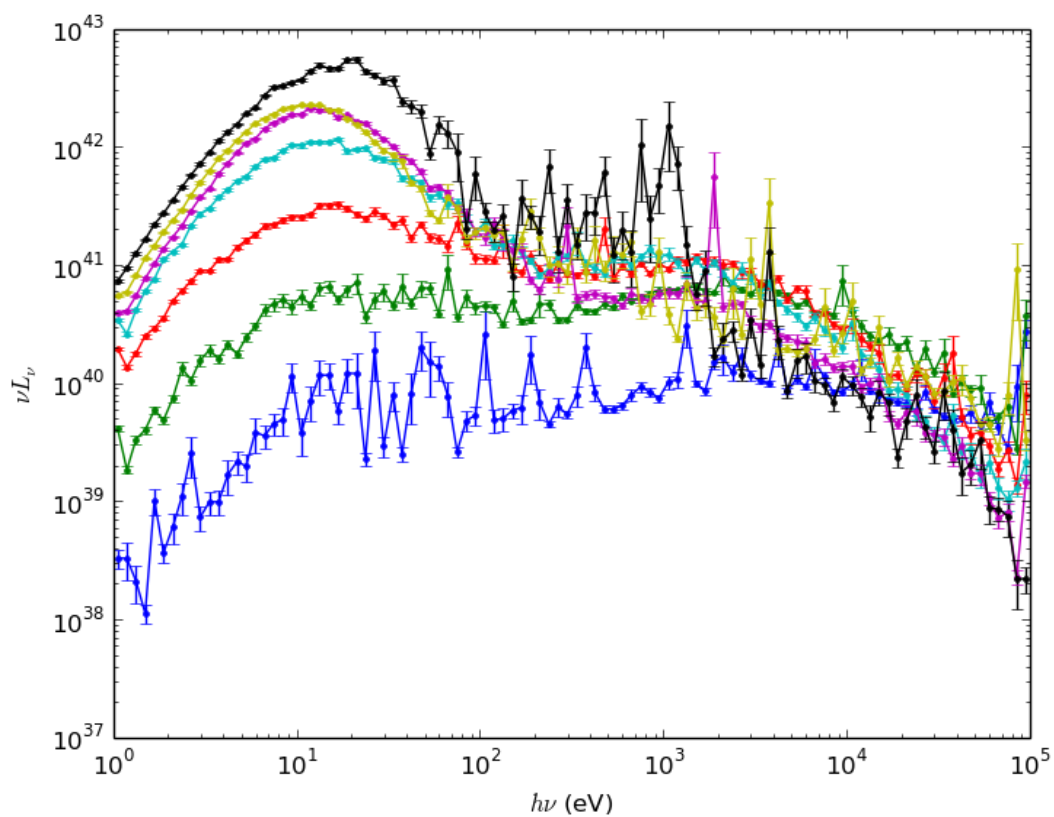

Figure 5.1: Luminosity distribution at $4 \leq r < 6$ (blue), $6 \leq r < 8$ (green), $8 \leq r < 10$ (red), $10 \leq r < 12$ (cyan), $12 \leq r < 15$ (magenta), $15 \leq r < 25$ (yellow), and $25 \leq r < 35$ (black) for the $L \sim 0.08 L_{\mathrm{Edd}}$ simulation computed with the global Monte Carlo code. The luminosity distribution is defined such that the total luminosity in a range $dr$ is $L(r)dr$.

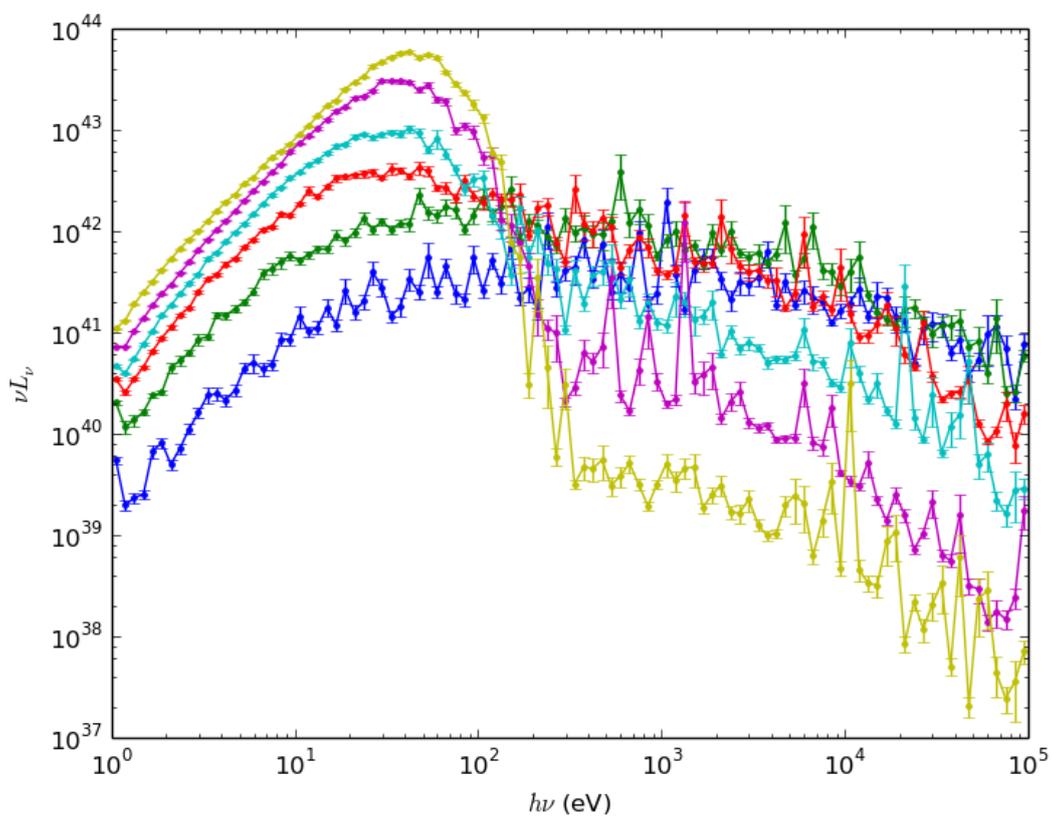

Figure 5.2: Luminosity distribution at $4 \leq r < 6$ (blue), $6 \leq r < 8$ (green), $8 \leq r < 12$ (red), $12 \leq r < 16$ (cyan), $16 \leq r < 25$ (magenta), and $25 \leq r < 40$ (yellow) for the $L \sim 0.2 L_{\mathrm{Edd}}$ simulation computed with the global Monte Carlo code. The luminosity distribution is defined such that the total luminosity in a range $dr$ is $L(r)dr$.

To explore the effect of turbulent Comptonization on the spectra, we compute spectra both including and excluding the turbulent velocities. But to do this, we must distinguish the turbulent velocities from the shear velocities. For a shearing box, the shear velocities are simply given by the background Keplerian flow. But for a global simulation this is inadequate for two reasons. First, this definition only applies to the midplane, i.e. where $z = 0$. For larger values of $|z|$ it is not accurate, although for a sufficiently thin disc it may be a good approximation. Second, it assumes what the shear flow is ahead of time instead of describing whatever happens to arise self-consistently in the simulation. Because of these concerns, we define the shear flow as the density weighted, azimuthal average of the velocity field. Defined this way, it is clear that the part of the velocity field that remains once the shear is subtracted off is actually turbulence, i.e. random fluctuations.

We find that there is no statistically significant difference between spectra computed with and without the turbulent velocities included, either overall or in any narrow radial range, for both the $L = 0.08L_{\text{Edd}}$ and $L = 0.2L_{\text{Edd}}$ simulations. This is not surprising given our shearing box results, which indicate that turbulent Comptonization becomes relevant closer to $L = L_{\text{Edd}}$. On the other hand, we find that for large radii bulk velocities are comparable to thermal velocities just inside the photosphere for the $L = 0.2L_{\text{Edd}}$ simulation, so it is somewhat surprising that we see no effect. We discuss how to continue this analysis in section 5.4.1.

In order for radiation MHD simulations to self-consistently include viscous energy exchange between the radiation and the gas due to bulk Comptonization, it is important that simulations correctly calculate the traceless components of the radiation pressure tensor (Chapter 2). To explore this, we compute the frequency-integrated radiation pressure tensor with the global code and compare it with the tensor computed in the original simulation in Figure 5.3. We see that overall the agreement is good. The discrepancy in the midplane around $r \sim 100$ is due to the fact that it is an extremely optically thick region and the Monte Carlo code does not include stimulated scattering (see section 5.4.2). This does not, however, impact the emitted spectrum since photons from this region are absorbed before they can escape the grid. The other notable discrepancy is at large $r$; the greater total area of darker blue colors suggests that the moment values are smaller for the original simulation in these regions. This discrepancy, however, is statistical, not mathematical; because it is an absolute value plot, the statistical error of the Monte Carlo simulations in these regions leads to greater prevalence of lighter colors. This error could, of course, be reduced by running the Monte Carlo simulation longer.

## 5.4   Future work

We now outline future work, both in bulk Comptonization and in other areas that can be explored with the global Monte Carlo code. In section 5.4.1 we discuss future work with the current global Monte Carlo code, in section 5.4.2 we discuss how the code can be improved, and in section 5.4.3 we discuss how to extend the theoretical analysis of Chapter 2.

### 5.4.1   Global Monte Carlo spectral calculations

**Separate high energy signal from noise**

For Monte Carlo shearing box simulations, the spectrum is unphysical above a certain photon energy value. The reason for this is that in each grid cell photon packets are drawn from a distribution that is uniform in log space over a fixed energy range. As a result, all energy bins end up with non-zero values, even if the simulation is not run long enough for the spectrum to converge at some energies. Typically as one moves from low energy bins to higher energy bins the errors bars increase, and this is consistent with the increasing lack of smoothness seen in the spectrum as one moves in this direction. Above a certain point, however, the spectrum suddenly changes; the error bars suddenly decrease and the spectrum suddenly becomes smooth again. It is this region that is usually unphysical.

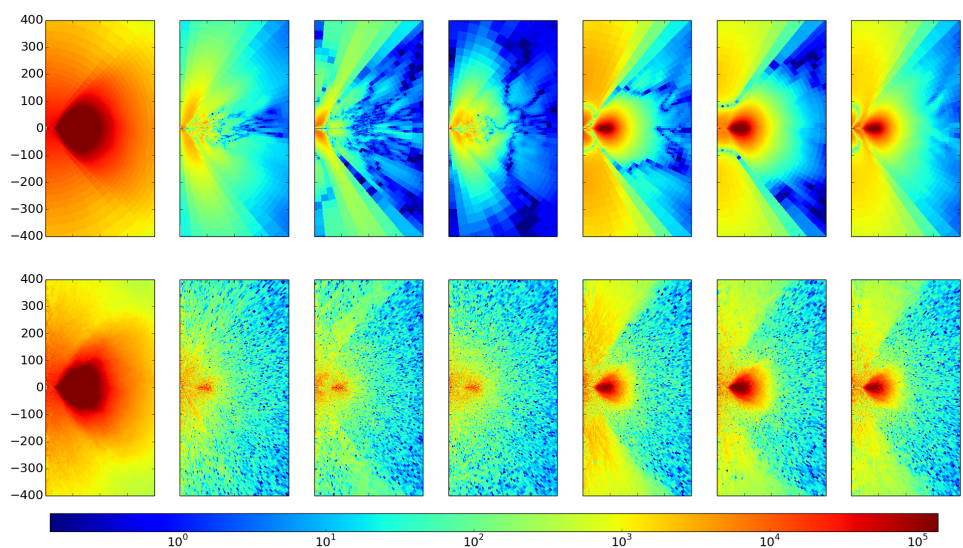

Figure 5.3: Plots of the azimuthally averaged, frequency-integrated radiation pressure tensor calculated in the original $L = 0.08L_{\rm Edd}$ radiation MHD simulation (top row) and with the global Monte Carlo code (bottom row). From left to right, the plots are the scalar radiation pressure $P$ (i.e. 1/3 the trace of the pressure tensor), $P_{r\phi}$, $P_{r\theta}$, $P_{\phi\theta}$, $P_{rr} - P$, $P_{\phi\phi} - P$, and $P_{\theta\theta} - P$. The numbers on the axes indicate the radii in units of gravitational radii. The numbers on the colorbar indicate the absolute value of the pressure in Ba.

For Monte Carlo shearing box simulations it is usually obvious from the shape of the spectrum what part of it should be trusted and what part should be discarded. But for global simulations it may be more difficult to make this distinction. Since the resulting spectrum is composed of photons that emanate from multiple regions of the disc, a sudden change in the spectrum above a certain energy may either be due to the effect we have just described or just be a contribution from a different region of the disc. For the results of global Monte Carlo simulations to be trusted, therefore, it is important to figure out how to make this distinction.

For example, the spectra plotted in Figures 5.1 and 5.2 clearly have both low energy and high energy components. It seems that the high energy component is due to the high gas temperature region immediately outside the disc. But in the cases where the high energy component has significantly less power than the low energy component, such as the yellow line in Figure 5.2, the high energy component looks somewhat similar to the unphysical component seen in shearing box simulations. In fact, if it were not for the other curves and the direct knowledge of the high temperature region outside the disc, we would probably assume that the high energy component in the yellow curve is unphysical. It is important, therefore, to sort out these difficulties before proceeding.

**Bulk Comptonization by turbulence**

Since for the $L = 0.02L_{\text{Edd}}$ simulation bulk velocities are comparable to thermal velocities near the photosphere at large radii (see section 5.3), it is worth investigating why the spectra show no turbulent Comptonization. Either the optical depth of this region is insufficiently large or there is some other effect that is suppressing it. For example, it could be that the shear flow is reducing the number of photon scatterings in such regions with advection. If there is a component of the shear flow in the $\theta'$ direction, it could advect photons vertically (recall that for the purpose of studying bulk Comptonization we generalized the definition of shear to refer to the axisymmetric component of the velocity field). Alternatively, if there are significant density inhomogeneities in the $\phi$ direction, then the azimuthal component of the shear flow may transport photons out of potential Comptonization regions into low density regions where they can escape before scattering appreciably. For example, we ran one global Monte Carlo simulation with turbulent velocities but not shear velocities. We expected that by omitting bulk Comptonization by shear we would obtain a less energetic spectrum, but instead the high energy part of the spectrum was significantly more energetic! It is worth exploring whether this is because the main effect of the shear is to transport photons into low density regions or some other factor. Perhaps this could be checked by running simulations with an azimuthally averaged density field and seeing whether omitting the shear flow has the same effect on the resulting spectra.

A broader question is whether such simulations are even able to adequately capture turbulent Comptonization at all. Since turbulent Comptonization arises from velocity differences between subsequent scatterings, the simulation grid resolution must be sufficiently high to capture this effect. Since shearing box simulations zoom in on a small portion of the disc they are capable of achieving the necessary resolution, whereas global simulations of thin discs do not necessarily resolve the turbulence in great detail. This raises an even greater concern, namely whether the magneto-rotational instability (MRI) itself is sufficiently well resolved in thin disc simulations to trust their results more broadly. This is critical since inward accretion of gas is entirely dependent on this effect.

Once the bulk Comptonization results of the $L = 0.08L_{\mathrm{Edd}}$ and $L = 0.2L_{\mathrm{Edd}}$ simulations are better understood, simulations with higher Eddington ratios and other initial magnetic field configurations should be explored.

**Bulk Comptonization by shear**

Bulk Comptonization is due not only to turbulence but also to the shear flow. We did not explore this effect in a shearing box in much detail because there are significant obstacles to self-consistently doing so.

To start, we observe that we cannot simply place the shear flow in a shearing box with a periodic boundary condition, because when crossing the periodic boundary the photons will see sudden, large changes in the shear flow that are unphysical. Since bulk Comptonization depends on velocity differences, this will overestimate bulk Comptonization in optically thick regions. On the other hand, if one analytically continues the shear flow across the boundary then this solves the problem in optically thick regions but will eventually lead to velocities that exceed the speed of light in optically thin regions (a photon traveling horizontally will eventually see velocities greater than $c$ for a linear shear flow). The most clever solution to the problem is to literally implement the shearing periodic boundary condition, but this too is insufficient. In this solution, the boundary is periodic in the photon fluid frame (rather than lab frame) energy and wave vector. Right before the photon crosses the boundary, therefore, a Lorentz transformation is performed to find the fluid frame values, and right after it crosses the boundary the new lab frame values are calculated based on the shear velocity in the new grid cell. This solution is also correct in optically thick regions, but in optically thin regions a photon traveling nearly horizontally sees a shear flow of infinite length (even though it is always non-relativistic). This corresponds to an infinitely large, spatially uniform flow, so that such a photon scatters many more times than it would have in an actual accretion disc. Furthermore, since bulk Comptonization depends on velocity differences, photons undergo massive energy changes per scattering in optically thin regions in this scenario.

It becomes clear that the reason why it is difficult to study bulk Comptonization by shear in a shearing box is that it is really a global phenomenon. It is easier to model shear analytically than it is to model turbulence, so in some ways as long as one realizes the problem must be approached globally it should be a somewhat easier problem to study. On the other hand, since shear can also advect photons through high density regions, the actual effect of the shear flow on spectra may depend on other aspects of the flow. To study this effect it may be helpful to run simulations with and without azimuthally averaging the density (see the above discussion on the effect of shear on turbulent Comptonization).

**Comparison of global disc spectra with observations**

Although the focus of this work is on bulk Comptonization, the global Monte Carlo simulations should be used to study spectra of radiation MHD simulations more broadly. The resulting spectra provide a basis for comparing the underlying simulations with observations. In particular, the fact that the spectra plotted in Figures 5.1 and 5.2 differ from classical thin disc theory should be explored in greater detail. To start, one should compare the resulting effective temperature profiles with that for $\alpha$ discs. In particular, one should account for a possible non-zero stress inner boundary condition as well as a mass accretion rate profile that varies somewhat with radius.

More broadly, these simulations can be used to explore whether magnetically domi-nated discs can explain why observed high Eddington accretion discs appear to be thermally stable. In the original $\alpha$ model there appears to be a thermal instability (Shakura & Sunyaev, 1976), and shearing box simulations seem to confirm this (Jiang, Stone & Davis, 2013). In the magnetically dominated global simulations, on the other hand, the strong magnetic field appears to stabilize the disc vertical structure. By comparing spectra for such simulations with observed spectra we can explore to what extent magnetically dominated discs are candidates for observed thermally stable discs.

## 5.4.2 Modifications to the global Monte Carlo code

In this section we describe useful future modifications to the global code in order from most important to least important.

### Output the effective luminosity

The shearing box code outputs the specific intensity $I(\nu, \mu_k)$ as a function of frequency $\nu$ and the $z$ component of the photon unit wave vector $k_z = \mu_k$. The reason it outputs this and not a luminosity is that the latter would be dependent on the size of the shearing box, which is not of physical significance. For the global simulation code, on the other hand, in order to compare with observation we are interested in the observed flux $F(\nu, i)$ as a function of the inclination angle $i$. The problem with this quantity, however, is that it depends on the observation distance, which is not of physical significance. We instead define the effective luminosity $L_{\text{eff}}(\nu, i) = 4\pi r^2 F(\nu, i)$, where $r$ is the observation distance. Since $F(\nu, i) \propto r^{-2}$, $L_{\text{eff}}(\nu, i)$ is independent of $r$. Note that the effective luminosity is the luminosity an observer would naively infer by assuming that the source is spherically symmetric. Currently the code outputs the actual luminosity (section 5.2) rather than the effective luminosity. While the actual luminosity is still useful for studying the physics in the disc, an important next step is to modify the code to output the effective luminosity in order to compare with observations.

We note that we have overlooked one relevant subtlety. Because global simulations are not necessarily axisymmetric, $F$ and hence $L_{\text{eff}}$ are really a function of two observation angles, the inclination from the vertical $i$ and the azimuthal angle $\phi_{\text{obs}}$. One could leave $L_{\text{eff}}$ as a function of these two angles, which may be useful for the purpose of studying departures from axisymmetry, for example. On the other hand, one can get better statistics by averaging over $\phi_{\text{obs}}$ since this includes more emitted photons. This gives $L_{\text{eff}}$ as a function of only one observation angle $i$ and is probably more useful for comparing with observations.

**Parallelize the code**

Currently the code can be run on multiple cores by simply running multiple copies of the code at the same time (Appendix G3). The problem with this approach is that it loads an unnecessarily large amount of data in RAM, which significantly limits either how many copies can be run or how detailed each individual copy can be. The code should be modified so that it can run on multiple cores efficiently.

**Include bound-free emission and absorption**

Currently the code includes only free-free emission and absorption and Compton scattering. For AGN at lower temperatures bound-free processes can be significant. Even at high temperatures bound-free emission and absorption of metals can be important. The code can be modified to include these processes.

**Track polarization**

The original shearing box code has the capability to track polarization. It should be straightforward to modify the global code to do this as well.

**Adjusting NT**

In order to speed up Compton scattering, the code computes two arrays before beginning to sample photon packets, each of size $NT \times n_r \times n_\phi \times n_{\theta'}$, where $NT = 20$ by default. For large grid sizes, this along with other arrays may take up a significant portion of RAM, especially if there are multiple copies running simultaneously. In the original shearing box code $NT = 40$, but we set $NT = 20$ to free up memory. Since lowering $NT$ reduces the precision of Compton scattering calculations (which will not show up in the estimated error in the luminosity distribution), it is worth exploring more carefully the trade off between losing precision and using up RAM.

**Include stimulated scattering**

The code does not include stimulated Compton scattering. This is significant at only comparative low frequencies, so it should not matter for calculating, for example, the peak and high energy tail of a spectrum. But it will give rise to the wrong distribution in thermal equilibrium at low energies if the atmosphere is very scattering dominated. In particular, instead of yielding the $\nu^2$ Rayleigh-Jeans law at low energies, it will result in a $\nu^3$ power law. Since the spectrum from a global simulation is the sum of spectra from different regions of the disc, it is not necessarily obvious what part of the spectrum we can be sure is immune to this problem. For example, a part that is high energy relative to the lowest energy part of the disc spectrum may have a low energy contribution from another region of the disc.

Even in cases in which omitting stimulating scattering has no effect on the spectrum emitted from the disc, it may result in the wrong frequency-integrated moments (Appendix G3) inside extremely optically thick regions. We found that the moments were too large in such regions relative to the correct thermal equilibrium values.

Including stimulated Compton scattering rigorously is probably not worthwhile. Since this effect depends on the spectrum at each point, it can only be computed iteratively, which is probably not a good use of computing time. On the other hand, if one implements stimulated scattering by assuming that the spectrum at each point is a black body in local thermal equilibrium with the gas, then at least the code will yield the correct result in the optically thick limit. In other words, at least in a situation in which one expects to get a black body one will indeed get a black body. Furthermore, because the local spectrum is often at least roughly approximated by a black body, this assumption may adequately treat stimulating scattering in more general situations. At the very least, it seems that it should be a significant improvement relative to not including it at all. Unfortunately, even though this approach is physically simple it will require non-trivial modifications to the code and therefore take significant time to implement. The reason for this is that the code currently uses tailor-made algorithms to sample from the necessary distributions, and these will have to be rewritten.

### 5.4.3 Complete analytical work on the curl-free component

We have done significant analytical work on bulk Comptonization by the divergence-less component of a velocity field in Chapter 2. While we have discussed the optically thin and thick limits of bulk Comptonization by the curl-free component, we have fewer closed-form results because order $v/c$ effects are intertwined non-trivially with the viscous, order $v^2/c^2$ effect for this component. We have not even, for example, derived the bulk viscosity coefficient in the optically thick limit. We have described these effects heuristically in enough detail that it may not be necessary to find additional analytic solutions for the purpose of understanding the impact of bulk Comptonization on accretion disc spectra. On the other hand, it would be intellectually satisfying to give a more complete analysis of these effects.

To study this it is important to solve for both the first and second order effects simultaneously, and since the first order effect is inherently time-dependent this problem is best approached by looking for specific, self-consistent solutions under certain limiting conditions. For example, one could start by looking for the viscous stress tensor components for a longitudinal, time-dependent, traveling (or standing) sinusoidal wave in the optically thick limit. Because this is a time-dependent problem, we expect that solutions will be functions not only of wavelength but of frequency as well (that is, the frequency of the traveling or standing wave mode).

# Chapter 6

# Conclusion

We have studied bulk Comptonization by turbulence in accretion discs from several different angles. Our principal results are as follows.

In Chapter 2 we examined the physical processes underlying bulk Comptonization in detail. Bulk Comptonization energy exchange is due to both ordinary work done by radiation pressure and radiation viscous dissipation. These effects are due to terms that are first and second order in the velocity field, respectively. Since in general these effects are intertwined non-trivially, we used the Helmholz theorem to decompose a velocity field into a divergenceless component and curl-free (compressible) component. For the divergenceless component, bulk Comptonization is due to radiation viscous dissipation alone and can be treated as thermal Comptonization with an equivalent "wave" temperature. If we decompose the velocity field into sinusoidal modes with wave vectors $\mathbf{k}$, then for statistically homogeneous turbulence the wave temperature is given by equation (2.28):

$$k_{\mathrm{B}} T_{\mathrm{w}} = \sum_{\mathbf{k}} \frac{1}{3} m_{\mathrm{e}} \left\langle v_{\mathbf{k}}^2 \right\rangle f(k). \tag{6.1}$$

The function $f(k)$ is a weighting function given by equation (2.32) which goes to unity for optically thin modes and downweights optically thick modes by a factor $2/9\tau_k^2$. For statistically homogeneous turbulence, therefore, the wave temperature is simply a weighted sum over the power present at each scale in the turbulent cascade. Scales with wavelengths that are short relative to the photon mean free path contribute fully to the wave temperature, while scales with wavelengths that are long relative to the photon mean free path are significantly downweighted and contribute negligibly.

The fact that the wave temperature downweights modes with wavelengths longer than the photon mean free path is physically intuitive because for these modes electron velocity differences between subsequent photon scatterings are significantly smaller. To confirm our physical intuition, we also define a heuristic wave temperature by equation (2.38):

$$\frac{3}{2} k_{\mathrm{B}} T_{\mathrm{w,heur}} = \frac{1}{4} m_{\mathrm{e}} \left\langle (\Delta \mathbf{v})^2 \right\rangle. \tag{6.2}$$

Here, $\left\langle (\Delta \mathbf{v})^2 \right\rangle$ is the average square velocity difference between subsequent photon scatterings. We find that $T_{\mathrm{w,heur}}$ is also given by equation (6.1) but with a slightly different weighting function $f_{\mathrm{heur}}(k)$ given by equation (2.39). The function $f_{\mathrm{heur}}(k)$ goes to unity for optically thin modes and downweights optically thick modes by a factor $1/3\tau_k^2$. Both $f(k)$ and $f_{\mathrm{heur}}(k)$ are plotted in Figure 2.1. The function $f_{\mathrm{heur}}(k)$ well approximates $f(k)$, which confirms that the wave temperature can be intuitively understood in terms of the electron velocity differences between subsequent photon scatterings.

Bulk Comptonization by the curl-free (compressible) component of the velocity field is due to both radiation viscous dissipation and ordinary work done by radiation pressure. Although these processes affect each other, we use the physical intuition we developed from studying the divergenceless component to gain physical insight into radiation viscous dissipation here. If the minimum turbulent wavelength is significantly larger than the photon mean free path, i.e. in the optically thick limit, radiation viscous dissipation is suppressed since electron velocity differences between subsequent photon scatterings are small. In this limit, therefore, the first order effect is dominant, and whether photons are upscattered or downscattered depends simply on whether the gas is converging (compressing) or diverging (expanding), respectively. These in turn depend on whether the sign of $-\nabla \cdot \mathbf{v}$ is positive or negative, respectively. The effect of this process on the emergent spectrum, however, depends on how effectively photons are able to escape from such regions to the observer. In the limit in which the maximum turbulent wavelength is significantly shorter than the photon mean free path, i.e. in the optically thin limit, the full turbulent power contributes to viscous dissipation. In this limit work done by radiation pressure is negligible, so bulk Comptonization can be treated as thermal Comptonization with $(3/2)k_B T_w = (1/2)m_e \langle v^2 \rangle$.

In vertically stratified accretion disc atmospheres we expect that the first order, pressure work effect will be subdominant to the viscous, second order effect in determining the emergent spectrum. For statistically homogeneous turbulence the former effect should be small on average since it can result in either upscattering or downscattering depending on the sign of $-\nabla \cdot \mathbf{v}$. Moreover, this effect has the potential to be greatest in optically thick regions, but in accretion discs the turbulence is mostly incompressible (divergenceless) in these regions.

In an accretion disc, equations (6.1) and (6.2) for the wave temperature should be applied to a local region in which the turbulence is statistically homogeneous. A single wave temperature should not be associated with the entire vertical structure since it is spatially stratified and therefore cannot be regarded as homogeneous. We expect that the wave temperature will be negligible near the midplane and increase as we move toward the photosphere since it increases as the photon mean free path increases. On the other hand, optically thin regions at or outside the photosphere cannot contribute to bulk Comptonization even if the wave temperature is substantial because the average number of photon scatterings in such regions is near zero. We therefore expect bulk Comptonization to be dominated by a region of moderate optical depth just inside the photosphere.

In Chapter 3 we modeled the contribution of bulk Comptonization to the soft X-ray excess in AGN. To do this, we calculated disc spectra both taking into account and not taking into account bulk velocities with data from radiation MHD simulations. Because our simulation data was limited, we developed a scheme to scale the disc vertical structure to different values of radius, mass, and accretion rate. For each parameter set, we characterized our results by a temperature and optical depth in order to facilitate comparisons with other warm Comptonization models of the soft excess. We chose our fiducial mass, $M = 2 \times 10^6 M_\odot$, and accretion rate, $L/L_{\mathrm{Edd}} = 2.5$, to correspond to the values fit by D12 to the super-Eddington narrow-line Seyfert 1 REJ1034+396, which has an unusually large soft excess. The temperatures, optical depths, and Compton $y$ parameters that we found broadly agree with those fit to REJ1034+396. Unlike in our Monte Carlo simulations, we expect that in a real disc bulk Comptonization at a given radius will be accompanied by a decrease in the gas temperature at the effective photosphere in order to leave the flux unchanged, which is required by energy conservation.

In Chapter 4 we used ideas developed in Chapter 2 to simplify and generalize the bulk Comptonization model of Chapter 3. Rather than fit the temperature and optical depth to spectra computed with Monte Carlo post-processing simulations, we developed a procedure to calculate the Comptonization temperature and optical depth directly from the underlying vertical structure data. Using this, we plotted the dependence of the Comptonization temperature, optical depth, and $y$ parameters on the shearing box parameters.

In particular, we showed that the shearing box parameter space can be reduced to two parameters, the surface density $\Sigma$ and the vertical epicyclic frequency $\Omega_z$. The Comptonization optical depth and $y$ parameter increase strongly with increasing $\Sigma^{-1}$ and weakly with increasing $\Omega_z^{-1}$. The Comptonization temperature also increases strongly with increasing $\Sigma^{-1}$, but decreases weakly with increasing $\Omega_z^{-1}$. We plotted the dependence of the bulk Comptonization temperature, optical depth, and $y$ parameter on $\Sigma$ and $\Omega_z$ (Figure 4.12). We showed how these results can be intuitively understood by analyzing the one dimensional temperature profiles and velocity scalings.

We then showed how we can analytically determine the dependence of bulk Comptonization on each accretion disc parameter individually. Since $\Omega_z^{-1}$ is directly proportional to mass, and $\Sigma$ is independent of mass, the dependence of bulk Comptonization on mass is identical to its dependence on $\Omega_z^{-1}$. Similarly, since $\Sigma^{-1}$ is directly proportional to luminosity, and $\Omega_z^{-1}$ is independent of luminosity, the dependence of bulk Comptonization on luminosity is identical to its dependence on $\Sigma^{-1}$. That is, the Comptonization optical depth and $y$ parameter increase strongly with increasing luminosity and weakly with increasing mass. The Comptonization temperature also increases strongly with increasing luminosity, but decreases weakly with increasing mass. This dependence is summarized in Figure 4.45.

The dependence of bulk Comptonization on the other accretion disc parameters is inferred by analyzing how they affect $\Sigma^{-1}$ since bulk Comptonization depends much more strongly on $\Sigma^{-1}$ than it does on $\Omega_z^{-1}$. Since $\Sigma^{-1}$ is proportional to the flux $F$, we showed that the dependence of bulk Comptonization on the other disc parameters can be understood intuitively in terms of how they effect $F$. In particular, at large radii bulk Comptonization always decreases with increasing radius. At small radius, whether bulk Comptonization increases or decreases with radius depends on the inner boundary condition. Using the same line of reasoning, we showed that bulk Comptonization increases with both spin and the inner boundary condition parameter $\Delta\epsilon$ at small radii, and decreases with those parameters at large radii.

We also studied bulk Comptonization for an entire accretion disc by examining how it varies when the radius is fixed to the region of maximum luminosity. The dependence of bulk Comptonization on mass and luminosity is unchanged from above since the radius of maximum luminosity does not vary with these parameters. But since this radius does depend on the spin and inner boundary condition parameter $\Delta\epsilon$, the dependence of bulk Comptonization on these parameters required a new treatment. We showed that in this case bulk Comptonization always increases with spin and $\Delta\epsilon$.

An important result of Chapter 4 is that bulk Comptonization is strongly dependent on the disc inner boundary condition. In particular, we showed that bulk Comptonization is negligible in black hole X-ray binaries unless the disc inner boundary condition parameter is very large ($\Delta\epsilon \sim 0.1$) or the luminosity greatly exceeds Eddington.

Our results agree with the expectations outlined in Chapter 2. In particular, the effect of bulk Comptonization on spectra is dominated by radiation viscous dissipation, which corresponds to terms that are second order, not first order, in the velocity field. Because our model connects the bulk Comptonization parameters to the disc vertical structure one dimensional temperature profiles in a way that is physically intuitive, it provides a useful framework for understanding bulk Comptonization even in situations in which some of our specific results may not hold, such as shearing box or global radiation MHD simulations run in new regimes.

Finally, in Chapter 5 we developed a global Monte Carlo post-processing code to compute spectra for global radiation MHD simulations. We computed spectra for two simulations of black hole accretion discs with $M = 5 \times 10^8 M_\odot$, both overall and as a function of radius. For one, $L \sim 0.08 L_{\rm Edd}$, and for the other $L \sim 0.2 L_{\rm Edd}$. For the purpose of studying bulk Comptonization in global simulations we defined the shear flow as the azimuthally averaged velocity field and the turbulence as the difference between the total velocity field and the shear flow. To study the effect of turbulent Comptonization we ran Monte Carlo simulations both including and excluding the turbulent velocities and found that there was no difference in the resulting spectra, either overall or at any radius. Although this was consistent with our results in Chapter 4, which indicate that this effect becomes relevant around $L \sim L_{\mathrm{Edd}}$, it was somewhat surprising given that bulk velocities are comparable to thermal velocities at large radii near the photosphere in the $L \sim 0.2L_{\mathrm{Edd}}$ simulation.

We discussed future work on bulk Comptonization with the global Monte Carlo simulations as well as how to use them more broadly to compare spectra of radiation MHD simulations with observed spectra of real systems. In particular, these simulations can be used to explore whether magnetically dominated discs can explain why observed high Eddington accretion discs appear to be thermally stable.

In this work we have provided an in-depth analysis of the equations underlying bulk Comptonization and developed significant physical intuition into many aspects of this phenomenon. We have described how to apply these ideas to the vertical structure of real accretion discs and estimated their effects on spectra using preliminary data from radiation MHD simulations. A complete understanding of bulk Comptonization will require global simulation data that correspond to a large space of accretion disc parameters. Our results will be useful for self-consistently resolving and interpreting this effect in future simulations.

# Appendix A

# A very brief overview of Compton scattering in astrophysics

In astrophysics Compton scattering refers to the process in which photons change energy by scattering off of electrons. In general physics, by contrast, Compton scattering refers to the process in which photons change energy by scattering off of electrons at rest. This is a subtle but crucial difference. To understand why it exists, we first examine scattering of photons by a single electron. In the classical (non-quantum) picture, this process is called Thomson scattering, and corresponds to the scattering of an electromagnetic plane wave by an electron. In steady state, the electron oscillates in place in response to the incoming wave and generates an outgoing wave. It does not, however, gain or lose energy over time. In other words, it continues to oscillate about the same point without recoiling backwards in response to the incoming wave. In this picture, therefore, a single photon does not gain or lose energy when scattering off of a single electron at rest; it only changes direction. In the quantum picture, on the other hand, this process is called Compton scattering (in general physics), and photons must be treated as particles with energy $\epsilon = h\nu$ and momentum $p = E/c = h\nu/c$. Conservation of energy and momentum then require that a photon lose energy when scattering off of an electron at rest. In this process the electron recoils backwards, gaining energy. The fractional energy change is $\sim \epsilon/m_{\mathrm{e}}c^2$. Since $m_{\mathrm{e}}c^2 = 511\,\mathrm{keV}$, which for photons corresponds to gamma rays, it is usually small. We see that in general physics, therefore, Compton scattering refers specifically to the process in which quantum mechanics plays a role.

In astrophysics, by contrast, Compton scattering refers to the process in which photons change energy by scattering off of electrons, whether or not quantum mechanics plays a role. Even in the classical (non-quantum) picture, it turns out that photons change energy when scattering off of electrons not at rest. Because in this case the energy change is simply a consequence of changing reference frames, it is often called a Doppler change. In other words, observers in the electron rest frame see no energy change but observers in other frames do. The energy change due to quantum effects, on the other hand, is often referred to as electron recoil. In astrophysics, therefore, Compton scattering includes energy changes due to both the Doppler change and electron recoil. Scattering in which photons do not change energy is referred to as coherent scattering. The term Thomson scattering depends on context; it often confusingly refers to coherent scattering rather than classical (non-quantum) scattering.

Now that we have sketched out the basics of Compton scattering by photons off of a single electron, we turn to the scattering of photons by a distribution of electrons. For non-relativistic, thermal electrons, for example, the average photon energy change is proportional to the difference between the average electron kinetic energy and the average photon energy. The average fractional photon energy change is given by

$$\frac{\langle \Delta \epsilon \rangle}{\epsilon} = \frac{4 k_{\mathrm{B}} T_{\mathrm{e}} - \epsilon}{m_{\mathrm{e}} c^2}. \tag{A1}$$

The term $4 k_{\mathrm{B}} T_{\mathrm{e}}/m_{\mathrm{e}}c^2$ is due to the Doppler change and the term $\epsilon/m_{\mathrm{e}}c^2$ is due to electron recoil. As long as the average photon energy is significantly less than the average electron kinetic energy, Compton scattering is well-approximated by the Doppler change alone. But in order for photons to actually come to thermal equilibrium with the electrons both the Doppler change and electron recoil are required. Roughly speaking, these ideas generalize to the scattering of photons by any isotropic distribution of velocities.

If the electrons are all moving in the same direction, by contrast, then a photon usually gains energy when colliding from the front (i.e. in head-on collisions) and loses energy when approaching from behind. This is true even when the photon energy is greater than the electron kinetic energy, for example. The notable exception to this is when the magnitude of the photon momentum is also greater than the magnitude of the electron momentum, in which case a photon will lose energy even when colliding in front. But the photon energy can greatly exceed the electron kinetic energy while at the same time having a momentum whose magnitude is much smaller than that of the electron. For example, for a photon whose energy is equal to the electron kinetic energy, $\epsilon = (1/2)m_e v^2$. Its momentum, therefore, is $p = \epsilon/c = (1/2)m_e v^2/c$, which is a factor of $v/2c$ smaller than the magnitude of the momentum of the electron!

We return to Comptonization by thermal electrons to discuss several important parameters. When low energy photons scatter off of high energy, thermal electrons, the effect on the spectrum depends on two parameters, the optical depth $\tau$ and the electron temperature $T_e$. The electron temperature determines the average fractional energy change per photon scattering, which we recall is given by $4k_B T_e/m_e c^2$. The optical depth determines the average number of scatterings, which depends somewhat on the geometry but is approximately given by $\tau^2$ (for optically thick conditions). The overall effect on the spectrum is characterized by the Compton $y$ parameter, given by $y_p = (4k_B T_e/m_e c^2)\tau^2$, which is simply the product of the fractional energy change per scattering and the number of scatterings. Note that we have approximated the fractional energy change per scatter-ing as $4k_{\mathrm{B}}T_{\mathrm{e}}/m_{\mathrm{e}}c^2$ since for low energy photons scattering off of high energy electrons the electron recoil term $\epsilon/m_{\mathrm{e}}c^2$ is negligible. If $y_{\mathrm{p}}$ is sufficiently larger than unity, however, the resulting average photon energy will be comparable to the thermal energy of the electrons, and so increasing $y_{\mathrm{p}}$ beyond this point has no further effect on the spectrum. For values of $y_{\mathrm{p}}$ this large we say the photon spectrum has saturated.

# Appendix B

# Derivation of the occupation number second moment due to a divergenceless velocity field

Beginning with equation (2.60), we prove that the steady-state occupation number second moment for a divergenceless velocity field of uniform density to first order in velocity is given by equation (2.62). First we find the solution for a single mode given by

$$\mathbf{v} = \sqrt{2}v_{\mathrm{rms}}\sin\left(\frac{2\pi z}{\lambda}\right)\hat{\mathbf{x}} = v(z)\hat{\mathbf{x}}.$$  (B1)

For this mode, the transfer equation is

$$\lambda_{\mathrm{p}}\ell^z\partial_z n_{0,1}(\hat{\ell}, z) = -n_{0,1}(\hat{\ell}, z) - \ell^x v(z)\epsilon\partial_\epsilon n_{0,0} + \frac{3}{2}\ell^x\ell^z n_{0,1}^{xz},$$  (B2)

where we assume $n_{0,1} = 0$ and $n_{0,1}^{zz} = 0$, which we can check later. Then, the transfer equation is

$$\lambda_{\mathrm{p}} \ell^z \frac{\partial n_{0,1}}{\partial z}(\hat{\ell}, z) = -n_{0,1}(\hat{\ell}, z) - \ell^x \sqrt{2} v_{\mathrm{rms}} \sin\left(\frac{2\pi z}{\lambda}\right) \epsilon \frac{\partial n_{0,0}}{\partial \epsilon} + \frac{3}{2} \ell^x \ell^z \frac{1}{4\pi} \oint d\Omega' \ell'^x \ell'^z n_{0,1}(\hat{\ell}', z). \tag{B3}$$

First we address the $z$-dependence. Because this equation is linear, it must be that $n_{0,1}(\hat{\ell}, z)$ is a superposition of a sine and a cosine,

$$n_{0,1}(\hat{\ell}, z) = A(\hat{\ell}) \cos\left(\frac{2\pi z}{\lambda}\right) + B(\hat{\ell}) \sin\left(\frac{2\pi z}{\lambda}\right). \tag{B4}$$

This gives two coupled integral equations for $A$ and $B$,

$$-\frac{\ell^z}{\tau_k} A = -B - \ell^x \sqrt{2} v_{\mathrm{rms}} \epsilon \frac{\partial n_{0,0}}{\partial \epsilon} + \frac{3}{2} \ell^x \ell^z \frac{1}{4\pi} \oint d\Omega' \ell'^x \ell'^z B(\hat{\ell}') \tag{B5}$$

and

$$\frac{\ell^z}{\tau_k} B = -A + \frac{3}{2} \ell^x \ell^z \frac{1}{4\pi} \oint d\Omega' \ell'^x \ell'^z A(\hat{\ell}'). \tag{B6}$$

It seems that both $A$ and $B$ are proportional to one power of $\ell^x$. Writing $A = \ell^x \tilde{A}$ and $B = \ell^x \tilde{B}$, and $\ell^z = \cos\theta = \mu$, we then obtain

$$-\frac{\mu}{\tau_k} \tilde{A} = -\tilde{B} - \sqrt{2} v_{\mathrm{rms}} \epsilon \frac{\partial n_{0,0}}{\partial \epsilon} + \frac{3}{8} \mu \int_{-1}^{1} d\mu'(1 - \mu'^2)\mu' \tilde{B}(\mu') \tag{B7}$$

and

$$\frac{\mu}{\tau_k} \tilde{B} = -\tilde{A} + \frac{3}{8} \mu \int_{-1}^{1} d\mu'(1 - \mu'^2)\mu' \tilde{A}(\mu'). \tag{B8}$$

Observing that letting $\tilde{A}$ and $\tilde{B}$ be odd and even, respectively, is a consistent solution, the $\mu'$ integral of $\tilde{B}$ vanishes, and the two equations can be combined to give a single equation for $\tilde{A}$,

$$-\frac{\mu}{\tau_k}\sqrt{2}v_{\text{rms}}\epsilon\frac{\partial n_{0,0}}{\partial \epsilon} + \frac{\mu^2}{\tau_k^2}\tilde{A} = -\tilde{A} + \frac{3}{8}\mu \int_{-1}^{1} d\mu'(1-\mu'^2)\mu'\tilde{A}(\mu'). \tag{B9}$$

We then solve this equation with a series expansion of odd powers of $\mu$:

$$\tilde{A} = \sum_{n=0}^{\infty} a_{2n+1}\mu^{2n+1} \tag{B10}$$

gives

$$a_{2n+1} = (-1)^n \frac{a_1}{\tau_k^{2n}}, \tag{B11}$$

which is just the expansion of $(1+\mu^2/\tau_k^2)^{-1}$, so that

$$\tilde{A} = \frac{a_1\mu}{1+\mu^2/\tau_k^2}. \tag{B12}$$

Substituting this back into the integral equation gives $a_1$, which completes the solution. So far then, we have

$$n_{0,1}(\hat{\ell},z) \;\; = \sqrt{2}v_{\text{rms}}\epsilon\frac{\partial n_{0,0}}{\partial \epsilon}\sin\theta\cos\phi\Bigg(\left(\frac{1}{Q}\right)\frac{\tau_k^2\cos\theta}{\tau_k^2+\cos^2\theta}\cos\left(\frac{2\pi z}{\lambda}\right)$$
$$+ \left(\frac{\tau_k\cos^2\theta}{Q(\tau_k^2+\cos^2\theta)}-1\right)\sin\left(\frac{2\pi z}{\lambda}\right)\Bigg), \tag{B13}$$

where

$$Q \equiv \tau_k - \frac{3}{4}\tau_k^3\int_0^1 d\mu\frac{\mu^2-\mu^4}{\tau_k^2+\mu^2} = \tau_k - \frac{3}{4}\tau_k^3\left(\frac{2}{3}+\tau_k^2-\tau_k(1+\tau_k^2)\tan^{-1}\left(\frac{1}{\tau_k}\right)\right). \tag{B14}$$

Note that $Q \to \tau_k - \tau_k^3/2$ in the optically thin limit, and $Q \to 9\tau_k/10$ in the optically thick limit. This solution is consistent with our assumptions and solves equation (2.60) to first order in velocity. The second moment is

$$n_{0,1}^{ij} = \frac{\lambda_{\mathrm{p}}\epsilon\partial_\epsilon n_{0,0}}{3c}\tau_k^2 f(k)\left(\partial_i v_j + \partial_j v_i\right),\tag{B15}$$

where $f(k)$ is given by equation (2.32). Since equation (2.60) is linear, the solution for an arbitrary, divergenceless velocity field of uniform density is then given by equation (2.62).

# Appendix C

# Monte Carlo implementation of bulk Compton scattering

We incorporated bulk velocities into the Monte Carlo code used by Davis et al. (2009), which is based on the statistically weighted photon packet method described in Pozdniakov et al. (1983). Although the applications in this work are non-relativistic, we use exact Lorentz transforms. To test our code, we ran simulations with relativistic velocity fields and checked that spectra resulting from Lorentz transforming the emissivity were the same as spectra from simulations with a Lorentz-boosted field. We also ran simulations of Comptonization by divergenceless velocity fields and checked that the results were in agreement with the results of Chapter 2.

The modifications we made in order to take bulk velocites into account are as follows. Photon packets are sampled from an emission function defined in the fluid frame, $\eta_0(\epsilon_0, \hat{\ell}_{\mathbf{0}})$, such as thermal brehmsstrahlung. The variables $\epsilon_0$ and $\hat{\ell}_{\mathbf{0}}$ denote the fluid frame photon energy and angle, respectively. Since the density grid is defined in the lab frame, we transform the density at a given point to the fluid frame before evaluating $\eta_0(\epsilon_0, \hat{\ell}_{\mathbf{0}})$. In this frame, the number of photons with energies between $\epsilon_0$ and $\epsilon_0 + d\epsilon_0$ within a solid angle $d\Omega_0$ per unit time per unit volume is $f_0(\epsilon_0, \hat{l}_0)d\epsilon_0 d\Omega_0 = (\eta_0(\epsilon_0, \hat{l}_0)/\epsilon_0)d\epsilon_0 d\Omega_0$. The photon packet is then assigned a fluid frame statistical weight proportional to $f_0(\epsilon_0, \hat{l}_0)$. Lab frame energies and directions are calculated with standard Lorentz transforms, but calculating the correct lab frame statistical weight is more subtle. Since we want to sample from the lab frame photon number emissivity (i.e., per unit time, per unit volume) distribution $f(\epsilon, \hat{l})$,

$$
\begin{aligned}
f(\epsilon, \hat{l}) &= \frac{\eta(\epsilon, \hat{l})}{\epsilon} \\
&= \left(\frac{\epsilon}{\epsilon_0}\right)\frac{\eta_0(\epsilon_0, \hat{l}_0)}{\epsilon_0} \\
&= \left(\frac{\epsilon}{\epsilon_0}\right)f_0(\epsilon_0, \hat{l}_0),
\end{aligned}
\tag{C1}
$$

it may seem that the fluid frame statistical weight should be multiplied by $\epsilon/\epsilon_0$, but this is in fact incorrect. To see why, note that even without changing the statistical weight, simply boosting the energy and direction already results in a new distribution,

$$
f_0(\epsilon_0(\epsilon, \hat{l}), \hat{l}_0(\epsilon, \hat{l}))\frac{\partial(\epsilon_0, \hat{l}_0)}{\partial(\epsilon, \hat{l})},
\tag{C2}
$$

which differs from the original distribution by the change of measure factor. Since the evaluation of this factor yields

$$
\frac{\partial(\epsilon_0, \hat{l}_0)}{\partial(\epsilon, \hat{l})} = \frac{\epsilon}{\epsilon_0},
\tag{C3}
$$

it so happens that the new distribution is already the lab frame photon number emissivity:

$$f_0(\epsilon_0(\epsilon, \hat{\ell}), \hat{\ell}_0(\epsilon, \hat{\ell})) \frac{\partial(\epsilon_0, \hat{\ell}_0)}{\partial(\epsilon, \hat{\ell})} = f_0(\epsilon_0(\epsilon, \hat{\ell}), \hat{\ell}_0(\epsilon, \hat{\ell})) \frac{\epsilon}{\epsilon_0}$$

$$= f(\epsilon, \hat{\ell}). \tag{C4}$$

Therefore, the fluid frame and lab frame statistical weights are equal. Once a photon packet's lab frame energy, direction, and statistical weight are assigned, the method used to evolve it is in essence the same as in Davis et al. (2009). Fluid frame parameters are self-consistently used in scattering events, and lab frame parameters are used to calculate changes in photon position between events. Fluid frame absorption coefficients are evaluated with densities transformed to the fluid frame.

We also attempted to upgrade the periodic boundary condition in the $x$ direction to a shearing periodic boundary condition. In principle this is required even if the effect of the bulk velocities on the spectrum is negligible since it applies to the density and temperature fields, not just the velocity field. In particular, for a box of width $L_x$ the periodic boundary condition in the $x$ direction assumes that

$$\rho(L_x, y, z) = \rho(0, y, z) \tag{C5}$$

and

$$T(L_x, y, z) = T(0, y, z). \tag{C6}$$

But since the boundary condition in the underlying radiation MHD simulation changes with time due to the background shear, the correct boundary condition is the shearing periodic boundary condition, which gives

$$\rho\left(L_x, y, z\right) = \rho\left(0, y + \Delta y\left(t_i\right), z\right) \tag{C7}$$

and

$$T\left(L_x, y, z\right) = T\left(0, y + \Delta y\left(t_i\right), z\right). \tag{C8}$$

The value $\Delta y$ is the change in the $y$ value at the $x$ boundary, and it depends on the time $t_i$ at the $i$th timestep in the underlying radiation MHD simulation. We note that the shearing periodic boundary condition must be implemented in order for the density and temperature fields to be continuous in the $x$ direction.

Since for most applications of interest the simulation data is statistically homogeneous in the $x$ and $y$ directions, it seems unlikely that using periodic boundary conditions instead of shearing periodic boundary conditions would impact the emitted spectrum. However, this choice definitely can impact the spectrum once the effects of bulk velocities are included. To show this, we first consider the effect of including only the turbulent velocities $\mathbf{v}_{\mathrm{turb}}$, which are defined by subtracting off the background shear $\mathbf{v}_{\mathrm{s}}$. In this case the periodic and shearing periodic boundary conditions in the $x$ direction are analogous to those for the density and temperatures fields. The periodic boundary condition is

$$\mathbf{v}_{\mathrm{turb}}\left(L_x, y, z\right) = \mathbf{v}_{\mathrm{turb}}\left(0, y, z\right), \tag{C9}$$

and the shearing periodic boundary condition is

$$\mathbf{v}_{\mathrm{turb}}\left(L_x, y, z\right) = \mathbf{v}_{\mathrm{turb}}\left(0, y + \Delta y\left(t_i\right), z\right). \tag{C10}$$

Since applying the periodic boundary condition to data from radiation MHD simulations that use the shearing periodic boundary condition results in a velocity field that is discontinuous at the boundary in the $x$ direction, this overestimates velocity differences between subsequent photon scatterings at this boundary. And since bulk Comptonization depends on velocity differences between subsequent photon scatterings, this overestimates bulk Comptonization. We initially calculated the spectra in Chapter 3 using the periodic boundary condition and found that it led to a small but not insignificant increase in bulk Comptonization.

It turns out that, ironically, there is no straightforward way to self-consistently apply shearing periodic boundary conditions to the background shear flow in a way that is physically meaningful. We discuss why this is so in detail in section 5.4.1. Correctly calculating bulk Comptonization by the background shear therefore requires global simulations. In Chapter 3 we simply added the shear flow to the turbulent velocity field. This is unphysical since it results in a shear flow that is discontinuous at the boundary in the $x$ direction, but it is still useful for giving a preliminary estimate of the effects of shear. It overestimates velocity differences between subsequent photon scatterings in optically thick regions but it may underestimate such differences in optically thin regions.

# Appendix D

# Shearing box scalings

## D1 Derivation of the radiation pressure profile scaling

The hydrostatic equilibrium equation is

$$\frac{dP}{dz} = -\rho z \Omega_z^2, \tag{D1}$$

where $\Omega_z$ is the vertical epicyclic frequency. The pressure profile is

$$
\begin{aligned}
P(z) =& P_{\mathrm{ph,in}} + \Omega_z^2 \int_z^{z_{\mathrm{ph}}} \rho\left(z'\right) z' dz' \\
=& P_{\mathrm{ph,in}} + \Omega_z^2 \left(\frac{\Sigma}{\Sigma_0}\right) \left(\frac{h}{h_0}\right)^{-1} \int_z^{z_{\mathrm{ph}}} \rho_0 \left(h_0 z'/h\right) z' dz' \\
=& P_{\mathrm{ph,in}} + \Omega_z^2 \left(\frac{\Sigma}{\Sigma_0}\right) \left(\frac{h}{h_0}\right)^{-1} \left(\int_z^{h z_{\mathrm{ph},0}/h_0} - \int_{z_{\mathrm{ph}}}^{h z_{\mathrm{ph},0}/h_0}\right) \rho_0 \left(h_0 z'/h\right) z' dz', \quad \text{(D2)}
\end{aligned}
$$

where the subscript "ph" denotes a value at the photosphere. Therefore,

$$P_0 \left( h_0 z / h \right) = P_{\mathrm{ph,in},0} + \Omega_{z,0}^2 \left( \frac{h}{h_0} \right)^{-2} \int_z^{h z_{\mathrm{ph},0} / h_0} \rho_0 \left( h_0 z' / h \right) z' dz' \tag{D3}$$

and

$$P_0 \left( h_0 z_{\mathrm{ph}} / h \right) = P_{\mathrm{ph,in},0} + \Omega_{z,0}^2 \left( \frac{h}{h_0} \right)^{-2} \int_{z_{\mathrm{ph}}}^{h z_{\mathrm{ph},0} / h_0} \rho_0 \left( h_0 z' / h \right) z' dz'. \tag{D4}$$

Substitution of equations (D3), (D4), and (3.2) into equation (D2) gives equation (3.17).

## D2   Shearing box scalings in terms of flux, shear, and vertical epicyclic frequency

The surface density scaling is

$$\left( \frac{\Sigma}{\Sigma_0} \right) = \left( \frac{\alpha}{\alpha_0} \right)^{-1} \left( \frac{\kappa}{\kappa_0} \right)^{-2} \left( \frac{\Omega_z}{\Omega_{z,0}} \right)^2 \left( \frac{\partial_x v_y}{\partial_x v_{y,0}} \right)^{-1} \left( \frac{F}{F_0} \right)^{-1}. \tag{D5}$$

The scale height scaling is

$$\left( \frac{h}{h_0} \right) = \left( \frac{\kappa}{\kappa_0} \right) \left( \frac{\Omega_z}{\Omega_{z,0}} \right)^{-2} \left( \frac{F}{F_0} \right). \tag{D6}$$

The density profile scaling is

$$\rho \left( z \right) = \left( \frac{\alpha}{\alpha_0} \right)^{-1} \left( \frac{\kappa}{\kappa_0} \right)^{-3} \left( \frac{\Omega_z}{\Omega_{z,0}} \right)^4 \left( \frac{\partial_x v_y}{\partial_x v_{y,0}} \right)^{-1} \left( \frac{F}{F_0} \right)^{-2} \rho_0 \left( h_0 z / h \right). \tag{D7}$$

The scalings for the pressure and gas temperature profiles are given by equations (3.17), (3.19), and (3.21). The turbulent velocity profile scaling is

$$v(z) = \left(\frac{\alpha}{\alpha_0}\right)^{1/2} \left(\frac{\beta}{\beta_0}\right)^{1/2} \left(\frac{\kappa}{\kappa_0}\right) \left(\frac{\Omega_z}{\Omega_{z,0}}\right)^{-1} \left(\frac{F}{F_0}\right) v_0(h_0 z/h). \tag{D8}$$

The shear velocity profile scaling is

$$v_{\rm s}(x) = \left(\frac{\kappa}{\kappa_0}\right) \left(\frac{\Omega_z}{\Omega_{z,0}}\right)^{-2} \left(\frac{\partial_x v_y}{\partial_x v_{y,0}}\right) \left(\frac{F}{F_0}\right) v_{\rm s,0}\left(h_0 x/h\right). \tag{D9}$$

# D3   Scalings for flux, shear and vertical epicyclic frequency in terms of radius, mass, and accretion rate

## D3.1   Newtonian scalings

Let $M$ and $\dot{M}$ be the mass and mass accretion rate, respectively. Define $r = R/R_{\rm g}$ and $\dot{m} = \dot{M}/\dot{M}_{\rm Edd}$. We also define

$$\eta = \frac{1}{2r_{\rm in}}. \tag{D10}$$

The flux, derived from energy and angular momentum conservation, is given by Agol & Krolik (2000) equation (11):

$$F = \frac{3GM\dot{M}}{8\pi R^3} \left(1 - \sqrt{r_{\rm in}/r} + \left(\sqrt{r_{\rm in}/r}\right) r_{\rm in}\Delta\epsilon\right), \tag{D11}$$

where $\Delta\epsilon$ is the change in efficiency due to a non-zero stress-free inner boundary condition. The flux scaling is

$$
\left(\frac{F}{F_0}\right) = \left(\frac{r}{r_0}\right)^{-3} \left(\frac{M}{M_0}\right)^{-1} \left(\frac{\dot{m}}{\dot{m}_0}\right) \left(\frac{\eta + \Delta\epsilon}{\eta_0 + \Delta\epsilon_0}\right)^{-1}
$$
$$
\left(\frac{1 - \sqrt{r_{\rm in}/r} + \left(\sqrt{r_{\rm in}/r}\right) r_{\rm in}\Delta\epsilon}{1 - \sqrt{r_{\rm in,0}/r_0} + \left(\sqrt{r_{\rm in,0}/r_0}\right) r_{\rm in,0}\Delta\epsilon_0}\right). \tag{D12}
$$

The vertical epicyclic frequency is

$$
\Omega_z = \sqrt{\frac{GM}{R^3}}. \tag{D13}
$$

The scaling for the vertical epicyclic frequency is

$$
\left(\frac{\Omega_z}{\Omega_{z,0}}\right) = \left(\frac{M}{M_0}\right)^{-1} \left(\frac{r}{r_0}\right)^{-3/2}. \tag{D14}
$$

The strain rate is

$$
\partial_x v_y = \frac{3}{2}\sqrt{\frac{GM}{R^3}}. \tag{D15}
$$

The strain rate scaling is

$$
\left(\frac{\partial_x v_y}{\partial_x v_{y,0}}\right) = \left(\frac{M}{M_0}\right)^{-1} \left(\frac{r}{r_0}\right)^{-3/2}. \tag{D16}
$$

## D3.2   Kerr scalings

Let $M$ and $\dot{M}$ be the mass and mass accretion rate, respectively. Let $R$ be the Boyer-Linquist radial coordinate and $a$ be the dimensionless spin parameter. Define $r = R/R_{\rm g}$ and $\dot{m} = \dot{M}/\dot{M}_{\rm Edd}$. The expressions for $A$, $B$, $C$, $D$, and $E$ are given by Riffert & Herold

(1995) (hereafter, RH95) equation (6). In terms of the dimensionless variables, they are

$$A = 1 - \frac{2}{r} + \frac{a^2}{r^2}, \tag{D17}$$

$$B = 1 - \frac{3}{r} + \frac{2a}{r^{3/2}}, \tag{D18}$$

$$C = 1 - \frac{4a}{r^{3/2}} + \frac{3a^2}{r^2}, \tag{D19}$$

$$D = \frac{1}{2\sqrt{r}} \int_{r_{\rm in}}^{r} \frac{x^2 - 6x + 8a\sqrt{x} - 3a^2}{\sqrt{x}\left(x^2 - 3x + 2a\sqrt{x}\right)} \, dx, \tag{D20}$$

$$E = 1 - \frac{6}{r} + \frac{8a}{r^{3/2}} - \frac{3a^2}{r^2}, \tag{D21}$$

where $r_{\rm in}$ is given by $E(r_{\rm in}) = 0$. We also define

$$\eta = 1 - \left(1 - \frac{2}{3r_{\rm in}}\right)^{1/2}, \tag{D22}$$

the efficiency parameter assuming a stress-free inner boundary condition. The flux is given by the thermal equilibrium equation, RH95 equation (19), modified by the non-zero stress inner boundary term in Agol & Krolik (2000) equation (8):

$$F = \frac{3\dot{M}M}{8\pi R^3} B^{-1} \left( r_{\rm in}^{3/2} B(r_{\rm in})^{1/2} \Delta\epsilon r^{-1/2} + D \right), \tag{D23}$$

where $\Delta\epsilon$ is the change in efficiency due to a non-zero stress-free inner boundary condition. The flux scaling is

$$\left(\frac{F}{F_0}\right) = \left(\frac{r}{r_0}\right)^{-3} \left(\frac{M}{M_0}\right)^{-1} \left(\frac{\dot{m}}{\dot{m}_0}\right) \left(\frac{\eta + \Delta\epsilon}{\eta_0 + \Delta\epsilon_0}\right)^{-1}$$
$$\left(\frac{B}{B_0}\right)^{-1} \left(\frac{r_{\text{in}}^{3/2} B(r_{\text{in}})^{1/2} \Delta\epsilon r^{-1/2} + D}{r_{\text{in,0}}^{3/2} B(r_{\text{in,0}})^{1/2} \Delta\epsilon_0 r_0^{-1/2} + D_0}\right). \tag{D24}$$

The vertical epicyclic frequency, inferred from RH95 equation (12), is

$$\Omega_z = \sqrt{\frac{GM}{R^3} C B^{-1}}. \tag{D25}$$

The scaling for the vertical epicyclic frequency is

$$\left(\frac{\Omega_z}{\Omega_{z,0}}\right) = \left(\frac{M}{M_0}\right)^{-1} \left(\frac{r}{r_0}\right)^{-3/2} \left(\frac{C}{C_0}\right)^{1/2} \left(\frac{B}{B_0}\right)^{-1/2}. \tag{D26}$$

The strain rate, inferred from RH95 equation (14), is

$$\partial_x v_y = \frac{3}{2} \sqrt{\frac{GM}{R^3}} A B^{-1}. \tag{D27}$$

The strain rate scaling is

$$\left(\frac{\partial_x v_y}{\partial_x v_{y,0}}\right) = \left(\frac{M}{M_0}\right)^{-1} \left(\frac{r}{r_0}\right)^{-3/2} \left(\frac{A}{A_0}\right) \left(\frac{B}{B_0}\right)^{-1}. \tag{D28}$$

# D4    Shearing box scalings in terms of radius, mass, and accretion rate

## D4.1    Newtonian scalings

In this section we substitute the results of Appendix D3.1 into the results of Appendix D2. The density profile scaling is

$$\rho\left(z\right) = \left(\frac{\alpha}{\alpha_0}\right)^{-1} \left(\frac{\kappa}{\kappa_0}\right)^{-3} \left(\frac{r}{r_0}\right)^{3/2} \left(\frac{M}{M_0}\right)^{-1} \left(\frac{\dot{m}}{\dot{m}_0}\right)^{-2} \left(\frac{\eta + \Delta\epsilon}{\eta_0 + \Delta\epsilon_0}\right)^2$$
$$\left(\frac{1 - \sqrt{r_{\mathrm{in}}/r} + \left(\sqrt{r_{\mathrm{in}}/r}\right) r_{\mathrm{in}}\Delta\epsilon}{1 - \sqrt{r_{\mathrm{in},0}/r_0} + \left(\sqrt{r_{\mathrm{in},0}/r_0}\right) r_{\mathrm{in},0}\Delta\epsilon_0}\right)^{-2} \rho_0\left(h_0 z/h\right). \tag{D29}$$

The pressure profile scaling is given by equation (3.17), and the gas temperature profile scaling is given by equations (3.19), (3.20), (3.21), where

$$\left(\frac{P_{\mathrm{c}}}{P_{\mathrm{c},0}}\right) = \left(\frac{\alpha}{\alpha_0}\right)^{-1} \left(\frac{\kappa}{\kappa_0}\right)^{-1} \left(\frac{r}{r_0}\right)^{-3/2} \left(\frac{M}{M_0}\right)^{-1}, \tag{D30}$$

and

$$P_{\mathrm{ph}} = \left(\frac{f_{\mathrm{cor}}}{f_{\mathrm{cor},0}}\right)^4 \left(\frac{r}{r_0}\right)^{-3} \left(\frac{M}{M_0}\right)^{-1} \left(\frac{\dot{m}}{\dot{m}_0}\right) \left(\frac{\eta + \Delta\epsilon}{\eta_0 + \Delta\epsilon_0}\right)^{-1}$$
$$\left(\frac{1 - \sqrt{r_{\mathrm{in}}/r} + \left(\sqrt{r_{\mathrm{in}}/r}\right) r_{\mathrm{in}}\Delta\epsilon}{1 - \sqrt{r_{\mathrm{in},0}/r_0} + \left(\sqrt{r_{\mathrm{in},0}/r_0}\right) r_{\mathrm{in},0}\Delta\epsilon_0}\right) P_{\mathrm{ph},0}. \tag{D31}$$

The turbulent velocity profile scaling is

$$v(z) = \left(\frac{\alpha}{\alpha_0}\right)^{1/2} \left(\frac{\beta}{\beta_0}\right)^{1/2} \left(\frac{\kappa}{\kappa_0}\right) \left(\frac{r}{r_0}\right)^{-3/2} \left(\frac{\dot{m}}{\dot{m}_0}\right) \left(\frac{\eta + \Delta\epsilon}{\eta_0 + \Delta\epsilon_0}\right)^{-1}$$
$$\left(\frac{1 - \sqrt{r_{\text{in}}/r} + \left(\sqrt{r_{\text{in}}/r}\right) r_{\text{in}}\Delta\epsilon}{1 - \sqrt{r_{\text{in},0}/r_0} + \left(\sqrt{r_{\text{in},0}/r_0}\right) r_{\text{in},0}\Delta\epsilon_0}\right) v_0(h_0 z/h). \tag{D32}$$

The shear velocity profile scaling is

$$v_{\text{s}}(x) = \left(\frac{\kappa}{\kappa_0}\right) \left(\frac{r}{r_0}\right)^{-3/2} \left(\frac{\dot{m}}{\dot{m}_0}\right) \left(\frac{\eta + \Delta\epsilon}{\eta_0 + \Delta\epsilon_0}\right)^{-1}$$
$$\left(\frac{1 - \sqrt{r_{\text{in}}/r} + \left(\sqrt{r_{\text{in}}/r}\right) r_{\text{in}}\Delta\epsilon}{1 - \sqrt{r_{\text{in},0}/r_0} + \left(\sqrt{r_{\text{in},0}/r_0}\right) r_{\text{in},0}\Delta\epsilon_0}\right) v_{\text{s},0}(h_0 x/h). \tag{D33}$$

The surface density profile scaling is

$$\left(\frac{\Sigma}{\Sigma_0}\right) = \left(\frac{\alpha}{\alpha_0}\right)^{-1} \left(\frac{\kappa}{\kappa_0}\right)^{-2} \left(\frac{r}{r_0}\right)^{3/2} \left(\frac{\dot{m}}{\dot{m}_0}\right)^{-1} \left(\frac{\eta + \Delta\epsilon}{\eta_0 + \Delta\epsilon_0}\right)$$
$$\left(\frac{1 - \sqrt{r_{\text{in}}/r} + \left(\sqrt{r_{\text{in}}/r}\right) r_{\text{in}}\Delta\epsilon}{1 - \sqrt{r_{\text{in},0}/r_0} + \left(\sqrt{r_{\text{in},0}/r_0}\right) r_{\text{in},0}\Delta\epsilon_0}\right)^{-1}, \tag{D34}$$

and the scale height scaling is

$$\left(\frac{h}{h_0}\right) = \left(\frac{\kappa}{\kappa_0}\right) \left(\frac{M}{M_0}\right) \left(\frac{\dot{m}}{\dot{m}_0}\right) \left(\frac{\eta + \Delta\epsilon}{\eta_0 + \Delta\epsilon_0}\right)^{-1}$$
$$\left(\frac{1 - \sqrt{r_{\text{in}}/r} + \left(\sqrt{r_{\text{in}}/r}\right) r_{\text{in}}\Delta\epsilon}{1 - \sqrt{r_{\text{in},0}/r_0} + \left(\sqrt{r_{\text{in},0}/r_0}\right) r_{\text{in},0}\Delta\epsilon_0}\right). \tag{D35}$$

## D4.2   Kerr scalings

In this section we substitute the results of Appendix D3.2 into the results of Appendix D2. The density profile scaling is

$$
\begin{aligned}
\rho\left(z\right) = {} & \left(\frac{\alpha}{\alpha_0}\right)^{-1} \left(\frac{\kappa}{\kappa_0}\right)^{-3} \left(\frac{r}{r_0}\right)^{3/2} \left(\frac{M}{M_0}\right)^{-1} \left(\frac{\dot{m}}{\dot{m}_0}\right)^{-2} \\
& \left(\frac{\eta + \Delta\epsilon}{\eta_0 + \Delta\epsilon_0}\right)^{2} \left(\frac{A}{A_0}\right)^{-1} \left(\frac{B}{B_0}\right) \left(\frac{C}{C_0}\right)^{2} \\
& \left(\frac{r_{\mathrm{in}}^{3/2} B(r_{\mathrm{in}})^{1/2} \Delta\epsilon r^{-1/2} + D}{r_{\mathrm{in},0}^{3/2} B(r_{\mathrm{in},0})^{1/2} \Delta\epsilon_0 r_0^{-1/2} + D_0}\right)^{-2} \rho_0\left(h_0 z/h\right).
\end{aligned} \tag{D36}
$$

The pressure profile scaling is given by equation (3.17), and the gas temperature profile scaling is given by equations (3.19), (3.20), (3.21), where

$$
\left(\frac{P_{\mathrm{c}}}{P_{\mathrm{c},0}}\right) = \left(\frac{\alpha}{\alpha_0}\right)^{-1} \left(\frac{\kappa}{\kappa_0}\right)^{-1} \left(\frac{r}{r_0}\right)^{-3/2} \left(\frac{M}{M_0}\right)^{-1} \left(\frac{A}{A_0}\right)^{-1} \left(\frac{C}{C_0}\right), \tag{D37}
$$

and

$$
\begin{aligned}
P_{\mathrm{ph}} = {} & \left(\frac{f_{\mathrm{cor}}}{f_{\mathrm{cor},0}}\right)^{4} \left(\frac{r}{r_0}\right)^{-3} \left(\frac{M}{M_0}\right)^{-1} \left(\frac{\dot{m}}{\dot{m}_0}\right) \left(\frac{\eta + \Delta\epsilon}{\eta_0 + \Delta\epsilon_0}\right)^{-1} \\
& \left(\frac{B}{B_0}\right)^{-1} \left(\frac{r_{\mathrm{in}}^{3/2} B(r_{\mathrm{in}})^{1/2} \Delta\epsilon r^{-1/2} + D}{r_{\mathrm{in},0}^{3/2} B(r_{\mathrm{in},0})^{1/2} \Delta\epsilon_0 r_0^{-1/2} + D_0}\right) P_{\mathrm{ph},0}. \tag{D38}
\end{aligned}
$$

The turbulent velocity profile scaling is

$$
\begin{aligned}
v(z) = {} & \left(\frac{\alpha}{\alpha_0}\right)^{1/2} \left(\frac{\beta}{\beta_0}\right)^{1/2} \left(\frac{\kappa}{\kappa_0}\right) \left(\frac{r}{r_0}\right)^{-3/2} \left(\frac{\dot{m}}{\dot{m}_0}\right) \left(\frac{\eta + \Delta\epsilon}{\eta_0 + \Delta\epsilon_0}\right)^{-1} \left(\frac{B}{B_0}\right)^{-1/2} \left(\frac{C}{C_0}\right)^{-1/2} \\
& \left(\frac{r_{\mathrm{in}}^{3/2} B(r_{\mathrm{in}})^{1/2} \Delta\epsilon r^{-1/2} + D}{r_{\mathrm{in},0}^{3/2} B(r_{\mathrm{in},0})^{1/2} \Delta\epsilon_0 r_0^{-1/2} + D_0}\right) v_0(h_0 z/h). \tag{D39}
\end{aligned}
$$

The shear velocity profile scaling is

$$v_{s}\left(x\right) = \left(\frac{\kappa}{\kappa_0}\right)\left(\frac{r}{r_0}\right)^{-3/2}\left(\frac{\dot{m}}{\dot{m}_0}\right)\left(\frac{\eta+\Delta\epsilon}{\eta_0+\Delta\epsilon_0}\right)^{-1}\left(\frac{A}{A_0}\right)\left(\frac{B}{B_0}\right)^{-1}\left(\frac{C}{C_0}\right)^{-1}$$
$$\left(\frac{r_{\text{in}}^{3/2}B(r_{\text{in}})^{1/2}\Delta\epsilon r^{-1/2}+D}{r_{\text{in},0}^{3/2}B(r_{\text{in},0})^{1/2}\Delta\epsilon_0 r_0^{-1/2}+D_0}\right)v_{s,0}\left(h_0 x/h\right). \tag{D40}$$

The surface density scaling is

$$\left(\frac{\Sigma}{\Sigma_0}\right) = \left(\frac{\alpha}{\alpha_0}\right)^{-1}\left(\frac{\kappa}{\kappa_0}\right)^{-2}\left(\frac{r}{r_0}\right)^{3/2}\left(\frac{\dot{m}}{\dot{m}_0}\right)^{-1}\left(\frac{A}{A_0}\right)^{-1}\left(\frac{\eta+\Delta\epsilon}{\eta_0+\Delta\epsilon_0}\right)$$
$$\left(\frac{B}{B_0}\right)\left(\frac{C}{C_0}\right)\left(\frac{r_{\text{in}}^{3/2}B(r_{\text{in}})^{1/2}\Delta\epsilon r^{-1/2}+D}{r_{\text{in},0}^{3/2}B(r_{\text{in},0})^{1/2}\Delta\epsilon_0 r_0^{-1/2}+D_0}\right)^{-1}, \tag{D41}$$

and the scale height scaling is

$$\left(\frac{h}{h_0}\right) = \left(\frac{\kappa}{\kappa_0}\right)\left(\frac{M}{M_0}\right)\left(\frac{\dot{m}}{\dot{m}_0}\right)\left(\frac{\eta+\Delta\epsilon}{\eta_0+\Delta\epsilon_0}\right)^{-1}\left(\frac{C}{C_0}\right)^{-1}$$
$$\left(\frac{r_{\text{in}}^{3/2}B(r_{\text{in}})^{1/2}\Delta\epsilon r^{-1/2}+D}{r_{\text{in},0}^{3/2}B(r_{\text{in},0})^{1/2}\Delta\epsilon_0 r_0^{-1/2}+D_0}\right). \tag{D42}$$

# D5   Including vertical radiation advection in shearing box scalings

## D5.1   Derivation of shearing box scalings without assuming radiation diffusion

Here we derive the shearing box scalings presented in Chapter 3 without assuming that the flux is carried by radiation diffusion. We give scalings for $\rho$, $T_{\text{g}}$, $v_{\text{turb}}$, and $v_{\text{s}}$ in terms of $\Sigma$, $\Omega_z$, $\partial_x v_y$, $\alpha$, $\beta$, $f_{\text{col}}$, and $h$. The result, therefore, of not assuming radiation diffusion is to leave the scale height $h$ as a free parameter.

To begin, the scaling for the density profile is still given by equation (3.13), except that $h$ is now a free parameter:

$$\rho(z) = \left(\frac{\Sigma}{\Sigma_0}\right) \left(\frac{h}{h_0}\right)^{-1} \rho_0 \left(h_0 z/h\right).$$ (D43)

The flux scaling is determined by equations (3.2), (3.3), and (3.5), which give

$$\left(\frac{F}{F_0}\right) = \left(\frac{\alpha}{\alpha_0}\right) \left(\frac{\Omega_z}{\Omega_{z,0}}\right)^2 \left(\frac{\partial_x v_y}{\partial_x v_{y,0}}\right) \left(\frac{\Sigma}{\Sigma_0}\right) \left(\frac{h}{h_0}\right)^2.$$ (D44)

The scaling for the turbulent velocity profile is derived from equations (3.2), (3.5), (3.10), and (3.11), which give

$$v(z) = \left(\frac{\alpha}{\alpha_0}\right)^{1/2} \left(\frac{\beta}{\beta_0}\right)^{1/2} \left(\frac{\Omega_z}{\Omega_{z,0}}\right) \left(\frac{h}{h_0}\right) v_0(h_0 z/h).$$ (D45)

The scalings for the shear velocity, pressure and gas temperature profiles are unchanged, except that $h$ is now a free parameter. The scaling for the shear velocity profile is given by equation (3.22),

$$v_s(x) = \left(\frac{\partial_x v_y}{\partial_x v_{y,0}}\right) \left(\frac{h}{h_0}\right) v_{s,0}\left(h_0 x/h\right).$$ (D46)

The pressure profile is given by equation (3.17),

$$P(z) = P_{\text{ph,in}} + \left(\frac{P_c}{P_{c,0}}\right) \left(P_0\left(h_0 z/h\right) - P_0\left(h_0 z_{\text{ph}}/h\right)\right),$$ (D47)

where

$$\left(\frac{P_c}{P_{c,0}}\right) = \left(\frac{\Omega_z}{\Omega_z}\right)^2 \left(\frac{\Sigma}{\Sigma_0}\right) \left(\frac{h}{h_0}\right)$$ (D48)

and

$$P_{\text{ph,in}} = \left( \frac{f_{\text{cor}}}{f_{\text{cor},0}} \right)^4 \left( \frac{F}{F_0} \right) P_{\text{ph,in},0}. \tag{D49}$$

The scaling for the gas temperature profile is given by equations (3.19), (3.20), and (3.21):

$$T_{\text{g,in}}^4(z) = T_{\text{g,ph}}^4 + \left( \frac{P_{\text{c}}}{P_{\text{c},0}} \right) \left( T_{\text{g},0}^4 \left( h_0 z/h \right) - T_{\text{g},0}^4 \left( h_0 z_{\text{ph}}/h \right) \right) \tag{D50}$$

$$T_{\text{g,ph}}^4 = \left( \frac{P_{\text{ph,in}}}{P_{\text{ph,in},0}} \right) T_{\text{g,ph},0}^4 \tag{D51}$$

$$T_{\text{g,out}}^4(z) = \left( \frac{P_{\text{ph,in}}}{P_{\text{ph,in},0}} \right) T_{\text{g},0}^4 \left( z_{\text{ph},0} + h_0(z - z_{\text{ph}})/h \right). \tag{D52}$$

## D5.2   Modelling radiation advection with an effective $\kappa$

The scaling for $h$ depends on how the radiation is vertically transported. For radiation diffusion, for example, the scaling for $h$ is given by equation (3.7):

$$\left( \frac{h}{h_0} \right) = \left( \frac{\alpha}{\alpha_0} \right)^{-1} \left( \frac{\kappa}{\kappa_0} \right)^{-1} \left( \frac{\partial_x v_y}{\partial_x v_{y,0}} \right)^{-1} \left( \frac{\Sigma}{\Sigma_0} \right)^{-1}. \tag{D53}$$

In section 4.3.5 we pointed out that including radiation advection is equivalent to simply decreasing $\kappa$ as far as the shearing box scalings are concerned. Now we also see that including radiation advection (at fixed surface density $\Sigma$) is therefore equivalent to simply *increasing* the scale height relative to the value set by radiation diffusion alone.

We note that although including advection *increases* the scale height in a shearing box, it has the opposite effect at a fixed radius in an accretion disc. This is because in a shearing box the surface density $\Sigma$ is fixed and the total flux $F$ is allowed to vary. In this case, including advection increases both the scale height and therefore the flux (equation D44). But if we substitute in equation (D44) everywhere for $\Sigma$, we can instead regard $F$ as a free parameter instead of $\Sigma$. Since $F$ is a function of accretion disc parameters such as the mass, mass accretion rate, radius, etc., this procedure gives the shearing box scalings in terms of accretion disc parameters rather than shearing box parameters. The scalings that result from this procedure, assuming radiation diffusion, are given in Appendix D3. The scaling for the scale height, for example, is given by

$$\left(\frac{h}{h_0}\right) = \left(\frac{\kappa}{\kappa_0}\right) \left(\frac{\Omega_z}{\Omega_{z,0}}\right)^{-2} \left(\frac{F}{F_0}\right). \tag{D54}$$

We see, therefore, that including advection *decreases* the scale height at a fixed radius in the disc. In this process, the flux is held constant and the surface density, therefore, increases (equation D44).

# Appendix E

# Applying the wave temperature definition to simulation data

It is problematic to apply equation (4.7) directly to simulation data for two reasons. First, a straightforward implementation runs too slowly for our purposes, since it requires computing an entire volume average for each grid cell. To speed up the computation for the applications in this work, we take the density in the entire region over which the spatial average is defined to be the density at position $\mathbf{r}$. With this approximation, equation (4.7) can be implemented in Python without explicitly using "for" loops to traverse the grid. In most of the simulation domain this approximation is sufficient since the probability that a photon scatters far from position $\mathbf{r}$ is negligible anyway. This approximation is less valid in the scattering photosphere, where the photon mean free path is large. But directly implementing equation (4.7) in this region is problematic for an entirely different reason, which is that we do not have access to the velocity function above the top of the simulation domain. As a result, for values of $\mathbf{r}$ near the top of the simulation domain the spatial average underestimates $\langle (\Delta \mathbf{v})^2 \rangle_{\mathbf{r}}$ . To compensate for this as well as our original approximation, we define an additional parameter $\tau_{\mathrm{break}}$ as follows. At each pair of $x$ and $y$ coordinates we set $T_\mathrm{w} = T_\mathrm{bulk}$ for values of $z$ where $\tau_\mathrm{s} \leq \tau_\mathrm{break} < 1$, since $T_\mathrm{w}$ approaches $T_\mathrm{bulk}$ in the optically thin limit. We set $\tau_\mathrm{break} = 0.5$. However, because the number of photon scatterings in a given region scales with $\tau_\mathrm{s}^2$ (section 4.2.2), it turns out that for our bulk Comptonization model the value of $T_\mathrm{w}$ does not matter for $\tau_\mathrm{s} < 1$ anyway, and so the approximations we make to define $T_\mathrm{w}$ in this region have no impact on our results. For example, we repeated the spectral calculations plotted in Figure 4.3 for $\tau_\mathrm{break} = 0$ and found that the results were unchanged.

# Appendix F

# Additional figures

In this section we show the plots of spectra at multiple radii omitted from section 4.2.

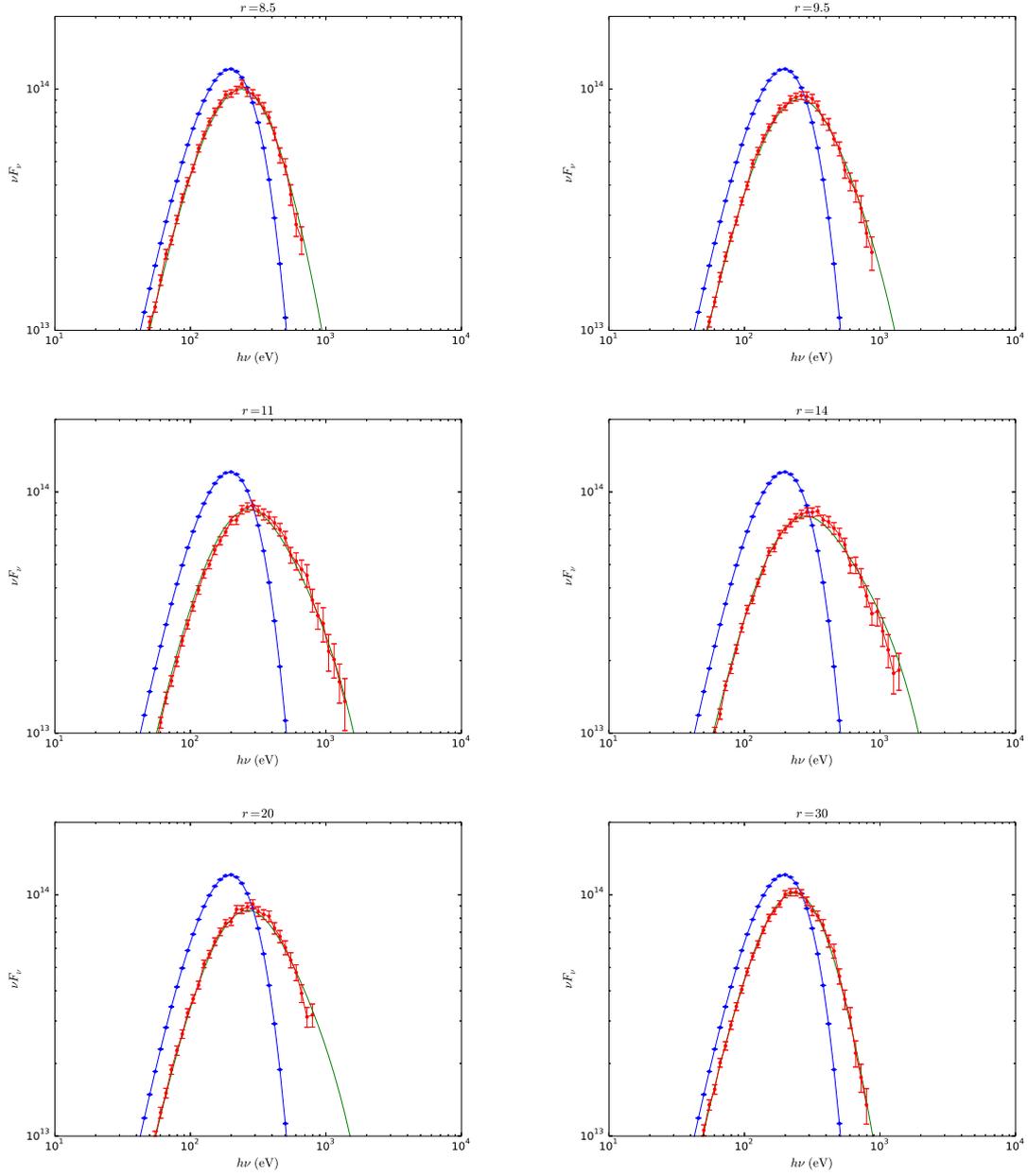

Figure F1: Normalized spectra (red) computed by passing a 50eV Planck source (blue) through vertical structure data truncated at $\tau_s = 10$ at multiple radii for the $M = 2 \times 10^6 M_\odot$, $L/L_{\mathrm{Edd}} = 5$ parameter set (Table 4.1). In all cases the velocities are zeroed and the wave temperatures are added to the gas temperatures. The green curves are calculated by using the Kompaneets equation to pass the 50eV Planck source through a homogeneous medium with temperature $T_{1D}$, given in Table 4.2. The spectra for only $r = 14$ were originally plotted in Figure 4.7.

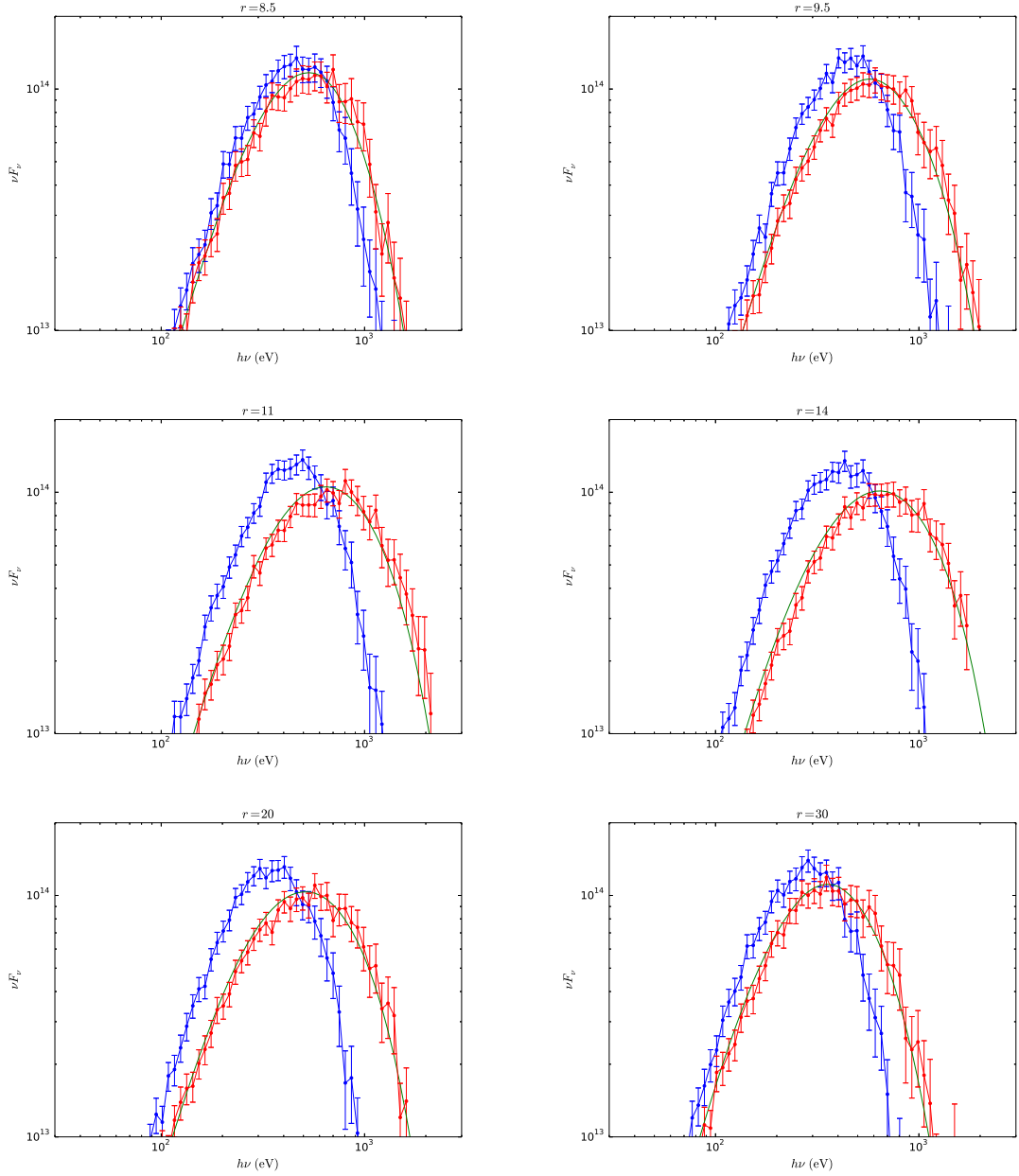

Figure F2: Normalized spectra at multiple radii for the $M = 2 \times 10^6 M_\odot$, $L/L_{\mathrm{Edd}} = 5$ parameter set (Table 4.1) computed with (red) and without (blue) velocities. The green curves are calculated by using the Kompaneets equation to pass the blue curves through a homogeneous Comptonizing medium with parameters $T_C$ and $\tau_C$, given in Table 4.3. The spectra for only $r = 14$ were originally plotted in Figure 4.8.

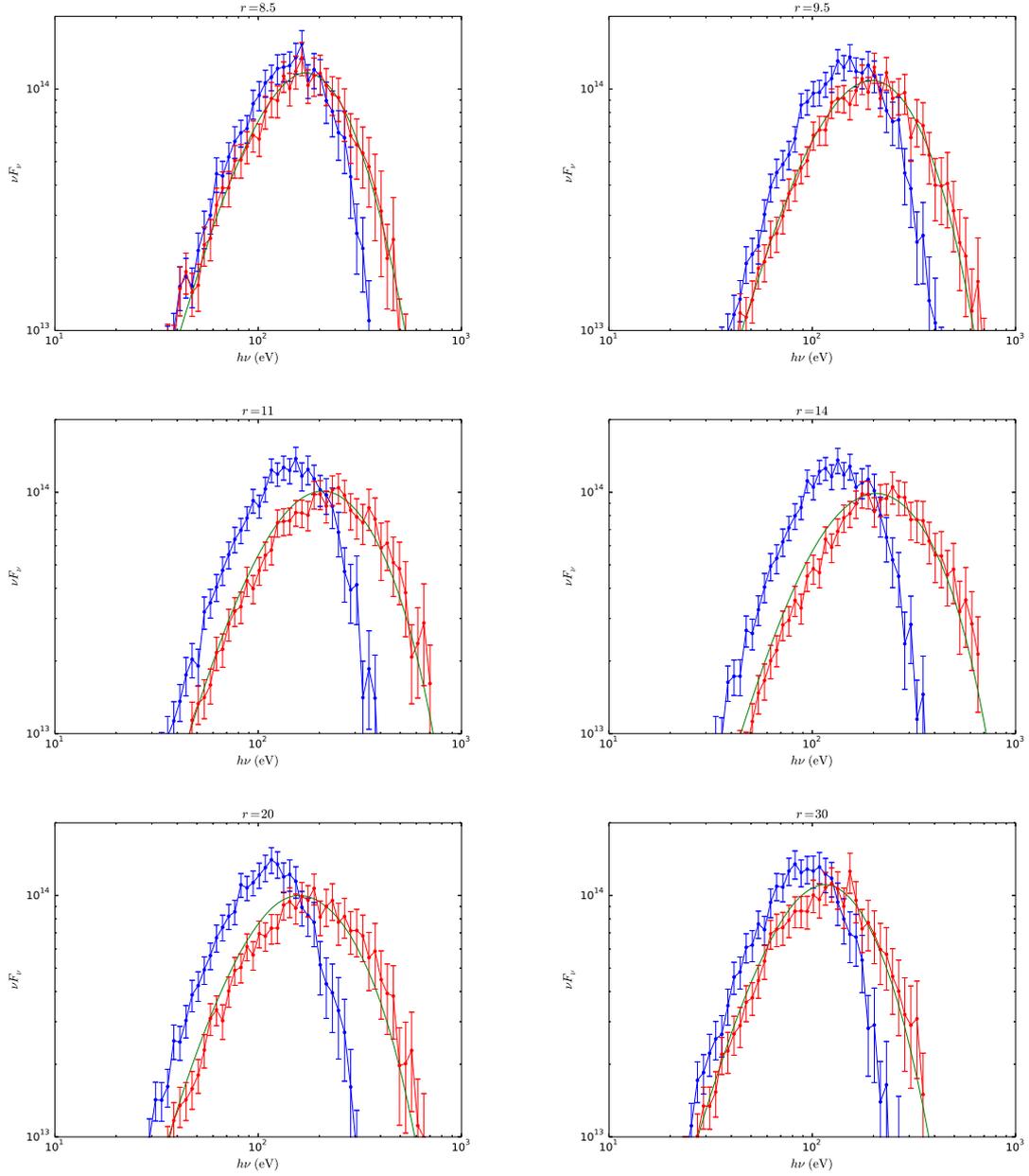

Figure F3: Normalized spectra at multiple radii for the $M = 2 \times 10^8 M_\odot$, $L/L_{\mathrm{Edd}} = 5$ parameter set (Table 4.1) computed with (red) and without (blue) velocities. The green curves are calculated by using the Komaneets equation to pass the blue curves through a homogeneous Comptonizing medium with parameters $T_{\mathrm{C}}$ and $\tau_{\mathrm{C}}$, given in Table 4.4. The spectra for only $r = 14$ were originally plotted in Figure 4.9.

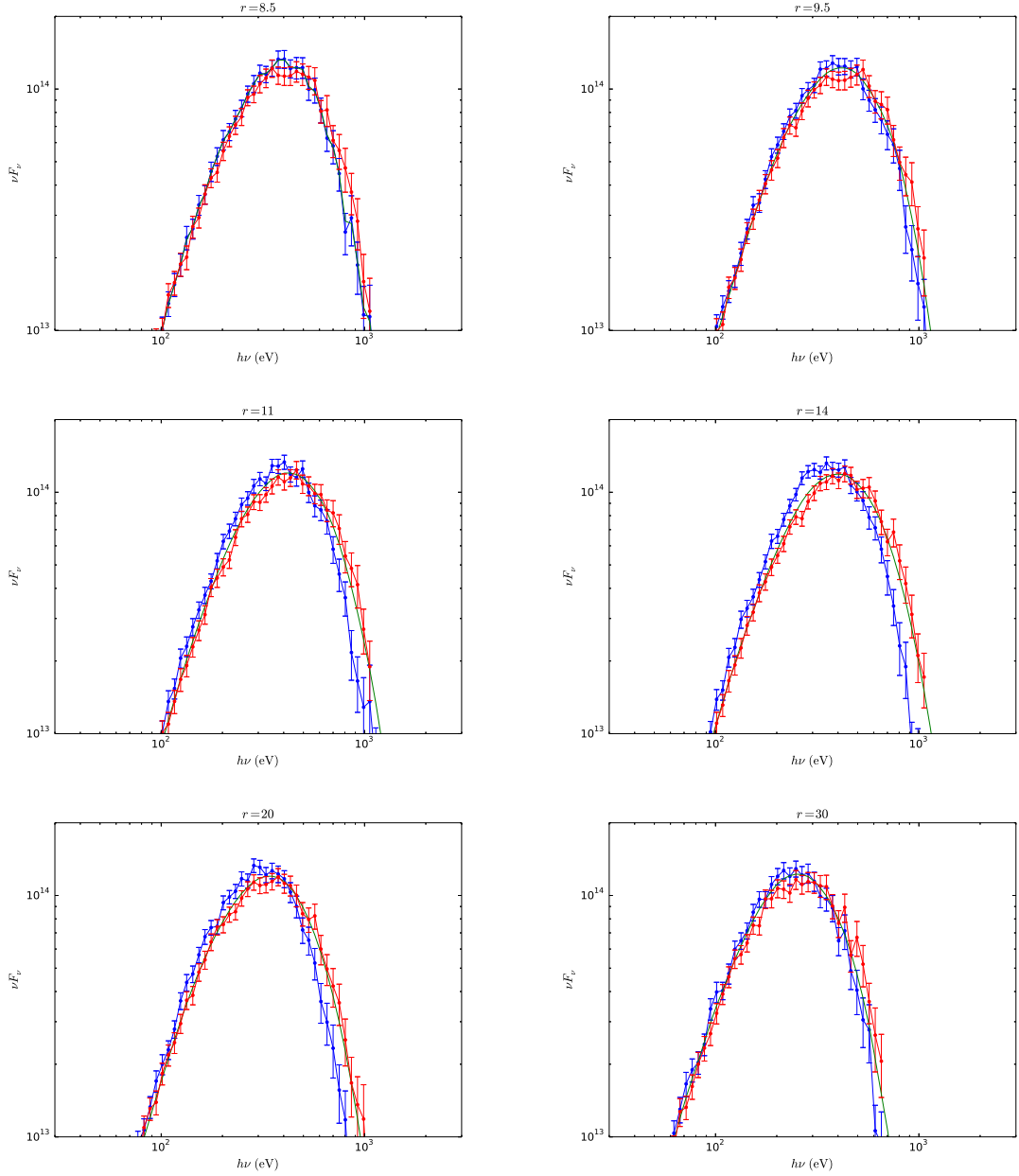

Figure F4: Normalized spectra at multiple radii for the $M = 2 \times 10^6 M_\odot$, $L/L_{\mathrm{Edd}} = 2.5$ parameter set (Table 4.1) computed with (red) and without (blue) velocities. The green curves are calculated by using the Kompaneets equation to pass the blue curves through a homogeneous Comptonizing medium with parameters $T_{\mathrm{C}}$ and $\tau_{\mathrm{C}}$, given in Table 4.5. The spectra for only $r = 14$ were originally plotted in Figure 4.10.

# Appendix G

# The global Monte Carlo code details

## G1  Photon propagation details

In this section we describe in detail how the global code propagates photons in the grid. It is this aspect of the code that was most difficult to implement and differs most from the corresponding part in the shearing box version.

First we review how photons are propagated in the shearing box grid. For a photon in a cell with indices $(i, j, k)$, the next cell that the photon enters depends on whether it hits an $x$, $y$, or $z$ cell boundary first. To determine this, we calculate the distance to each boundary separately. The distance to the $x$ boundary is $(x_{i+1} - x)/k_x$ if $k_x > 0$ and $(x_i - x)/k_x$ if $k_x < 0$, where $x_i$ is the $x$ coordinate of the left cell boundary. The distances to the $y$ and $z$ boundaries are determined analogously. If the distance to the $x$ boundary is the minimum, then the photon next enters the cell with indices $(i + 1, j, k)$ if $k_x > 0$ and $(i - 1, j, k)$ if $k_x < 0$, and the analogous statement is true for the $y$ and $z$ boundaries.

To propagate photons in the global grid, we generalize the procedure from the shearing box grid. For a photon in a cell with (global) indices $(i, j, k)$, the next cell that the photon enters depends on whether it hits an $r$, $\phi$, or $\theta'$ cell boundary first. To determine this, we calculate the distance to each boundary separately. But because the cells are non-rectangular, this requires us to solve quadratic equations and construct decision trees that choose the correct root depending on the situation. Several aspects of this procedure are non-trivial. For example, for the shearing box grid the $x$ coordinate of the nearest $x$ boundary is $x_{i+1}$ if $k_x > 0$ and $x_i$ if $k_x < 0$. But the analogous statement is not true for the global grid; we cannot know whether the nearest $r$ boundary is $r_{i+1}$ or $r_i$ based on the sign of $k_r$ alone. For the $\phi$ and $\theta'$ coordinates, there may not even be a nearest boundary (for example, if the photon is traveling in approximately the $r$ direction then it will travel infinitely far before crossing a $\phi$ boundary)! Therefore, to determine the distance to the nearest $r$ boundary (before even comparing with the distances to the nearest $\phi$ and $\theta'$ boundaries), for example, we must compute the distance to both the upper boundary with coordinate $r_{i+1}$ and the lower boundary with coordinate $r_i$ and then choose the boundary that is closest. Once the distance to the nearest boundary for each coordinate separately is calculated, the next step is analogous to the case of the shearing box. If the distance to the $r$ boundary is the minimum, then the photon next enters the cell with indices $(i+1, j, k)$ if the nearest $r$ boundary coordinate is $r_{i+1}$ and otherwise enters the cell with indices $(i-1, j, k)$ if the nearest $r$ boundary coordinate is $r_i$. The analogous statement is true for the $\phi$ and $\theta'$ boundaries.

But there is an additional problem, unlike in the case of the shearing box code. Due to rounding errors, the procedure for updating the photon's indices does not perfectly interface with the part of the code that modifies the photon position coordinates. The problem is not that rounding errors increase the statistical error of the code; the rounding errors are too small to have any effect on this. The problem, rather, is that these errors on occasion lead to inconsistencies between the photon's indices and its position coordinates that cause the code to crash. If, for example, the code calculates that a photon will cross a boundary that it does not end up crossing, then the indices it assigns the photon will not correspond to the photon's actual position coordinates.

This is a tricky problem to solve because it may seem that one should simply update the indices to conform to the position coordinates. But if the discrepancy is due to the fact that the photon has the wrong position coordinates then simply changing the indices may not be effective. For example, if the photon is supposed to cross into the next cell but the position coordinates place it immediately outside the cell boundary due to a rounding error, the photon position coordinates should be adjusted, not its indices. Adjusting its indices could cause it to stay stuck right outside the border. If one is not careful, therefore, attempts to correct errors only lead to new problems.

To deal with rounding errors we could instead try to adjust the position coordinates to correspond to the indices rather than the other way around. But this is problematic, too, since the position coordinates are obviously not uniquely defined by the indices. The trick is to recognize that the position coordinates corresponding to indices that change at a cell boundary are specified by the indices. For example, if a photon crosses an $r$ boundary so that its $r$ index changes from $i$ to $i+1$, the new $r$ coordinate is given by $r_{i+1}$. The other coordinates cannot be inferred from the indices, but it is safe to change the other indices to conform to the coordinates. If, again, a photon crosses an $r$ boundary, for example, then once its $r$ coordinate is updated to $r_{i+1}$ we can change the indices for $\phi$ and $\theta'$ to conform to the values of those coordinates. Changing only the $\phi$ and $\theta'$ indices therefore avoids the danger of getting stuck at the $r$ boundary.

In other words, once the code computes which boundary the photon will cross, all that matters is that it successfully crosses this boundary and that it does so self-consistently. To do this, the coordinate corresponding to this boundary must be corrected to make sure it conforms to the new index, and the other indices must be corrected to make sure they conform to the new coordinates.

On top of these issues the code also takes into account scattering events, boundary conditions, photons crossing boundaries for multiple coordinates at once, etc. The implementation of these effects carries over from the shearing box code in a fairly straightforward way, and we have omitted these details from this discussion for clarity.

## G2   Basic use

The code requires two input files, one to input the simulation parameters and another to input the simulation grid. It writes to two output files, one for photons with $k_z > 0$ and the other for photons with $k_z < 0$.

### The parameter input file

The parameter input file is a text file that should be formatted as follows:

```
nen nmu nphi
en0 emin emax
nph1 nph2
idum
ir1min ir2max
stepsize
filename
```

The first line contains the number of frequency bins, $\mu_k$ bins for $\mu_k > 0$ (so that the total number of $\mu_k$ bins is $2 \times nmu$), and $\phi_k$ bins, respectively. The number of $\phi_k$ bins is irrelevant and should therefore be set to 1, since the code takes the sum over these bins in order to calculate $L(\nu, \mu_k)$ (section 5.2). These bins may be useful if the code is modified to, for example, output the luminosity distribution as a function of both $\mu_k$ and $\phi_k$, or output the effective luminosity as a function of observation angle (section 5.4.2).

The parameter $en0$ is the energy unit for photon energy variables. In theory it can be set to any value, but to minimize numerical errors it should be set to the typical photon energy of the simulation. For example, for simulations of high Eddington black hole accretion disc spectra we set this to 1keV in cgs units. The parameters *emin* and *emax* are the lower and upper bounds on the photon energies of the frequency bins in units of *en0*. For example, if the desired lower and upper bounds of the frequency bins are $\nu_1$ and $\nu_2$, respectively, then *emin* and *emax* should be set to $h\nu_1/en0$ and $h\nu_2/en0$, respectively, where $h$ is Planck's constant.

The parameters *ir1min* and *ir2max* correspond to the minimum and maximum values of the index of the $\phi$ coordinate for which photon packets are sampled. Unless the code is running on multiple cores (Appendix G3), they should be set to 0 and $n_\phi - 1$, respectively, where $n_\phi$ is the number of cells in the $\phi$ direction.

The parameter *nph1* is the number of photon packets sampled per grid cell when $stepsize = 1$, which should be the default value. To speed up the code in exchange for increasing statistical error one can set it to larger values (Appendix G3). If the number of cells in the $r$ and $\theta'$ directions are given by $n_r$ and $n_{\theta'}$, respectively, then for $stepsize = 1$ the total number of photon packets sampled is $nph1 \times n_r \times (ir2max - ir1min + 1) \times n_{\theta'}$. Assuming $ir1min = 0$ and $ir1max = n_\phi - 1$, the total number of photon packets sampled is $nph1 \times n_r \times n_\phi \times n_{\theta'}$ The parameter *nph2* corresponds to a currently inactive feature. In the original shearing box version, the intermediate photon bin contents were outputted to a file for every $nph2 \times n_x \times n_y \times n_z$ photon packets sampled. In other words, the code would give $nph1/nph2 - 1$ intermediate outputs. The global code could easily be modified to include this feature.

The parameter *idum* is the seed for the random number generator. Generally we always set this to 1 so that if we run the code multiple times, each with slightly different settings, then we can be sure that any differences in the output are not directly due to the random number generator. On the other hand, it may be useful to run the code with different values of *idum* while holding everything else fixed in order to determine, for example, whether a fluctuation in the output is physical or due to statistical error. The parameter *idum* must also be adjusted when running the code on multiple cores (Appendix G3).

The parameter *filename* should be set to the name of the file that contains the simulation grid.

## The grid input file

The grid input file is a binary file named *filename* (see above). The first three items are the integers $n_r$, $n_\phi$ and $n_{\theta'}$, which are the number of cells in the $r$, $\phi$ and $\theta'$ directions, respectively. The next five items are the double precision floats $r_{\min}$, $r_{\max}$, $\phi_{\max}$, $\theta'_{\min}$, and $\theta'_{\max}$, respectively, for the simulation grid. Without loss of generality $\phi_{\min}$ is set to 0. The values $r_{\min}$ and $r_{\max}$ should be given in centimeters (i.e. cgs units). If the grid corresponds to the whole sphere, for example, then $\phi_{\max} = 2\pi$, $\theta'_{\min} = -\pi/2$, and $\theta'_{\max} = \pi/2$.

The next three items are one dimensional arrays of double precision floats. The first two arrays are the density and temperature grids, respectively. They each have size $n_r \times n_\phi \times n_{\theta'}$ and should be created by incrementing the $\theta'$ index first (i.e. it should be the inner "for" loop) and the $r$ index last (i.e. it should be the outer "for" loop). The last array is the velocity array. It has size $3 \times n_r \times n_\phi \times n_{\theta'}$ and its order of indices is the same as for the density and temperature arrays. For each grid cell, the three velocity components $v_r$, $v_\phi$, and $v_{\theta'}$ are stored contiguously in the array in that order. For example, if the velocity array is denoted $v[i]$, then the $r$, $\phi$, and $\theta'$ components of the velocity at a given grid cell corresponding to index $i$ are given by $v_r[i] = v[3i]$, $v_\phi[i] = v[3i+1]$, and $v_{\theta'}[i] = v[3i+2]$, respectively.

### The luminosity output files

The code writes to two output files, one for photons with $k_z > 0$ and the other for photons with $k_z < 0$. Each output file is formatted as follows:

```
nen nmu
freq
mu L Lerr
mu L Lerr
...
freq
mu L Lerr
mu L Lerr
...
...
```

The first line contains the number of bins in frequency (i.e. energy) and $k_z$, respectively (see above). In all that follows, $freq$ is the frequency of each frequency bin. For each frequency, $L$ and $Lerr$ are the values of the luminosity distribution and luminosity distribution error, respectively, at $|k_z| = mu$.

## G3   Additional options

Here we describe the options we have made the greatest use of. The code has others as well that we do not list here.

## Number of cells sampled

The number of photon packets sampled is $nph1 \times n_r \times (ir2max - ir2min + 1) \times n_{\theta'}/stepsize$, so if $nph1 = 1$ then to make the code run faster in exchange for greater statistical error $stepsize$ can be set to a value greater than 1. Because for an accretion disc we expect to see the greatest symmetry with respect to changes in $\phi$, if $stepsize > 1$ then the code traverses all indices for the $r$ and $\theta'$ coordinates but only samples indices for the $\phi$ coordinate. Therefore, $1 \leq stepsize \leq (ir2max - ir2min + 1)$, and furthermore $(ir2max - ir2min + 1)$ should be divisible by $stepsize$.

## Boundary conditions

The code has many options for alternative boundary conditions. If the simulation is of a comparatively thin disc then there are advantages to truncating the simulation grid so that $-f\pi/2 \leq \theta' \leq f\pi/2$ for some factor $f < 1$. We discuss these below. In such a case, the boundary condition for the $\theta'$ coordinate should be changed to escape.

## Outputting moments

The code can output the first three moments of the frequency-integrated specific intensity in every grid cell to a binary file. These are the energy density $E$, the flux $F^i$, and the radiation pressure tensor $P^{ij}$. The first five items in the file are the double precision floats $r_{\min}$, $r_{\max}$, $\phi_{\max}$, $\theta'_{\min}$, and $\theta'_{\max}$. The next three items are the integers $n_r$, $n_\phi$, and $n_{\theta'}$. Afterwards is a one dimensional array of double precision floats. It has size $20 n_r \times n_\phi \times n_{\theta'}$ and is created by incrementing the $\theta'$ index first (i.e. it is the inner "for" loop) and the $r$ index last (i.e. it is the outer "for" loop). For each grid cell, the 20 moment components and the corresponding errors $E$, $E_{\mathrm{err}}$, $F^x$, $F^x_{\mathrm{err}}$, $F^y$, $F^y_{\mathrm{err}}$, $F^z$, $F^z_{\mathrm{err}}$, $P^{xx}$, $P^{xx}_{\mathrm{err}}$, $P^{yy}$, $P^{yy}_{\mathrm{err}}$, $P^{zz}$, $P^{zz}_{\mathrm{err}}$, $P^{xy}$, $P^{xy}_{\mathrm{err}}$, $P^{xz}$, $P^{xz}_{\mathrm{err}}$, $P^{yz}$, and $P^{yz}_{\mathrm{err}}$ are stored contiguously in the array in that order. For example, if the moments array is denoted $m[i]$, then the $E$, $E_{\mathrm{err}}$, and $F^x$ components at a given grid cell corresponding to index $i$ are given by $E[i] = m[20i]$, $E_{\mathrm{err}}[i] = m[20i + 1]$, and $F^x[i] = m[20i + 2]$, respectively. Note that even though the position indices correspond to the coordinates $(r, \phi, \theta')$, the moment components are cartesian. The reason for this is that the moments are computed from the photon unit wave vectors, which we recall (section 5.2) are stored as $(k_x, k_y, k_z)$, not $(k_r, k_\phi, k_{\theta'})$.

## Outputting dependence of spectra on $\theta'$

The code can output the luminosity distribution $L(\nu)$ at each value of $\theta'$ to $n_{\theta'}$ different text files. Each file is formatted as follows:

```
nen
freq
0.5 L Lerr
freq
0.5 L Lerr
...
```

The first line contains the number of bins in frequency. In all that follows, $freq$ is the frequency of each frequency bin. For each frequency, $L$ and $L_{\mathrm{err}}$ are the values of the luminosity distribution and luminosity distribution error, respectively. Note that in this case the dependence of the distribution on $k_z$ has been integrated out. The code could of course be changed to include this. The 0.5 factors reflect that there is therefore only a single $k_z$ bin with nominal value $|k_z| = 0.5$.

Note that for an accretion disc it is necessary to use this option simply to make sure only photons emitted from the disc photosphere and not the disc midplane are included in the spectrum. To do this one should omit all photons emitted within some bounds $\theta'_1 < \theta' < \theta'_2$.

## Outputting dependence of spectra on $r$

If the simulation is of a comparatively thin disc then one can truncate the simulation grid before inputting it to the code so that $-f\pi/2 \leq \theta' \leq f\pi/2$ for some factor $f < 1$. One advantage of this is that one can then find the luminosity distribution as a function of $r$. To do this, one must both change the boundary condition for $\theta'$ to escape (see above) and set the code to output the luminosity distribution $L(\nu)$ at each value of $r$ to $n_r$ different text files. Each file is formatted as follows:

```
nen
freq
0.5 L Lerr
freq
0.5 L Lerr
...
```

The first line contains the number of bins in frequency. In all that follows, $freq$ is the frequency of each frequency bin. For each frequency, $L$ and $L_{\mathrm{err}}$ are the values of the luminosity distribution and luminosity distribution error, respectively. Note that in this case the dependence of the distribution on $k_z$ has been integrated out. The code could of course be changed to include this. The 0.5 factors reflect that there is therefore only a single $k_z$ bin with nominal value $|k_z| = 0.5$.

We note that for the file that corresponds to the largest value of $r$ it is possible that there are contributions to the spectrum from photons emitted from the $r = r_{\mathrm{max}}$ surface rather than the $\theta' = \theta'_{\mathrm{min}}$ or $\theta' = \theta'_{\mathrm{max}}$ surface. In particular, this spectrum will include photons emitted not only from the photosphere but from the disc midplane. When comparing spectra at different radii, therefore, the spectrum from this file should be omitted. On the other hand, it should be included if one wants to, for example, check that the total emitted spectrum is invariant under various conditions.

## Running on multiple cores

To run the code on multiple cores, one can simply run multiple copies of the code. The value of $idum$ in the parameter file should be different for each copy since it is the random number generator seed (Appendix G2). In addition, one can choose different ranges in the coordinate $\phi$ for each copy by modifying $ir2min$ and $ir2max$ (Appendix G2). For example, if $n_\phi = 128$ then to run on two cores one can set $ir2min = 0$, $ir2max = 63$ for the first copy and $ir2min = 64$, $ir2max = 127$ for the second copy.

## Discarding stuck photons

If the density in a grid cell is sufficiently large then a photon can get stuck in that cell and prevent the code from running properly. One can set the code to discard a photon if it scatters greater than a certain number of times. If this number is sufficiently large then it should not effect the resulting spectra, for two reasons. First, if this number is sufficiently large then such photons are rare. Second, such photons are likely to be discarded anyway after being sufficiently downweighted.

## Cylindrical grid

The code can be set to read in a cylindrical grid rather than a spherical grid. Since we have used this setting rarely it should be thoroughly tested before being put to use.

# G4  Tests

In this section we describe several ways we tested the code.

## Photon propagation

We first checked that in optically thin conditions photons travel in straight lines by outputting and plotting their trajectories. This test was important because the hardest part of creating the global code was to modify the propagation algorithm to work for non-rectangular cells (section 5.2, Appendix G1).

## Comptonization

First we checked that spectra obtained with homogeneous thermal Comptonization were given by numerical solutions to the Kompaneets equation. We checked this for multiple geometries. For example, $r_{min} \ll r_{max}$ corresponds to a spherical geometry, whereas $r_{min} \approx r_{max}$ corresponds to a (locally) plane parallel geometry. In this limit we also checked that bulk Comptonization by transverse modes was correctly described by numerical solutions to the Kompaneets equation with the "wave" temperature given by our closed-form solution.

## Shearing box limit

The shearing box limit corresponds to $r_{min} \approx r_{max}$, $\phi_{max} \ll 2\pi$, $\theta'_{min} = 0$, and $\theta'_{max} \ll \pi/2$. In this limit we checked that spectra were the same as for the shearing box code.

## Moments

We computed the frequency-integrated energy and flux described in Appendix G3 and checked that they approximately agree with those computed in the underlying radiation MHD simulations. The only notable differences were in extremely optically thick regions, which were due to the fact that the code does not include stimulated scattering (see section 5.4.2). This does not impact the emitted spectrum, however, since photons from those regions are absorbed before they have the chance to escape.